\documentclass[prd,aps,preprint,nofootinbib,endfloats]{revtex4}
\usepackage{graphics}
\usepackage{epsfig}

\def\g{$\gamma$}
\def\gr{$\gamma$-rays~}
\def\G{\Gamma}
\def\Ga{\Gamma}
\def\ga{\gamma}
\def\xr{X-ray~}

\def\araa{Ann. Rev. Astron. \& Astroph.,}
\def\apj{Ap. J.,}
\def\aj{Astron. J.,}
\def\apjl{Ap. J. Lett.,}
\def\apjs{Ap. J. Supp.,}
\def\apss{Astrophys. \& Space Sci.}
\def\mnras{Mon. Not. RAS}
\def\physrep{Physics Reports}
\def\aap{Astron. \& Astrophys.,}
\def\aaps{Astron. \& Astrophys. Supp.,}
\def\pasp{Publ. Astron. Soc. Pacific}
\def\pasj{Publ. Astron. Soc. Japan}

\def\etal,{{\it et al., }}
\def\msun{M_\odot}
\def\ee{\end{equation}}
\def\be{\begin{equation}}
\begin{document}
\ifx\href\undefined\else\hypersetup{linktocpage=true}\fi
\title{The Physics of Gamma-Ray Bursts}

\author{Tsvi Piran}
\email{tsvi@phys.huji.ac.il} \affiliation{Racah Institute for
Physics, The Hebrew University, Jerusalem, 91904, Israel}

\begin{abstract}
Gamma-Ray Bursts (GRBs),  short and intense pulses of low energy
\gr, have fascinated astronomers and astrophysicists since their
unexpected discovery in the late sixties. During the last decade,
several space missions: BATSE (Burst and Transient Source
Experiment) on Compton Gamma-Ray Observatory, BeppoSAX and now
HETE II (High-Energy Transient Explorer), together with ground
optical, infrared  and radio observatories have revolutionized our
understanding of GRBs showing that they are cosmological, that
they are accompanied by long lasting afterglows and that they are
associated with core collapse Supernovae. At the same time a
theoretical understanding has emerged in the form of the fireball
internal-external shocks model. According to this model GRBs are
produced when the kinetic energy of an ultra-relativistic flow is
dissipated in internal collisions. The afterglow arises when the
flow is slowed down  by shocks with the surrounding circum-burst
matter. This model has numerous successful predictions like the
prediction of the afterglow itself, the prediction of jet breaks
in the afterglow light curve and of an optical flash that
accompanies the GRBs themselves. In this review I focus on
theoretical aspects and on physical processes believed to take
place in GRBs.
\end{abstract}

\maketitle \tableofcontents

\section{INTRODUCTION }
\label{sec:intro}

Gamma-Ray Bursts (GRBs) are short and intense pulses of soft \gr.
The bursts last  from a fraction of a second to several hundred
seconds. GRBs  arrive from cosmological distances from random
directions in the sky. The overall observed fluences range from
$10^{-4}$ergs/cm$^2$ to $10^{-7}$ergs/cm$^2$ (the lower limit
depends, of course, on the characteristic of the detectors and not
on the bursts themselves). This corresponds to isotropic
luminosity of $10^{51}-10^{52}$ergs/sec, making GRBs the most
luminous objects in the sky. However, we know today that most GRBs
are narrowly beamed and the corresponding energies are ``only"
around $10^{51}$ergs \cite{Frail01,PanaitescuK01,Piranetal01},
making them comparable to Supernovae in the total energy release.

The GRBs are followed by afterglow - lower energy, long lasting
emission in the X-ray, optical and radio. The radio afterglow was
observed in some cases several years after the bursts. The
accurate afterglow positions enabled the identification of host
galaxies in almost all cases when afterglow was detected and this
in turn enabled the determination of the corresponding redshifts
that range from 0.16 (or possibly even down to 0.0085) to 4.5.
Within the host galaxies  there is evidence that (long duration)
GRBs arise within star forming regions and there is evidence that
they follow the star formation rate.

While not all observed features are understood there is an
overall agreement between the observations and the fireball
model. According to the fireball model GRBs are produced when the
kinetic energy of an ultra-relativistic flow is dissipated. The
GRB itself is produced by internal dissipation within the flow
while the afterglow is produced via external shocks with the
circum-burst medium. I will focus in this review on this model.

The numerous observations of the GRB and the observations of the
afterglow constrain the fireball model that describes the emitting
regions. The evidence on nature of the inner engine that powers
the GRB and produces the ultra-relativistic flow is however,
indirect. The energetic requirements and the time scales suggest
that GRB involve the formation of the black hole via a
catastrophic stellar collapse event or possibly a neutron star
merger. Additional indirect evidence arises from the requirement
of the fireball model of long (several dozen seconds) activity of
the inner engine. This hints towards an inner engine built on an
accreting black hole. On the other hand, the evidence of
association of GRBs with star forming regions indicates that GRBs
progenitors are massive stars. Finally, the appearance of
Supernova bumps in the afterglow light curve (most notably in GRB
030329) suggest association with Supernovae and stellar collapse.

I review here  the theory of GRB, focusing as mentioned earlier on
the fireball internal-external shocks model. I begin in \S
\ref{sec:obs} with a brief discussion of the observations. I turn
in \S \ref{sec:accepted} to some generally accepted properties of
GRB models - such as the essential ultra-relativistic nature of
this phenomenon. Before turning to a specific discussion of the
fireball model I review in \S \ref{sec:rel} several relativistic
effects and in \S \ref{sec:physical-Processes} the physical
process, such as synchrotron emission or particle acceleration in
relativistic shocks that are essential ingredients of this model.
In \S \ref{sec:PROMPT} I turn to  a discussion of the prompt
emission and the GRB. In \S \ref{sec:afterglow} I discuss
modelling the afterglow emission. I consider other related
phenomenon - such as TeV \gr emission, high energy neutrinos,
Ultra High energy cosmic rays and gravitational radiation in \S
\ref{sec:Other}. Finally, I turn in \S \ref{sec:inner-engine} to
examine different `inner engines' and various aspects related to
their activity. I conclude with a discussion of open questions and
observational prospects.

While writing this review I realized how large is the scope of
this field and how difficult it is to cover all aspects of this
interesting phenomenon. Some important aspects had to be left out.
I also did not attempt to give a complete historical coverage of
the field. I am sure that inadvertently I have missed many
important references.  I refer the reader to several other recent
review papers
\cite{Fishman1995,P99,ParadijsARAA00,P00,Meszaros01,Hurleyetal02,Meszaros02a,Galama_sari}
that discuss these and other aspects of GRB theory and
observations from different points of view.

\section{OBSERVATIONS }
\label{sec:obs}

I begin with a short review of  the basic observed properties of
GRBs. This review is brief as a complete review requires  a whole
paper by itself. I refer the reader to several review papers for a
detailed summary of the observations
\cite{Fishman1995,ParadijsARAA00,Hurleyetal02,Galama_sari}. I
divide this section to three parts. I begin with the prompt
emission - the GRB itself. I continue with properties of the
afterglow. I conclude with a discussion of the rates of GRBs, the
location of the bursts within their host galaxies and the
properties of the host galaxies.

\subsection{Prompt Emission }
\label{sec:prompt-obs}

I begin with a discussion of the GRB itself, namely the \gr and
any lower-energy emission that occurs simultaneously with them.
This includes \xr emission that generally accompanies the
$\gamma$-ray emission as a low energy tail. In some cases, called
\xr flashes (XRFs), the \gr signal is weak and all that we have is
this \xr signal. Prompt (operationally defined as the time period
when the \g-ray detector detects a signal above background)
longer-wavelength emission may also occur at the optical and radio
but it is harder to detect. However, so far optical flashes was
observed In three cases \cite{Akerlof99,Foxetal03,LiEtal03}
simultaneously with the \g-ray emission.

\subsubsection{Spectrum }
\label{sec:spec-obs}

The spectrum is non thermal. The energy flux peaks at a few
hundred keV and in many bursts there is a long high energy tail
extending in cases up to GeV. The spectrum varies strongly from
one burst to another. An excellent phenomenological fit for the
spectrum was introduced by  \textcite{Band93} using  two power
laws joined smoothly at a break energy
$(\tilde\alpha-\tilde\beta)E_0$:
\begin{equation}
N(\nu) = N_0 \cases { \big({h\nu})^{\tilde \alpha} \exp (-{h \nu
\over E_0}) & for $ h\nu < (\tilde\alpha-\tilde\beta)E_0$ ;\cr
\big[{(\tilde \alpha-\tilde \beta) E_0 } \big]^{(\tilde
\alpha-\tilde\beta)} \big({h \nu }\big)^{\tilde\beta} \exp (\tilde
\beta-\tilde \alpha), & for $h \nu
> (\tilde\alpha-\tilde\beta)E_0,$ \cr} \ .
\end{equation}
 I denote the
spectral indices here as $\tilde \alpha$ and $\tilde \beta$ to
distinguish them from the afterglow parameters ($\alpha$ and
$\beta$) discussed later.  There is no particular theoretical
model that predicts this spectral shape. Still, this function
provides an excellent fit to most of the observed spectra. For
most observed values of $\tilde\alpha$ and $\tilde\beta$, $\nu
F_\nu \propto \nu^2 N(\nu)$ peaks at $E_p = (\tilde\alpha+2)E_0$.
For about 10\% of the bursts the upper slope is larger than -2 and
there is no peak for $\nu F_\nu$ within the observed spectrum.
Another group of bursts, NHE bursts, (no high energy)
\cite{Pendleton_NHE97} does not have a hard component (which is
reflected by a very negative value of $\tilde \beta$). The
``typical'' energy of the observed radiation is $E_p$. $E_p$
defined in this way should not be confused with the commonly used
hardness ratio which is the ratio of photons observed in two BATSE
\footnote{BATSE is the Burst and Transient Source Experiment on
the CGRO (Compton Gamma-Ray Observatory), see e.g.
http://cossc.gsfc.nasa.gov/batse/. It operated for almost a decade
detecting several thousand bursts, more than any satellite before
or after it. The BATSE data was published in several catalogues
see \textcite{Paciesas99,Paciesas00} for the most recent one}
channels: Channel 3 (100-300keV) counts divided by Channel 2
(50-100keV) counts. The break frequency and the peak flux
frequencies are lower on average for bursts with lower observed
flux \cite{Mallozi95,Mallozzi98}.

\textcite{Band93} present a small catalogue of the spectra of 52
bright bursts which they analyze in terms of the Band function.
\textcite{PreeceEtal00} present a larger catalogue with 156 bursts
selected for either high flux or fluence. They consider several
spectral shape including the Band function.

\begin{figure}[htb]
\begin{center}
\epsfig{file=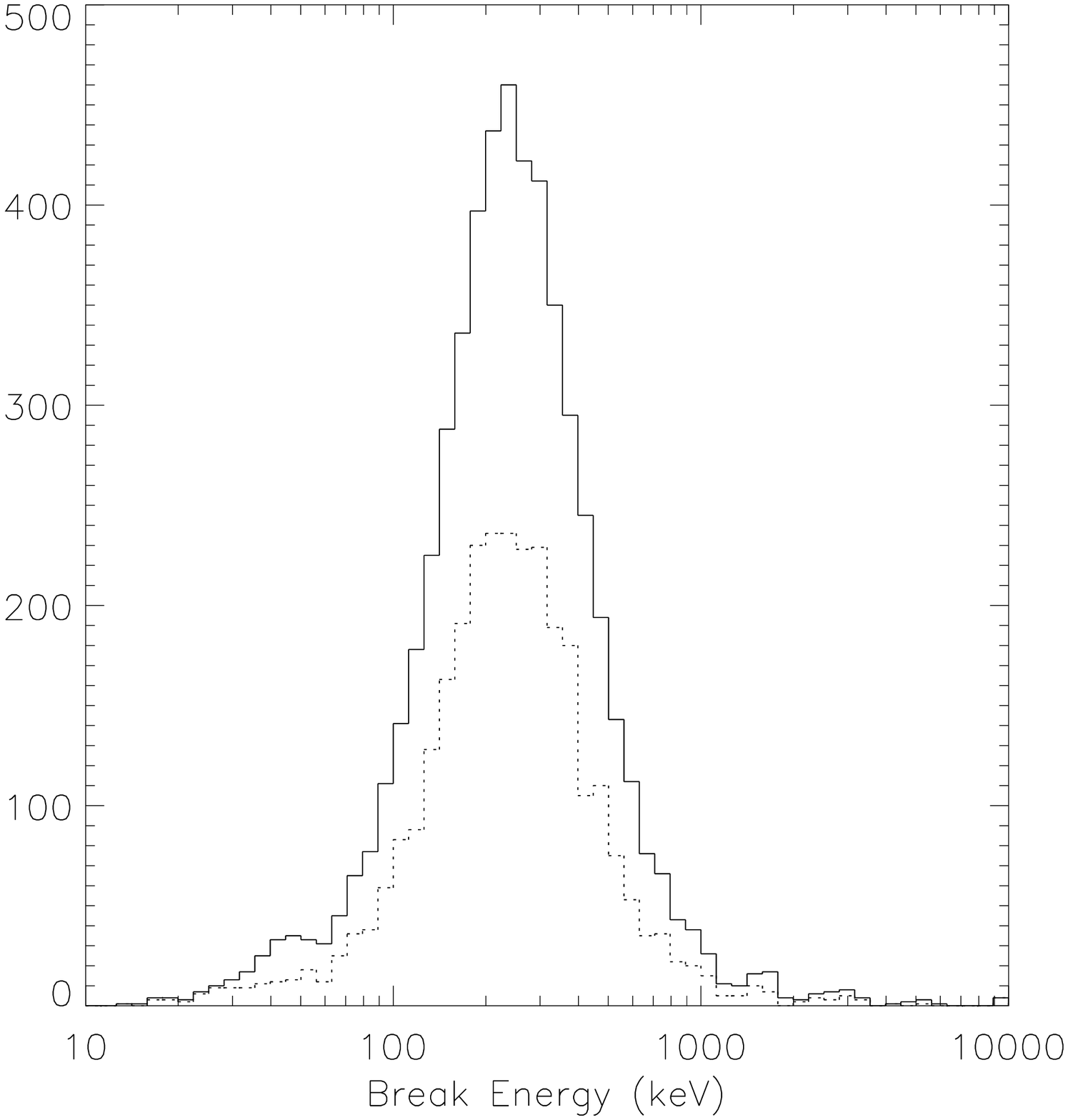,height=2in} \caption{Distribution of
the observed values of the break energy
$(\tilde\alpha-\tilde\beta)E_0$ in a sample of bright bursts
\cite{PreeceEtal00}}. The solid line represents the whole sample
while the dashed line represent a subset of the data.
\label{fig:spectrum_distribution}
\end{center}
\end{figure}

Fig. \ref{fig:spectrum_distribution} shows the distribution of
observed values of the break energy,
$(\tilde\alpha-\tilde\beta)E_0$,  in a sample of bright bursts
\cite{PreeceEtal00}. Most of the bursts are the range $100\,{\rm
keV}<(\tilde\alpha-\tilde\beta)E_0<400\,{\rm keV}$, with a clear
maximum in the distribution around
$(\tilde\alpha-\tilde\beta)E_0\sim 250$keV. There are not many
soft GRBs - that is, GRBs with peak energy in the tens of keV
range. However, the discovery \cite{XRF} of XRFs - \xr flashes
with similar temporal structure to GRBs but lower typical energies
- shows that the low peak energy cutoff is not real and it
reflects the lower sensitivity of BATSE in this range
\cite{BATSE_XRF}.

Similarly,  it is debatable whether there is a real paucity in
hard GRBs and there is an upper cutoff to the GRB hardness or it
just happens that the detection is optimal in this (a few hundred
keV) band. BATSE triggers, for example, are based mostly on the
count rate between 50keV and 300keV. BATSE is, therefore, less
sensitive to harder bursts that emit most of their energy in the
MeV range. Using BATSE's observation alone one cannot rule out the
possibility that there is a population of harder GRBs that emit
equal power in total energy which are not observed because of this
selection effect
\cite{PN96,CohenKatzP98,Llyod_Petrosian99,Lingen97}.  More
generally, a  harder burst with the same energy as a soft one
emits fewer photons. Furthermore, the spectrum is generally flat
in the high energy range and it decays quickly at low energies.
Therefore it is intrinsically more difficult to detect a harder
burst. A study of the SMM (Solar Maximum Mission) data
\cite{Harris97} suggests that there is a deficiency (by at least a
factor of 5) of GRBs with hardness above 3MeV, relative to GRBs
peaking at $\sim$0.5MeV, but this data is consistent with a
population of hardness that extends up to 2MeV.

Overall the narrowness of the hardness distribution is very
puzzling.  First, as I stressed earlier it is not clear whether it
is real and not a result of an observational artifact. If it is
real then on one hand there is no clear explanation to what is the
physical process that controls the narrowness of the distribution
(see however \textcite{Guetta_Spada_Waxman01}). On the other hand
cosmological redshift effects must broaden this distribution and
it seem likely (but not demonstrated yet) that if the GRB
distribution extends to z=10 as some suggest
\cite{ReichartLamb00,CiardiLoeb00,BrommLoeb02,Lloyd-RonningFryerRamirez-Ruiz02}
then such a narrow distribution requires an intrinsic correlation
between the intrinsic hardness of the burst and its redshift,
namely that the intrinsic hardness increases with the redshift.
There is some evidence for such a correlation between $E_p$ and
the observed peak flux \cite{Mallozi95,Mallozzi98}. More recently
\textcite{AmatiEtal02} reported on a correlation between $E_p$ and
the isotropic equivalent energy seen in 12 BeppoSAX bursts that
they have analyzed. They also report on a correlation between
$E_p$ and the redshift as, the bursts with higher isotropic
equivalent energy are typically more distant.  These three
different correlations are consistent with each other if the
observed peak flux of bursts is determined by their intrinsic
luminosity more than by the distance of the bursts. In such a case
(because of the larger volume at larger distances)  the observed
more distant bursts are on average brighter than nearer ones (see
also \S \ref{sec:hosts-distribution}).

 Even though the burst hardness distribution shows a single population a plot of the
hardness vs temporal duration shows that short bursts (see Fig.
\ref{fig:hardness-duration}) are typically harder
\cite{Dezalay96,Kouveliotou96}. The correlation is significant.
Another interesting sub-group of bursts is the NHE (no high
energy) bursts - bursts with no hard component that is no emission
above 300keV \cite{Pendleton_NHE97}. This group is characterized
by a large negative value of $\beta$, the high energy spectral
slope.  The NHE bursts have luminosities about an order of
magnitude lower than regular bursts and they exhibit an
effectively homogeneous intensity distribution with $\langle  V
/V_{max} \rangle=  0.53 \pm 0.029$. As I discuss later in \S
\ref{sec:temp-obs} most GRB light curves are composed of many
individual pulses. It is interesting that in many bursts there are
NHE pulses combined with regular pulses.

EGRET (The Energetic Gamma Ray Experiment Telescope) the high
energy \gr detector on Compton - GRO detected seven GRBs with
photon energies ranging from 100 MeV to 18 GeV \cite{EGRET_GRB}.
In some cases this very high energy emission  is delayed more than
an hour after the burst \cite{Hurley94,Sommer94}. No high-energy
cutoff above a few MeV has been observed in any GRB spectrum.
Recently, \cite{Gonzalez03} have combined the BATSE (30keV -2Mev)
data with the EGRET data for 26 bursts. In one of these bursts,
GRB 941017 (according to the common notation GRBs are numbered by
the date), they have discovered a high energy tail that extended
up to 200 MeV and looked like a different component. This high
energy component appeared 10-20 sec after the beginning of the
burst and displayed a roughly constant flux with a relatively hard
spectral slope ($F_\nu \propto \nu^0$) up to 200 sec. At late time
(150 after the trigger)  the very high energy (10-200 MeV) tail
contained 50 times more energy than the ``main" \gr energy
(30keV-2MeV) band. The TeV detector, Milagrito, discovered (at a
statistical significance of 1.5e-3 or so, namely at 3$\sigma$) a
TeV signal coincident with GRB 970417
\cite{Milagrito_970417,Atkins03}. If true this would correspond to
a TeV fluence that exceeds the low energy \gr fluence. However no
further TeV signals were discovered from other 53 bursts observed
by Milagrito \cite{Milagrito_970417} or from several bursts
observed by the more sensitive Milagro \cite{Milagro_GRB}. One
should recall however, that due to the attenuation of the IR
background TeV photons could not be detected from $z>0.1$. Thus
even if most GRBs emit TeV photons those photons won't be detected
on Earth.

Another puzzle is the low energy tail. \textcite{CohenKatzP98}
analyze several strong bursts and find that their low energy slope
is around 1/3 to -1/2. However,  \textcite{PreeceEtal98,Preece02}
suggest that about 1/5 of the bursts have a the low energy power
spectrum, $\alpha$, steeper than  1/3 (the  synchrotron slow
cooling low energy slope). A larger fraction is steeper than -1/2
(the fast cooling synchrotron low energy slope). However, this is
not seen in any of the HETE spectrum whose low energy resolution
is somewhat better. All HETE bursts have a low energy spectrum
that is within the range 1/3 and -1/2 \cite{BarraudEtal03}. As
both BATSE and HETE use NaI detectors that have a poor low energy
resolution \cite{CohenKatzP98}, this problem might be resolved
only when a better low energy spectrometer  will be flown.

\subsubsection{Temporal Structure }
\label{sec:temp-obs}

\begin{figure}[htb]
\begin{center}
\epsfig{file=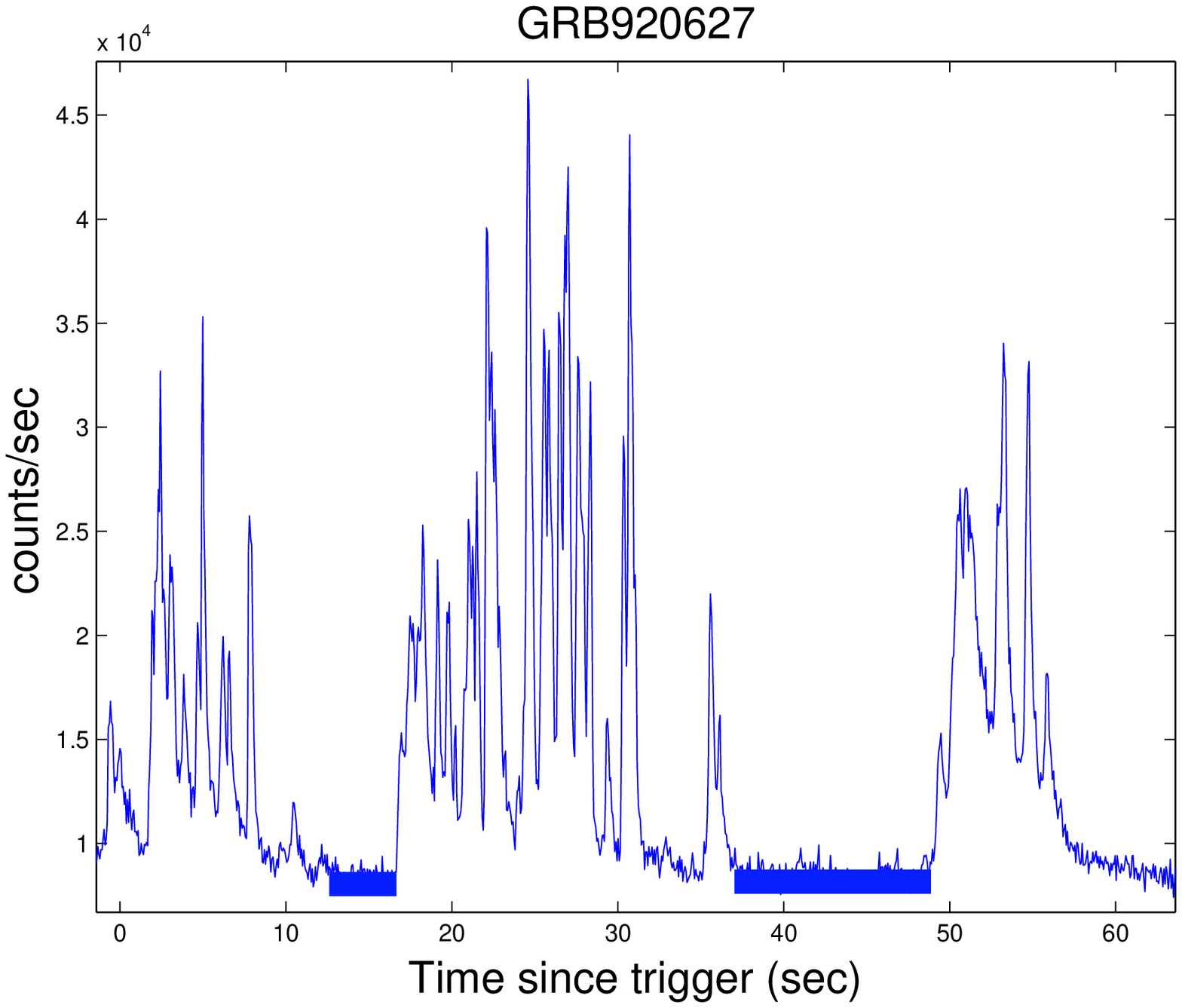,height=2in} \caption{The light curve
of GRB 920627. The total duration of the burst is 52sec, while
typical pulses are 0.8sec wide. Two quiescent periods lasting
$\sim$10 seconds are marked by horizontal solid bold lines. }
\label{fig:variable}
\end{center}
\end{figure}

The duration of the bursts spans five orders, ranging from less
than 0.01sec to more than 100sec. Common measures for the duration
are $T_{90}$ ($T_{50}$) which correspond to the time in which 90\%
(50\%) of the counts of the GRB arrives. As I discuss below (see
\S \ref{sec:pop}) the bursts are divided to long and short bursts
according to their $T_{90}$. Most GRBs are highly variable,
showing 100\% variations in the flux on a time scale much shorter
than the overall duration of the burst. Fig \ref{fig:variable}
depicts the light curve of a typical variable GRB (GRB 920627).
The variability time scale, $\delta t$, is determined by the width
of the peaks. $\delta t$ is much shorter (in some cases by a more
than a factor of $10^4$) than $T$, the duration of the burst.
Variability on a time scale of milliseconds has been  observed in
some long bursts \cite{NakarPiran02a,McBreenEtal01short}. However,
only $\sim 80$\% of the bursts show substantial substructure in
their light curves. The rest are rather smooth, typically with a
FRED (Fast Rise Exponential Decay) structure.

\textcite{Fenimore_Ramirez-Ruiz01} (see also
\textcite{Reichartetal01}) discovered a correlation between the
variability and the luminosity of the bursts. This correlation (as
well as the lag-luminosity relation discussed later) allow us to
estimate the luminosity of bursts that do not have a known
redshift.

 The bursts
seem to be composed of individual pulses, with a pulse being the
``building blocks" of the overall light curve. Individual pulses
display a hard to soft evolution with the peak energy decreasing
exponentially with the photon fluence
\cite{Liang96,Norrisetal96,Ford95}. The pulses have the following
temporal and spectral features. (i) The light curve of an
individual pulse is a FRED - fast rise exponential decay -  with
an average rise to decay ratio of 1:3 \cite{Norrisetal96}. (ii)
The low energy emission is delayed compared to the high energy
emission\footnote{Low/high energy implies the low vs. the high
BATSE channels. The four BATSE channels at 20-50keV, 50-100keV,
100-300keV and $>300$keV.} \cite{Norrisetal96}.
\textcite{Norris_lags00} have found that these spectral lags are
anti-correlated with the luminosity of the bursts: Luminous bursts
have long lags. This lag luminosity relation provides another way
to estimate the luminosity of a burst from its (multi-spectra)
light curve. (iii) The pulses' low energy light curves are wider
compared to the high energy light curves. The width goes as $ \sim
E^{-0.4} $\cite{Fenimoreetal95}. (iv) There is a
Width-Symmetry-Intensity correlation. High intensity pulses are
(statistically) more symmetric (lower decay to rise ratio) and
with shorter spectral lags \cite{Norrisetal96}. (v) There is a
Hardness-Intensity correlation. The instantaneous spectral
hardness of a pulse is correlated to the instantaneous intensity
(the pulse become softer during the pulse decay)
\cite{Borgonovo01}.

Both the pulse widths, $\delta t$, and the pulse
separation,$\Delta t$ , have a rather similar log-normal
distributions. However, the pulse separation, distribution,
reveals has an excess of long intervals \cite{NakarPiran02a}.
These long intervals can be classified as quiescent periods
\cite{Ramirez-Ruiz_Merloni01}, relatively long periods of several
dozen seconds with no activity. When excluding these  quiescent
periods both distributions are log-normal with a comparable
parameters \cite{NakarPiran02a,QuilliganEtal02}. The average pulse
interval, $\bar \Delta t = 1.3sec$ is larger by a factor 1.3 then
the average pulse width $\bar \delta t= 1sec$. One also finds that
the pulse widths are correlated with the preceding interval
\cite{NakarPiran02a}. \textcite{Ramirez-Ruiz_Fenimore00} found
that the pulses' width does not vary along the bursts.

One can also analyze the temporal behavior using the traditional
Fourier transform method to analyze. The power density spectra
(PDS) of light curves shows a power law slope of $ \sim -5/3 $ and
a sharp break at 1Hz \cite{Beloborodov_pds_00}.

\begin{figure}[htb]
\begin{center}
\epsfig{file=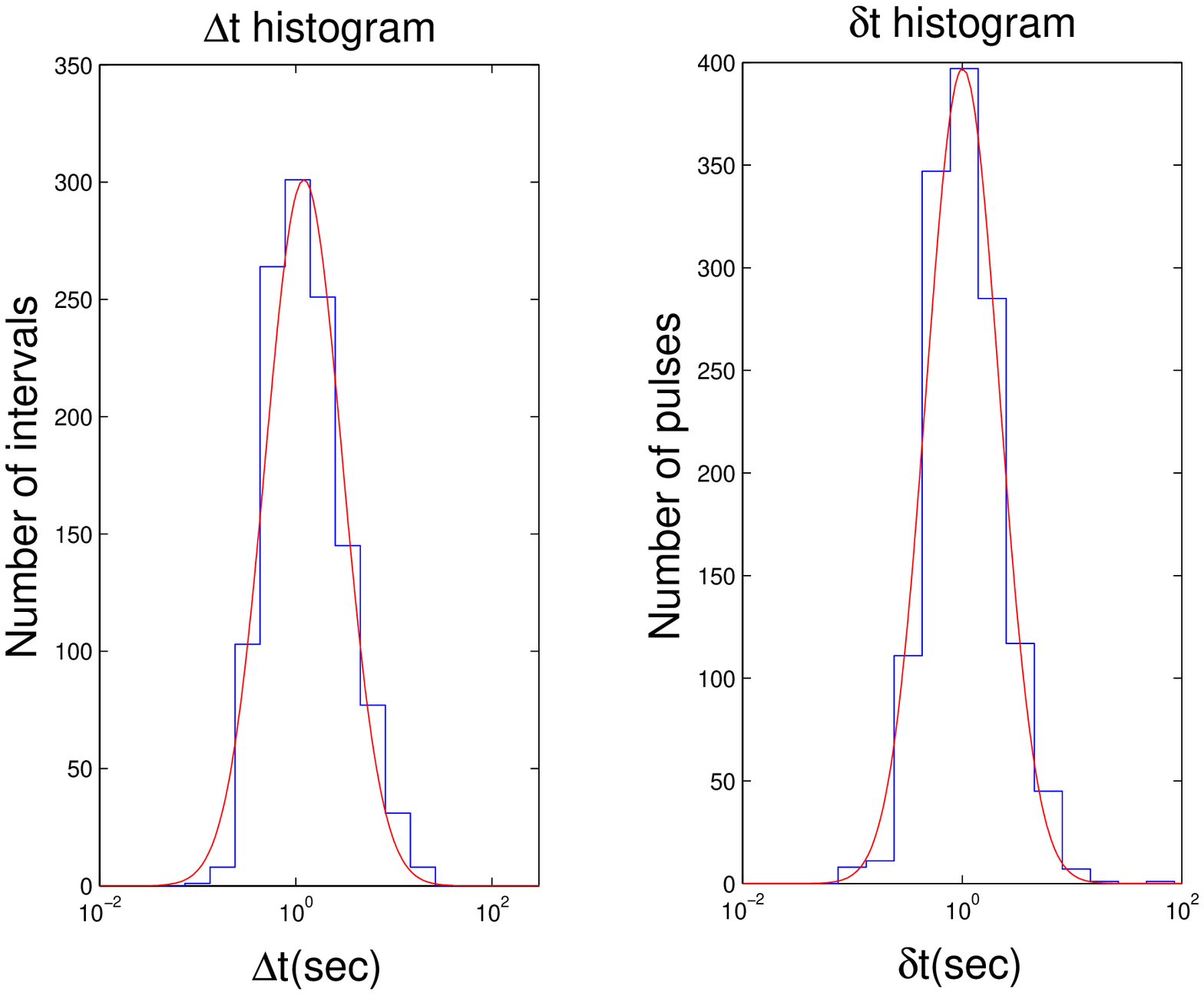,height=2in} \caption{The pulse width
distribution (right) and the distribution of intervals between
pulses (left) (from \cite{NakarPiran02a}).}
\label{fig:distributions}
\end{center}
\end{figure}

The results described so far are for long bursts.  the
variability of short ($T<2$sec) bursts is more difficult to
analyze. The duration of these bursts is closer to the limiting
resolution of the detectors. Still  most ($\sim 66\%$) short
bursts are variable with $\delta t/T < 0.1$ \cite{NakarPiran02b}.
These variable bursts are composed of multiple subpulses.

\subsubsection{Populations}
\label{sec:pop}

{\bf Long and Short Bursts} The clearest classification of bursts
is based on their duration. \textcite{Kouveliotou_2pop_93} have
shown that GRB can be divided to two distinct groups: long burst
with $T_{90}>2$sec and short bursts with $T_{90}< 2$sec. Note that
it was suggested \cite{Mukherjee98,Horvath98} that there is a
third intermediate class with $2.5{\rm sec} <T_{90}< 7$sec.
However, it is not clear if this division to three classes is
statistically significant \cite{Hakkila00}.

\begin{figure}[htb]
\begin{center}
\epsfig{file=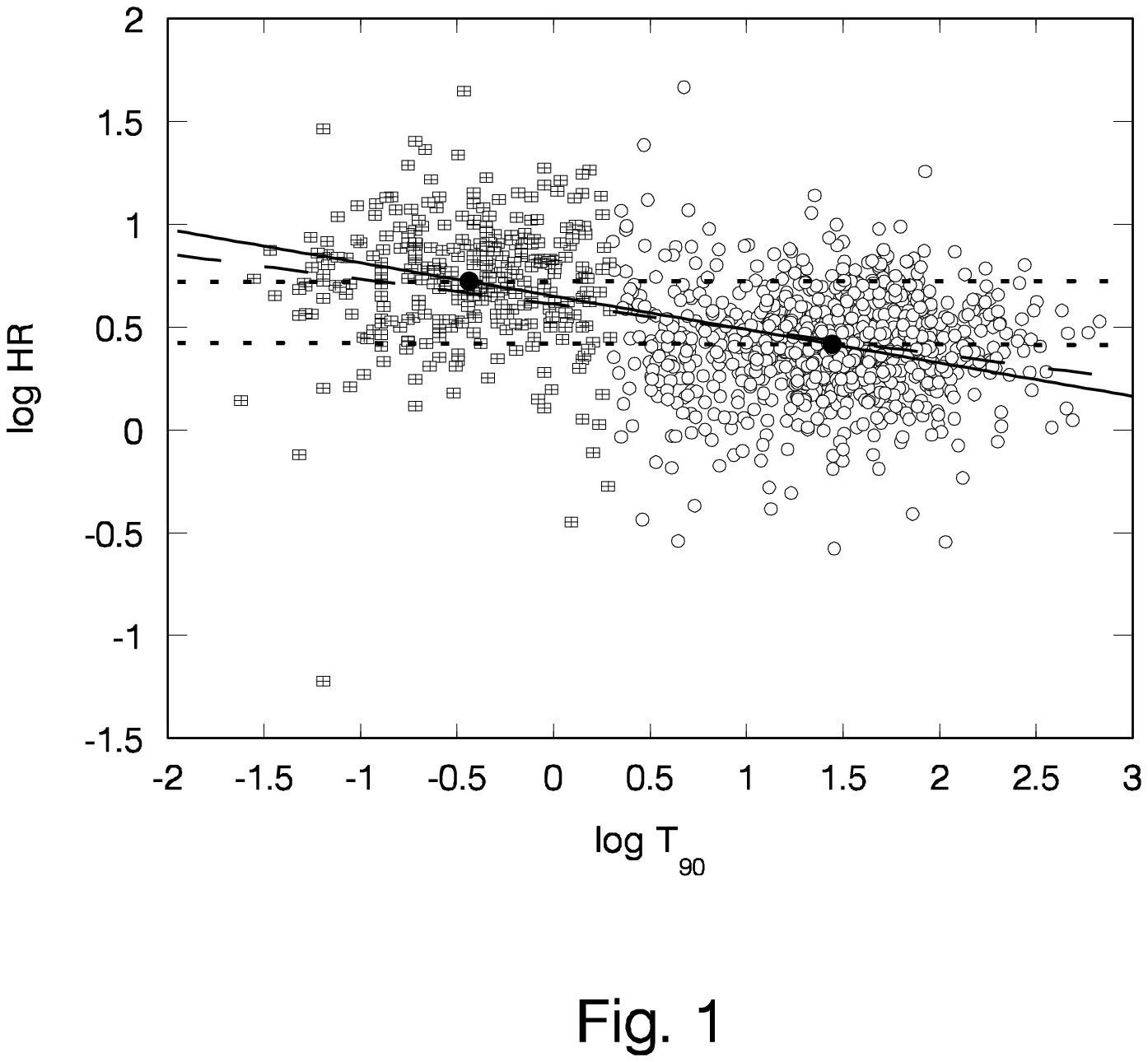,height=2in} \caption{The
hardness-duration correlation for BATSE bursts. HR is the ratio of
fluence between BATSE channels 3 and 2. Squares represent short
burst and circle long bursts. The solid line is a regression line
for the whole sample. The two dotted lines are the regresions
lines for the short and long samples respectively (from
\cite{Qinetal00}.)} \label{fig:hardness-duration}
\end{center}
\end{figure}

An interesting question is whether short bursts could arise from
single peaks of long bursts in which the rest of the long burst is
hidden by noise. \textcite{NakarPiran02b} have shown that in
practically all long bursts the second highest peak is comparable
in height to the first one. Thus, if the highest peak is above the
noise so should be the second one. Short bursts are a different
entity. This is supported by the observation that short bursts are
typically harder \cite{Dezalay96,Kouveliotou96}. The
duration-hardness distribution (see Fig.
\ref{fig:hardness-duration}) shows clearly that there are not soft
short bursts.

The spatial distribution of the {\bf observed} short bursts is
clearly different from the distribution of the {\bf observed} long
one. A measure of the spatial distribution is the average ratio $
\langle V/V_{max} \rangle \equiv \langle (C/C_{min})^{-3/2}
\rangle $, where $C$ is the count rate and $C_{min}$ is the
minimal rate required for triggering. In a uniform Eucleadian
sample this ratio equals $0.5$ regardless of the luminosity
function. One of the first signs of a cosmological origin of GRBs
was the deviation of this value from 0.5 for the BATSE sample
\cite{Meeganetal92Nat}. The the $ \langle V/V_{max} \rangle $ of
the BATSE short bursts sample
\cite{Mao_Narayan_P94,P96IAU,Katz_Canel96}  is significantly
higher than $ \langle V/V_{max} \rangle  $ of the long bursts
sample. Note that more recently \cite{Schmidt01} suggested that
the two values are similar and the distribution of long and short
bursts is similar. However, \textcite{GuettaPiran03} finds
$\langle V/V_{max} \rangle_{long} = 0.282$ and $\langle V/V_{max}
\rangle_{long} = 0.390$ (I discuss this point further in \S
\ref{sec:rates}). This implies that the population of {\bf
observed} short bursts is nearer on average than the population of
the observed long ones. This is not necessarily a statement on the
location of short vs. the location of long bursts. Instead it
simply reflects the fact that it is more difficult to detect a
short burst. For a short burst one has to trigger on a shorter
(and hence noisier) window the detector (specifically BATSE that
triggers on 64 ms for short bursts and on 1 sec for long ones) is
less sensitive to short bursts. I discuss later, in \S
\ref{sec:rates}, the question of rates of long vs. short bursts.

So far afterglow was detected only from long bursts. It is not
clear whether this is an observational artifact or a real feature.
However, there was no \xr afterglow observed for the only well
localized short hard burst: GRB020531 \cite{Hurley020531}. Chandra
observations show an intensity weaker by at least a factor of
100-300 than the intensity of the \xr afterglow from long bursts
at a similar time \cite{Butler020531}. Afterglow was not observed
in other wavelength as well \cite{Klotz03}

As identification of hosts and redshifts depend on the detection
of afterglow this implies that nothing is known about the
distribution, progenitors, environment etc.. of short burst.
These bursts are still waiting for their afterglow revolution.

{\bf \xr Flashes}  (XRFs) are \xr bursts with a similar temporal
structure to GRBs but lower typical energies. \textcite{XRF}
discovered these flushes by comparing GRBM (GRB Monitor) with
sensitivity above 40 keV and WFC (Wide Field Camera) triggering on
BeppoSAX\footnote{see http://www.asdc.asi.it/bepposax/ for
information on BeppoSAX and its the different instruments.}.
 In 39 cases the WFCs were triggered
without GRBM triggering implying that these flashed do not have
any hard component and most of their flux is in \xr. The duration
of 17 of these transients (out of the 39 transients), denoted \xr
flashes (XRFs), is comparable to the duration of the X-ray
emission accompanying GRBs.   The peak fluxes of the XRFs are
similar to the \xr fluxes observed during GRBs in the WFCs ($\sim
10^{-8}$ergs/sec/cm$^2$) but their peak energy is clearly below 40
keV. These finding confirmed the detection of
\textcite{StrohmayerEtal98} of 7 GRBs with $E_p < 10$keV and 5
additional GRBs with $E_p< 50$keV in the GINGA data.

\textcite{BarraudEtal03} analyze 35 bursts detected on HETE
II\footnote{HETE II is a dedicated GRB satellite that aims at
locating quickly bursts with high positional accuracy. See
http://space.mit.edu/HETE/ for a description of HETE II and its
instruments}. They find that XRFs lie on the extension of all the
relevant GRB distributions. Namely there is a continuity from
GRBs to XRFs. Detailed searches in the BATSE data revealed that
some of these bursts have also been detected by BATSE
\cite{BATSE_XRF}.
 Using the
complete search in 90\% of the WFC data available,
\textcite{Heise03} find that the observed frequency of XRFs is
approximately half of the GRB-frequency: In 6 years of BeppoSAX
observations they have observed 32 XRFs above a threshold
peak-luminosity of $5 \times 10^{-9}$erg/s/cm$^2$ in the 2-25 keV
range compared with 54 GRBs (all GRBs above BATSE thresholds are
observed if in the field of view).

By now \textcite{Soderberg02} discovered optical afterglow from
XRF 020903 and they suggest that the burst was at $z=0.25$. They
also suggest a hint of an underlying SN signal (see \S
\ref{sec:obs-SN})  peaking between ~7-24 days after the initial
XRF trigger. Afterglow was discovered from XRF 030723as well
\cite{Fox030723}.

\subsubsection{Polarization}
\label{sec:prompt-polarization}

Recently, \textcite{CoburnBoggs03} reported on a detection of a
very high ($80\%\pm 20\%$) linear  polarization  during the prompt
$ \gamma $-ray emission of GRB 021206.  This burst was extremely
powerful. The observed fluence of GRB 021206 was $1.6 \cdot
10^{-4}ergs/cm^2$ at the energy range of 25-100Kev
\cite{Hurley02GCN1727,Hurley02GCN1728}. This puts GRB 021206 as
one of the most powerful bursts, and the most powerful one (a
factor of 2-3 above GRB990123) after correcting for the fact that
it was observed only in a narrow band (compared to the wide BATSE
band of 20-2000keV). \textcite{CoburnBoggs03} analyzed the data
recorded by the  Reuven Ramaty High Energy Solar Spectroscopic
Imager (RHESSI). The polarization is measured in this detector by
the angular dependence of the number detection of simultaneous
pairs of events that are most likely caused by a scattering of the
detected \gr within the detector. The data data analysis is based
on 12 data points which are collected over 5sec. Each of these
points is a sum of several independent observations taken at
different times. Thus the data is some kind of convolution of the
polarization over the whole duration of the burst.

\textcite{CoburnBoggs03} test two hypothesis. First they test the
null hypothesis of no polarization. This hypothesis is rejected at
a confidence level of $ 5.7\sigma  $.  Second they estimate the
modulation factor assuming a constant polarization  during the
whole burst. The best fit to the data is achieved with $ \Pi
=(80\pm 20)\% $.  However, \textcite{CoburnBoggs03} find that the
probability that $\chi^2$ is greater than the value obtained with
this fit is 5\%, namely the model of constant polarization is
consistent with the analysis and observations only at the 5\%
level.

\textcite{Rutledge03Polarization} reanalyzed this data and pointed
out several inconsistencies within the methodology of
\textcite{CoburnBoggs03}. Their upper limit on the polarization
(based on the same data) is $\sim 4\%$. In their rebuttle
\textcite{BoggsCoburn03} point out that the strong upper limit
(obtained by \textcite{Rutledge03Polarization} is inconsistent
with the low S/N estimated by these authors. However, they do not
provide a clear answer to the criticism of the methodology raised
by \textcite{Rutledge03Polarization}. This leaves the situation,
concerning the prompt polarization from this burst highly
uncertain.

\subsubsection{Prompt Optical Flashes }
\label{sec:prompt-optical}

The Robotic telescope ROTSE (Robotic Optical Transient Search
Experiment) detected a 9th magnitude optical flash that was
concurrent with the GRB emission from GRB 990123 \cite{Akerlof99}.
The six snapshots begun 40sec after the trigger and lasted until
three minutes after the burst. The second snapshot that took place
60sec after the trigger recorded a 9th magnitude flash. While the
six snapshots do not provide a ``light curve" it is clear that the
peak optical flux does not coincide with the peak \gr emission
that takes place around the first ROTSE snapshot. This suggests
that the optical flux is not the ``low energy tail" of the \gr
emission. Recently, \textcite{Foxetal03} reported on a detection
of 15.45 magnitude optical signal from GRB 021004 193 sec after
the trigger. This is just 93 seconds after the 100 sec long burst
stopped being active. Shortly afterwards \textcite{LiEtal03}
reported on a detection of 14.67 magnitude optical signal from GRB
021211 105 sec after the trigger. Finally, \textcite{Price030329}
detected a  12th magnitude prompt flash, albeit this is more than
1.5 hours after the trigger. Similar prompt signal was not
observed from any other burst in spite of extensive searches that
provided upper limits. \textcite{Kehoe01} searched 5 bright bursts
and found single-image upper limits ranging from 13th to 14th
magnitude around 10 sec after the initial burst detection and from
14 to 15.8 magnitudes one hour later. These upper limits are
consistent with the two recent detections  which are around 15th
mag.  The recent events of rapid detection suggest that we should
expect many more such discoveries in the near future.

\subsubsection{The GRB-Afterglow Transition - Observations }
\label{sec:transition-obs}

There is no direct correlation between the $\gamma$-ray fluxes and
the \xr (or optical) afterglow fluxes. The extrapolation of the
\xr afterglow fluxes backwards generally does not fit the
$\gamma$-ray fluxes. Instead they fit the late prompt \xr signal.
These results are in nice agreement with the predictions of the
Internal - External shocks scenario in which the two phenomena are
produced by different effects and one should not expect a simple
extrapolation to work.

The expected GRB afterglow transition have been observed in
several cases. The first observation took place (but was not
reported until much latter) already in 1992 \cite{BureninEtal99}.
BeppoSAX data shows a rather sharp transition in the hardness
that takes place several dozen seconds after the beginning of the
bursts.
This transition is seen clearly in  the different energy bands
light curves of GRB990123 and in GRB980923 \cite{GiblinEtal99}.
\textcite{Connaughton02} have averaged the light curves of many
GRBs and discovered  long and soft tails: the early \xr
afterglow. Additional evidence for the transition from the GRB to
the afterglow can be observed in the observations of the
different spectrum within the GRB \cite{Preece02}.

\subsection{The Afterglow }
\label{sec:obs-afterglow}

Until 1997 there were no known counterparts to GRBs in other
wavelengths. On Feb 28 1997 the Italian-Dutch satellite  BeppoSAX
detected \xr afterglow from GRB 970228 \cite{Costa_970228}. The
exact position given by BeppoSAX led to the discovery of optical
afterglow \cite{vanParadijs970228}. Radio afterglow was detected
in GRB 970508 \cite{Frail970508}. By now more than forty \xr
afterglows have been observed (see
http://www.mpe.mpg.de/$\sim$jcg/grb.html for a complete up to date
tables of well localized GRBs with or without afterglow. Another
useful page is: http://grad40.as.utexas.edu/grblog.php). About
half of these have optical and radio afterglow (see Fig
\ref{fig:Venn}). The accurate positions given by the afterglow
enabled the identification of the host galaxies of many bursts. In
twenty or so cases the redshift has been measured. The observed
redshifts range from 0.16 for GRB 030329 (or 0.0085 for GRB
980425) to a record of 4.5 (GRB 000131).  Even though the
afterglow is a single entity I will follow the astronomical
wavelength division and I will review here the observational
properties of \xr, optical and radio afterglows.

\subsubsection{The \xr afterglow }
\label{sec:obs-xr}

The \xr afterglow is the first and strongest, but shortest
signal. In fact it seems to begin already while the GRB is going
on (see \S \ref{sec:transition-obs} for a discussion of the
GRB-afterglow transition). The light curve observed several hours
after the burst can usually be extrapolated to the late parts of
the prompt emission.

\begin{figure}[htb]
\begin{center}
\epsfig{file=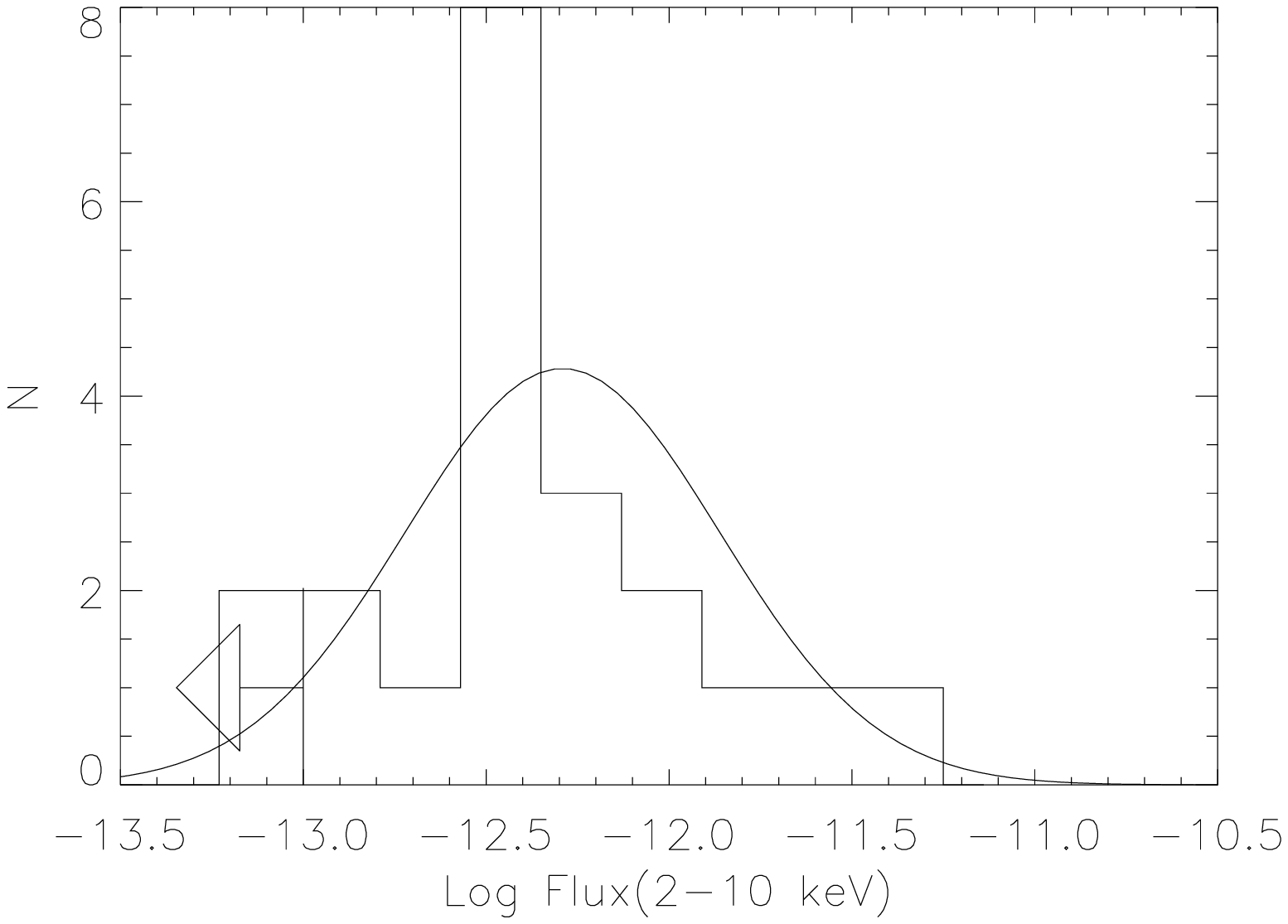,height=2in} \caption{The distribution
of X-ray fluxes (2-10 keV) at t=11 hours after the GRB in  21
afterglows observed by BeppoSAX (from \textcite{Piranetal01}). The
sample includes all the fast observations performed by BeppoSAX on
GRB from January 1997 to October 1999.  No X-ray afterglow was
detected in GB990217  to the limiting instrumental sensitivity of
$10^{-13}$ erg cm$^{-2}$ s$^{-1}$, 6 hours after the burst. In the
case of GB970111  a candidate was detected, but evidence of fading
behaviour is marginal, so both cases are considered as upper
limits (indicated by the arrow).} \label{fig:x-rays1}
\end{center}
\end{figure}

The \xr afterglow fluxes from GRBs have a power law dependence on
$\nu$ and on the observed time $t$ \cite{Piro01}: $f_\nu(t)
\propto \nu^{-\beta} t^{-\alpha}$ with $\alpha \sim 1.4$ and
$\beta \sim 0.9$. The flux distribution, when normalized to a
fixed hour after the burst has a rather narrow distribution. A
cancellation of the  k corrections and the temporal decay makes
this flux, which is proportional to $(1+z)^{\beta-\alpha}$
insensitive to the redshift. Using 21 BeppoSAX bursts
\cite{Piro01} \textcite{Piranetal01} find that the 1-10keV flux,
11 hours after the burst is $ 5 \times 10^{-13}$ergs/cm$^{-2}$sec.
The distribution is log-normal with $\sigma_{f_{x}}\approx 0.43
\pm 0.1$ (see fig. \ref{fig:x-rays1}). \textcite{Pasquale03} find
a similar result for a larger sample. However, they find that the
\xr afterglow of GRBs with optical counterparts is on average 5
times brighter than the \xr afterglow of dark GRBs (GRBs with no
detected optical afterglow).  The overall energy emitted in the
\xr afterglow is generally a few percent of the GRB energy.
\textcite{BergerKulkarniFrail03} find that the \xr luminosity is
indeed correlated with the opening angle and when taking the
beaming correction into account they find that  $L_X=f_b
L_{X,iso}$, is approximately constant, with a dispersion of only a
factor of 2.

\xr lines were seen in 7 GRBs: GRB 970508 \cite{Piro970508}, GRB
970828 \cite{Yoshida970828}, GRB 990705 \cite{Amati990705}, GRB
991216 \cite{Piro991216}, GRB 001025a \cite{Watson010220}, GRB
000214 \cite{Antonelli000214} and GRB 011211 \cite{Reeves011211}.
The lines were detected using different instruments: BeppoSAX,
ASCA (Advanced Satellite for Cosmology and Astrophysics) , Chandra
and XMM-Newton. The lines were detected around 10 hours after the
burst. The typical luminosity in the lines is around
$10^{44}-10^{45}$ergs/sec, corresponding to a total fluence of
about $10^{49}$ergs. Most of the lines are interpreted as emission
lines of Fe K$\alpha$. However, there are also a
radiative-recombination-continuum line edge and K$\alpha$ lines of
lighter elements like Si, S, Ar and Ca (all seen in the afterglow
of GRB 011211 \cite{Reeves011211}). In one case  (GRB 990705,
\textcite{Amati990705}) there is a transient absorption feature
within the prompt \xr emission, corresponding also to Fe
K$\alpha$. The statistical significance of the detection of these
lines is of some concern (2-5 $\sigma$), and even thought the late
instruments are much more sensitive than the early ones all
detections remain at this low significance level.
\textcite{Rutledge03} and \textcite{Sako03HEAD} expressed concern
about the statistical analysis of the data showing these lines and
claim that none of the observed lines is statistically
significant. The theoretical implications are far reaching. Not
only the lines require, in most models, a very large amount of
Iron at rest (the lines are quite narrow), they most likely
require \cite{Ghisellinietal02A&A} a huge energy supply ($>
10^{52}$ergs), twenty time larger than the typical estimated \gr
energy ($\sim 5 \cdot 10^{50}$ergs).

\subsubsection{Optical and IR afterglow }
\label{sec:Obs-opt}
\begin{figure}[htb]
\begin{center}
\epsfig{file=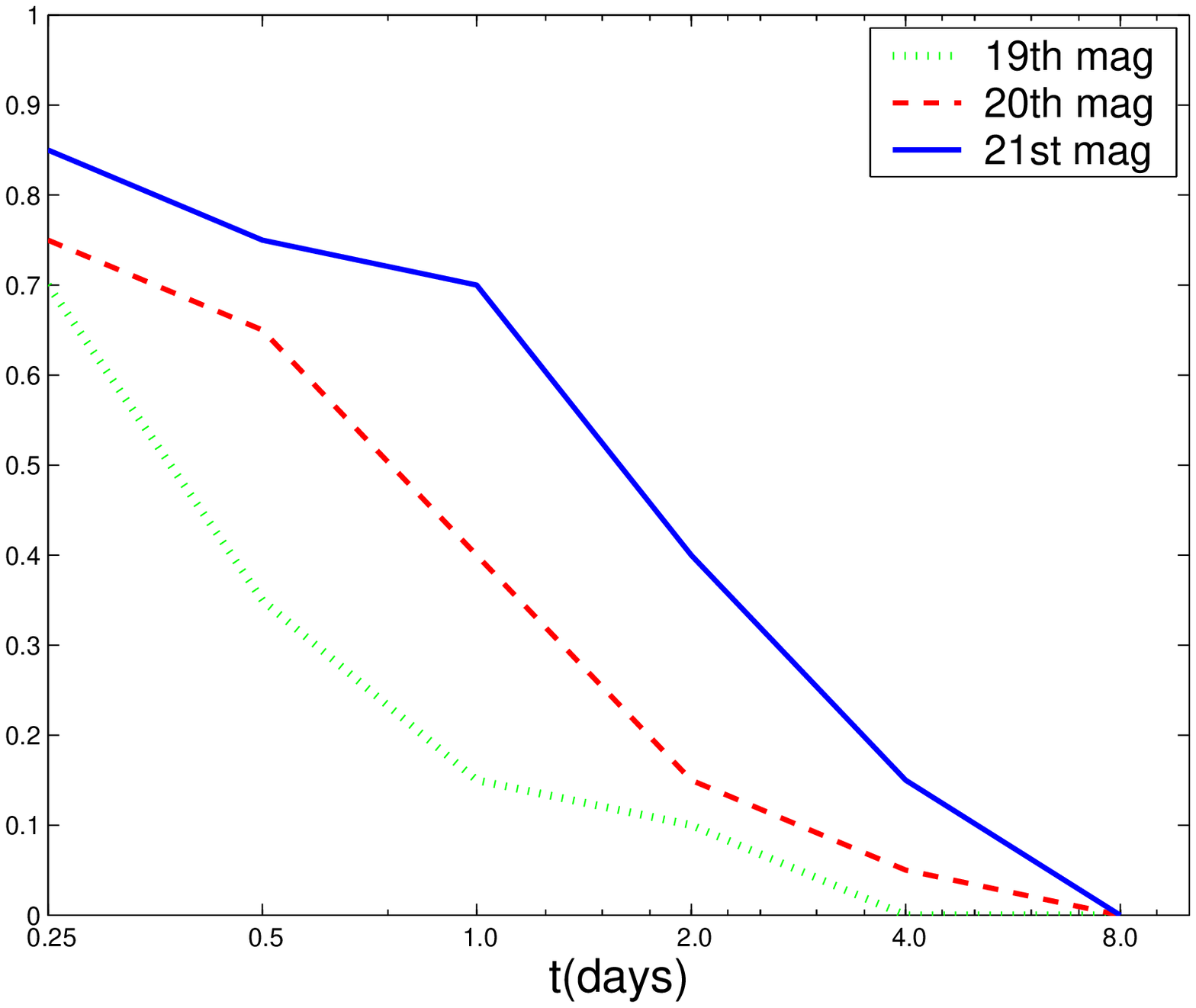,height=2in} \caption{The fraction of
bursts with optical afterglow above three limiting magnitudes as a
function of time (compared to the total number of bursts with
optical afterglow).} \label{fig:optical_one_day}
\end{center}
\end{figure}

 About 50\% of well localized GRBs show optical
and IR afterglow. The observed optical afterglow is typically
around 19-20 mag one day after the burst (See fig
\ref{fig:optical_one_day}). The signal decays, initially, as a
power law in time, $t^{-\alpha}$ with a typical value of $\alpha
\approx 1.2$ and large variations around this value. In all cases
the observed optical spectrum is also a power law $\nu^{-\beta}$.
Generally absorption lines are superimposed on this power law. The
absorption lines correspond to absorption on the way from the
source to earth. Typically the highest redshift lines are
associated with the host galaxy,  providing a measurement of the
redshift of the GRB. In a few cases emission lines, presumably
from excited gas along the line of site were also observed.

Technical difficulties led a gap of several hours between the
burst and the detection of the optical afterglow, which could be
found only after an accurate position was available. The rapid
localization provided by HETE II helped to close this gap and an
almost complete light curve from 193 sec after the trigger
($\approx 93$ sec after the end of the burst) is available now for
GRB021004 \cite{Foxetal03}.

\begin{figure}[htb]
\begin{center}
\epsfig{file=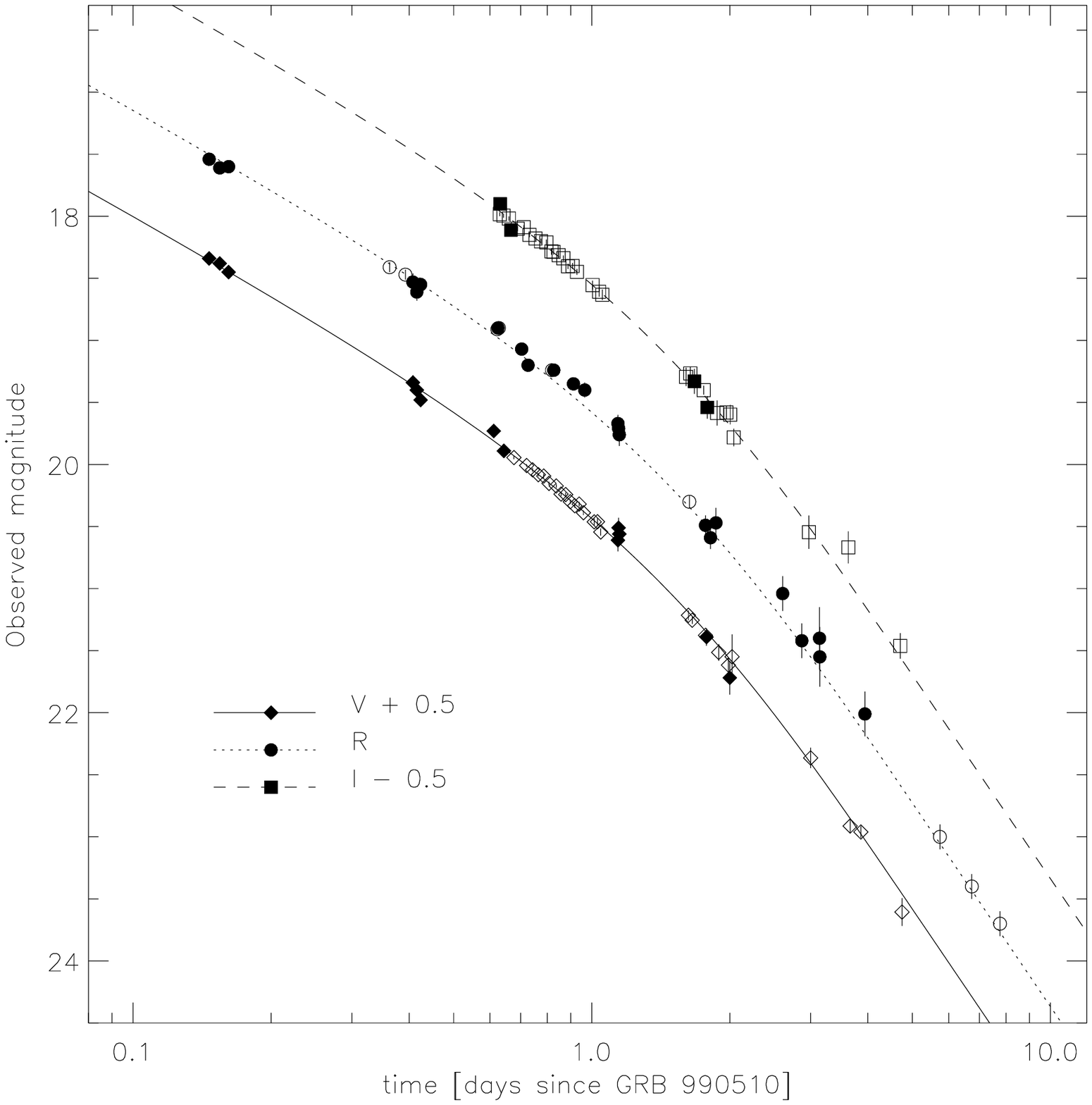,height=2in} \caption{The optical
light curves of GRB990510. A fit for the observed optical light
curves is obtained with $\alpha_1 = 0.82 \pm 0.02$, $\alpha_2 =
2.18 \pm 0.05$  and $t_* = 1.2 \pm 0.08$ days (from
\textcite{Harrisonetal99}).} \label{fig:990510}
\end{center}
\end{figure}

Many afterglow light curves show an achromatic break  to a steeper
decline with $\alpha \approx 2$. The classical example of such a
break was seen in GRB 990510 \cite{Harrisonetal99,Staneketal99}
and it is shown here in Fig. \ref{fig:990510}. It is common to fit
the break with the phenomenological formula: $F_\nu (t) = f_*
(t/t_*)^{-\alpha_{1}}\{
1-\exp[-(t/t_*)^{(\alpha_{1}-\alpha_{2})}](t/t_*)^{(\alpha_{1}-\alpha_{2})}
\}$. This break is commonly interpreted as a jet break that allows
us to estimate the opening angle of the jet \cite{Rhoads99,SPH99}
or the viewing angle within the standard jet model \cite{Rossi02}
(see \S \ref{sec:Energetics} below).

The optical light curve of the first detected afterglow (from GRB
970228) could be seen for more than half a year
\cite{Fruchteretal98}. In most cases the afterglow fades faster
and cannot be followed for more than several weeks.  At  this
stage the afterglow becomes significantly dimer than its host
galaxy and the light curve reaches a plateau corresponding to the
emission of the host.

In a several cases: e.g. GRB 980326 \cite{Bloom99}, GRB 970228
\cite{Reichart99} GRB 011121 \cite{Bloometal02,Garnavichetal03}
red bumps are seen at late times (several weeks to a month). These
bumps are usually interpreted as evidence for an underlying SN. A
most remarkable Supernova signature was seen recently in GRB
030329 \cite{Stanek03SN,Hjorth03SN}. This supernova had the same
signature as SN98bw that was associated with GRB 990425 (see \S
\ref{sec:obs-SN}).

Finally, I note that  varying polarization at optical wavelengths
has been observed in GRB afterglows at the level of a few to ten
percent
\cite{CovinoEtal99,WijersEtal99,RolEtal00,CovinoEtal02,Bersier03,Greineretal03}.
These observations are in agreement with rough predictions
(\cite{Sari99,GhiselliniLazzati99}) of the synchrotron emission
model provided that there is a deviation from spherical symmetry
(see \S \ref{sec:pol_theory} below).

\subsubsection{Dark GRBs}

Only $\sim 50\%$ of well-localized GRBs show optical transients
(OTs) successive to the prompt gamma-ray emission, whereas an \xr
counterpart is present in 90\% of cases (see Fig. \ref{fig:Venn}).
Several possible explanations have been suggested for this
situation. It is possible that late and shallow observations could
not detect the OTs in some cases; several authors argue that dim
and/or rapid decaying transients could bias the determination of
the fraction of truly obscure GRBs \citep{Fyn01a,Ber02}. However,
recent reanalysis of optical observations
\citep{Rei01,Ghi00,Laz00} has shown that GRBs without OT detection
(called dark GRBs, FOAs Failed Optical Afterglows, or GHOSTs,
Gamma ray burst Hiding an Optical Source Transient) have had on
average weaker optical counterparts, at least 2 magnitudes in the
R band, than GRBs with OTs. Therefore, they appear to constitute a
different class of objects, albeit there could be a fraction
undetected for bad imaging.

The nature of dark GRBs is not clear. So far three hypothesis have
been put forward to explain the behavior of dark GRBs. First, they
are similar to the other bright GRBs, except for the fact that
their lines of sight pass through large and dusty molecular
clouds, that cause high absorption \cite{ReichartPrice02}. Second,
they are more distant than GRBs with OT, at $ z \ge 5 $
\citep{Fruchter_970228,ReichartLamb00}, so that the Lyman break is
redshifted into the optical band. Nevertheless, the distances of a
few dark GRBs have been determined and they do not imply high
redshifts \citep{Djo02,Ant00,Pir02}. A third possibility is that
the optical afterglow of dark GRBs is intrinsically much fainter
(2-3 mag below) than that of other GRBs.

\textcite{Pasquale03} find that GRBs with optical transients show
a remarkably narrow distribution of flux ratios, which corresponds
to an average optical-to-x spectral index $0.794\pm 0.054$. They
find that, while 75\% of dark GRBs have flux ratio upper limits
still consistent with those of GRBs with optical transients, the
remaining 25\% are 4 - 10 times weaker in optical than in X-rays.
This result suggests  that the afterglows of most  dark GRBs are
intrinsically fainter in all wavelength relative to the afterglows
of GRBs with observed optical transients.  As for the remaining
25\% here the spectrum (optical to X-ray ratio) must be different
than the spectrum of other afterglows with a suppression of the
optical band.

\subsubsection{Radio afterglow}

\begin{figure}[htb]
\begin{center}
\epsfig{file=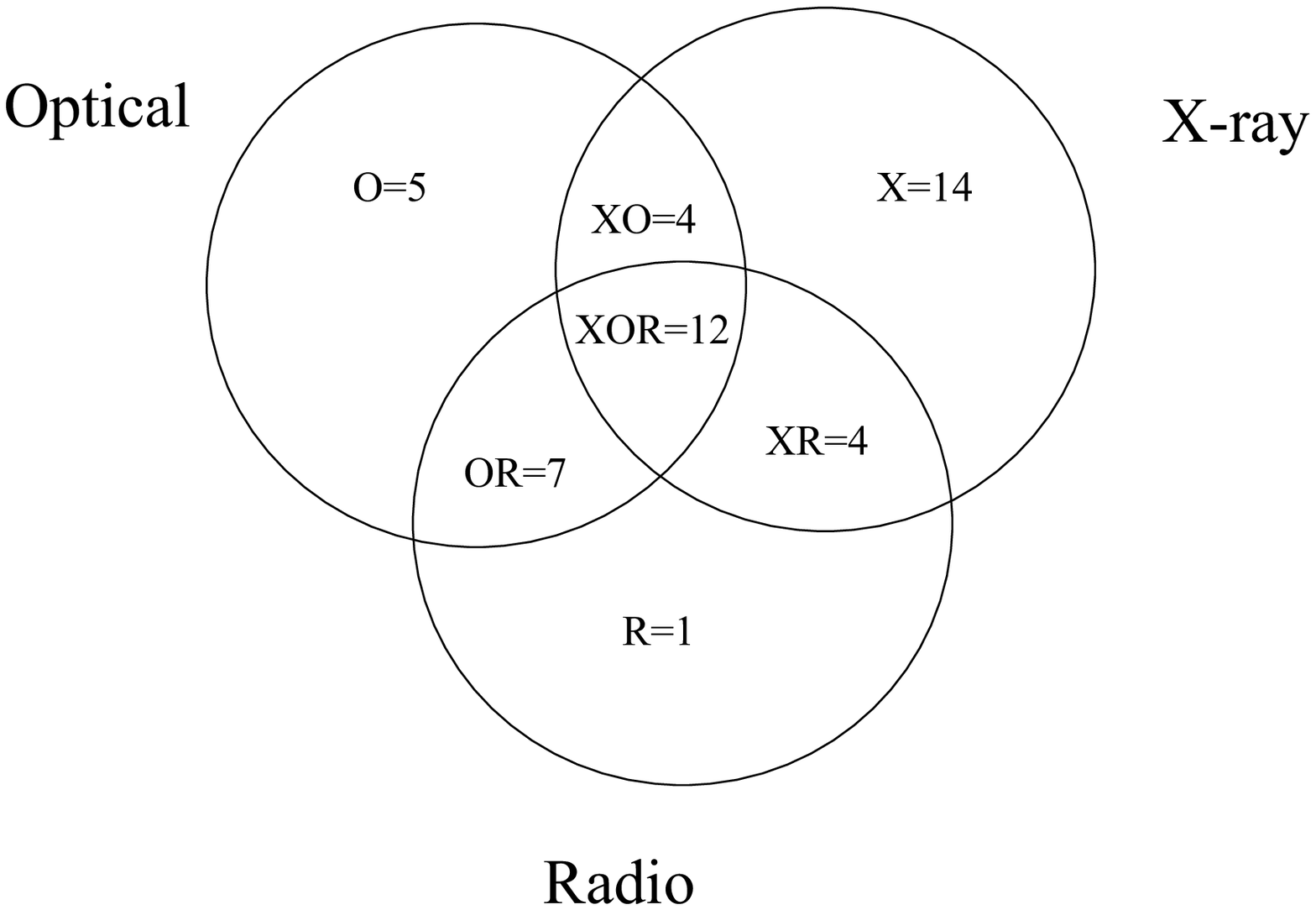,height=2in} \caption{Venn diagram of
the distribution of 47 Afterglows observed in different wavelength
between 1997-2001 (from a talk given by D. Frail at the Sackler
GRB workshop, Harvard, May 2001}
\label{fig:Venn}
\end{center}
\end{figure}

Radio afterglow was detected in $\sim 50$ \% of the well localized
bursts.  Most observations are done at about 8 GHz since the
detection falls off drastically at higher and lower frequencies.
The observed peak fluxes are at the level of 2 mJy. A turnover is
seen around $0.2$ mJy and  the undetected bursts have upper
limits of the order of 0.1 mJy. As the localization is based on
the \xr afterglow (and as practically all bursts have \xr
afterglow) almost all these bursts were detected in \xr. $\sim
80$ \% of the radio-afterglow bursts have also optical afterglow.
The rest are optically dark. Similarly $\sim 80$\% of the
optically observed afterglow have also a radio component (see fig
\ref{fig:Venn}).

Several bursts (GRBs 980329, 990123, 91216, 000926, 001018,
010222, 011030, 011121) were detected at around one day. Recent
radio observations begin well before that but do not get a
detection until about 24 hrs after a burst. The earliest radio
detection took place in GRB 011030 at about 0.8 days after the
burst \cite{TaylorFrailFox01}. In several cases (GRBs 990123,
990506, 991216,  980329 and  020405) the afterglow was detected
early enough to indicate emission from the reverse shock and a
transition from the reverse shock to the forward shock.

\begin{figure}[htb]
\begin{center}
\epsfig{file=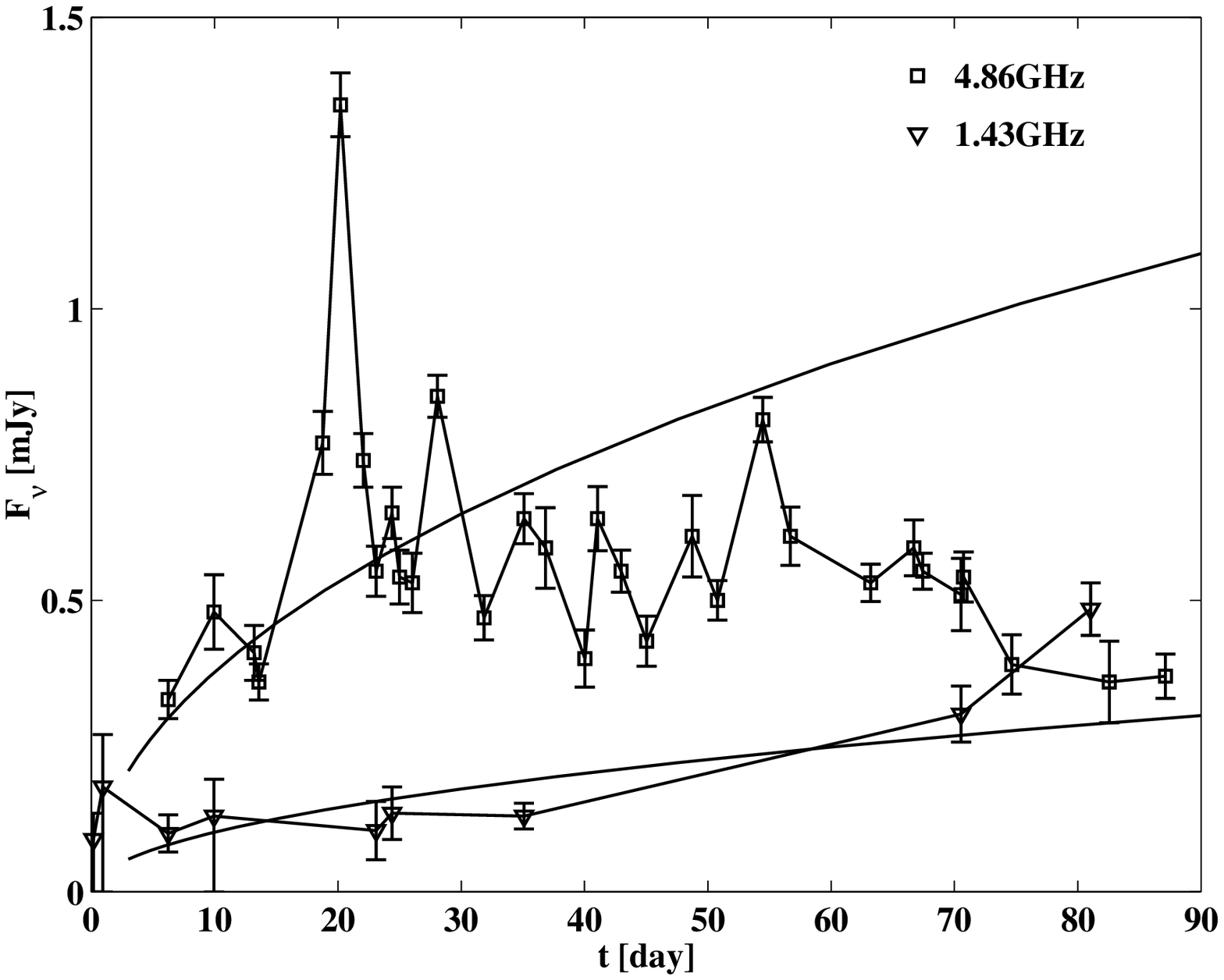,height=2in} \caption{Light curves of
the radio afterglow of GRB970508 at 4.86GHz and 1.43GHz, compared
with the predictions of the adiabatic fireball model (from
\textcite{Frail970508}). } \label{fig:radio970508}
\end{center}
\end{figure}

The radio light curve of GRB 970508 (see fig
\ref{fig:radio970508}) depicts early strong fluctuations (of order
unity) in the flux \cite{Frail970508}. \textcite{Goodman97}
suggested that these fluctuations arise due to scintillations and
the decrease (with time)  in the amplitude of the   fluctuations
arises from a transition from strong to weak scintillations.
\textcite{Frail970508} used this to infer the size of the emitting
region of GRB 970508 at $\sim 4$ weeks after the burst as $\sim
10^{17}$cm. This observations provided the first direct proof of
relativistic expansion in GRBs.

The self-absorbed frequencies fall in the centimeter to meter wave
radio regime and hence the lower radio emission is within the
self-absorption part of the spectrum (see \S \ref{Sec:self-abs}
later). In this case the spectrum rises as $\nu^2$ \cite{KP97}.
The spectral shape that arises from a the fact that the system is
optically thick enables us (using similar arguments to those of a
simple black body emission) to determine the size of the emitting
region. In GRB 990508 this has lead to $\sim 10^{17}$cm. A
comparable estimate to the one derived from scintillations.

The long-lived nature of the radio afterglow allows for
unambiguous calorimetry of the blast wave to be made when its
expansion has become sub-relativistic and quasi-spherical. The
light curves evolves on a longer time scale in the radio. Some GRB
afterglows have been detected years after the burst even after the
relativistic-Newtonian transition (see \S \ref{sec:Newtonian}). At
this stage the expansion is essentially spherical and this enables
a direct "calorimetric" estimate of the total energy within the
ejecta \cite{Waxmanetal98}.

\subsection{Hosts and Distribution }
\label{sec:hosts-distribution}

\subsubsection{Hosts }
\label{sec:hosts}
 By now (early 2004) host
galaxies have been observed for all but 1 or 2 bursts with
optical, radio or \xr afterglow localization with arcsec precision
\cite{Hurleyetal02}. The no-host problem which made a lot of noise
in the nineties has disappeared. GRBs are located within host
galaxies (see \textcite{Djorgovski02a,Djorgovski02b} and
\textcite{Hurleyetal02} for detailed reviews). While many
researchers believe that the GRB host population seem to be
representative of the normal star-forming field galaxy population
at a comparable redshifts, others argue that GRB host galaxies are
significantly bluer than average and their star formation rate is
much higher than average.

The host galaxies are faint with median apparent magnitude
$R\approx 25$. Some faint hosts are at $R\approx 29$. Down to
$R\approx 25$ the observed distribution is consistent with deep
field galaxy counts. \textcite{Jimenezetal01} find that the
likelihood of finding a GRB in a galaxy is proportional to the
galaxy's luminosity.

The magnitude and redshift distribution of GRB host galaxies are
typical for normal, faint field galaxies, as are their
morphologies
\cite{Odewahn98,Holland01,Bloom02,Hurleyetal02,Djorgovski02b}.
While some researchers argue that  the broad band optical colors
of GRB hosts are not distinguishable from those of normal field
galaxies at comparable magnitudes and redshifts
\cite{Bloom02,Sokolov01},  others \cite{Fruchter_970228} asserts
that the host galaxies are  unusually blue and that they are
strongly star forming. \textcite{LeFlochetal03}  argues that R-K
colors of GRB hosts are unusually blue and the hosts may be of low
metallicity and luminosity. This suggests \cite{LeFloch04} that
the hosts of GRBs might be different from the cites of the
majority of star forming galaxies that are  luminous, reddened and
dust-enshrouded infrared starbursts (\textcite{ElbazCesarsky03}
and references therein). \textcite{LeFloch04} also suggests that
this difference might rise due to an observational bias and that
GRBs that arise in dust-enshrouded infrared starbursts are dark
GRBs whose afterglow is not detectable due to obscuration. Whether
this is tru or not is very relevant to the interesting question to
which extend GRBs follow the SFR and to which extend they can be
used to determine the SFR at high redshifts.

\textcite{Totani97}, \textcite{Wijersetal_SFR98} and
\textcite{Pac98} suggested that GRBs follow the star formation
rate. As early as 1998  \textcite{Fruchter_970228} noted that  all
four early GRBs with spectroscopic identification or deep
multicolor broadband imaging of the host (GRB 970228 GRB 970508,
GRB 971214, and GRB 980703) lie in rapidly star-forming galaxies.
Within the host galaxies the distribution of GRB-host offset
follows the light distribution of the hosts \cite{Bloom02}. The
light is roughly proportional to the density of star formation.
Spectroscopic measurements suggest that GRBs are within Galaxies
with a higher SFR. However, this is typical for normal field
galaxy population at comparable redshifts \cite{Hurelyetal02}.
There are some intriguing hints, in particular the flux ratios of
[Ne III] 3859 to [OII] 3727 are on average a factor of 4 to 5
higher in GRB hosts than in star forming galaxies at low redshifts
\cite{Djorgovski02b}. This may represent an indirect evidence
linking GRBs with massive star formation. The link between GRBs
and massive stars has been strengthened with the centimeter and
submillimeter discoveries of GRB host galaxies
\cite{BergerKF01,FrailBMetal02}  undergoing prodigious star
formation (SFR$\sim 10^3$ M$_\odot$ yr$^{-1}$), which remains
obscured at optical wavelengths.

Evidence for a different characteristics of GRB host galaxies
arise from the work of \textcite{Fynbo02,Fynbo03} who find that
GRB host galaxies ``always" show Lyman alpha emission in cases
where a suitable search has been conducted. This back up the claim
for active star formation and at most moderate metallicity in GRB
hosts.  It clearly distinguishes GRB hosts from the Lyman break
galaxy population, in which only about 1/4 of galaxies show strong
Lyman alpha.

\subsubsection{The Spatial Distribution}
\label{sec:spatial}

BATSE's discovery that the bursts are distributed uniformly on the
sky \cite{Meeganetal92Nat} was among the first indication of the
cosmological nature of GRBs. The uniform distribution indicated
that GRBs are not associated with the Galaxy or with ``local"
structure in the near Universe.

Recently there have been several claims that sub-groups of the
whole GRB population shows a deviation from a uniform
distribution. \textcite{MeszarosA00a,MeszarosA00b}, for example,
find that the angular distribution of the intermediate sub-group
of bursts (more specifically of the weak intermediate sub-group)
is not random.  \textcite{Celottietal02} reported that the
two-point angular correlation function of 407 short BATSE GRBs
reveal a $\sim 2\sigma$ deviation from isotropy on angular scales
$2^o-4^o$. This results  is consistent with the possibility that
observed short GRBs are nearer and the angular correlation is
induced by the large scale structure correlations on this scale.
These claims are important as they could arise only if these
bursts are relatively nearby. Alternatively this indicates
repetition of these sources \cite{Celottietal02}.  Any such
deviation would imply that these sub-groups are associated with
different objects than the main GRB population at least that these
subgroup are associated with a specific feature, such as a
different viewing angle.

\textcite{Clineetal03} studied the shortest GRB population, burst
with a typical durations several dozen ms. They find that there is
a significant angular asymmetry and the $\langle V/V_{max} \rangle
$ distribution provides evidence for a homogeneous sources
distribution.  They suggest that these features  are best
interpreted as sources of a galactic origin. However, one has to
realize that there are strong selection effects that are involved
in the detection of this particular subgroup.

\subsubsection{GRB rates and the {\it isotropic} luminosity function}
\label{sec:rates}

There have been many attempts to determine the GRB luminosity
function and rate from  the BATSE peak flux distribution. This was
done by numerous using different levels of statistical
sophistication and different physical assumptions on the evolution
of the rate of GRBs with time and on the shape of the luminosity
function.

Roughly speaking the situation is the following. There are now
more than 30 redshift measured. The median redshift is $z\approx
1$ and the redshift range is from 0.16 (or even 0.0085 if the
association of GRB 980425 with SN 98bw should be also considered)
to 4.5 (for GRB 000131). Direct estimates from the sample of GRBs
with determined redshifts are contaminated by observational biases
and are insufficient to determine the rate and luminosity
function. An alternative approach is to estimates these quantities
from  the BATSE peak flux distribution. However, the observed
sample with a known redshifts clearly shows that the luminosity
function is wide. With a wide luminosity function, the rate of GRB
is only weakly constraint by the peak flux distribution. The
analysis is further complicated by the fact that the observed peak
luminosity, at a given detector with a given observation energy
band depends also on the intrinsic spectrum. Hence different
assumptions on the spectrum yield different results. This
situation suggest that there is no point in employing
sophisticated statistical tools (see however,
\cite{LoredoWasserman95,P99} for a discussion of these methods)
and a simple analysis is sufficient to obtain an idea on the
relevant parameters.

I will not attempt to review the various approaches here. A
partial list  of calculations includes
\cite{Piran92,Cohen_Piran95,FenimoreBloom95,LoredoWasserman95,HorackHakkila97,LoredoWasserman98,P99,Schmidt99,Schmidt01,Schmidt01a,SethiBhargavi01}.
 Instead I will just quote
results of some estimates of the rates and luminosities of GRBs.
The simplest approach is to fit $\langle V/V_{max} \rangle$, which
is the first moment of the peak flux distribution.
\textcite{Schmidt99,Schmidt01,Schmidt01a} finds using $\langle
V/V_{max} \rangle$ of the long burst distribution and assuming
that the bursts follow the \cite{PorcianiMadau01} SFR2,  that the
present local rate of long observed GRBs is $\approx ~0.15 {\rm
Gpc}^{-3} {\rm yr}^{-1}$ \cite{Schmidt01}. Note that this rate
from \cite{Schmidt01} is smaller by a factor of ten than the
earlier rate of \cite{Schmidt99}!  This estimate corresponds to a
typical (isotropic) peak luminosity of $\sim 10^{51}$ergs/sec.
These are the observed rate and the isotropic peak luminosity.

Recently \textcite{GuettaPiranWaxman03} have repeated these
calculations . They use both the  \cite{Rowan-Robinson99} SFR
formation rate:
\begin{equation}
\label{RR} R_{GRB}(z) = \rho_0 \left\{ \begin{array}{ll}
10^{0.75 z} & z<1 \nonumber \\
10^{0.75 z_{\rm peak}} & z>1.
\end{array}
\right. \; ,
\end{equation}
and SFR2 from  \cite{PorcianiMadau01}. Their best fit luminosity
function (per logarithmic luminosity interval, $d\log L $) is:
\begin{equation}
\label{Lfun} \Phi_o(L) =c_o \left\{ \begin{array}{ll}
(L/L^*)^{\alpha}\qquad & L^*/30  <  L < L^*  \\
(L/L^*)^{\beta} \qquad &  L^* erg/sec < L < 30L^*
\end{array}
\right. ;,
\end{equation}
and 0 otherwise with a typical luminosity, $L^*=1.1
\times10^{51}$ergs/sec, $\alpha=-0.6$ and $\beta=-2$, and $c_o$ is
a normalization constant so that the integral over the luminosity
function equals unity.  The corresponding local GRB rate is
$\rho_0=0.44$Gpc$^{-1}$yr$^{-1}$. There is an uncertainty of a
factor of $\sim 2$ in the typical energy, $L^*$, and in the local
rate. I will use these numbers as the ``canonical" values in the
rest of this review.

The observed (BATSE) rate of short GRBs is smaller by a factor of
three than the rate of long ones. However, this is not the ratio
of the real  rates as :(i) The BATSE detector is less sensitive to
short bursts than to long ones; (ii) The true rate depends on the
spatial distribution of the short bursts.  So far no redshift was
detected for any short bursts and hence this distribution is
uncertain. For short bursts  we can resort only to estimates based
on the peak flux distribution. There are indications that $\langle
V/V_{max} \rangle$ of short burst is larger (and close to the
Eucleadian value of 0.5) than the $\langle V/V_{max} \rangle$
value of  long ones (which is around 0.32). This implies that the
observed short bursts are nearer to us that the long ones
\cite{Mao_Narayan_P94,Katz_Canel96,Tavani98} possible with  all
observed short bursts are at $z<0.5$. However,
\textcite{Schmidt01} finds for short bursts $\langle V/V_{max}
\rangle= 0.354$, which is rather close to the value of long
bursts. Assuming that short GRBs also follow the SFR he obtains a
local rate of $0.075 {\rm Gpc}^{-3} {\rm yr}^{-1}$  - a factor of
two below the rate of long GRBs! The (isotropic) peak luminosities
are comparable. This results differs from a recent calculation of
\textcite{GuettaPiran03} who find for short bursts $\langle
V/V_{max} \rangle= 0.390$ and determine from this a local rate of
$1.7 {\rm Gpc}^{-3} {\rm yr}^{-1}$ which is about four times the
rate of long bursts. This reflects the fact that the {\bf
observed} short GRBs are significantly nearer than the {\bf
observed} long ones.

These rates and luminosities are assuming that the bursts are {\bf
isotropic}. Beaming reduces the actual peak luminosity increases
the implied rate by a  factor $f_b^{-1}=2 / \theta^2$.  By now
there is evidence that GRBs are beamed and moreover the total
energy  in narrowly distributed \cite{Frail01,PanaitescuK01}.
There is also a good evidence that the corrected peak luminosity
is much more narrowly distributed than the isotropic peak
luminosity \cite{vanPuttenRegimbau03,GuettaPiranWaxman03}. The
corrected peak luminosity is $L_{peak} (\theta^2/2) \sim const$.
\textcite{Frail01} suggest that the true rate is larger by a
factor of 500 than the observed isotropic estimated rate. However,
\textcite{GuettaPiranWaxman03} repeated this calculation
performing a careful average over the luminosity function and find
that that true rate is only a factor of $\sim 75 \pm25 $ times the
isotropically estimate one. Over all the true rate is: $ 33 \pm 11
h_{65}^{3} {\rm Gpc}^{-3} {\rm yr}^{-1}$.

With increasing number of GRBs with redshifts it may be possible
soon to determine the GRB redshift distribution directly from this
data. However, it is not clear what are the observational biases
that influence this data set and one needs a homogenous data set
in order to perform this calculation.  Alternatively one can try
to determine luminosity estimators
\cite{Norris_lags00,Fenimore_Ramirez-Ruiz01,SchaeferDengBand01,Schaefer03a}
from the subset with known redshifts and to obtain, using them a
redshift distribution for the whole GRB sample.
\textcite{Lloyd-RonningFryerRamirez-Ruiz02} find using the
\textcite{Fenimore_Ramirez-Ruiz01} sample that this method implies
that (i) The rate of GRBs increases backwards with time even for
$z>10$, (ii) The Luminosity of GRBs increases with redshift as
$(1+z)^{1.4\pm 0.5}$; (iii) Hardness and luminosity are strongly
correlated. It is not clear how these features, which clearly
depend on the inner engine could depend strongly on the redshift.
Note that in view of the luminosity-angle relation (see \S
\ref{sec:Energetics} below) the luminosity depends mostly on the
opening angle. An increase of the luminosity with redshift would
imply that GRBs were more narrowly collimated at earlier times.

\subsubsection{Association with Supernovae}
\label{sec:obs-SN}

The association of GRBs with star forming regions and the
indications that GRBs follow the star formation rate suggest that
GRBs are related to stellar death, namely to Supernovae
\cite{Pac98}. Additionally there is some direct evidence of
association of GRBs with Supernovae.

{\bf GRB 980425 and SN98bw:} The first indication of an
association between GRBs and SNes was found when SN 98bw was
discovered within the error box of  GRB 980425 \cite{Galama98bw}.
This was an usual type Ic SN which was much brighter than most
SNs. Typical ejection velocities in the SN were larger than usual
($ \sim 2\cdot 10^4 km/sec$) corresponding to a kinetic energy of
$2-5 \time 10^{52}$ ergs, more than ten times  than previously
known energy of SNes, \cite{IwamotoEtal98}. Additionally radio
observations suggested a component expanding sub relativistically
with $v \sim 0.3 c$ \cite{Kulkarnietal98}. Thus, 1998bw was an
unusual type Ic supernovae, significantly more powerful than
comparable SNes. This may imply that SNs are associated with more
powerful SNes. Indeed all other observations of SN signature in
GRB afterglow light curves use a SN 98bw templates.  The
accompanying GRB,  980425 was also unusual. GRB 980425 had a
smooth FRED light curve and no high energy component in its
spectrum. Other bursts like this exist but they are rare. The
redshift of SN98bw was 0.0085 implying an isotropic equivalent
energy of $\sim 10^{48}$ergs. Weaker by several orders of
magnitude than a typical GRB.

The BeppoSAX Wide Field Cameras had localized GRB980425 with a 8
arcmin radius accuracy.  In this circle, the BeppoSAX NFI (Narrow
Field Instrument) had detected two sources, S1 and S2. The NFI
could associate with each of these 2 sources an error circle of
1.5 arcmin radius.  The radio and optical position of SN1998bw
were consistent only with the NFI error circle of S1, and was out
of the NFI error circle of S2.  Therefore, \textcite{Pianetal00}
identified S1 with X-ray emission from SN1998bw, although this was
of course no proof of association between SN and GRB.  It was
difficult, based only on the BeppoSAX NFI data, to characterize
the behavior and variability of S2 and it could not be excluded
that S2 was the afterglow of GRB980425. The XMM observations of
March 2002 \cite{Pianetal03} seem to have brought us closer to the
solution. XMM detects well S1, and its flux is lower than in 1998:
the SN emission has evidently decreased. Concerning the crucial
issue, S2: XMM, having a better angular resolution than BeppoSAX
NFIs, seems to resolve S2 in a number of sources. In other words,
S2 seems to be not a single source, but a group of small faint
sources.  Their random variability (typical fluctuations of X-ray
sources close to the level of the background) may have caused the
flickering detected for S2.  This demolishes the case for the
afterglow nature of S2, and strengthens in turn the case for
association between GRB980425 and SN1998bw.

{\bf Red Bumps:}  The late red bumps (see \S \ref{sec:Obs-opt})
have been  discovered in several GRB light curves
\cite{Bloom99,Reichart99,Bloometal02,Garnavichetal03}. These bumps
involve both a brightening (or a flattening) of the afterglow as
well as a transition  to a much redder spectrum. These bumps have
been generally interpreted as due to an underlining SN
\cite{Bloom99}. In all cases the bumps have been fit with a
template of SN 1998bw, which was associated with GRB 980425.
\textcite{EsinBlandford00} proposed that these bumps are produced
by  light echoes on surrounding dust (but see \cite{Reichart01}).
\textcite{Waxman-Draine00} purposed another alternative
explanation based on dust sublimation.

\begin{figure}[htb]
\begin{center}
\epsfig{file=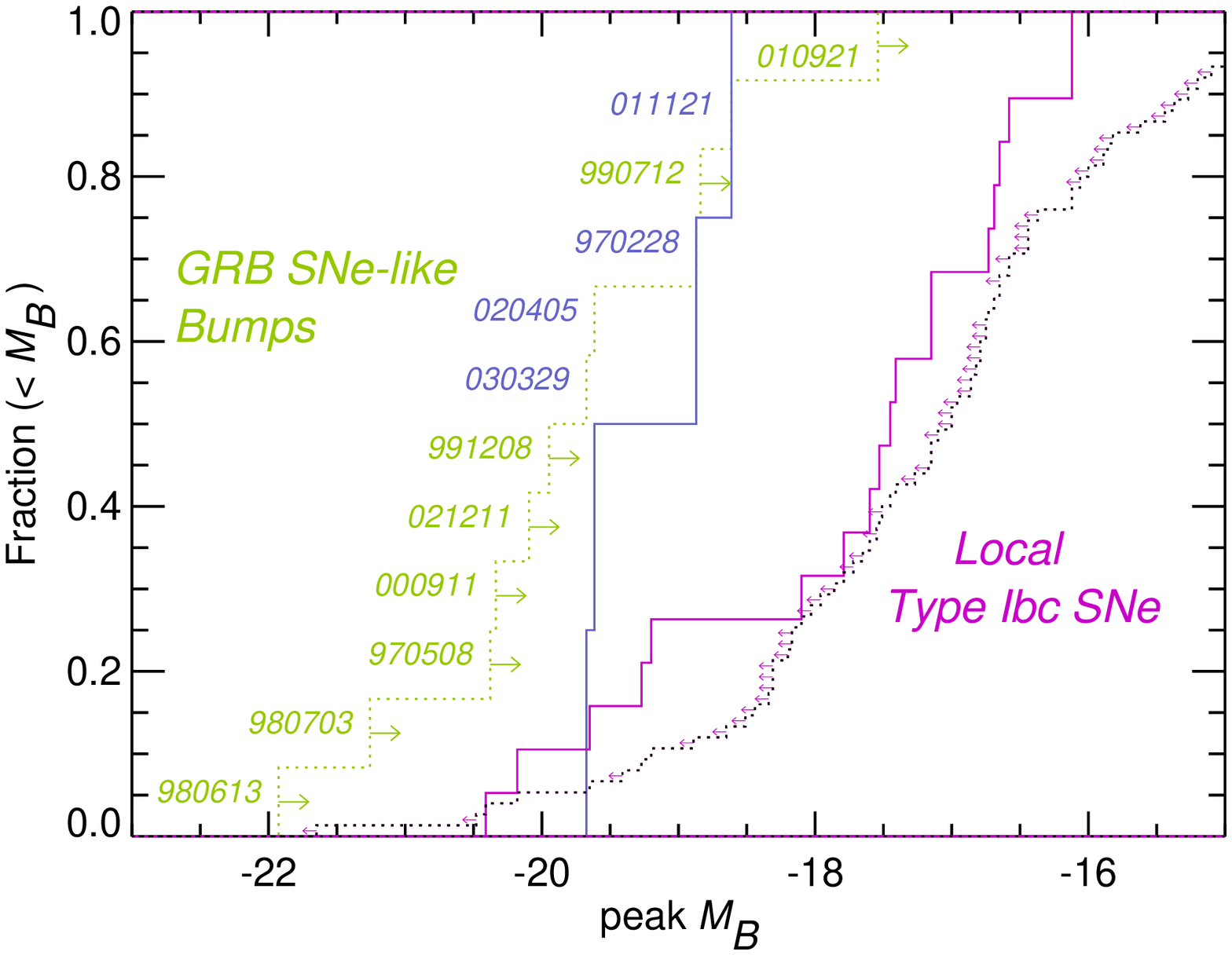,height=5in} \caption{Comparison of the
peak magnitudes of GRB-SNe with those of local Type Ib/Ic SNe. The
solid cumulative histogram to the left is for those GRBs with a
believable detection of a SN bump; the brightness of all other
claimed GRB-SN peaks or reported upper-limits are shown as a
dotted histogram. The solid histogram to the right is for those
local Ibc SNe for which the SN was observed before peak; all
others were discovered after peak. The faintest GRB-SN
non-detection (010921) only probes the top $\sim$40th-percentile
of local Type Ib/Ic SNe.  It is clear that the current GRB-SNe
population may have only revealed the tip of the iceberg;
plausibly, then, SNe could accompany all long-duration GRBs, from
\cite{Bloom03}} \label{fig:SNbump}
\end{center}
\end{figure}

For most GRBs there is only an upper limit to the magnitude of the
bump in the light curve.  A comparison of these upper limits (see
Fig. \ref{fig:SNbump}) with the maximal magnitudes of type Ibc SNe
shows that the faintest GRB-SN non-detection (GRB 010921) only
probes the top $\sim$40th-percentile of local Type Ib/Ic SNe.  It
is clear that the current GRB-SNe population may have only
revealed the tip of the iceberg; plausibly, then, SNe could
accompany all long-duration GRBs.

\begin{figure}[htb]
\begin{center}
\epsfig{file=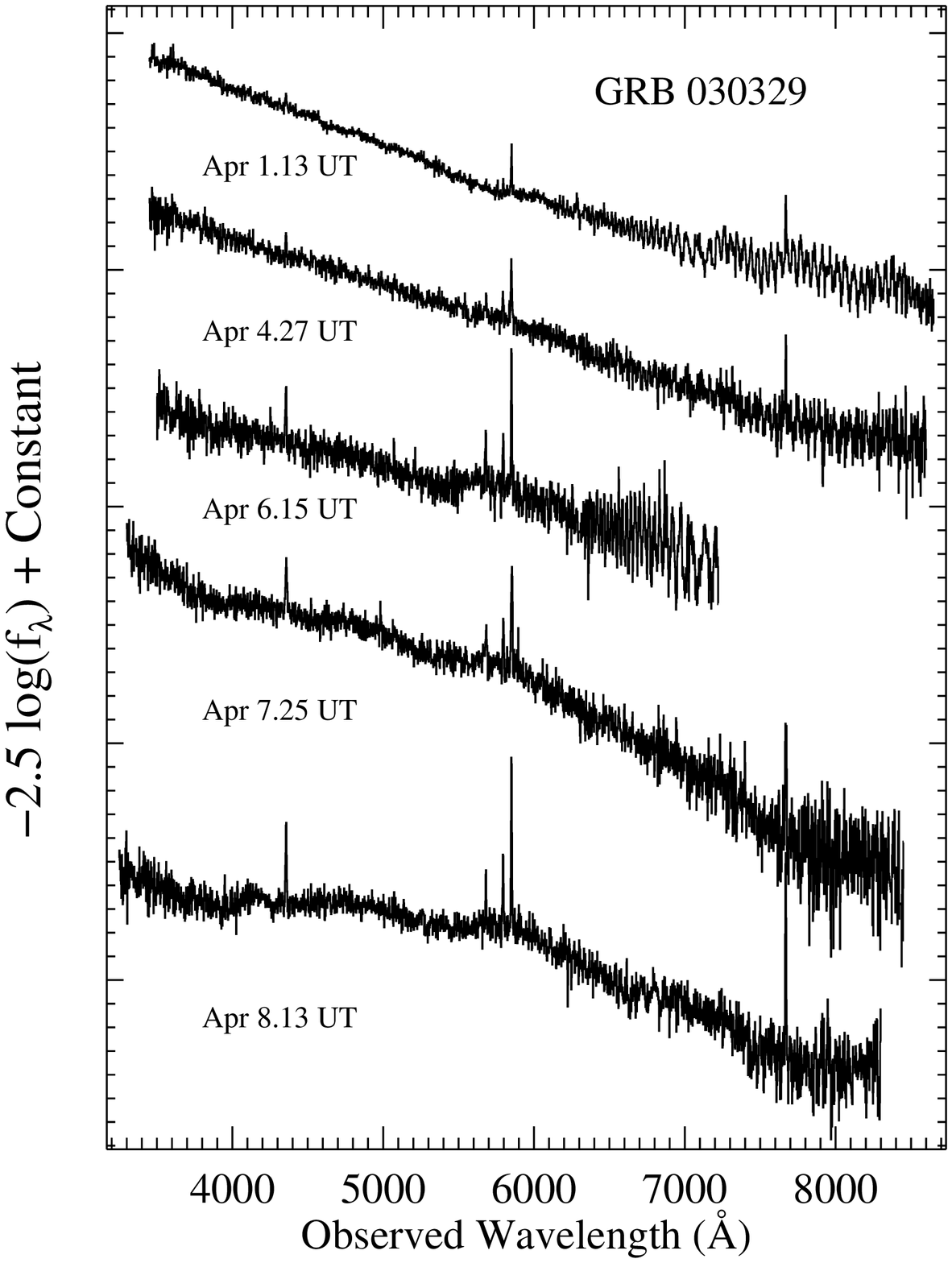,height=5in} \caption{Evolution of the
GRB 030329/SN2004dh spectrum, from April 1.13 UT (2.64 days after
the burst), to April 8.13 UT (6.94 days after the burst). The
early spectra consist of a power law continuum ($F_\nu \propto
\nu^{-0.9}$) with narrow emission lines originating from HII
regions within the host galaxy at redshift z=0.168. Spectra taken
after April 5 show the development of broad peaks characteristic
of a supernova (from \cite{StanekEtal03}.} \label{SN_2003dh}
\end{center}
\end{figure}

{\bf GRB 030329 and CN 2003dh:} The confirmation of SN 98bw like
bump and the confirmation of the GRB-SN association was
dramatically seen recently \cite{StanekEtal03,Hjorth03SN} in the
very bright GRB 030329 that is associated with SN 2003dh
\cite{ChornockEtal03}. The bump begun to be noticed six days after
the bursts and the SN 1999bw like spectrum dominated the optical
light curve at later times (see Fig. \ref{SN_2003dh}. The spectral
shapes of 2003dh and 1998bw were quite similar, although there are
also differences. For example \ref{sec:Energetics} estimated a
somewhat larger expansion velocity for 2003dh. Additionally the
\xr signal was much brighter (but this could be purely afterglow).

For most researchers in the field this discovery provided the
final conclusive link between SNe and GRBs (at least with long
GRBs). As the SN signature coincides with the GRB this
observations also provides evidence against a Supranova
interpretation, in which the GRB arises from a collapse of a
Neutron star that takes place sometime after the Supernova in
which the Neutron star was born - see \ref{sec:Supranova} .
(unless there is a variety of Supranova types, some with long
delay and others with short delay between the first and the second
collapses) the spectral shapes of 2003dh and 1998bw were quite
similar, although there are also differences. For example there is
a slightly larger expansion velocity for 2003dh. It is interesting
that while not as week as GRB 990425, the accompanying GRB
99030329 was significantly weaker than average. The implied
opening angle reveals that the prompt $\gamma$-ray energy output,
$E_\gamma$, and the X-ray luminosity at $10\;$hr, $L_X$, are a
factor of $\sim 20$ and $\sim 30$, respectively, below the average
value around which most GRBs are narrowly clustered (see
\ref{sec:Energetics} below).

It is interesting to compare SN 1999bw and SN 2003dh. Basically,
at all epochs \textcite{Mathesonetal03} find that  the best fit to
spectra of 2003dh is given by 1998bw at about the same age . The
light curve is harder, as the afterglow contribution is
significant, but using spectral information they find that 2003dh
had basically the same light curve as 1998bw.
\textcite{Mazzalietal03} model the spectra  and find again that it
was very similar to 1998bw. They find some differences, but some
of that might be due to a somewhat different approach to spectral
decomposition, which gives somewhat fainter supernova.

{\bf \xr lines:} The appearance of iron \xr lines (see \S
\ref{sec:obs-xr}) has been  interpreted as additional evidence for
SN. One has to be careful with this interpretation as the iron \xr
lines are seen as if emitted  by matter at very low velocities and
at rather large distances. This is difficult to achieve if the
supernova is simultaneous with the GRB, as the SN bumps imply.
This \xr lines might be consistent with the Supranova model
\cite{VietriStella98} in which the SN takes place month before the
GRB. However, in this case there won't be a SN bump in the light
curve! \textcite{MR00,MR01} and \textcite{KumarNarayan03} suggest
alternative interpretations which do not require a Supranova.

\subsection{Energetics }
\label{sec:Energetics}

Before redshift measurements were available the  GRB energy  was
estimated from the BATSE catalogue by fitting an (isotropic)
luminosity function to the flux distribution (see e.g
\textcite{Cohen_Piran95,LoredoWasserman98,Schmidt99,Schmidt01,Schmidt01a,GuettaPiranWaxman03}
and many others). This lead to a statistical estimate of the
luminosity function of  a distribution of bursts.

These estimates were  revolutionized with the direct determination
of the redshift for individual bursts. Now the energy could be
estimated directly for specific bursts. Given an observed
$\gamma$-ray fluence and the redshift to a burst one can easily
estimate the energy emitted in $\gamma$-rays, $E_{\gamma,iso}$
assuming that the emission is isotropic (see
\textcite{Bloom_Frail_Sari01} for a detailed study including k
corrections). The inferred energy, $E_{\gamma,iso}$ was the
isotropic energy, namely the energy assuming that the GRB emission
is isotropic in all directions. The energy of the first burst with
a determined redshift, GRB 970508, was around $10^{51}$ergs.
However, as afterglow observations proceeded, alarmingly large
values ({\it e.g.} $3.4 \times 10^{54}$ergs for GRB990123) were
measured for $E_{\gamma,iso}$. The variance was around three
orders of magnitude.

However, it turned out \cite{Rhoads99,SPH99} that GRBs are beamed
and $E_{\gamma,iso}$ would not then be a good estimate for the
total energy emitted in $\gamma$-rays. Instead: $E_\gamma \equiv
(\theta^2/2)E_{\gamma,iso}$. The angle, $\theta$, is the effective
angle of $\gamma$-ray emission. It can be estimated from $t_{b}$,
the time of the break in the afterglow light curve \cite{SPH99}:
\begin{equation}
\theta =0.16 (n/E_{k,iso,52})^{1/8} t_{b,days}^{3/8} = 0.07
(n/E_{k,\theta,52})^{1/6} t_{b,days}^{1/2},
\end{equation}
where $t_{b,days}$ is the break time in days. $E_{k,iso,52}$ is
``isotropic equivalent" kinetic energy, discussed below, in units
of $10^{52}$ergs, while $E_{k,\theta,52}$ is the real kinetic
energy in the jet i.e: $E_{k,\theta,52}=(\theta^2/2)
E_{k,iso,52}$. One has to be careful which of the two energies
one discusses. In the following I will usually consider, unless
specifically mentioned differently, $E_{k,iso,52}$, which is also
related to the energy per unit solid angle as: $E_{k,iso,52}/4
\pi$. The jet break is observed both in the optical and in the
radio frequencies. Note that the  the observational signature in
the radio differs from that at optical and \xr
\cite{SPH99,Harrisonetal99} (see Fig. \ref{fig:990510_radio}) and
this provides an additional confirmation for this interpretation.

\begin{figure}[htb]
\begin{center}
\epsfig{file=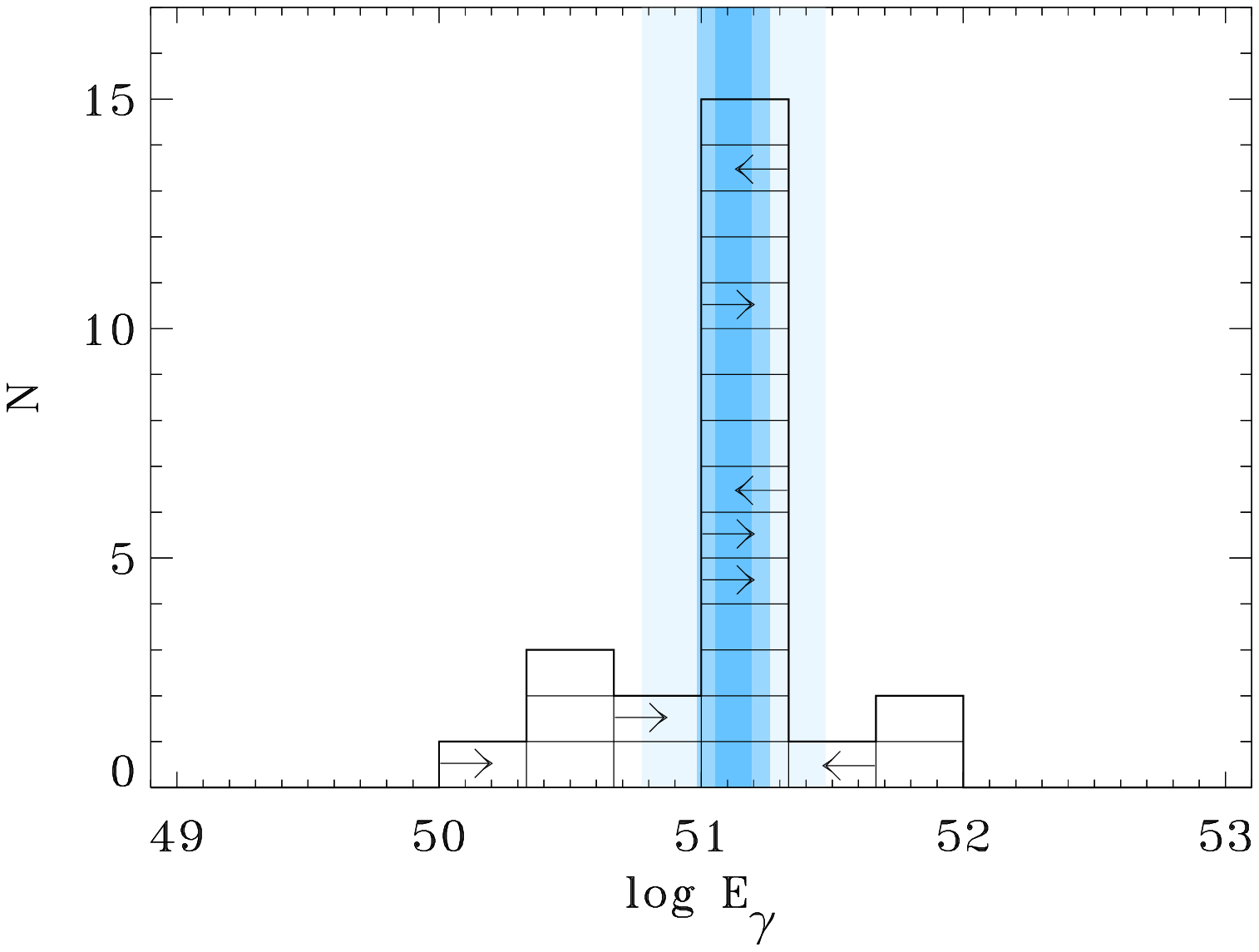,height=2in} \caption{A histogram of
GRB energies ($E_\gamma$) with three equal logarithmic spacings
per decade. The histogram shows a narrow distribution of GRB
energies about the standard energy ${E}_\gamma = 1.33 \cdot
10^{51}$ ergs , with an error of $\sigma = 0.07$\,dex. The
observed burst--to--burst rms spread is $0.35$\,dex (a factor of
2.23) about this value. Bands of 1, 2, and 5 $\sigma$ about the
standard energy are shown. There are five identifiable outliers,
which lie more than 5 $\sigma$ from the mean, however, there is
currently no basis other than discrepant energy to exclude these
bursts from the sample (\cite{BloomFrailKulkarni03}.}
\label{fig:energy1}
\end{center}
\end{figure}

\begin{figure}[htb]
\begin{center}
\epsfig{file=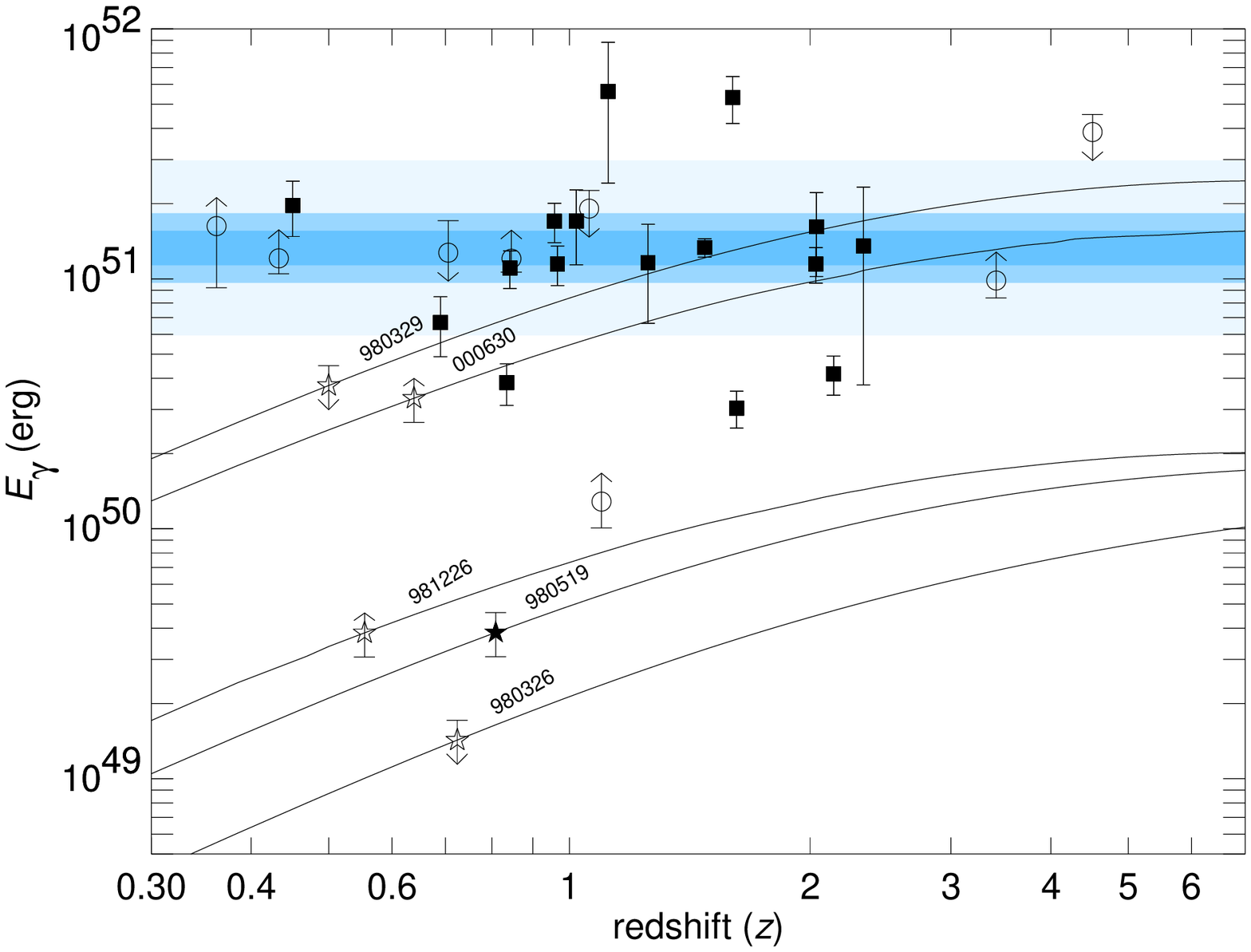,height=2in} \caption{GRB energy
release versus redshift. Bands of 1, 2, and 5 $\sigma$ about the
mean energy ${E}_\gamma = 1.33 \cdot 10^{51}$ ergs  are shown.
Plotted are the trajectories of five GRBs with no known
spectroscopic redshift (labeled with star symbols). While the
energies of GRB\,980329 and GRB\,000630 could be consistent with
the standard value at redshifts beyond $z \sim 1.5$, at no
redshift could the energies of GRB\,980326 and GRB\,980519 be
consistent cite{BloomFrailKulkarni03}.} \label{fig:energy2}
\end{center}
\end{figure}

\textcite{Frail01} estimated $E_\gamma$ for 18 bursts, finding
typical values around $10^{51}$ergs (see also
\textcite{PanaitescuK01}). \textcite{BloomFrailKulkarni03} find
${E}_\gamma = 1.33 \times 10^{51}\,h_{65}^{-2}$ erg and a
burst--to--burst variance about this value  $\sim 0.35$ dex, a
factor of 2.2. This is three orders of magnitude smaller than the
variance in the isotropic equivalent $E_\gamma$. A compilation of
the beamed energies from \cite{BloomFrailKulkarni03}, is shown in
Figs \ref{fig:energy1} and \ref{fig:energy2}. It demonstrates
nicely this phenomenon. The constancy of $E_\gamma$ is remarkable,
as it involves a product of a factor inferred from the GRB
observation (the \gr flux) with a factor inferred from the
afterglow observations (the jet opening angle).  However,
$E_\gamma$ might not be a good estimate for $E_{tot}$, the total
energy emitted by the central engine. First, an unknown conversion
efficiency of energy to $\gamma$-rays has to be considered:
$E_{tot} = \epsilon^{-1} E_\gamma =\epsilon^{-1} (\theta^2/2)
E_{\gamma,iso}$. Second, the large Lorentz factor during the
$\gamma$-ray emission phase, makes the observed $E_\gamma$ rather
sensitive to angular inhomogeneities of the relativistic ejecta
\cite{KP00b}. The recent early observations of the afterglow of
GRB 021004 indicate that indeed a significant angular variability
of this kind exists \cite{NakarPiranGranot03,NakarPiran03b}.

The kinetic energy of the flow during the adiabatic afterglow
phase, $E_k$ is yet another energy measure that arises. This
energy (per unit solid angle) can be estimated from the afterglow
light curve and spectra. Specifically it is rather closely
related to the observed afterglow \xr flux
\cite{Kumar00,Waxman_Friedman,Piranetal01}. As this energy is
measured when the Lorentz factor is smaller it is less sensitive
than $E_\gamma$ to angular variability. The constancy of the \xr
flux \cite{Piranetal01} suggest that  this energy is also
constant. Estimates of $E_{k,\theta}$ \cite{PanaitescuK01} show
that $\bar E_\gamma \approx 3 \bar E_{k,\theta}$, namely the
observed ``beamed" GRB energy is larger than the estimated
``beamed" kinetic energy of the afterglow. \textcite{Frail01},
however, find that $\bar E_\gamma \approx \bar E_{k,\theta}$,
namely that the two energies are comparable.

An alternative interpretation to the observed  breaks is that we
are viewing a ``universal" angle dependent, namely, ``structured"
jet - from different viewing angles
\cite{Lipunov_Postnov_Pro01,Rossi02,Zhang02}.  The observed break
corresponds in this model to the observing angle $\theta$ and not
to the opening angle of the jet. This interpretation means that
the GRB beams are wide and hence the rate of GRBs is  smaller than
the rate implied by the usual beaming factor. On the other hand it
implies that GRBs are more energetic.
\textcite{GuettaPiranWaxman03} estimate that this factor (the
ratio of the fixed energy of a ``structured" jet relative to the
energy of a uniform jet to be $\sim 7$. However they find that the
observing angle distribution is somewhat inconsistent with the
simple geometric one  that should arise in universal structured
jets (see also \textcite{Pernaetal03,NakarGranotGuetta03}). The
energy-angle relation discussed earlier require (see  \S
\ref{sec:structured} below) an angle dependent jet with $E(\theta)
\propto \theta^{-2}$.

Regardless of the nature of the jet (universal structured jet or
uniform with a opening angle that differs from one burst to
another) at late time it becomes non relativistic and spherical.
With no relativistic beaming every observer detects emission from
the whole shell. Radio observations at this stage enable us to
obtain a direct calorimetric estimate of the total kinetic energy
of the ejecta at late times  \cite{FrailWaxmanKulkarni00}
Estimates performed in several cases yield a comparable value for
the total energy.

If GRBs are beamed  we should expect orphan afterglows (see \S
\ref{sec:orphan}): events in which we will miss the GRB but we
will  observe the late afterglow that is not so beamed. A
comparison of the rate of orphan afterglows to GRBs will give us a
direct estimate of the beaming of GRBs (and hence of their
energy). Unfortunately there are not even good upper limits on the
rate of orphan afterglows. \textcite{Veerswijk03} consider the
observations within the Faint Sky Variability Survey (FSVS)
carried out on the Wide Field Camerea on teh 2.5-m Isacc Newton
Telescope on La Palma. This survey mapped 23 suare degree down to
a limiting magnitude of about V=24. They have found one object
which faded and was not detected after a year. However, its colors
suggest that it was a supernova and not a GRB. Similarly,
\textcite{VandenBerketal02} find a single candidate within the
Sloan Digital Sky Survey. Here the colors were compatible with an
afterglow. However, later it was revealed that this was a variable
AGN and not an orphan afterglow. As I discuss later this limits
are still far to constrain the current beaming estimates (see \S
\ref{sec:orphan}).

One exception is for late radio emission for which there are some
limits \cite{PernaLoeb98,Levinsonetal02}.
\textcite{Levinsonetal02} show that the number of orphan radio
afterglows associated with GRBs that should be detected by a
flux-limited radio survey  is smaller for a smaller jet opening
angle $\theta$. This might seen at first sight contrary to
expectation as narrower beams imply more GRBs. But, on the other
hand, with narrower beams each GRB has a lower energy and hence
its radio afterglow is more difficult to detect. Overall the
second factor wins. Using the results of FIRST and NVSS surveys
they find nine afterglow candidates. If all candidates are
associated with GRBs then there is a lower limit on the beaming
factor of $f^{-1}_b \equiv (\theta^2/2)> 13$. If none are
associated with GRBs they find $f^{-1}_b > 90$. This give
immediately a corresponding upper limit on the average energies of
GRBs. \textcite{GuettaPiranWaxman03} revise this values in view of
a recent estimates of the correction to the rate of GRBs to:
$f^{-1}_b = 40$.

When considering  the energy of GRBs one has to remember the
possibility, as some models suggest,  that an additional energy is
emitted which is not involved in the GRB itself or in the
afterglow. \textcite{vanPuttenLevinson01}, for example, suggest
that a powerful Newtonian wind collimates the less powerful
relativistic one. The ``standard jet" model also suggests a large
amount of energy emitted sideways with a lower energy per solid
angle and a lower Lorentz factors.
 It is interesting to note that the
calorimetric estimates mentioned earlier limit the total amount of
energy ejected regardless of the nature of the flow. More
generally, typically during the afterglow matter moving with a
lower Lorentz factor emits lower frequencies. Hence by comparing
the relative beaming of afterglow emission in different wavelength
one can estimate the relative beaming factors, $f^{-1}_b(E)$, at
different wavelength and hence at different energies.
\textcite{NakarPiran03a} use various \xr searches for orphan \xr
afterglow to limit the (hard) \xr energy to be at most comparable
to the \gr energy. This implies that the total energy of matter
moving at a Lorentz factor of $\sim 40$ is at most comparable to
the energy of matter moving with a Lorentz factor of a few hundred
and producing the GRB itself. At present limits on optical orphan
afterglow are insufficient to set significant limits on matter
moving at slower rate, while as mentioned earlier radio
observations already limit the overall energy output.

These observations won't limit, of course, the energy emitted in
gravitational radiation, neutrinos, Cosmic Rays or very high
energy photons that may be emitted simultaneously by the source
and influence the source'e energy budget without influencing the
afterglow.

\section{THE GLOBAL PICTURE -  GENERALLY ACCEPTED INGREDIENTS }
\label{sec:accepted}

 There are several generally accepted ingredients in practically all
 current GRB models.

{\bf Relativistic Motion:} Practically all current GRB models
 involve a relativistic motion with a Lorentz factor,
$\Gamma > 100$. This is essential to overcome the compactness
problem (see \S \ref{sec:comp} below). At first this understanding
was based only on theoretical arguments. However, now there are
direct observational proofs of this concept: It is now generally
accepted that both the radio scintillation \cite{Goodman97} and
the lower frequency self-absorption \cite{KP97} provide
independent estimates of the size of the afterglow, $\sim
10^{17}$cm, two weeks after the burst. These observations imply
that the afterglow has indeed expanded relativistically.
\textcite{SP99a} suggested that the optical flash accompanying GRB
990123 provided a direct evidence for ultra-relativistic motion
with $\Gamma \sim 100$. \textcite{SoderbergRamirezRuiz03} find a
higher value: $1000 \pm 100$. However, these interpretations are
model dependent.

The relativistic motion implies that we observe blue shifted
photons which are significantly softer in the moving rest frame.
It also implies that when the objects have a size $R$ the observed
emission arrives on a typical time scale of $R/c \G^2$ (see \S
\ref{sec:Temporal}).  Relativistic beaming also implies that we
observe only a small fraction ($1/\Gamma$) of the source. As I
discussed earlier (see \S \ref{sec:Energetics} and also
\ref{sec:patchy-shell}) this has important implications on our
ability to estimate the total energy of GRBs.

While all models are based on ultra-relativistic motion, none
explains convincingly (this is clearly a subjective statement) how
this relativistic motion is attained. There is no agreement even
on the nature of the relativistic flow. While in some models the
energy is carried out in the form of kinetic energy of baryonic
outflow in others it is a Poynting dominated flow or both.

{\bf Dissipation} In most models the  energy of the relativistic
flow is dissipated and this provides the energy needed for the
GRB and the subsequent afterglow.  The dissipation is in the form
of (collisionless) shocks, possibly via plasma instability. There
is a general agreement that the afterglow is produced via
external shocks with the circumburst matter (see
\ref{sec:afterglow}). There is convincing evidence (see {\it e.g.}
\textcite{Fenimoreetal96,SP97,Ramirez-Ruiz_Fenimore00,Piran_Nakar02}
and \S \ref{sec:ex-int} below) that in most bursts the
dissipation during the GRB phase takes place via internal shocks,
namely shocks within the relativistic flow itself. Some (see e.g.
\textcite{Dermer_Mitman99,Begelman99,RuffiniEtal01,Dar03})
disagree with this statement.

{\bf Synchrotron Radiation:} Most models (both of the GRB and the
afterglow)  are based on Synchrotron emission from relativistic
electrons accelerated within the shocks. There is a reasonable
agreement between the predictions of the synchrotron model and
afterglow observations
\cite{Wijers_Galama98,GPS99a,PanaitescuK01}. These are also
supported by measurements of linear polarization in  several
optical afterglows (see \S \ref{sec:Obs-opt}). As for the GRB
itself there are various worries about the validity of this model.
In particular there are some inconsistencies between the observed
special slopes and those predicted by the synchrotron model (see
\cite{Preece02} and \S \ref{sec:spec-obs}). The main alternative
to Synchrotron emission is synchrotron-self Compton
\cite{Waxman97a,Ghisellini_Celotti99} or inverse Compton of
external light
\cite{Shemi94,Brainerd94,ShavivDar95,Lazzatietal03}. The last
model requires, of course a reasonable source of external light.

{\bf Jets and Collimation:} Monochromatic breaks appear in many
afterglow light curves. These breaks are interpreted as ``jet
breaks" due to the sideways beaming of the relativistic emission
\cite{PanaitescuMeszaros99,Rhoads99,SPH99}  (when the Lorentz
factor drops below $1/\theta_0$ the radiation is beamed outside of
the original jet reducing the observed flux) and due to the
sideways spreading of a beamed flow \cite{Rhoads99,SPH99}. An
alternative interpretation is of  a viewing angles of a
``universal structured jet"
\cite{Lipunov_Postnov_Pro01,Rossi02,Zhang02} whose energy varies
with the angle. Both interpretations suggest that GRBs are beamed.
However, they give different estimates of the overall rate and the
energies of GRBs (see \S \ref{sec:structured} below). In either
case the energy involved with GRBs is smaller than the naively
interpreted isotropic energy and the rate is higher than the
observed rate.

{\bf A (Newborn) Compact Object} If one accepts the beaming
interpretation of the breaks in the optical light curve the total
energy release in GRBs is $\sim 10^{51}$ergs
\cite{Frail01,PanaitescuK01}. It is higher if, as some models
suggest, the beaming interpretation is wrong or if a significant
amount of additional energy (which does not contribute to the GRB
or to the afterglow) is emitted from the source. This energy,
$\sim 10^{51}$ergs, is comparable to the energy released in a
supernovae.  It indicates that the process must involve a compact
object. No other known source can release  so much energy within
such a short time scale. The process requires a dissipation of
$\sim 0.1 m_\odot$ within the central engine over a period of a
few seconds. The sudden appearance of so much matter in the
vicinity of the compact object suggest a violent process, one that
most likely involves the birth of the compact object itself.

{\bf Association with Star Formation and SNe:} Afterglow
observations, which exist for a subset of relatively bright long
bursts, show that GRBs arise within galaxies with a high star
formation rate (see \cite{Djorgovski01b} and \S \ref{sec:hosts}).
Within the galaxies the bursts distribution follows the light
distribution \cite{Bloom02}.  This has lead to the understanding
that (long) GRB arise from the collapse of massive stars (see \S
\ref{sec:Collapsar}). This understanding has been confirmed by the
appearance of SN bumps in the afterglow light curve (see \S
\ref{sec:obs-SN} earlier) and in particular by the associations of
SN 1999bw with GRB 980425 and of  SN 2003dh with GRB 030329.

 {\bf Summary:} Based on these generally accepted ideas one can
sketch the following generic GRB model: GRBs are a rare phenomenon
observed within star forming regions and associated with the death
of massive stars and the birth of compact objects. The \gr
emission arises from internal dissipation within a relativistic
flow. This takes place at a distances of $\sim 10^{13}-10^{15}$cm
from the central source that produces the relativistic outflow.
Subsequent dissipation of the remaining energy due to interaction
with  the surrounding circumburst matter produces the afterglow.
The nature of the ``inner engine" is not resolved yet, however, a
the association with SN (like 1998bw and 2003dh) shows that long
GRBs involve a a collapsing star. Much less is known on the origin
of short GRBs.

\section{RELATIVISTIC EFFECTS }
\label{sec:rel}

\subsection{Compactness and relativistic motion }
\label{sec:comp}

The first theoretical clues to the necessity of relativistic
motion in GRBs arose from the Compactness problem
\cite{Ruderman75}. The conceptual argument is simple. GRBs show a
non thermal spectrum with a significant high energy tail (see \S
\ref{sec:spec-obs}).  On the other hand a naive calculation
implies that the source is optically thick.  The fluctuations on a
time scale $\delta t$ imply that the source is smaller than $c
\delta t$. Given an observed flux $F$, a duration $T$, and an
distance $d$ we can estimate the energy $E$ at the source. For a
typical photon's energy $\bar E_\gamma$ this yields a photon
density $ \approx 4 \pi d^2 F / \bar E_\gamma c^3 \delta t^2$.
Now, two \gr can annihilate and produce e$^+$e$^-$ pairs, if the
energy in their CM frame is larger than $2 m_e c^2$. The optical
depth for pair creation is:
\begin{equation}
\tau_{\gamma\gamma} \approx { f_{e^\pm} \sigma_T 4 \pi d^2 F
\over \bar E_\gamma c^2 \delta t}
\end{equation}
where, $f_{e^\pm}$ is a numerical factor denoting the average
probability that photon will collide with another photon whose
energy is sufficient for pair creation.  For typical values and
cosmological distances, the resulting optical depth is extremely
large $\tau_{e^\pm} \sim 10^{15}$ \cite{Piran95}. This is, of
course, inconsistent with the non-thermal spectrum.

The compactness problem can be resolved if the emitting matter is
moving relativistically towards the observer. I denote the Lorentz
factor of the motion by $\G$. Two corrections appear in this
case. First, the observed photons are blue shifted and therefore,
their energy at the source frame is lower by a factor $\G$.
Second, the implied size of a source moving towards us with a
Lorentz factor $\G$ is $c \delta t \G^2$ (see \S
\ref{sec:Temporal} below). The first effect modifies $f_{e^\pm}$
by a factor $\G^{-2 \alpha}$ where $\alpha$ is the photon's index
of the observed \gr (namely the number of observed photons per
unit energy is proportional to $E^{-\alpha}$.). The second effect
modifies the density estimate by a factor $\G^{-4}$ and it
influences the optical depth as $\G^{-2}$. Together one finds that
for $\alpha \sim 2$ one needs $\G \gtrsim 100$ to obtain an
optically thin source.

The requirement that the source would be optically thin can be
used to obtain direct limits from  specific bursts on the minimal
Lorentz factor within those bursts
\cite{Krolik_Pier91,FenimoreEpsteinHo93,Woods_Loeb95,Piran95,Baring_Harding97,P99,Lithwick_Sari01}.
A complete calculation requires a detailed integration over
angular integrals and over the energy dependent pair production
cross section. The minimal Lorentz factor depends also on the
maximal photon energy, $E_{\rm max}$, the upper energy cutoff of
the spectrum. \textcite{Lithwick_Sari01} provide a detailed
comparison of the different calculations and point our various
flaws in some of the previous estimates. They find that:
\begin{equation}
\tau_{\gamma\gamma} = {11\over 180} {\sigma_T d^2  (m_e
c^2)^{-\alpha+1}{\cal F}\over c^2 \delta T (\alpha-1)} (
           {E_{\rm max}\over m_ec^2})^{\alpha-1} \G^{2\alpha + 2}
           (1+z)^{2\alpha-2} \ ,
           \label{opt}
\end{equation}
where the high end of the observed photon flux is given by ${\cal
F} E^{-\alpha}$ (photons per cm$^2$ per sec per unit photon
energy). A lower limit on $\G$ is obtained by equating Eq.
\ref{opt} to unity.

\subsection{Relativistic time effects }
\label{sec:Temporal}

Consider first a source moving relativistically with a constant
velocity  along a line towards the observer and two photons
emitted at $R_1$ and $R_2$. The first photon (emitted at $R_1$)
will reach the observer at time $(R_2-R_1)/v-(R_2-R_1)/c$ before
the second photon (emitted at $R_2$). For $\G \gg 1$ this equals
$\approx (R_2-R_1)/2 c \G^2$. This allows us to associate an
``observer time" $R/2 c \G^2$ with the distance $R$ and for this
reason I have associated a scale  $c \delta t \G^{-2}$ with
fluctuations on a time scale $\delta t$  in the optical depth
equation earlier (see \S \ref{sec:comp}). This last relation
should be modified if the source moves a varying velocity
(v=v(R)). Now
\begin{equation}
\delta t_{12} \approx \int_{R_1}^{R_2}\frac{ dR}{ 2 c \G^2(R)} \ ,
\end{equation}
which reduces to
\begin{equation}
T_R \approx R/ 2 c \G^2 \ , \label{Rt}
\end{equation}
for motion with a constant velocity. The difference between a
constant velocity source and a decelerating source introduces a
numerical factor of order eight  which is important during the
afterglow phase \cite{Sari97}.

\begin{figure}[htb]
\begin{center}
\epsfig{file=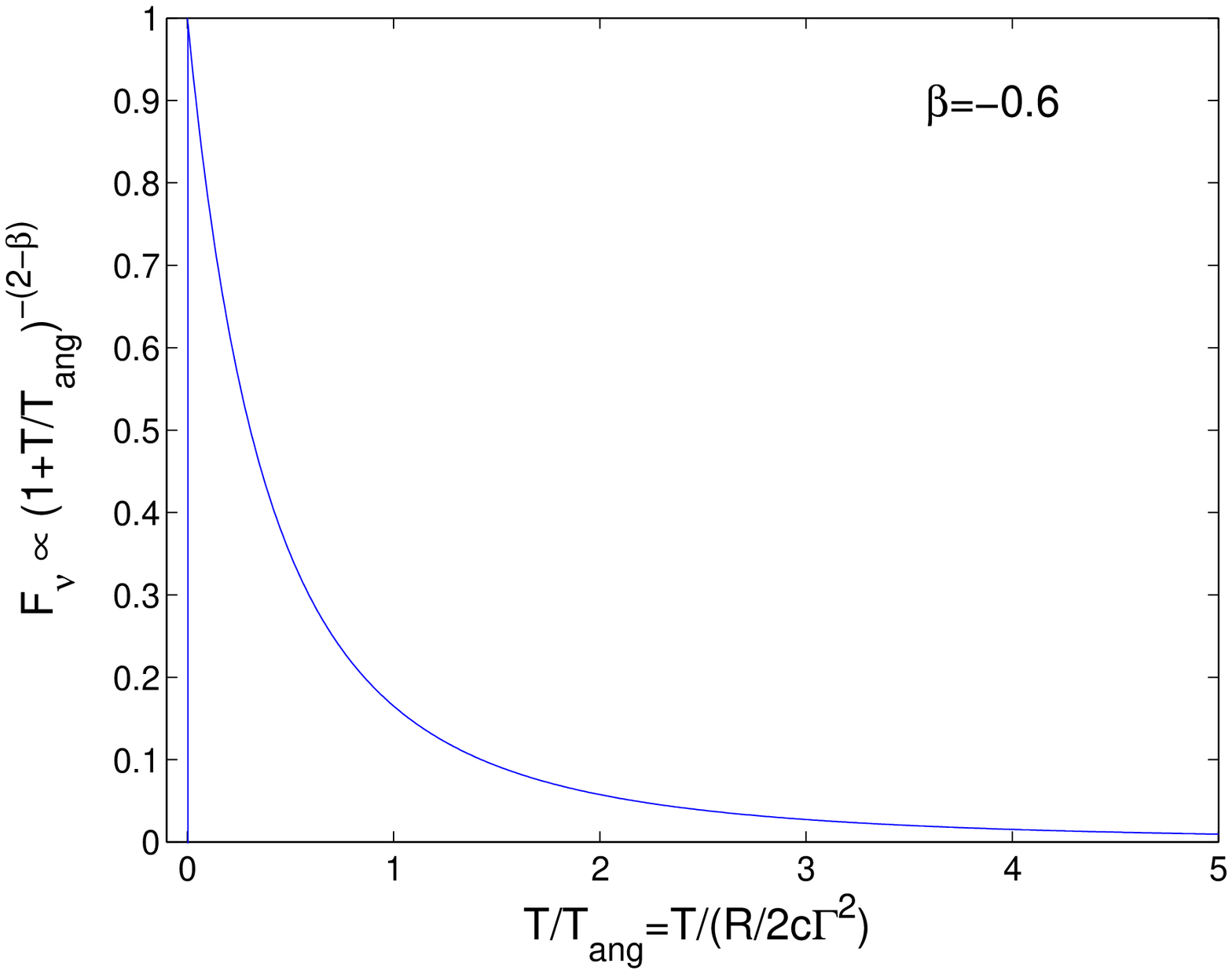,height=2in} \caption{The observed
pulse from an instantaneous flash from a spherical relativistic
thin shell moving relativistically and emitting emitting with a
power low $\nu^{-0.6}$.} \label{fig:thinshell}
\end{center}
\end{figure}

Consider now a relativistically expanding spherical shell, or at
least a shell that is locally spherical (on a scale larger than
$1/\G$). Emission from parts of the shell moving at angle $\theta$
relative to the line of sight to the observer will arrive later
with a time delay $R(1-cos \theta)/c$. For small angles this time
delay equals $R \theta^2/2 c$.  As the radiation is beamed with an
effective beaming angle $\approx 1/\G$ most of the radiation will
arrive within a typical angular time scale:
\begin{equation}
T_{ang} \equiv R /2 c \G^2 \ . \label{tang}
\end{equation}
The combination of time delay and blueshift implies that if  the
emitted spectrum is a power law spectrum with a spectral index
$\alpha$ then the observed signal from the instantaneous emission
of a thin shell will decay at late time as a power law with
$t^{-(2 - \alpha)}$ \cite{Fenimoreetal96,NakarPiran03a}. The
observed pulse from an instantaneous flash from a thin shell is
shown in Fig. \ref{fig:thinshell}.

As I discuss later (see \S \ref{sec:ex-int}) the similarity
between the angular time scale and the radial time scale plays a
crucial role in GRB models.

\subsection{Relativistic Beaming and the Patchy Shell Model }
\label{sec:patchy-shell}

The radiation from a relativistic source is beamed with a typical
beaming angle $1/\Gamma$. This implies that if the source that is
expanding radially with an ultra-relativistic speed a given
observer ``sees" radiation only from a region that is within
$\Gamma^{-1}$ from its line of sight to the source. If the radius
of the emitting region is $R$ the observer will see radiation from
a region of size $R/\Gamma$. Since $\Gamma$ is extremely large
during the GRB we observe emission only from a small fraction of
the emitting shell. It is possible, and even likely, that the
conditions within the small region that we observe will be
different from the average ones across the shell. This means that
 the conditions that we infer won't reflect the true average
conditions within this particular GRB.

An interesting point related to the internal shocks (discussed
later) model in this context is the following. According to the
internal shocks model individual pulses are obtained by
collisions between individual shells. Here the inhomogeneity of
individual shells could be wiped out when the contributions of
different hot spots from different shells is added. Alternatively
the ``inner engine" may produce  a consistent angular pattern in
which the hot spot is in the same position in all shells and in
this case averaging won't lead to a cancellation of the patchy
shell structure.

Within the internal-external model the GRB is produced by internal
shocks in which only the relative motion within the flow is
dissipated. The bulk Lorentz factor remains unchanged. During the
afterglow the shell is slowed down by external shocks. As the
Lorentz factor decreases with time (see Eq. \ref{RGamma}) we
observe a larger and larger fraction of the emitting region until
$\Gamma \approx \theta^{-1}$, where $\theta$ is the angular size
of the whole emitting region -  the GRB jet, see \S
\ref{sec:jets}. This has several inevitable implications. If the
initial relativistic flow is inhomogenous on a small angular scale
then different observers looking at the same GRB (from different
viewing angles)  will see different \gr light curves. A strong
burst to one observer might look weak to another one if it is
located at an angle larger than $1/\Gamma$ from the first. The two
observers will see similar conditions later on, during the
afterglow, as then they will observe the same angular regions.
This has the following implications: (i) Given that the GRB
population originate from some `typical' distribution we expect
that fluctuation between different bursts at early time during the
GRB will be larger than fluctuations observed at late times during
the afterglow \cite{KP00b}. A direct consequence of this behaviour
is the appearance of a bias in the observations of GRBs. As we are
more likely to detect stronger events we will tend to identify
bursts in which a `hot spot` was pointing towards us during the
GRB phase. If the original GRB shells are inhomogenous this would
inevitably lead to a bias in the estimates of the GRB emission as
compared to the kinetic energy during the afterglow. (ii) As the
afterglow slows down we observe a larger and larger region. The
angular structure would produces a variability in the light curve
with a typical time scale of t, the observed time. These
fluctuations will decay later as the Lorentz factor decreases and
the observations are averaged over a larger viewing angle.
\textcite{NakarPiranGranot03} have suggested that this is the
source of the early fluctuations in the light curve of GRB 021004.
\textcite{NakarOren03} modelled this process with a numerical
simulation. They find that the flucutation light curve of GRB
021004 can be nicely fitted by this model and that it also
explains the correlated fluctuations in the polarization (see also
\cite{Granot03}).

\section{PHYSICAL PROCESSES }
\label{sec:physical-Processes}

The observed prompt emission must be generated by energetic
particles that have been accelerated within the collisionless
shocks. The most likely process is synchrotron emission, even
though there is some evidence that a simple synchrotron spectra
does not fit all bursts \cite{Preece02} (but see however,
\cite{BarraudEtal03} who finds consistency with the synchrotron
model). I consider here, the different physical ingredient that
determine the emission process: particle acceleration, magnetic
field amplification, synchrotron emission and inverse Compton
emission that could be relevant in some cases.

\subsection{Relativistic Shocks}
\label{sec:shocks}

Shocks involve sharp jumps in the physical conditions.
Conservation of mass, energy and momentum determine the Hugoniot
shock jump conditions across the relativistic shocks for the case
when the upstream matter is cold (see e.g. \textcite{BLmc1}):
\begin{eqnarray}
n_2 = 4 \Gamma n_1 \\
\nonumber e_2 = 4 \Gamma n_1 m_p c^2 \\
\Gamma_{sh}^2 = 2 \Gamma^2  \nonumber \label{jump}
\end{eqnarray}
where $n_{1,2}$,$e_{1,2}$ are the number density and the energy
density (measured in the local rest frame) of the matter upstream
(region 1) and downstream (region 2). I have assumed that the
energy density in region 1 is very small compared to the rest mass
density. $\Ga$ is the Lorentz factor of the fluid just behind the
shock  and $\Ga_{sh}$ is the Lorentz factor of the shock front
(both measured in the rest frame of the upstream fluid). The
matter is compressed by a factor $\Gamma$ across a relativistic
shock. The pressure, or the internal energy density behind the
shock is of order $\Gamma^2 n_1 m_p c^2$. Thus,  in the shock's
rest frame the relative ``thermal" energy per particle
(downstream) is of the same order of the kinetic energy per
particle (ahead of the shock) upstream. Put differently the shock
converts the `ordered' kinetic energy to a comparable random
kinetic energy. In an ultra-relativistic shock the downstream
random velocities are ultra-relativistic.

Similar jump conditions can be derived for the  Magnetic fields
across the shock. The parallel magnetic field (parallel to the
shock front) $B_{||}$ is compressed and amplified:
\begin{equation}
B_{||2}  = \Gamma B_{||1}
\end{equation}
The perpendicular magnetic field $B_\bot$ remains unchanged.

The energy distribution of the (relativistic) electrons and the
magnetic field behind the shock are needed to calculate  the
Synchrotron spectrum.  In principle these parameters should be
determined from the microscopic physical processes that take place
in the shocks. However, it is difficult to estimate them from
first principles. Instead I define two dimensionless parameters,
$\epsilon_B$ and $\epsilon_e$, that incorporate our ignorance and
uncertainties \cite{PacRho93,Pi94,SNP96}. It is commonly assumed
that these energies are a constant fraction of the internal energy
behind the shock (see however, \textcite{DaigneMochkovitch03}). I
denote by $\epsilon_e$ and by $\epsilon_B$ the ratio between these
energies and the total internal energy:
\begin{eqnarray}
e_e \equiv \epsilon_e e = 4 \Gamma^2_{sh} \epsilon_e n_1 m_p c^2 \\
\nonumber e_B = B^2 /8 \pi \equiv \epsilon_B e = 4 \Gamma^2_{sh}
\epsilon_B n_1 m_p c^2 \label{epsilons}
\end{eqnarray}

One usually assumes that these factors, $\epsilon_{e,B}$, are
constant through out the burst evolution. One may even expect
that they should be constant from one burst to another (as they
reflect similar underlying physical processes). However, it seems
that a simple model that assumes that these parameters are
constant during the prompt burst cannot reproduce the observed
spectrum \cite{DaigneMochkovitch03}. This leads to explorations
of models in which the equipartition parameters $\epsilon_{e,B}$
depend on the physical conditions within the matter.

In GRBs, as well as in SNRs the shocks are  collisionless. The
densities are so small so that mean free path of the particles for
collisions is larger than the typical size of the system. However,
one expects that ordered or random magnetic fields or
alternatively plasma waves will replace in these shocks the role
of particle collisions. One can generally use in these cases the
Larmour radius as a typical shock width. A remarkable feature of
the above shock jump conditions is that as they arise from general
conservation laws they are independent of the detailed conditions
within the shocks and hence are expected to hold within
collisionless shocks as well. See however \cite{Mitra96} for a
discussion of the conditions for collisionless shocks in GRBs.


\subsection{Particle Acceleration }
\label{sec:acc}

It is now generally accepted that Cosmic rays (more specifically
the lower energy  component below $10^{15}$eV) are accelerated
within shocks in SNRs is the Galaxy (see e.g.
\textcite{Gaisser91}). A beautiful demonstration of this effect
arises in the observation of synchrotron emission from Supernova
remnants, which shows \xr emission from these accelerated
particles within the shocks.

The common model for  particle shock acceleration is the Diffuse
Shock Acceleration (DSA) model. According to this model the
particles are accelerated when they repeatedly cross a shock.
Magnetic field irregularities keep scattering the particles back
so that they keep crossing the same shock. The competition
\cite{Fermi49} between the average energy gain, $E_f/E_i$ per
shock crossing cycle (upstream-downstream and back) and the escape
probability per cycle, $P_{esc}$ leads to a power-law spectrum
$N(E) dE \propto E^{-p} dE $ with
\begin{equation}
p =  1+ ln[1/(1-P_{esc})]/ln[\langle E_f /E_i \rangle].
\label{acce}
\end{equation}
Note that within the particle acceleration literature this index
$p$ is usually denoted as $s$.  Our notation follows the
 common notation within the GRB literature.

\textcite{Blandford_Eichler87} review the theory of DSA in
non-relativistic shocks. However, in GRBs the shocks are
relativistic (mildly relativistic in internal shocks and extremely
relativistic in external shocks).  Acceleration in ultra
relativistic shocks have been discussed by several groups
\cite{Heavens_Drury88,Bednarz98,Gallant99,Achterberg_etal01,Kirk00,Vietri02}
In relativistic shocks the considerations are quite different from
those in non-relativistic ones. Using the relativistic shock jump
conditions (Eq. \ref{jump} and kinematic considerations one can
find (see \textcite{Vietri95,Gallant_99,Achterberg_etal01}) that
the energy gain in the first shock crossing is of the order
$\Ga^2_{sh}$. However, subsequent shock crossing are not as
efficient and the energy gain is of order unity $\langle E_f /E_i
\rangle \approx 2$ \cite{Gallant_99,Achterberg_etal01}.

The deflection process in the upstream region is due to a large
scale smooth background magnetic field perturbed by MHD
fluctuations. A tiny change of the particle's momentum in the
upstream region is sufficient for the shock to overtake the
particle. Within the downstream region the momentum change should
have a large angle before the particle overtakes the shock and
reaches the upstream region. As the shock moves with a
sub-relativistic velocity ($\approx c/\sqrt 3$) relative to this
frame it is easy for a relativistic particle to overtake the
shock.  A finite fraction of the particles reach the upstream
region. Repeated cycles of this type (in each one the particles
gain a factor of $\sim 2$ in energy) lead to a power-law spectrum
with $p \approx 2.2-2.3$ (for $\Gamma_{sh} \gg 1$). Like in
non-relativistic shock this result it fairly robust and it does
not depend on specific assumptions on the scattering process. It
was obtained by several groups using different approaches,
including both numerical simulations and analytic considerations.
The insensitivity of this result arises, naturally from the
logarithmic dependence in equation \ref{acce} and from the fact
that both the denominator and the numerator are of order unity.
This result agrees nicely with what was inferred from GRB spectrum
\cite{Sari_Piran_97MNRAS} or with the afterglow spectrum
\cite{PanaitescuK01}. \textcite{Ostrowski02} point out, however,
that this result requires highly turbulent conditions downstream
of the shock. If the turbulence is weaker the resulting energy
spectrum could be much steeper. Additionally as internal shocks
are only mildly relativistic the conditions in these shocks might
be different.

The maximal energy that the shock accelerated particles can be
obtained by comparing the age of the shock $R/c$ (in the upstream
frame) with the duration of an  acceleration cycle. For a simple
magnetic deflection, this later time is just half of the Larmour
time, $ E/Z q_{e}  B$ (in the same frame). The combination yields:
\begin{equation}
E_{max} \approx Z q_{e} B  R =  10^{20} {\rm eV} B_3 R_{15} \ ,
\label{emax_acc}
\end{equation}
where the values that I have used in the last equality reflect
the conditions within the reverse external shocks where UHECRs
(Ultra High Energy Cosmic Rays) can be accelerated (see \S
\ref{sec:UHECRs} below). For particle diffusion in a random
upstream field (with a diffusion length $l$) one finds that $R$
in the above equation is replaced by $\sqrt{R l /3}$.

The acceleration process has to compete with radiation losses of
the accelerated particles. Synchrotron losses are inevitable as
they occur within the same magnetic field that is essential for
deflecting the particles. Comparing the energy loss rate with the
energy gain one obtain a maximal energy of:
\begin{equation}
E_{max} \approx m c^2 \left( {4 \pi q_{e} \Ga_{sh} \over \sigma_T
B } \right )^{1/2} \approx 5 \cdot 10^{17} {\rm eV} (m/m_p)
\G_{100}^{1/2} B^{-1/2} \label{Emax_syn} .
\end{equation}
The corresponding Lorentz factor is of the order of $10^8$ for
$\Ga_{sh}=100$ and $B=1$ Gauss. Note that this formula assumes
that the acceleration time is the Larmour time and hence that the
synchrotron cooling time is equal to the Larmour time. Obviously
it should be modified by a numerical factor which is mostly likely
of order unity.

\subsection{Synchrotron }
\label{sec:sync}

Synchrotron radiation play,  most likely, an important role in
both the GRB and its afterglow. An important feature of
synchrotron emission is its polarization (see \S
\ref{sec:pol_theory}). Observations of polarization in GRB
afterglows and in one case in the prompt emission support the idea
that synchrotron emission is indeed taking place there (note
however that IC also produces polarized emission). I review here
the basic features of synchrotron emission focusing on aspects
relevant to GRBs. I refer the reader to
\textcite{Rybicki_Lightman79} for a more detailed discussion.

\subsubsection{Frequency and Power}

The typical energy of synchrotron photons as well as the
synchrotron cooling time depend on the Lorentz factor $\gamma_e$
of the relativistic electron under consideration and on the
strength of the magnetic field . If the emitting material moves
with a Lorentz factor $\Gamma$ the photons are blue shifted. The
characteristic photon energy in the observer frame is given by:
\begin{equation}
\label{syn_obs} (h\nu_{syn})_{obs}=\frac{\hbar q_eB}{m_ec}\gamma
_e^2\Ga ,
\end{equation}
where $q_e$ is the electron's charge.

The power emitted, in the local frame, by a single electron due
to synchrotron radiation is:
\begin{equation}
P_{syn} = \frac 4 3 \sigma_T c U_B \gamma_e^2 \ ,
\label{syn_power}
\end{equation}
where $U_B \equiv B^2/8 \pi \equiv \epsilon_B e$ is the magnetic
energy density and $\sigma_T$ is the Thompson cross section. The
cooling time of the electron in the fluid frame is then $\gamma_e
m_e c^2/P$. The observed cooling time $t_{syn}$ is shorter by a
factor of $\Ga$:
\begin{equation}
\label{cooling} t_{syn}(\gamma_e) =\frac{3 m_e c}{4\sigma _T
U_B\gamma_e\Ga}.
\end{equation}

Substituting the value of $ \gamma_e$ from equation \ref{syn_obs}
into the cooling rate Eq. \ref{cooling} one obtains the cooling
time scale as a function of the observed photon energy:
\begin{equation}
\label{tausyn2} t_{syn} (\nu) = \frac{3}{\sigma_T}  \sqrt{\frac{ 2
\pi c m_e q_e} {B^{3} \Ga}} \nu^{-1/2}
\end{equation}

Since $\ga_e$ does not appear explicitly in this equation
$t_{syn}$ at a given observed frequency is independent of the
electrons' energy distribution within the shock. This is
provided, of course, that there are electrons with the required
$\ga_e$ so that there will be emission in the frequency
considered.  As long as there is such an electron the cooling
time is ``universal''.   This equation shows a characteristic
scaling of $t_{syn} (\nu) \propto \nu^{-1/2}$. This is not very
different from the observed relation $\delta T \propto
\nu^{-0.4}$ \cite{Fenimore95}. However, it is unlikely that
cooling and not a physical process determines the temporal
profile.

The cooling time calculated above sets a lower limit to the
variability time scale of a GRB since the burst cannot possibly
contain spikes that are shorter than its cooling time.
Observations of GRBs typically show asymmetric spikes in the
intensity variation, where a peak generally has a fast rise and a
slower decay.  A plausible explanation of this observation is that
the shock heating of the electrons happens rapidly (though
episodically), and that the rise time of a spike is related to
the heating time.  The decay time is then set by the cooling, so
that the widths of spikes directly measure the cooling time.
However, it seems that there are problems with this simple
explanation. First when plugging reasonable parameters one finds
that the decay time as implied by this equation is too short.
Second, if the cooling time is long the shocked region would
suffer adiabatic losses and this would reduce the efficiency of
the process. Thus it is unlikely that the pulse shapes can be
explained by Synchrotron physics alone.

\subsubsection{The Optically thin Synchrotron Spectrum }
\label{sec:synch-spec}

The instantaneous synchrotron spectrum of a single relativistic
electron with an initial energy $\ga_e m_e c^2$ is approximately
a power law with $F_\nu \propto \nu^{1/3}$ up to
$\nu_{syn}(\ga_e)$ and an exponential decay above it.  The peak
power occurs at $\nu_{syn}(\gamma_e)$, where it has the
approximate value
\begin{equation}
\label{flux}
P_{\nu,max}\approx\frac{P(\gamma_e)}{\nu_{syn}(\gamma_e)}= \frac
{m_e c^2 \sigma_T} {3 q_e} \Gamma B .
\end{equation}
Note that $P_{\nu,max}$ does not depend on $\gamma_e$, whereas the
position of the peak does.

If the electron is energetic it will cool rapidly until it will
reach $\ga_{e,c}$,   the Lorentz factor of an electron that cools
on a hydrodynamic time scale.  For a rapidly cooling electron we
have to consider the time integrated spectrum. For an initial
Lorentz factor $\ga_e$: $F_\nu \propto \nu^{-1/2}$ for
$\nu_{syn}(\ga_{e,c}) < \nu < \nu_{syn}(\ga_e)$.

To calculate the overall spectrum due to the electrons one needs
to integrate over the electron's Lorentz factor distribution. I
consider first, following  \cite{SPN98a}, a power-law
distribution  a power index $p $ and a minimal Lorentz factor
$\ga_{e,min}$.  This is, of course, the simplest distribution and
as discussed  in \S \ref{sec:acc}  this is the expected
distribution of shock accelerated particles:
\begin{equation}
N(\gamma_e )\sim \gamma_e^{-p}\ \ \ {\rm for}\ \gamma_e
>\gamma _{e,min}\;.
\label{e_distribution}
\end{equation}
The condition $p>2$ is required so that the energy does not
diverge at large $\gamma_e$ (\textcite{DaiCheng01,Bhattacharya01}
consider also distributions with $2>p>1$ with a maximal energy
cutoff, see below). The minimum Lorentz factor, $\gamma _{e,min}$,
of the distribution is related to the electron's energy density
$e_e$ and the electron's number density $n_e$ as:
\begin{equation}
\gamma_{e,min}=  {p-2\over p-1}{e_e \over n_e m_e c^2}=
 {p-2\over p-1}
\langle\gamma_e\rangle. \label{gemin}
\end{equation}
The minimal Lorentz factor plays an important role as it
characterizes the `typical' electron's Lorentz factor and the
corresponding `typical' synchrotron frequency, $\nu_m \equiv
\nu_{syn}(\ga_{e,min})$. Interestingly the upper energy cutoff
(which essentially exists somewhere) does not play a critical
role in the spectrum for $p>2$. Of course it will lead to a high
frequency cutoff of the spectrum around $\nu_{syn}$ that
corresponds to this energy. However, quite generally, this happens
at the high energy tail far from where the peak flux or the peak
energy are emitted.

A simple modification of the above idea arises if only a fraction,
$\xi_e$, of the electrons is accelerated to high energies and the
rest of the electrons remain cold
\cite{BykovMeszaros96,Guetta_Spada_Waxman01}. If a small fraction
of electrons shares the energy $e_e$ then the typical Lorentz
factor would be $\xi_e^{-1} \ga_{e,min}$, where $\ga_{e,min}$ is
calculated from Eq. \ref{gemin} above. All the corresponding
places where $\ga_{e,min}$ is used should be modified according to
this factor. At the same time fewer electrons will be radiating.
This will introduce a factor $\xi_e$ that  should multiply the
total emitted flux.  In the following discussion I will not add
this factors into the analysis. Similarly in situations when
multiple pair are formed \cite{Ghisellini_Celotti99} the
electron's energy is shared by a larger number of electron. In
this case $\xi_e$ is larger than unity and  similar modifications
of the spectrum applies.

The lowest part of the spectrum (strictly speaking the lowest part
of the optically thin spectrum, as at very low frequencies self
absorption sets in, see \S \ref{Sec:self-abs} below) is always the
sum of the contributions of the tails of all the electron's
emission: $F_\nu \propto \nu^{1/3}$. This is typical to
synchrotron \cite{Meszaros-Rees93,Katz94,CohenKatzP98} and is
independent of the exact shape of the electron's distribution.
\textcite{Tavani96a,Tavani96b}, for example obtain such a low
energy spectrum both for a Gaussian or for a Gaussian and a high
energy power-law tail. The observation of bursts (about 1/5 of the
bursts) with steeper spectrum at the lower energy part, i.e. below
the ``synchrotron line of death" \cite{PreeceEtal98,Preece02} is
one of the problems that this model faces. The problem is even
more severe as in order that the  GRB will be radiating
efficiently, otherwise the efficiency will be very low, it must be
in the fast cooling regime and  the relevant low energy spectrum
will be $\propto \nu^{-1/2}$
\cite{CohenKatzP98,GhiselliniCelottiLazzati00}. However, as
stressed earlier (see \S ref{sec:spec-obs}) this problem is not
seen in any of the HETE spectrum whose low energy tail is always
in the proper synchrotron range with a slope \cite{BarraudEtal03}
and it might be an artifact of the low energy resolution of BATSE
in this energy range \cite{CohenKatzP98}.

On the other hand the most energetic electrons will always be
cooling rapidly (independently of the behavior of the ``typical
electron''). These electrons emit practically all their energy
$m_e c^2 \gamma$, at their synchrotron frequency. The number of
electrons with Lorentz factors $\sim\gamma$ is
$\propto\gamma^{1-p}$ and their energy $\propto\gamma^{2-p}$. As
these electrons cool, they deposit most of their energy into a
frequency range $\sim\nu_{syn}(\gamma)\propto\gamma^2$ and
therefore $F_{\nu}\propto\gamma^{-p}\propto\nu^{-p/2}$. Thus the
uppermost part of the spectrum will satisfy:
\begin{equation}
F_\nu = N[\gamma(\nu)] m_e c^2 \gamma(\nu) d\gamma /d\nu \propto
\nu^{-p/2}. \label{fastfnu}
\end{equation}

In the intermediate frequency region the spectrum differs between
a `slow cooling' if the `typical' electrons with $\ga_{e,min}$ do
not cool on a hydrodynamic time scale  and `fast cooling' if they
do. The critical parameter that determines if the electrons are
cooling fast or slow is  $\ga_{e,c}$, the Lorentz factor of an
electron that cools on a hydrodynamic time scale. To estimate
$\ga_{e,c}$ compare $t_{syn}$ (Eq. \ref{cooling}) with $t_{hyd}$,
the hydrodynamic time scale (in the observer's rest frame):
\begin{equation}
\ga_{e,c} = {{3m_{e} c }\over{4 \sigma _{T}  U_{B}\Ga  t_{hyd}}}
\label{ga_c}
\end{equation}
For fast cooling   $\ga_{e,min}< \ga_{e,c}$, while $\ga_{e,min} >
\ga_{e,c}$ for slow cooling. In the following discussion two
important frequencies play a dominant role:
\begin{eqnarray}
 \nu_{m} \equiv \nu_{syn}(\ga_{e,min}) \ ; \\ \nonumber
 \nu_{c}  \equiv \nu_{syn}(\ga_{e,c}) \ . \label{numc}
\end{eqnarray}
These are the synchrotron frequencies of electrons with
$\ga{e,min}$ and with $\ga_{e,c}$.

 {\bf Fast cooling   ($\ga_{e,c} < \ga_{e,min}$):}
The typical electron is cooling rapidly hence  $\nu_c < \nu_m$.
The low frequency spectrum  $F_\nu \propto \nu^{1/3}$ extends up
to $\nu_c$. In the intermediate range between, $\nu_c$ and
$\nu_m$, we observe the energy of all the cooling electrons. The
energy of an electron $\propto\gamma$, and its typical frequency
$\propto\gamma^2$ the flux per unit frequency is
$\propto\gamma^{-1}\propto \nu^{-1/2}$. Overall the observed
flux, $F_\nu$, is given by:
\begin{equation}\label{spectrumfast}
F_\nu \propto \cases{ ( \nu / \nu_c )^{1/3} F_{\nu,max}, &
$\nu<\nu_c$, \cr ( \nu / \nu_c )^{-1/2} F_{\nu,max}, &
$\nu_c<\nu<\nu_m$, \cr ( \nu_m / \nu_c )^{-1/2} ( \nu /
\nu_m)^{-p/2} F_{\nu,max}, & $\nu_m<\nu$, \cr }
\end{equation}
where $\nu_m \equiv \nu_{syn}(\gamma_{e,min}), \nu_{c} \equiv
\nu_{syn}(\gamma_{e,c})$ and $F_{\nu,max}$ is the observed peak
flux. The peak flux is at $\nu_c$ $F_{\nu,max}\equiv N_e
P_{\nu,max}/4\pi D^2$ (where D is the distance to the source and I
ignore cosmological corrections).  The power emitted is simply the
power given to the electrons, that is $\epsilon_e $ times the
power generated by the shock, $dE/dt$:
\begin{equation}
P_{fast} = \epsilon_e {dE \over dt} . \label{Pfast-cool}
\end{equation}
The peak energy emitted (which corresponds to the peak of $\nu
F_\nu$) is at $\nu_m$. The resulting spectrum is shown in Fig.
\ref{fig:full_spectrum}.


{\bf Slow cooling  ($\ga_{e,c} > \ga_{e,min}$):} Now only the high
energy tail of the distribution (those electrons above
$\ga_{e,c}$) cools efficiently. The electrons with
$\gamma_e\sim\gamma_{e,min}$, which form the bulk of the
population, do not cool. Now  $f_\nu \propto \nu^{1/3}$ up to
$\nu_m$, and $F_\nu \propto \nu^{-p/2}$ above $\nu_c$. In the
intermediate region between these two frequencies:
\begin{equation}
F_\nu = N[(\gamma(\nu)] P[( \gamma(\nu)] d\gamma /d\nu \propto
\nu^{-(p-1)/2} ,
\end{equation}
where $\gamma(\nu)$ is the Lorentz factor for which the
synchrotron frequency equals $\nu$, $N[\ga]$ is the number of
electrons with a Lorentz factor $\ga$ and $P[\ga]$ the power
emitted by an electron with $\ga$. Overall one finds:
\begin{equation}
\label{spectrumslow} F_\nu \propto \cases{  (\nu/\nu_m)^{1/3}
F_{\nu,max},
            & $\nu<\nu_m$, \cr
(\nu/\nu_m)^{-(p-1)/2} F_{\nu,max},
            & $\nu_m<\nu<\nu_c$, \cr
\left( \nu_c/\nu_m \right)^{-(p-1)/2} \left( \nu/\nu_c
\right)^{-p/2} F_{\nu,max},
            & $\nu_c<\nu$. \cr
}
\end{equation}
The peak flux is at $\nu_m$ while the peak energy emitted is at
$\nu_c$. The emitted power is determined by the ability of the
electrons to radiate their energy:
\begin{equation}
P_{slow} = N_e P_{syn} (\ga_{e,min}) \label{Pslow-cool}
\end{equation}
where, $N_e$ is the number of electrons in the emitting region and
$P_{syn} (\ga_{e,min}) $, the synchrotron power of an  electron
with $\ga_{e,min}$, is given by Eq. \ref{syn_power}.

Typical spectra corresponding to fast and slow cooling are shown
in Fig. \ref{fig:full_spectrum}. The light curve depends on the
hydrodynamic evolution, which in turn determines the time
dependence of $\nu_m,\nu_c$ and $F_{\nu,max}$. The spectra
presented here are composed of broken power laws.
\textcite{GranotSari02} present a more accurate spectra in which
the asymptotic power law segments are connected by smooth curves.
They fit the transitions by $ [(\nu/\nu_b)^{-n\beta_1}
+(\nu/\nu_b)^{-n \beta_2}]^{-1/n}$. The parameter $n$ estimates
the smoothness of the transition with $n \approx 1$ for all
transitions.

Fast cooling must take place during  the GRB itself: the
relativistic shocks  must emit their energy effectively -
otherwise there will be a serious inefficiency problem.
Additionally the pulse won't be  variable if the cooling time is
too long. The electrons must cool rapidly and release all their
energy. It is most likely that during the early stages of an
external shock (that is within the afterglow phase - provided that
it arises due to external shocks) there will be a transition from
fast to slow cooling
\cite{MR97,Waxman97a,MesReesWei97,Waxman97b,KP97}.


\textcite{Tavani96a,Tavani96b} discusses the synchrotron spectrum
from a Gaussian electron distribution and from a Gaussian electron
distribution with a high energy tail. As mentioned earlier the
Gaussian (thermal) distribution has a typical low frequency
$\nu^{1/3}$ spectrum. However, as expected,  there is a sharp
exponential cutoff at high frequencies. Without a high energy
tail this spectrum does not fit the observed GRB spectra of most
GRBs (see \S \ref{sec:spec-obs}). Note, however, that it may fit a
small subgroup with a NHE \cite{Pendleton_NHE97}.  With an
electron distribution composed of a Gaussian and an added high
energy tail the resulting spectra has the typical $\nu^{1/3}$
component and an additional high energy tail which depends on the
electrons power law index.
 Such a spectra fits several observed GRB
spectra \cite{Tavani96a,Tavani96b}.

Another variant is the synchrotron spectrum from a power-law
electron distribution with $1<p<2$
\cite{Bhattacharya01,DaiCheng01}. In this case there must be a
high energy cutoff $\gamma_{e,max}$ and the `typical' electron's
energy corresponds to this upper cutoff. A possible cutoff can
arise from Synchrotron losses at the energy where the
acceleration time equals to the energy loss time (see e.g.
\textcite{deJagerEtal96} and the discussion in \S \ref{sec:acc}):
\begin{equation}
\gamma_{e,Max} \approx 4 \times 10^7 B^{-1/2} \ .
\end{equation}
The resulting  ``typical" Lorentz factor $\gamma_{e,min}$ differs
now from the one given by Eq. \ref{gemin}.
\textcite{DaiCheng01,Bhattacharya01} find that it is replaced
with:
\begin{equation}
\gamma_{e,min}=
\left[\left(\frac{2-p}{p-1}\right)\left(\frac{m_p}{m_e}\right)\epsilon_e
\Gamma\gamma_{e,Max}^{p-2}\right]^{1/(p-1)} \ . \label{gemin1}
\end{equation}
The resulting spectrum is now similar to the one obtained for
fast or slow cooling with the new critical frequencies $\nu_m$
given by plugging the result of Eq. \ref{gemin1} into Eq.
\ref{numc}.

\subsubsection{Synchrotron Self-Absorption }
\label{Sec:self-abs}

At low frequencies synchrotron self-absorption may take place. It
leads to  a steep cutoff of the low energy spectrum, either as the
commonly known $\nu^{5/2}$ or as $\nu^2$. To estimate the self
absorption frequency one needs the optical depth along the line of
sight. A simple approximation is: $\alpha'_{\nu'}R/\Gamma$ where
$\alpha'_{\nu'}$ is the absorption coefficient
\cite{Rybicki_Lightman79}:
\begin{equation}
\label{alpha_nu} \alpha'_{\nu'} = {(p+2) \over 8 \pi m_e
\nu'^2}\int^{\infty}_{\gamma_{min}} d\gamma_e
P'_{\nu',e}(\gamma_e){n(\gamma_e) \over \gamma_e} \  \  .
\end{equation}
The self absorption frequency $\nu_a$ satisfies:
$\alpha'_{\nu'_0} R/\Gamma=1$. It can be estimates only once we
have a model for the hydrodynamics and how do $R$ and $\gamma$
vary with time \cite{Wijers_Galama98,GPS99b}.

The spectrum below the the self-absorption frequency depends on
the electron distribution. One obtains the well known
\cite{Rybicki_Lightman79}, $\nu^{5/2}$  when the synchrotron
frequency of the electron emitting the self absorbed radiation is
inside the self absorption range. One obtains  $\nu^2$  if the
radiation within the  self-absorption frequency range is due to
the low energy tail of electrons that are radiating effectively at
higher energies. For this latter case, which is more appropriate
for GRB afterglow (for slow cooling with $\nu_m < \nu_c$)
\cite{PacRho93,Meszaros-Rees93,Katz94,KP97}:
\begin{equation}
F_\nu \propto \nu^2  [k_B T_e / (\Gamma m_p c^2)] {R^2},
\label{sa_spec}
\end{equation}
where $R$ is the radius of the radiating shell and the factor
$k_B T_e / (\Gamma m_p c^2)$ describes the degree of electron
equipartition in the plasma shock-heated to an internal energy per
particle $m_p c^2$ and moving with Lorentz factor $\gamma$.

The situation is slightly different for a shock heated fast
cooling i.e. if $\nu_c < \nu_m$ \cite{GPS00}. In this case we
expect the electron's distribution to be inhomogeneous, as
electrons near the shock did not cool yet but electrons further
downstream are cool. This leads to a new spectral range $\nu_{sa}
< \nu < \nu_{sa'}$ with $F_\nu \propto \nu^{11/8}$ (see Fig.
\ref{fig:full_spectrum}).

Synchrotron self-absorption is probably irrelevant during the GRB
itself. Note, however, that  under extreme conditions the self
absorption frequency might be in the low \xr and this may explain
the steep low energy spectra seen in some bursts. These extreme
conditions are needed in order to make the system optically thick
to synchrotron radiation but keeping it optically thin to Thompson
scattering and pair creation \cite{GPS00}. Self absorption appears
regularly  during the afterglow and is observed typically in radio
emission \cite{Katz94,Waxman97b,KP97,Wijers_Galama98,GPS99b}. The
expected fast cooling self-absorbed spectra may arise in the early
radio afterglow. So far it was not observed.

\subsection{Inverse Compton}
\label{sec:IC}

Inverse Compton (IC) scattering may modify our analysis in several
ways.  IC can influence the spectrum even if the system is
optically thin (as it must be) to Compton scattering (see e.g.
\textcite{Rybicki_Lightman79}). In view of the high energies
involved  a photon is  IC scattered only once.  After a single IC
scattering the photon's energy is so high that in the electron's
rest frame it is above the Klein-Nishina energy ($m_e c^2 \sim
0.5$Mev), and the decrease in the Compton cross section in this
energy range makes a second scattering unlikely. Note that in some
cases ({\it e.g.} in forward external shocks) even the first
scattering may suffer from this problem. The effect of IC depends
on the Comptonization parameter $Y=\gamma^2 \tau_e$. For fast
cooling one can  show \cite{SNP96} that  $Y$ satisfies:
\begin{eqnarray}
Y= {\epsilon _e/U _B}~~~     & \ \ {\rm  if }\ \ & U_e \ll U _B\\
\nonumber Y= \sqrt{U _e/U_B} & \ \ {\rm  if }\ \ & U_e \gg U _B ,
\nonumber
\end{eqnarray}
where $U_e$ and $U_B$ are the energy densities of the electron's
and of the magnetic field respectively.  IC is unimportant if
$Y<1$ and in this case it can be ignored.

If $Y>1$, which corresponds to $U_e > U_B$ (or to $\epsilon_e >
\epsilon_{B}$) and to $Y=\sqrt{U_e/U_B}$, then a large fraction of
the low energy synchrotron radiation will be up scattered by IC
and a large fraction of the energy will be emitted via the IC
processes. Those IC photons might  be too energetic, that is their
energy may be far beyond the observed energy range. In this case
IC will not influence the observed spectra directly. However, as
IC will take a significant fraction of the energy of the cooling
electrons it will influence the observations in two ways: it will
shorten the cooling time (the emitting electrons will be cooled by
both synchrotron and IC process).  Second, assuming that the
observed $\ga$-ray photons results from synchrotron emission, IC
will influence the overall energy budget and reduce the efficiency
of the production of the observed radiation. I turn now to each of
this cases.

An IC scattering boosts the energy of the photon by a factor
$\gamma^2_e$. Typical synchroton photon that have been scattered
once by  IC will be observed at the energy:
\begin{equation}
\label{IC_obs} (h\nu_{IC})_{obs}=\frac{\hbar q_eB}{m_ec}\gamma
_e^4 \Ga .
\end{equation}
The electrons are cooled both by synchrotron and by IC.  The
latter is more efficient and the cooling is enhanced by the
Compton parameter $Y$.  The cooling time scale is:
\begin{equation} t_{IC}={\frac{6 \pi c^{3/4}
\sqrt{U_B/U_e} \hbar^{1/4} m_e^{3/4}{q_e}^{1/4}} {B^{7/4}
 (h \nu)^{1/4} \Ga^{3/4} \sigma_T}}
\label{tIC}
\end{equation}

The conditions needed to produce the observed emission using IC
are probably not fulfilled in either external or internal shocks
(see however \textcite{Ghisellini_Celotti99} and the discussion in
\S \ref{sec:QIC} below). However even if IC does not produce the
observed $\ga$-ray photons it still influences the process if $Y>
1$. First it will add an ultra high energy component to the GRB
spectrum. This component will typically be at around $\ga_e^2$
times the observed $\sim 100$KeV photons, namely at the GeV-TeV
range (see e.g. \textcite{Vietri97,Bottcher_Dermer98} and the
discussion in  \S \ref{sec:TeV}). This component might have been
already observed in some GRBs during the early afterglow (see \S
\ref{sec:spec-obs}). Inverse Compton will also speed up the
cooling of the emitting regions and shorten the cooling time,
$t_{syn}$ estimated earlier (Eq. \ref{tausyn2}) by a factor of
$Y$. At the same time this also reduces the efficiency (for
producing the observed $\gamma$-rays) by the same factor.

\subsection{Quasi-Thermal Comptonization}
\label{sec:QIC}

\textcite{Ghisellini_Celotti99} suggested that  the prompt GRB
emission arises in a quasi-thermal Comptonization process.  In
their model the optical depth within the emitting region (of
internal shocks) is of order unity  leading to a copious pair
production. The system is optically thick to synchrotron emission.
The self-absorbed synchrotron emission is the seed for an Inverse
Compton emission produced by the pairs. The effective Compton
paramter in the new system, $\tilde Y$, is:
\begin{equation}
\tilde Y \equiv 4 \tau ({ kT' \over m_e c^2}) (1 + \tau ) [ 1  + 4
( {kT' \over m_e c^2})] \label{tildeY},
\end{equation}
where $T'$ is the effective temperature of the pair and $\tau$ is
the total cross section for scattering. The pairs act as a
thermostat controlling the effective temperature within the
emitting region to 30-300kev \cite{Svensson82,Svensson84}. The
resulting spectrum from this model is a flat spectrum $F_\nu
\propto \nu^0$ between the $h \nu_{sa} \Ga $ and $k T' \Ga$
\cite{Ghisellini_Celotti99}. The spectrum will evolve rapidly
during the burst while the pairs are being created and the
effective temperature decreases.

\subsection{Polarization from Relativistically Moving Sources}
\label{sec:pol_theory}

Polarization can provide information on both the emission process
and on the geometry of the emitting regions. Usually the observed
polarization is obtained by first integrating the Stokes
parameters of the radiation emitted by the individual electrons
over the electron's distribution. This yields the local
polarization. Then we integrate over the emitting region to obtain
the global polarization. In GRBs (both in the prompt emission and
in the afterglow) the emitting regions move relativistically
towards the observed. The implied Lorentz transformations play a
very important role in the second integration as they change the
direction of propagation of the photons and hence the direction of
the local polarization. The final results are sometimes surprising
and counter intuitive. For example even if the intrinsic (local)
emission is 100\% polarized in the same direction the integration
over the emitting region would reduce this to ~70\% polarization.
I consider polarization from synchrotron emission here, but the
results can be easily applied to IC as well.  I apply the results
derived in this section to the possible polarization from the
prompt emission and from the afterglow in the corresponding
sections \S \ref{sec:pol_prompt} and \ref{sec:pol_after}.

As an example I consider synchrotron emission. Synchrotron
emission is polarized with and the intrinsic local polarization
level depends on the spectral index of the energy distribution of
the emitting electrons, $p$, \cite{Rybicki_Lightman79}. For
typical values ($2<p<3$) it can reach 75\%. The polarization
vector is perpendicular to the magnetic field  and, of course, to
the direction of the emitted radiation. The formalism can be
easily adopted also to Inverse Compton for which the intrinsic
local polarization is higher and  could reach 100\% when the
photons are scattered at 90$^o$.

\begin{figure}[htb]
\begin{center}
\epsfig{file=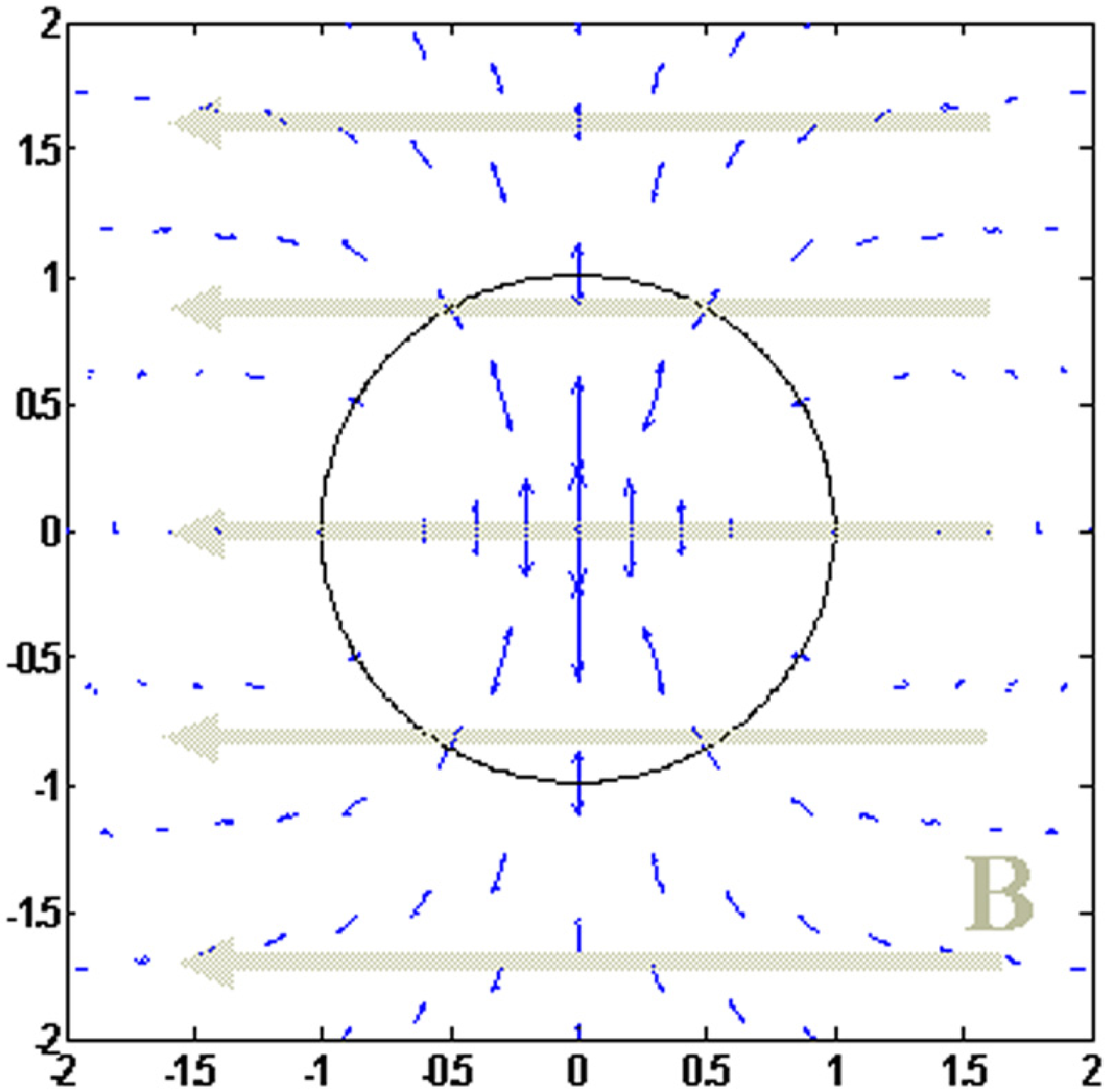,height=3in} \caption{ Polarization
from a uniform magentic field (following \cite{GranotKonigl03}).
The circle marks the and angle where the matter moves at an angle
$\Gamma^{-1}$ from the observer. Note that the polarization is
maximal along the line perpendicular to the uniform field but it
vanished in the other direction near the $\Gamma^{-1}$ circle.}
\label{fig:pol_uniform}
\end{center}
\end{figure}

Consider first a case where the magnetic field is uniform locally
(over a regions of angular size $\Gamma^{-1}$). This could happen,
for example, if we have an ordered magnetic field along the $\phi$
direction and the observer is more than $\Gamma^{-1}$ away from
the symmetry axis. This would be the case within internal shocks
if the magnetic field is dragged from the source or within several
Poynting flux dominated models. The locally emitted polarization
is uniform and is in the plane of the sky and perpendicular to the
direction of the magnetic field. In a Newtonian system it would
combine so that the observed polarization equals the emitted one.
However, the Lorentz transformations induce their own signature on
the observed polarization \cite{GranotKonigl03,Granot03}. This is
depicted in Fig. \ref{fig:pol_uniform}. It is clear from this
figure that the polarization vector varies along the observed
region (whose angular size is $1/\Gamma$. Consequently the
observed global polarization will be smaller than the local
polarization.

The observed stokes parameters are weighted averages of the local
stokes parameters at different regions of the shell. The
instantaneous polarization is calculated using the instantaneous
observed flux $ F_{\nu }(y,T)\propto (1+y)^{-(3+\alpha )} $, with
$\alpha$ the relevant spectral index at this segment,  as the
weights, where $ y \equiv (\Gamma \theta )^{2} $ and $ T $ is the
observer time. The time integrated polarization is calculated
using the fluences as weights: $ \int ^{\infty }_{0}F_{\nu
}(y,T)dT\propto (1+y)^{-(2+\alpha)} $.

The fluxes depend on how the intensity varies with the magnetic
field. For $ I_\nu \propto B^0 $, which is relevant for fast
cooling\footnote{see however \textcite{Granot03}.} (and the prompt
GRB), the time integrated stokes parameters (note that $ V=0 $ as
the polarization is linear) and polarization are given by:
\begin{equation} \label{Eq QU ordered} \frac{\left\{ \begin{array}{c}Q \\U \\
\end{array}\right\}}{I}=
\Pi _{synch} \frac{\int _{0}^{2\pi }\int _{0}^{\infty }(1+y)^{-(2+\alpha
)}\left\{ \begin{array}{c} \cos(2\theta _{p}) \\ \sin(2\theta _{p}) \\
\end{array}\right\}dyd\phi }{\int _{0}^{2\pi }\int _{0}^{\infty
}(1+y)
 ^{-(2+\alpha )}dyd\phi } ,
\end{equation}
and the relative polarization is given by
\begin{equation} \label{Eq Pi} \Pi
=\frac{\sqrt{U^{2}+Q^{2}}}{I},
\end{equation}
where $ \theta _{p}=\phi +\arctan(\frac{1-y}{1+y}\cot\phi ) $
\cite{GranotKonigl03} (see also \cite{LyutikovPB03}).   For $
\alpha =1 $ Eqs. \ref{Eq QU ordered}-\ref{Eq Pi} yield a
polarization level of $ \Pi /\Pi _{synch}\approx 60\% $. I.e. 60\%
of the maximal synchrotron polarization, or an overall
polarization of $\sim 45\%$.  Taking the exact values of $\alpha$
and the dependence of $I_\nu$ on $B$ for fast cooling and $p=2.5$
results in an overall polarization of $\sim 50\%$
\cite{GranotKonigl03,NakarPiranWaxman03}.

\begin{figure}[htb]
\begin{center}
\epsfig{file=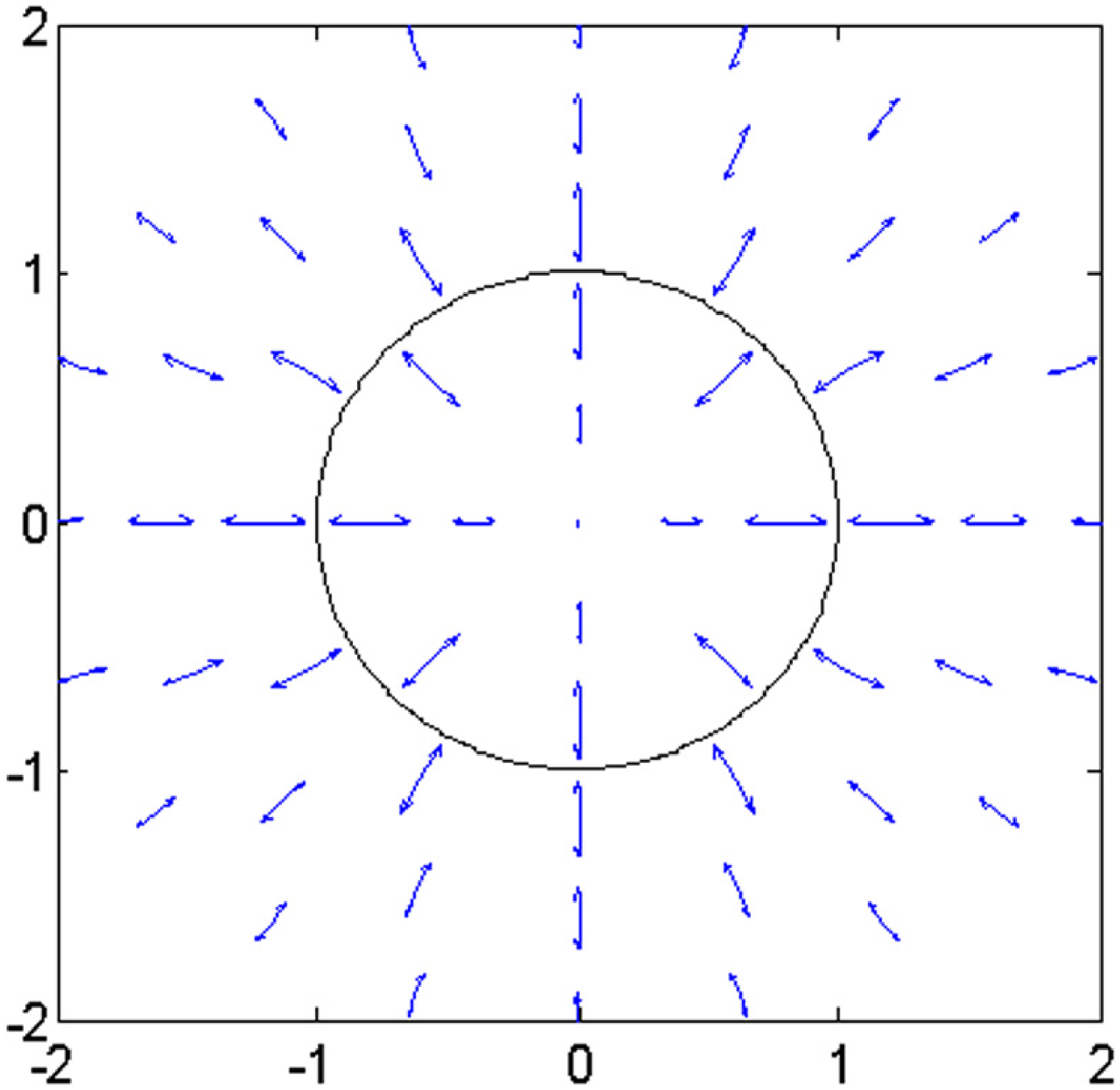,height=3in} \caption{Polarization
from a random magnetic field in the plane of the shock
\cite{NakarPiranWaxman03}. The solid circle marks the and angle
where the matter moves at an angle $\Gamma^{-1}$ from the
observer. Polarization can arise if we view a part of a jet
(dashed line), or if the emission is dominated by hot spot (dashed
- dotted regions) or if there is an overall gradient of the
emissivity as would arise in the standard jet model.}
 \label{fig:pol_random}
\end{center}
\end{figure}

It turns out that one can get a polarized emission even from
random magnetic field \textcite{GruzinovWaxman99} and
\textcite{MedvedevLoeb99}. This happens if the system has non
spherical geometry.  Consider a two dimensional random magnetic
field which is in the plane of the shock and assume that the
correlation length of this magnetic field is very short compared
to all other length scales in the system.  The  Lorentz
transformation induce in this case a radial polarization pattern
going out from the center (where the velocity of  the matter is
towards the observer and the polarization vanishes). This
polarization pattern is shown in Fig. \ref{fig:pol_random}. It is
clear that a simple integration over this pattern will lead to a
vanishing polarization.

However, a net polarization can arise in several cases if the
overall symmetry is broken. Polarization will arise if (see Fig.
\ref{fig:pol_random}):
\begin{itemize}
\item We observe a jet in an angle so that only a part of the jet
is within an angle of $\Gamma^{-1}$. \item If the emission is
nonuniform and there are stronger patches with angular size
smaller than $\Gamma^{-1}$from which most of the emission arise.
\item We observe a standard jet whose emission is angle dependent
and this dependence is of the order of $\Gamma^{-1}$.
\end{itemize}

\textcite{Gruzinov99,GhiselliniLazzati99,Sari99,Waxman03}
suggested that polarization  can arise from  a jet  even if the
magnetic field is random. \textcite{NakarPiranWaxman03} considered
a random magnetic field that remains planner in the plane of the
shock (for a three dimensional random magnetic field the
polarization essentially vanishes). For $I_\nu \propto B^{0} $ the
degree of observed polarization of the emission emitted from a
small region at angle $y$ is: $ \Pi (y)/\Pi _{synch}=min(y,1/y) $.
The overall time integrated stokes parameters are:
\begin{equation}
\label{Eq QU rand} \frac{\left\{ \begin{array}{c}Q \\U \\
\end{array}\right\}}{I}=\Pi _{synch}\frac{\int _{0}^{2\pi
}\int_{0}^{\infty}P'_{\nu',m}(1+y)^{-(2+\alpha)}\min(y,1/y)\left\{
\begin{array}{c}
\cos(2\phi) \\ \sin(2\phi)\\
\end{array}\right\}dyd\phi }{\int _{0}^{2\pi }\int_{0}^{\infty }P'_{\nu
',m}(1+y)^{-(2+\alpha )}dyd\phi},
\end{equation}
where $ P'_{\nu ',m}=P'_{\nu ',m}(y,\phi ) $ is the emitted power
at the synchrotron frequency in the fluid rest frame. For a
top-hat jet with sharp edges $ P'_{\nu ',m} $ is constant for any
$ y $ and $ \phi $ within the jet and zero otherwise. For a
structured jet $ P'_{\nu ',m} $ depends on the angle from the jet
axis.

The maximal polarization is observed when one sees the edge of the
jet. The probability to see the edge of a top-hat jet with sharp
edges and an opening angle $ \theta _{j} \Gamma \gg 1 $ is
negligible. On the other hand a jet with $ \theta _{j} \Gamma \ll
1 $ is not expected. Thus the only physical cases in which we can
expect a large polarization are $ 1 \lesssim \theta _{j}\Gamma
<{\textrm{a few }} $.
\begin{figure}
\resizebox*{0.45\columnwidth}{0.3\textheight}{\includegraphics{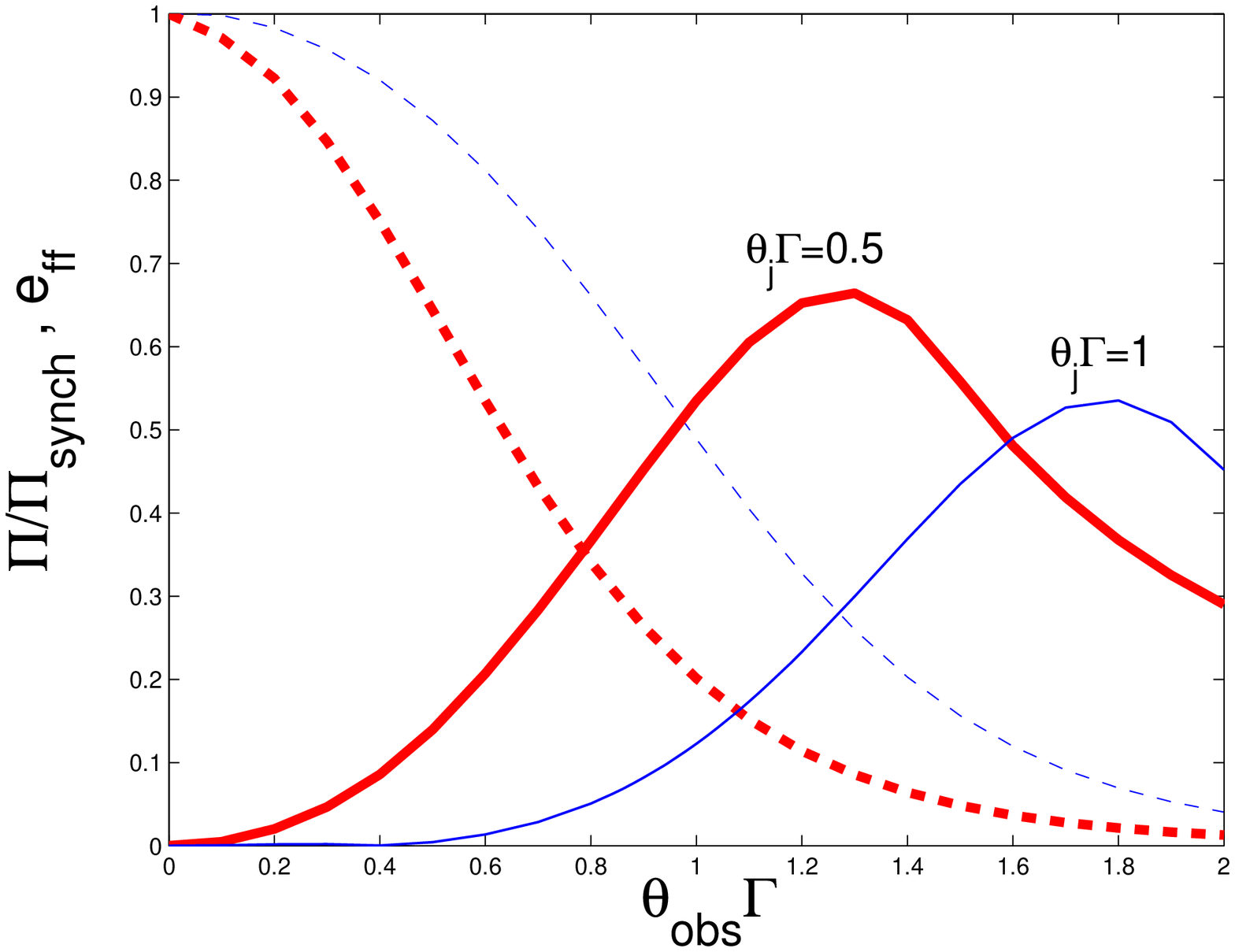}}
\resizebox*{0.45\columnwidth}{0.3\textheight}{\includegraphics{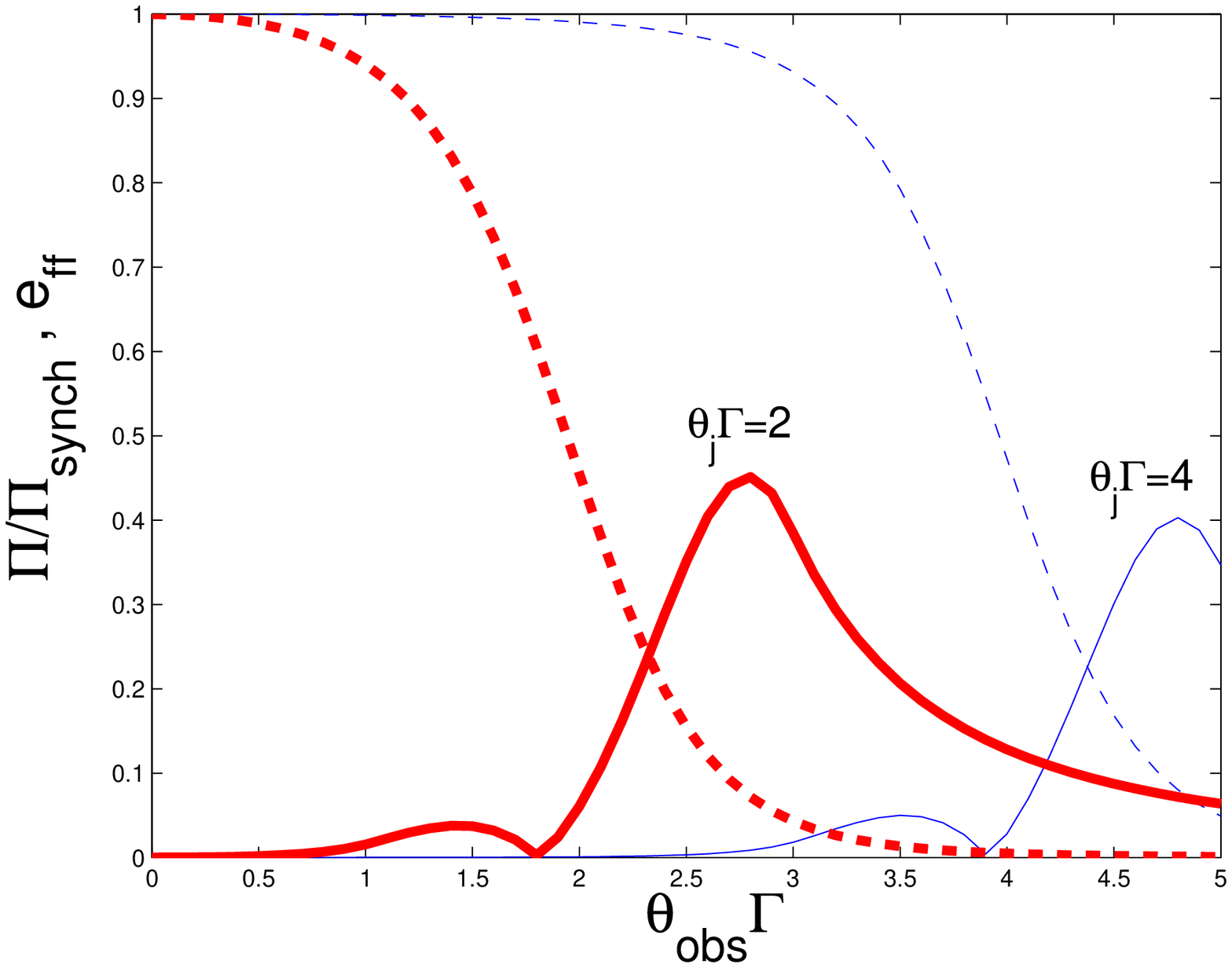}}
\caption{The time integrated polarization ({\it solid line}) and
the efficiency ({\it dashed line}) as a function of $\theta_{obs}
\Gamma$ for four different values of $\theta_j$ for a random
magnetic field.} \label{fig:pol_ran2}
\end{figure}

Fig. \ref{fig:pol_ran2} depicts the time integrated polarization
and the efficiency from sharp edged jets with different opening
angles as a function of the angle between the jet axis and the
line of sight, $ \theta _{obs} $. The efficiency, $ e_{ff} $ is
defined to be the ratio between the observed fluence at $
\theta_{obs} $ and the maximal possible observed fluence at $
\theta _{obs}=0 $. In all these cases the polarization is peaked
above 40\%, however the efficiency decrease sharply as the
polarization increase. Thus the probability to see high
polarization grows when $ \theta _{j} $ decrease. The probability
that $ \theta _{obs} $ is such that the polarization is larger
than $ 30\% $ ($\cdot \Pi_{synch}$) while $ e_{ff}>0.1 $ is 0.68,
0.41, 0.2 \& 0.08 for $ \theta _{j} \Gamma =0.5,1,2,4 $
respectively. In reality this probability will be smaller, as the
chance to observe a burst increases with its observed flux.

These later calculations also apply for IC emission
\cite{Lazzatietal03,DarDeRujula03}. However, in this case the
intrinsic local polarization is around 100\% and hence one can
reach a maximal polarization of $\sim 70$\%.

Polarization could also arise if the magnetic field is uniform
over random patches within a region of size $\Gamma^{-1}$. Here it
is difficult, of course to estimate the total polarization without
a detailed model of the structure of the jet
\cite{GruzinovWaxman99}.



\section{THE GRB AND THE PROMPT EMISSION}
\label{sec:PROMPT}

I turn now to discussion of the theory of the GRB and the prompt
emission. It is generally accepted that both the GRB and the
afterglow arise due to dissipation of the kinetic energy of the
relativistic flow. The relativistic motion can be dissipated by
either external \cite{MR92,RM92,Katz94} or internal shocks
\cite{NPP92,PaczynskiXu94,MR94b}. The first involve slowing down
by the external medium surrounding the burst. This would be the
analogue of a supernova remnant in which the ejecta is slowed
down by the surrounding ISM. Like in SNRs external shocks can
dissipate all the kinetic energy of the relativistic flow. On the
other hand internal shocks are shocks within the flow itself.
These take place when faster moving matter takes over a slower
moving shell.

\textcite{SP97} have shown that external shocks cannot produce
variable bursts (see also \textcite{Fenimoreetal96}). By variable
I mean here, following \cite{SP97} that  $\delta t \ll T $, where
$T$ is the overall duration of the burst (e.g. $T_{90}$) and
$\delta t$ is the duration of a typical pulse (see \S
\ref{sec:temp-obs}). As most GRBs are variable \textcite{SP97}
concluded that most GRBs are produced by internal shocks
\cite{MR94b}. Internal shocks can dissipate only a fraction of the
kinetic energy. Therefore, they must be accompanied by external
shocks that follow and dissipate the remaining energy. This leads
to the internal-external shocks scenario \cite{PiranSari98}. GRBs
are produced by internal shocks within a relativistic flow.
Subsequent external shocks between the flow and the circum-burst
medium produce a smooth long lasting emission - the afterglow.
Various observations (see \S \ref{sec:transition-obs}) support
this picture. I begin with the discussion with a comparison of
internal vs. external shocks. I review then the prompt emission
from internal shocks, then the prompt emission from external
shocks (which includes contributions to the late part of long GRBs
and the prompt optical flash). I also discuss the transition from
the observations of one shock to the other.

\subsection{Internal vs. External Shocks }
\label{sec:ex-int}
\subsubsection{General Considerations}
\begin{figure}[htb]
\begin{center}
\epsfig{file=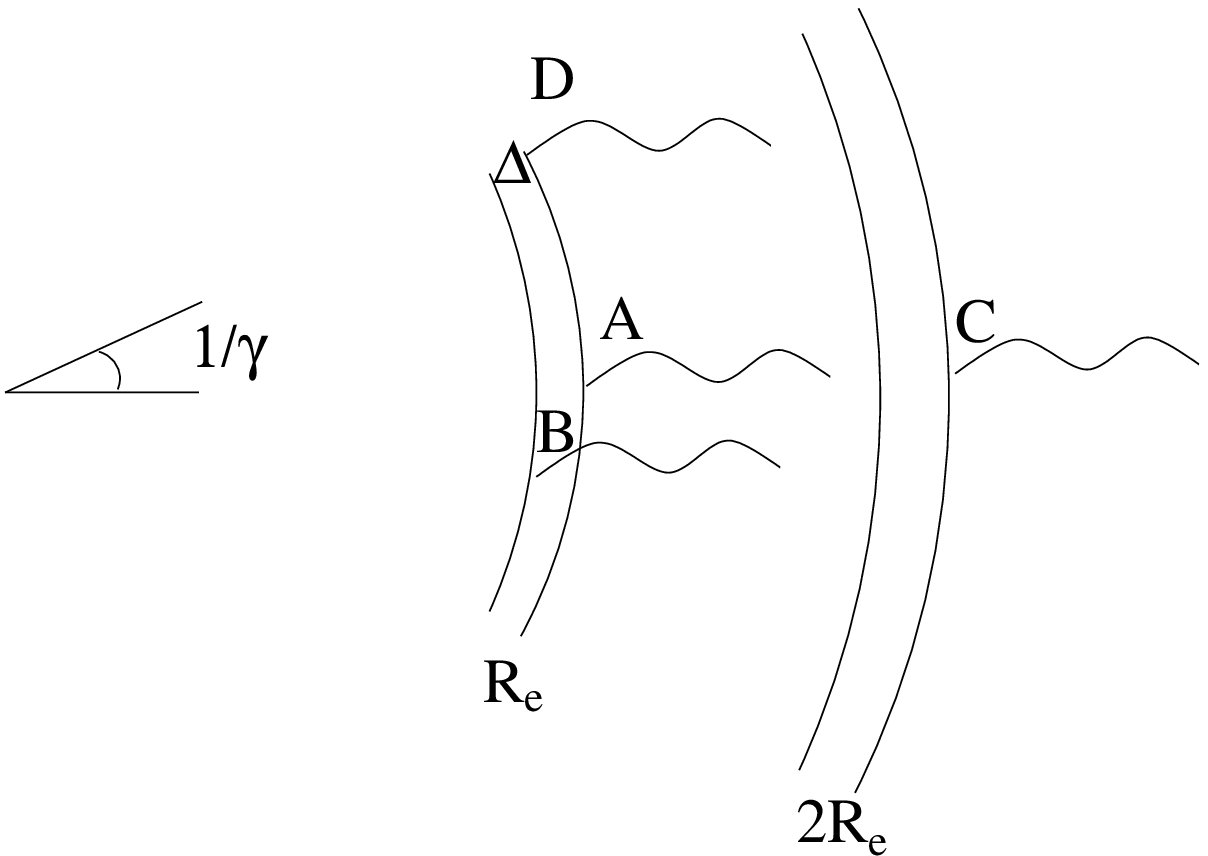,height=1.5in} \caption{Different time
scale from a relativistic expanding shell in terms of the arrival
times ($t_i$) of various photons: $t_{ang} = t_D-t_A$, $t_{R} =
t_C-t_A$ and $t_\Delta = t_B-t_A$.} \label{fig:times}
\end{center}
\end{figure}

Consider a ``quasi'' spherical relativistic emitting shell with a
radius $R$, a width $\Delta$  and a Lorentz factor $\Gamma$. This
can be a whole spherical shell or a spherical like section of a
jet whose opening angle $\theta$ is larger than $\Gamma^{-1}$.
Because of relativistic beaming an observer would observe
radiation only from a region of angular size $\sim \Gamma^{-1}$.
Consider now photons emitted at different points along the shock
(see Fig. \ref{fig:times}). Photons emitted by matter moving
directly towards the observer (point A in Fig. \ref{fig:times})
will arrive first. Photons emitted by matter moving at an angle
$\Gamma^{-1}$ (point D in Fig. \ref{fig:times}) would arrive after
$t_{ang} = R/2c\Gamma^2$. This is also, $t_{R}$, the time of
arrival of photons emitted by matter moving directly towards the
observer but emitted at $2R$ (point C in Fig. \ref{fig:times}).
Thus, $t_{R} \approx t_{ang}$ \cite{SP97,Fenimoreetal96}. This
coincidence is the first part of the argument that  rules out
external shocks in variable GRBs.

At a given point particles are continuously accelerated and emit
radiation as long as the shell with a width $\Delta$ is crossing
this point. The photons emitted at the front of this shell will
reach the observer a time  $t_\Delta = \Delta /c$ before those
emitted from the rear (point B in Fig. \ref{fig:times}). In fact
photons are emitted slightly longer as it takes some time for the
accelerated electrons to cool.  However, for most reasonable
parameters the cooling time is much shorter from the other time
scales \cite{SNP96} and I ignore it hereafter.

The emission from different angular points smoothes the signal on
a time scale $t_{ang}$.   If $t_\Delta \le t_{ang}\approx t_{R}$
the resulting burst will be smooth with a width $t_{ang}\approx
t_{R}$. The second part of this argument follows from the
hydrodynamics of external shocks. I show later in \S
\ref{sec:Ex-shocks} (see also \textcite{SP97}) that for external
shocks $\Delta/c \le R/c \Gamma^2 \approx t_{R} \approx t_{ang}$
and for a spreading shell $\Delta \approx R/c \Gamma^2$. Therefore
external shocks can produce only smooth bursts!

\begin{figure}[htb]
\begin{center}
\epsfig{file=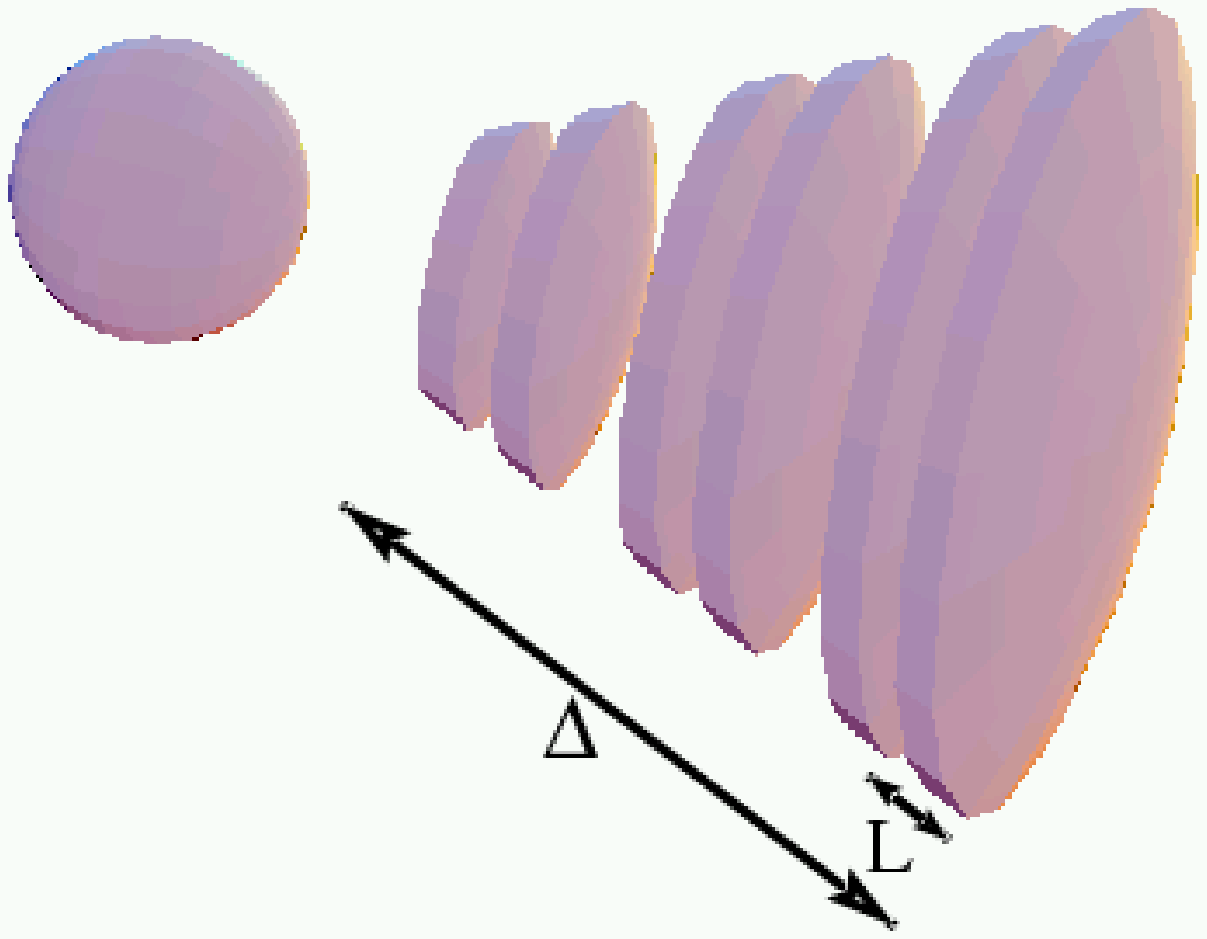,height=2.3in,width=3in} \caption{The
internal shocks model (from \cite{Sari_PhD}) Faster. shells
collide with slower ones and produce the observed $\gamma$ rays.
The variability time scale is $L/c$ while the total duration of
the burst is $\Delta/c$ .} \label{fig:internal_shocks}
\end{center}
\end{figure}

As we find only two time scales and as the emission is smoothed
over a time scale $t_{ang}$, a necessary condition for the
production of a variable light curve is that $t_\Delta = \Delta/c
> t_{ang}$. In this case $t_\Delta$ would be the duration of the burst
and $t_{ang}$ the variability time scale. This can be easily
satisfied within internal shocks (see Fig
\ref{fig:internal_shocks}). Consider an ``inner engine" emitting a
relativistic wind active over a time $t_\Delta =\Delta/c$
($\Delta$ is the overall width of the flow in the observer frame).
The source is variable on a scale $L /c$. Internal shocks will
take place at $R_s \approx L \Gamma^2$. At this place the angular
time and the radial time satisfy: $t_{ang} \approx t_{R} \approx
L/c $. Internal shocks continue as long as the source is active,
thus the overall observed duration $T = t_\Delta$ reflects the
time that the ``inner engine" is active. Note that now $t_{ang}
\approx L/c < t_\Delta$ is trivially satisfied. The observed
variability time scale in the light curve, $\delta t$,  reflects
the variability of the source $L/c$. While the overall duration of
the burst reflects the overall duration of the activity of the
``inner engine".

Numerical simulations  \cite{KPS97}  have shown that not only the
time scales are preserved but the source's temporal behavior is
reproduced on an almost  one to one basis in the observed light
curve. This can be explained now \cite{NakarPiran02c} by a simple
toy model (see \S \ref{sec:toy} below).

\subsubsection{Caveats and Complications}

Clearly the way to get around the previous argument is if $t_{ang}
< t_{R}$. In this case one can identify $t_{R}$ with the duration
of the burst and $t_{ang}$ as the variability time scale. The
observed variability would require in this case that:
$t_{ang}/t_{R} = \delta t /T$. For this the emitting regions must
be smaller than $R/\Gamma$.

\begin{figure}[htb]
\begin{center}
\epsfig{file=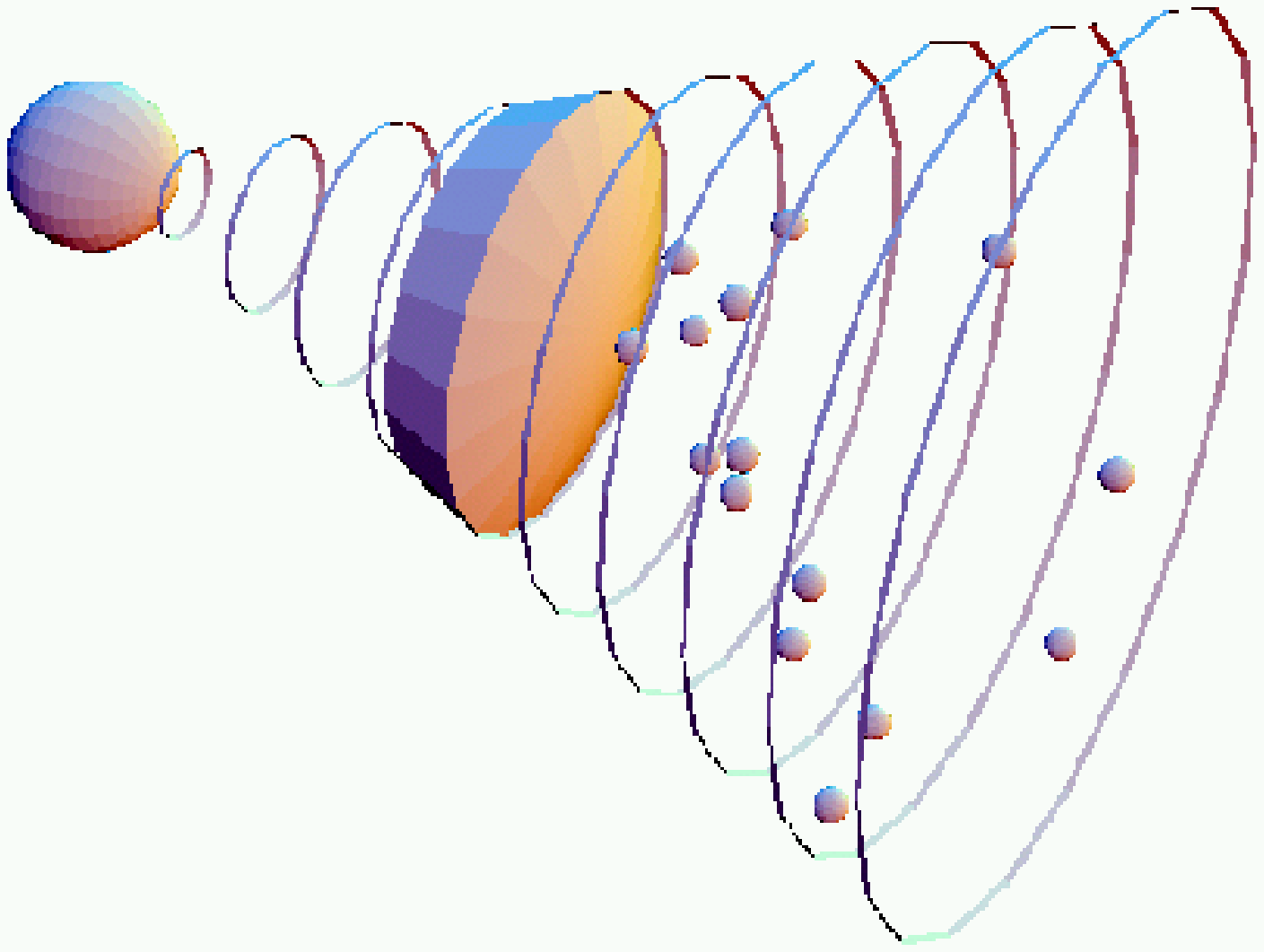,height=2.5in,width=3in} \caption{The
clumpy ISM model (from \cite{Sari_PhD}). Note the small covering
factor and the resulting ``geometrical" inefficiency.}
\label{fig:clumps}
\end{center}
\end{figure}

One can imagine an  inhomogenous external medium which is clumpy
on a scale $d \ll R/\Gamma$ (see Fig \ref{fig:clumps}). Consider
such a clump located at an angle $\theta \sim \Gamma^{-1}$ to the
direction of motion of the matter towards the observer. The
resulting angular time, which is the difference in arrival time
between the first and the last photons emitted from this clump
would be:$\sim d/c \Gamma $. Now $t_{ang} \sim {d/c\Gamma}< t_{R}$
and it seems that one can get around the argument presented
before.

However, Sari and Piran \cite{SP97} have shown that such a
configuration would be extremely inefficient. This  third part of
this argument rules out this caveat. The observations limit the
size of the clumps to $d \approx c \Gamma \delta t$ and the
location of the shock to  $R \approx c T \Gamma^2 $. The number
of clumps within the observed angular cone with an opening angle
$\Gamma^{-1}$ equals the number of pulses which is of the order
$T/\delta t$. The covering factor of the clumps can be directly
estimated in terms of the observed parameters by multiplying the
number of clumps ($T/\delta t$) times their area $d^2= (\delta t
\Gamma)^2$ and dividing by the cross section of the cone
$(R/\Gamma)^2$. The resulting covering factor equals $ \delta t
/T \ll 1$.  The efficiency of conversion of kinetic energy to
$\gamma$-rays in this scenario is smaller than this covering
factor which for a typical variable burst could be smaller than
$10^{-2}$.

I turn now to several attempts to find a way around this result.
I will not discuss here the feasibility of the suggested models
(namely is it likely that the surrounding matter will be clumpy
on the needed length scale \cite{Dermer_Mitman99}, or can an
inner engine eject ``bullets" \cite{Begelman99} with an angular
width of $\sim 10^{-2}$ degrees and what keeps these bullets so
small even when they are shocked and heated). I examine only the
question whether the observed temporal structure can arise within
these models.

\subsubsection{External Shocks on a Clumpy Medium {sec:Clumpy}}
\label{Clumpy} \textcite{Dermer_Mitman99} claim that the simple
efficiency argument of \textcite{SP97} was flawed. They point out
that if the direction of motion of a specific blob is almost
exactly towards the observer the corresponding angular time will
be of order $d^2/cR$ and not $d/{c\Gamma}$ used for a ``generic''
blob. This is narrower by a factor $d\Gamma/R$ than the angular
time across the same blob that is located at a typical angle of
$\Gamma^{-1}$. These special blobs would produce strong narrow
peaks and will form a small region along a narrow cone with a
larger covering factor. \textcite{Dermer_Mitman99} present a
numerical simulation of light curves produced by external shocks
on a clumpy inhomogeneous medium with $\delta t/ T \sim 10^{-2} $
and efficiency of up to $\sim 10$\%.

A detailed analysis of the light curve poses, however, several
problems for this model.  While this result is marginal for bursts
with $\delta t/T \sim 10^{-2}$ with a modulation of 50\% it is
insufficient for bursts with $\delta t /T \sim 10^{-3}$ or if the
modulation is $\sim 100\%$.   Variability on a time scale of
milliseconds has been  observed \cite{NakarPiran02b} in many long
GRBs (namely $\delta t / T $ can be as small as $10^{-4}$.).
Moreover, in this case one would expect that earlier pulses (that
arise from blobs along the direction of motion) would be narrower
than latter pulses. This is not seen in the observed bursts
\cite{Ramirez-Ruiz_Fenimore00}.

Finally the arrival time of individual pulses depends on the
position of the emitting clumps relative to the observers. Two
following pulses would arise from two different clumps that are
rather distant from each other. There is no reason why the pulses
and intervals should be correlated in any way. Recall (\S
\ref{sec:temp-obs}) that the duration of a pulse and the
subsequent interval are correlated \cite{NakarPiran02a}.

\subsubsection{The Shot-Gun Model }
\label{sec:shot-gun}

\begin{figure}[htb]
\begin{center}
\epsfig{file=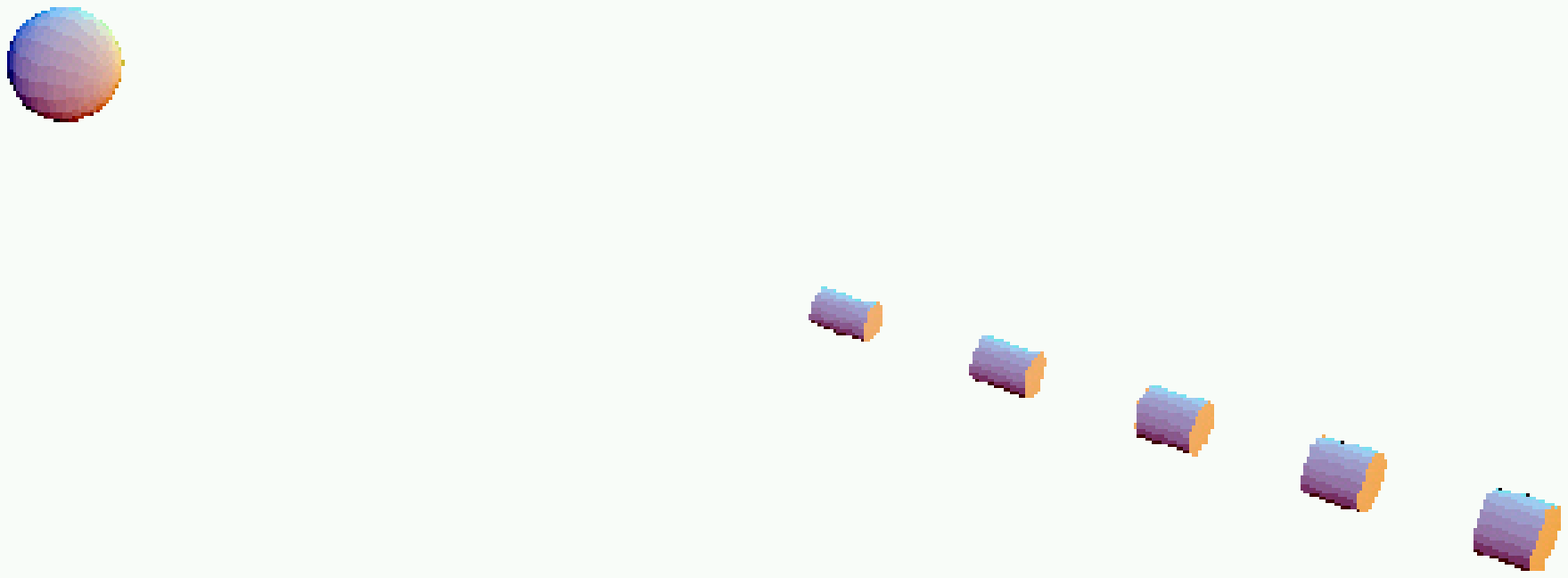,height=2in,width=4in} \caption{The
shot-gun  model (from \cite{Sari_PhD}). The inner engine emits
narrow ``bullets" that collide with the ISM.} \label{fig:bullets}
\end{center}
\end{figure}

\textcite{Begelman99} suggested  that the ``inner engine" operates
as a shot-gun emitting multiple narrow bullets with an angular
size much smaller than $\Gamma^{-1}$ (see Fig \ref{fig:bullets}).
These bullets do not spread while propagating and they  are slowed
down rapidly by an external shock with a very dense circumburst
matter.  The pulses width is given by $t_{ang}$ or by the slowing
down time.  The duration of the burst is determined by the time
that the ``inner engine" emits the bullets.

This model  can produce the observed variability and like in the
internal shocks model the  observed light curve represents also
here the temporal activity of the source. However, in this model
the width of the pulses is determined by the angular time or the
hydrodynamic time or the cooling time of the shocked material. On
the other hand the intervals between the pulses depend only on the
activity of the inner engine. Again, there is no reason why the
two distributions will be similar and why there should be a
correlation between them (see \S \ref{sec:temp-obs} and
\cite{NakarPiran02a}).

\subsubsection{Relativistic Turbulence}

An interesting alternative to shocks as a way to dissipate kinetic
energy is within plasma turbulence
\cite{SmolskyUsov96,SmolskyUsov00,LyutikovBlandford02,LyutikovBlandford03}.
It has been suggested that in this case the kinetic energy of the
shock is dissipated downstream to a combination of macroscopic
(relativistic) random motion of plasma blobs with a Lorentz factor
$\Gamma_b$. Within these blobs the particles have also a
(relativistic) random velocity with a Lorentz factor $\Gamma_p$,
such that: $\Gamma_s \approx \Gamma_b \Gamma_p$.

Relativistic turbulence may be the only way to produce
variability in a situation that the matter is slowed down by the
external medium and not by internal interaction. I stress that in
this case the process is not described by regular shocks and
hence some of the previous arguments do not hold. Two crucial open
questions are i)  Whether one can produce the observed
correlations between pulses and intervals. ii)  Why there is no
spreading of pulses at later times, as would be expected if the
emitting region is slowing down and increasing its radius.

\subsection{Internal Shocks }
\label{sec:In-shocks}

\subsubsection{Hydrodynamics of Internal Shocks }
\label{sec:Int-hydro}
 Internal shocks take place when a faster
shell catches a slower one, namely at:
\begin{equation}
R_{int} \approx  c \delta t \Gamma^2 =  3 \times 10^{14} {\rm cm}
\Ga_{100}^2 \tilde \delta t \label{Rint}
\end{equation}
where $\Ga_{100}$ is the typical Lorentz factor in units of
$10^{2}$ and $\tilde \delta t$ is the time difference between the
emission of the two shells. I show later that $\tilde \delta t$
defined here is roughly equal to the observed fluctuations in the
light curve of the burst $\delta t$. Clearly $R_{int}<R_{ext}$
must hold otherwise internal shocks won't take place. $R_{ext}$ is
defined as the location of  efficient extraction of energy by
external shocks (see \S \ref{sec:Ex-shocks}). If follows from the
discussion in \S \ref{sec:Ex-shocks} that the condition
$R_{int}<R_{ext}$ implies:
\begin{equation}
 \delta \Gamma^2 < {\rm max}( {l \over \Gamma^{2/3}}, l^{3/4}
 \Delta^{1/4})
\end{equation}
where $l$ is defined by Eq. \ref{Sedov} and it is typically of
the order of $10^{18}$cm, while $\Delta$ is the width of the
shell and it is of order $10^{12}$cm. Both conditions set upper
limits on $\Gamma$ (of the order of a few thousands) for internal
shocks. If the initial Lorentz factor is too large then internal
shocks will take place at large radii and external shocks will
take place before the internal shocks could take place. It is
possible that this fact plays an important role in limiting the
relevant Lorentz factors and hence the range of variability of
$E_p$, the peak energy observed in GRBs.

Internal shocks are characterized by a comparable Lorentz factor
of order of a few ($1 < \Ga < 10$) reflecting the relative motion
of the shells and by comparable densities $n$ in both shells. In
this case, for an adiabatic index (4/3), the Loretz factor of the
shocked region $\hat \Gamma$ satisfies:
\begin{equation}
\label{internal_conditions} \hat\Ga =\sqrt{(\Ga^2+1)/2}\ \ .
\end{equation}

The shocked density $\hat n$ and energy $\hat e$ are:

\begin{equation}
{\hat n} = (4 {\hat \Ga} +3 ) n \approx 4 {\hat \Ga} n \ \ ; \ \
{\hat e}  =  {\hat \Ga} {\hat n} m_p c^2 \ .
\end{equation}
Both shocks are mildly relativistic and their strength depends on
the relative Lorentz factors of the two shells.

\subsubsection{The Efficiency of Internal Shocks }
\label{sec:efficiency}

Consider collision between two shells with masses $m_{r}$ and
$m_{s}$ that are moving at different relativistic velocities:
$\Gamma_r \gtrsim \Gamma_s \gg 1$. The resulting bulk Lorentz
factor, $\Gamma_{m}$ in an elastic collision is:
\begin{equation}
\Ga_{m}\simeq
\sqrt{\frac{m_{r}\Ga_{r}+m_{s}\Ga_{s}}{m_{r}/\Ga_{r}+m_{s}/
\Ga_{s}}}. \label{gammam}
\end{equation}
The internal energy, ${\cal E}_{int}$, in the local frame and
$E_{int}$, in the frame of an external observer, of the merged
shell: $E_{int} =\Gamma_m{\cal E}_{int}$, is the difference of the
kinetic energies before and after the collision:
\begin{equation}
E_{int}=m_{r}c^{2}(\Ga_{r}-\Ga_{m})+m_{s}c^{2}(\Ga_{s}-\Ga_{m}).
\end{equation}
The conversion efficiency of kinetic energy into internal energy
is \cite{KPS97}:
\begin{equation}
\epsilon =1-{(m_{r}+m_{s})\Gamma _{m} \over (m_{r}\Gamma
_{r}+m_{s} \Gamma _{s})} . \label{two-shell-efficiency}
\end{equation}
As can be expected a conversion of a significant fraction of the
initial kinetic energy to internal energy requires that the
difference in velocities between the shells will be significant:
$\Ga_r \gg \Ga_s$ and that the two masses will be comparable $m_r
\approx m_s$ \cite{KPS97,DaigneMochkovitch98}.

\textcite{Beloborodov_efficiency_00} considered internal shocks
between shells with a lognormal distribution of
$(\Ga-1)/(\Ga_0-1)$, where $\Gamma_{0}$ is the average Lorentz
factor. The dimensionless parameter, $A$, measures the width of
the distribution.  He shows that the efficiency increases and
reached unity when $A$ is of order unity, that is typical
fluctuation in $\Ga$  are by a factor of 10 compared to the
average. Similarly numerical simulations of
\textcite{Guetta_Spada_Waxman01} show that a significant fraction
of the wind kinetic energy, on the order of 20\%, can be converted
to radiation, provided the distribution of Lorentz factors within
the wind has a large variance and the minimum Lorentz factor is
greater than $\approx 10^{2.5}L^{2/9}_{52}$, where $L_{52}$ is the
(isotropic) wind luminosity in units of $10^{52}$ergs/sec.

Another problem that involves the efficiency of GRBs is that not
all the internal energy generated is emitted. This depends further
on $\epsilon_e$, the fraction of energy given to the electron. If
this fraction is small and if there is no strong coupling between
the electrons and the protons the thermal energy of the shocked
particles (which is stored in this case mostly in the protons)
will not be radiated away. Instead it will be converted again to
kinetic energy by adiabatic cooling. \textcite{KobayashiSari01}
consider a more elaborated model in which colliding shells that do
not emit all their internal energy are reflected from each other,
causing subsequent collisions and thereby allowing more energy to
be emitted. In this case more energy is eventually emitted than
what would have been emitted if we considered only the first
collision. They obtain about 60\% overall efficiency even if the
fraction of energy that goes to electrons is small
$\epsilon_e=0.1$. This is provided that the shells' Lorentz factor
varies between 10 and 10$^4$.

\subsubsection{Light Curves from Internal Shocks }
\label{sec:toy}

 Both  the similarity between the pulse width and
the pulse separation distribution  and the correlation between
intervals and the subsequent pulses
\cite{NakarPiran02a,QuilliganEtal02} arise naturally within the
internal shocks model \cite{NakarPiran02c}. In this model both the
pulse duration and the separation between the pulses are
determined by the same parameter - the interval between the
emitted shells.  I outline here the main argument (see
\textcite{NakarPiran02c} for details). Consider two shells with a
separation $ L $. The Lorentz factor of the slower outer shell is
$\Gamma_{S}=\Gamma $ and of the Lorentz factor inner faster shell
is $ \Gamma_{F}=a\Gamma  $ ($ a>2 $ for an efficient collision).
Both are measured in the observer frame. The shells are ejected at
$ t_{1} $ and $ t_{2}\approx t_{1}+L/c$. The collision takes place
at a radius $ R_s\approx 2\Gamma ^{2}L $ (Note that $ R_s $ does
not depend on $ \Gamma _{2} $). The emitted photons from the
collision will reach the observer at time (omitting the photons
flight time, and assuming transparent shells):

\begin{equation}
\label{to} t_{o} \approx t_{1}+R_s/( 2c\Gamma ^{2})\approx
t_{1}+L/c \approx t_{2} \ .
\end{equation}
The photons from this pulse are observed almost simultaneously
with a (hypothetical) photon that was emitted from the ``inner
engine'' together with the second shell (at $ t_{2} $). This
explains why various numerical simulations
\cite{KPS97,DaigneMochkovitch98,PanaitescuSpadaMeszaros99} find
that for internal shocks the observed light curve replicates the
temporal activity of the source.

In order to determine the time between the bursts we should
consider multiple collisions. It turns out that there are just
three types of collisions, (i), (ii) and (iii), that characterize
the system and all combinations of multiple collisions can be
divided to these three types.  Consider  four shells emitted at
times $ t_{i} $ ($ i=1,2,3,4 $) with a separation of the order of
$ L $ between them. In type (i)  there are two collisions -
between the first and the second shells and between the third and
the fourth shells. The first collision will be observed at $ t_{2}
$ while the second one will be observed at $ t_{4} $. Therefore, $
\Delta t\approx t_{4}-t_{2}\approx 2L/c $. A different collision
scenario (ii) occurs if the second and the first shells collide,
and afterward the third shell takes over and collide with them
(the forth shell does not play any roll in this case). The first
collision will be observed at $ t_{2} $ while the second one will
be observed at $ t_{3} $. Therefore, $ \Delta t\approx
t_{3}-t_{2}\approx L/c. $ Numerical simulations
\cite{NakarPiran02c} show that more then 80\% of the efficient
collisions follows one of these two scenarios ((i) or (ii)).
Therefore one can estimate:
\begin{equation}
\Delta t\approx L/c \ . \label{separation}
\end{equation}
Note that this result is independent of the shells' masses.

A third type of a multiple collision (iii) arises if the third
shell collides first with the second shell. Then the merged shell
will collide with the first one (again the fourth shell does not
participate in this scenario). In this case the two pulses merge
and will arrive almost simultaneously, at the same time with a
(hypothetical) photon that would have been emitted from the inner
engine simultaneously with the third (fastest) shell. $t \sim
t_3$. Only a 20\% fraction exhibits this type of collision.

The pulse width is determined by the angular time (ignoring the
cooling time): $ \delta t=R_s/(2c\Gamma ^{2}_{s}) $ where $
\Gamma _{s} $ is the Lorentz factor of the shocked emitting
region. If the shells have an equal mass ($ m_{1}=m_{2} $) then $
\Gamma _{s}=\sqrt{a}\Gamma $ while if they have equal energy ($
m_{1}=am_{2} $) then $ \Gamma _{s}\approx \Gamma  $. Therefore:
\begin{equation}
\delta t \approx
  \left\{ \begin{array}{r@{\quad\quad}l}
    R_s/2a\Gamma^{2}c\approx L/ac & \rm{equal \  mass}, \\
    R_s/2\Gamma ^{2}c \approx L/c  & \rm {equal \ energy}.
    \end{array} \right .
    \label{width}
\end{equation}
The ratio of the  Lorentz factors $ a $, determines  the
collision's efficiency. For efficient collision the variations in
the shells Lorentz factor (and therefore  $ a $) must be large.

It follows from Eqs. \ref{separation} and \ref{width} that for
equal energy shells the $ \Delta t $-$ \delta t $ similarity and
correlation arises naturally from the reflection of the shells
initial separation in both variables. However, for  equal mass
shells $ \delta t $ is shorter by a factor of $a$ than $ \Delta t
$. This shortens the pulses relative to the intervals.
Additionally, the large variance of $a$ would wipe off the $\Delta
t $-$ \delta t$ correlation. This suggests that equal energy
shells are more likely to  produce the observed light curves.


\subsection{External Shocks }
\label{sec:Ex-shocks}

\subsubsection{Hydrodynamics }
\label{sec:Ex-hydro}

\begin{figure}[htb]
\begin{center}
\epsfig{file=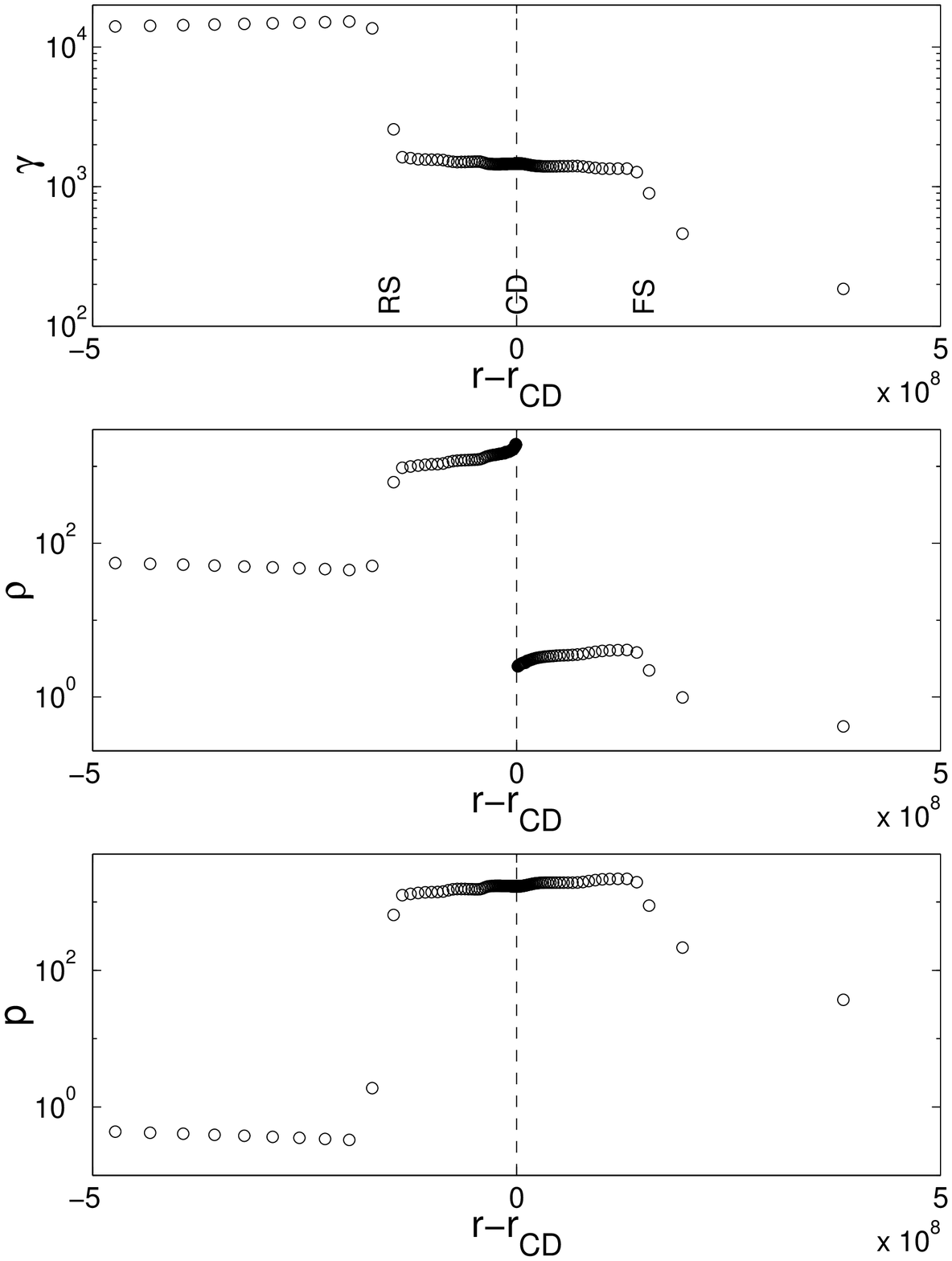,height=2in,width=4in} \caption{The
Lorentz factor $\Gamma$, the density $\rho$ and the pressure $p$
in the shocks.  There are four regions: the ISM (region 1), the
shocked ISM (region 2), the shocked shell (region 3) and the
un-shocked shell (region 4), which are separated by the forward
shock (FS), the contact discontinuity (CD) and the reverse shock
(RS). From \cite{KPS98}.} \label{shock_profile}
\end{center}
\end{figure}

Consider the situation when a cold relativistic shell (whose
internal energy is negligible compared to the rest mass) moves
into the cold ISM. Generally, two shocks form: an outgoing shock
that propagates into the ISM or into the external shell, and a
reverse shock that propagates into the inner shell, with a contact
discontinuity between the shocked material (see Fig.
\ref{shock_profile}).

There dual shocks system is divided to four distinct regions (see
Fig. \ref{shock_profile}): the ambient matter at rest (denoted by
the subscript 1), the shocked ambient matter which has passed
through the forward shock (subscript 2 or f), the shocked shell
material which has passed through the reverse shock (subscript 3
or r), and the unshocked material of the shell (subscript 4).  The
nature of the emitted radiation and the efficiency of the cooling
processes depend on the conditions in the shocked regions 2 and 3.
Both regions have the same energy density $e$. The particle
densities $n_2$ and $n_3$ are, however, different and hence the
effective ``temperatures,'' i.e. the mean Lorentz factors of the
random motions of the shocked protons and electrons, are
different.

Two quantities determine the shocks' structure: $\Ga$, the Lorentz
factor of the motion of the inner expanding matter (denoted 4)
relative to the outer matter (the ISM or the outer shell in the
case of internal collisions - denoted 1) , and the ratio between
the particle number densities in these regions, $n_4/n_1$.
Initially the density contrast between the spherically expanding
shell and the ISM is large. Specifically $n_4/n_1 > \Ga^2$. This
happens during the early phase of an external shock when the shell
is small and dense. This configuration is denoted ``Newtonian''
because the reverse shock is non-relativistic at most (or mildly
relativistic). In this case all the energy conversion takes place
in the forward shock. Only a negligible fraction of the energy is
converted to thermal energy in the reverse shock if it is
Newtonian \cite{SaP95}. Let $\Gamma_2$ be the Lorentz factor of
the motion of the shocked fluid relative to the rest frame of the
fluid at 1 and let $\bar \Gamma_3$ be the Lorentz factor of the
motion of this fluid relative to the rest frame of the
relativistic shell (4):
\begin{equation}
\Ga_2  \approx \Ga \ \ \ ; \ \ \ \bar \Gamma_{3} \approx 1 .
\label{nr1}
\end{equation}
The particle and energy densities $(n, e)$ in the shocked regions
satisfy:
\begin{equation}
n_2 \approx 4 \Ga n_1, \ \  ;  \ \  e \equiv e_2  = 4 \Ga^2 n_1
m_p c^2 \ \ ; \ \ n_3 = 7 n_4, \ \  ;  \ \   e_3 = e . \label{nr3}
\end{equation}

Later, the shell expands and the density ratio decreases (like
$R^{-2}$ if the width of the shell is constant and like $R^{-3}$
if the shell is spreading)  and $n_4/n_1 < \Ga^2$ (but
$n_4/n_1>1$). In this case both the forward and the reverse shocks
are relativistic. The shock equations between regions 1 and 2
combined with  the contact discontinuity between 3 and 2 yield
\cite{BLmc1,BLmc2,Pi94}:
\begin{equation}
\Gamma_2  = (n_4/n_1)^{1/4} \Ga^{1/2} /\sqrt 2 \ \ ; \ \ n_2 = 4
\Gamma_2 n_1 \ \ ; \ \ e \equiv e_2 = 4 \Gamma^2_2 n_1 m_p c^2 ,
\label{cond12}
\end{equation}
Similar relations hold for the reverse shock:
\begin{equation}
\bar \Gamma_3 = (n_4/n_1)^{-1/4} \Ga^{1/2} /\sqrt 2 \ \ ;  \ \ n_3
= 4 \bar \Gamma_{3} n_4. \label{cond34}
\end{equation}
Additinally,
\begin{equation}
  e_3=e \ \ ; \ \  \bar \Gamma_3\cong  (\Ga/\Ga _2+\Ga_2/\Ga)/2 \
  ,
\end{equation}
which follow from the equality of pressures and velocity on the
contact discontinuity. Comparable amounts of energy are converted
to thermal energy in both shocks when both shocks are
relativistic.

The interaction between a relativistic flow and an external medium
depends on the Sedov length that is defined generally as:
\begin{equation}
 E = m_p c^2 \int_0^l 4 \pi n(r) r^2 dr \ .
\end{equation}
The rest mass energy within the Sedov sphere equals the energy of
the explosion. For a homogeneous ISM:
\begin{equation}
 l \equiv ( { E \over
 (4 \pi /3) n_{ism} m_p c^2 } )^{1/3} \approx 10^{18} {\rm cm} E_{52}^{1/3}n_1^{1/3} \ .
\label{Sedov}
\end{equation}
Note that in this section $E$ stands for the isotropic equivalent
energy. Because of the very large Lorentz factor angular
structure on a scale larger than $\Gamma^{-1}$ does not influence
the evolution of the system and it behaves as if it is a part of
a spherical system.  A second length scale that appears in the
problem is $\Delta$, the width of the relativistic shell in the
observer's rest frame.

Initially the reverse shocks is Newtonian and only a negligible
amount of energy is extracted from the shell. At this stage the
whole shell acts ``together". Half of the shell's kinetic energy
is converted to thermal energy when the collected external mass is
$M/\G$, where $M$ is the shell's mass \cite{RM92,Katz94}. This
takes place at a distance:
\begin{equation}
R_\Ga = {l \over \Gamma^{2/3}} = \bigg({E  \over n_{ism} m_p c^2
\Gamma^2} \bigg)^{1/3} =  5.4 \times 10^{16}~{\rm cm }~
E_{52}^{1/3} n_{1}^{-1/3} \Ga_{100}^{-2/3} , \label{rm}
\end{equation}
where $E_{52}$ is the equivalent isotropic energy in
$10^{52}$ergs, $n_1= n_{ism}/ 1~ {\rm particle/ cm}^3$.

However, the reverse shock might become relativistic before
$R_\Ga$. Now energy extraction from the shell is efficient and
one passage of the reverse shock through the shell is sufficient
for complete conversion of the shell's energy to thermal energy.
The energy of the shell will be extracted during a single passage
of the reverse shock across the shell. Using the expression for
the velocity of the reverse shock into the shell (Eq.
\ref{cond34}) one finds that the reverse shock reaches the inner
edge of the shell at $R_\Delta$ \cite{SaP95}:
\begin{equation}
R_\Delta = l^{3/4} \Delta^{1/4} \approx 10^{15}{ \rm cm}
l_{18}^{3/4} \Delta_{12}^{1/4}\ . \label{Rdelta}
\end{equation}
The reverse shock becomes relativistic at $R_N$, where $n_4/n_1 =
\Ga^2$:
\begin{equation}
R_N = l^{3/2} /\Delta^{1/2} \Gamma^2
\end{equation}
Clearly, if $R_N > R_\Ga$ then the energy of the shell is
dissipated while the shocks are still ``Newtonain". If
$R_N<R_\Ga$ the reverse shock becomes relativistic. In this case
$R_\Ga$ looses its meaning as the radius where the energy is
dissipated.  The energy of the shell is dissipated in this
``relativistic" case at $r_\Delta$. The question which of the two
conditions is relevant depends on the parameter $\xi$
\cite{SaP95}:
\begin{equation}
\label{xi} \xi \equiv (l/ \Delta )^{1/2} \Ga^{-4/3} = 2
(l_{18}/\Delta_{12})^{1/2}\Ga_{100}^{-4/3} .
\end{equation}
I have used a canonical value for $\Delta$ as $10^{12}$cm. It
will be shown later that  within the internal-external scenario
$\Delta/c$ corresponds to the duration of the bursts and
$10^{12}$cm corresponds to a typical burst of $30$sec.

Using $\xi$ one can express the different radii as:
\begin{equation}
\label{order0} R_{int}/\zeta =     R_\Delta /\xi^{3/2}=  R_\ga
\xi^{2} = R_N /\xi^{3} \ .
\end{equation}
For completeness I have added to this equation $R_{Int}$, where
internal shocks take place (see Eq. \ref{Rint}). The dimensionless
quantity $\zeta$ : $\zeta \equiv \delta /\Delta$. Thus:
\begin{equation}
\cases   {R_\Delta  <   {\bf R_\Ga} < R_N
            &  $\xi>1$  { (Newtonian~reverse~shock)}\cr
R_N < R_\Ga < {\bf R_\Delta}
            & $\xi<1$ { (Relativistic~reverse~shock)}.
}
\end{equation}
I have marked in bold face the location where the effective energy
extraction does take place.  With typical values for $l$, $\Delta$
and $\Gamma$ $\xi$ is around unity. The radius where energy
extraction takes place is marked in bold face!

{\bf Expanding shell:} A physical shell is expected to expand
during  as it propagates with $\Delta = \Delta_0 + R\Gamma^2$
\cite{PShN93}. This will lead to a monotoneously decreasing
$\xi$. As the value of $R_\Gamma$ is independent of $\Delta$ it
does not vary. However, $R_\Delta$ and $R_N$ decrease from their
initial values. If $\Delta_0 < R_\Gamma \Gamma^2$ (corresponding
to $\xi_0
> 1$) then $\xi =1$ at $R_\Delta =R_\Gamma = R_N$ and all
three radii coincide. Given the fact that with typical parameters
$\xi$ is of order unity this seems to be the ``typical" case.  The
reverse shocks becomes mildly relativistic just when the energy
extraction becomes efficient.  However, if $\xi_0 \ll 1$ then the
shell won't expand enough and  still there will be a relativistic
reverse shock operating at $R_\Delta$.  It is useful to note that
in this case the effective energy extraction takes place at
$R_\Delta$ for all initial values of $\xi_0$. In the following I
denote by $\tilde \xi$ the value of $\xi$ at $R_\Delta$: $\tilde
\xi \approx \xi_0$ if $\xi_0 < 1$ and otherwise $\tilde \xi
\approx 1$.

Overall the external shocks take place at:
\begin{equation}
R_{ext}= \cases{  {\rm max}( {l /\Gamma^{2/3}}, l^{3/4}
 \Delta^{1/4}),
            & Non spreading shell, \cr
l / \Gamma^{2/3} \approx l^{3/4}
 \Delta^{1/4} \approx  5 \times 10^{16} {\rm cm}
E_{52}^{1/3}n_1^{1/3} \Ga^{-2/3}_{100},
            & Spreading shell. \cr
} \label{Rext}
\end{equation}
Usually I will use the second relation (the spreading shell one)
in the following discussion. Note that in the case of non
spreading shell one uses the maximum of the two possible radii.
For example in the Newtonian case where the extraction is at
$l/\Gamma^{2/3}$ the shocks pass the shall many times and hence
$l/Gamma^{2/3}>l^{3/4}\Delta^{1/4}$.

\subsubsection{Synchrotron Spectrum from External Shocks }
\label{sec:Ex-spec}

The bulk of the kinetic energy of the shell is converted to
thermal energy via the two shocks at around the time the shell
has expanded to the radius $R_\Delta$ (this would be the case in
either a thick shell with $\xi< 1$ or with an expanding shell that
begins with $\xi_0 > 1$ but reaches $\xi \approx 1$ due to
expansion of the shell around the time when $R_\Ga = R_\Delta$ and
efficient dissipation takes place . At this radius, the
conditions at the forward shock are:
\begin{equation}
\label{hydroforward} \Gamma _2  =  \Gamma \xi ^{3/4},   \ \ \
n_2  =  4\Gamma _2n_{1},  \ \ \ e_2  =  4\Gamma _2^2n_{1}m_pc^2,
\end{equation}
while at the reverse shock:
\begin{equation}
\label{hydroreverse} \bar \Gamma _3  =  \xi^{-3/4},   \ \ \
\Gamma_3        =  \Gamma\xi^{3/4}, \ \ \ n_3             =  4\xi
^{9/4}\Gamma ^2n_{1}, \ \ \ e_3            =  e_2.
\end{equation}

Substitution of $\Ga_{sh}=\Ga_2 = \Ga \xi^{3/4}$ in Eq.
\ref{epsilons} yields, for the equipartition magnetic field:
\begin{equation}
B= \sqrt{32 \pi} c \epsilon_B^{1/2} \Ga \xi^{3/4} m_p^{1/2}
n_{1}^{1/2} =(40~{\rm G})~\epsilon_B^{1/2}\xi^{3/4} {\G_{100}}
n_{1}^{1/2}.
\end{equation}
If the magnetic field in region 2 behind the forward shock is
obtained purely by shock compression of the ISM field, the field
would be very weak, with $\epsilon_B \ll 1$.  Such low fields are
incompatible with observations of GRBs. I consider, therefore, the
possibility that there may be some kind of a turbulent instability
which brings the magnetic field to approximate equipartition
\cite{Medvedevetal03,Frederiksenetal03}. In the case of the
reverse shock, i.e. in region 3, magnetic fields of considerable
strength might be present in the pre-shock shell material if the
original exploding fireball was magnetic. The exact nature of
magnetic field evolution during fireball expansion depends on
several assumptions. \textcite{Thompson94} found that the magnetic
field will remain in equipartition if it started off originally in
equipartition. M\'esz\'aros, Laguna \& Rees \cite{MLR} on the
other hand, estimated that if the magnetic field was initially in
equipartition then it would be below equipartition by a factor of
$10^{-5}$ by the time the shell expands to $R_\Delta$. It is
uncertain which, if either one, of the estimates is right. As in
the forward shock, an instability could boost the field back to
equipartition. Thus, while both shocks may have $\epsilon_B\ll 1$
with pure flux freezing, both could achieve
$\epsilon_B\rightarrow1$ in the presence of instabilities.  In
principle, $\epsilon_B$ could be different for the two shocks. For
simplicity I will consider the same value in the following
discussions.

Following the discussion in \S \ref{sec:acc}, I assume that in
both regions 2 and 3 the electrons have a power law distribution
with a minimal Lorentz factor $\gamma_{e,min}$ given by Eq.
\ref{gemin} with the corresponding Lorentz factors for the forward
and the reverse shocks.

{\bf Forward shock:} The typical energy of synchrotron photons as
well as the synchrotron cooling time depend on the Lorentz factor
$\gamma_e$ of the relativistic electrons under consideration and
on the strength of the magnetic field. Using Eq. \ref{gemin} for
$\gamma_{e,min} $ and Eq. \ref{syn_obs} for the characteristic
synchrotron energy for the forward shock:
\begin{equation}
\label{hnu_gemin}(h\nu_{syn})_{obs|\ga_{e,min}}= 160~ {\rm keV}~
\epsilon_B^{1/2} \epsilon_e^2 \Ga_{2,100}^4 n_1^{1/2} = 0.5~{\rm
keV}~ (\epsilon_B/0.1)^{1/2} (\epsilon_e/0.1)^2 \tilde \xi_0^{3}
\Ga_{100}^4
 n_1^{1/2} ,
\end{equation}
and
\begin{equation}
\label{cooling_gemin} t_{syn|\ga_{e,min}}= 0.085~ {\rm sec}
~\epsilon_B^{-1} \epsilon_e^{-1} \Ga_{2,100}^{-4} n_1^{-1} =
0.085~ {\rm sec} ~\epsilon_B^{-1} \epsilon_e^{-1} \tilde \xi^{-3}
\Ga_{100}^{-4} n_1^{-1}
\end{equation}
The characteristic frequency and the corresponding cooling time
for the ``typical'' electron are larger  and shorter by a factor
of $[(p-2)/(p-1)]^2$, correspondingly.

The electrons at the forward shock are fast cooling and w the
typical cooling frequency is \cite{SP99b}:
\begin{equation}
\nu_c = 6 keV  (\epsilon_B/0.1)^{-3/2} (\Ga_2/100)^{-4} n_1^{-3/2}
t_s^{-2} \ ,
\end{equation}
where $t_s$ is the time in seconds.  The photons from the early
forward shock are in the low \gr to \xr range, but this depends
strongly on the various parameters (note the strong $\Ga_2^4$
dependence in equation  \ref{hnu_gemin}). For this set of
canonical parameters $\nu_m < \nu_c$. However,  the ratio of these
two frequencies depends on $\Ga^8$! For $\Ga$ slightly larger then
100 the inequality will reverse and the system will be in the fast
cooling regime.

{\bf Reverse Shock:} The Lorentz factor of the reverse shock,
$\bar \Gamma_3$ is smaller by a factor of $\xi^{3/2}\Ga $ than
the Lorentz factor of the forward shock $\Ga_2$. Similarly the
Lorentz factor of
 a ``typical electron'' in the reverse shock is
lower by a factor $\xi^{3/2}\Ga $.  Therefore the observed energy
is lower by a factor $\xi^3 \Ga^2$. The typical synchrotron
frequency of the reverse shock is
\begin{equation}
\nu_{m|reverse~shock} =1.3 \times 10^{13} {\rm Hz} ~
(\epsilon_B/0.1)^{1/2} (\epsilon_e/0.1)^2 \Ga_{100}^2
\label{nu_syn_r} \ .
\end{equation}
This is in the IR regions but note again the strong dependence on
the Lorentz factor and on $\epsilon_e$, which could easily bring
this frequency up to the optical regime. The cooling frequency in
the reverse shock region is the same as the cooling frequency of
the forward shock (if both regions have the same $\epsilon_B$)
\cite{SP99b} hence:
\begin{eqnarray}
\nonumber \nu_{c|reverse~shock}  = 8 \times 10^{18}{\rm Hz}
(\epsilon_{B}/0.1)^{-3/2}(\Gamma_2/{100})^{-4}n_{1}^{-3/2}t_{s}^{-2} = \\
8.8 \times 10^{15} {\rm Hz} (\epsilon_B/0.1)^{-3/2} E_{52}^{-1/2}
n_1^{-1} t_s^{-1/2} \ . \label{nu_c_R}
\end{eqnarray}

In the forward shock $\nu_m$ is comparable or larger than $\nu_c$.
In the reverse shock $\nu_m < \nu_c$ and it is  usually in the
slow cooling regime. The reverse shocks exists for a short time
until it reaches the back of the relativistic shell. Then it
turns into a rarefraction wave that propagates forwards. After
some back and forth bounces of these wave all the matter behind
the forward shock organizes itself in the form of the
Blandford-McKee self similar solution discussed latter in \S
\ref{sec:Blast}.  This above estimates suggest
\cite{MeszarosRees97,SP99c,SP99a,SP99b} that during the short
phase in which the reverse shock exists it should produce a
powerful optical flash. This flash should coincide with the late
part of the GRB. \textcite{Kobayashi00} calculates the light
curves and typical frequencies of the reverse shock for a variety
of conditions.

\subsection{The Transition from Internal Shocks to External Shocks
} \label{sec:transition}

The internal shocks take place at a distance $R_{int} \sim c
\delta t \Gamma^2 \sim (\delta t/0.3sec)  \Gamma_2^2 10^{14}$cm.
These shocks last as long as the inner engine is active. The
typical observed time scale for this activity  $\sim 50 $sec (for
long bursts) and $\sim 0.5$sec (for short ones). External shocks
begin at $R_{ext} \sim 10^{16}$cm. If $R_{ext} /\Gamma^2 \le
T=\Delta/c $, namely if the burst is long,  the afterglows begins
while internal shocks are still going on and the initial part of
the afterglow overlaps the late part of the GRB \cite{Sari97}. At
the early time the afterglow emission (from the forward shock)
peaks in the high X-rays contributing also to the observed
$\gamma$-ray flux. One can expect, therefore,  a transition within
the GRB from hard (pure GRB) to softer and smoother (GRB and
afterglow) signal. Some observational evidence for this transition
was presented in \S \ref{sec:transition-obs}.

\subsection{Prompt Polarization}
\label{sec:pol_prompt}

In \S \ref{sec:prompt-polarization} I discussed the detection of
very high linear polarization from GRB 021206
\cite{CoburnBoggs03}. While the data analysis is uncertain several
papers claimed that this detection has strong implications. First
\textcite{CoburnBoggs03} suggest that this polarization indicates
that the emission mechanism is synchrotron.
\textcite{LyutikovPB03} and \textcite{Granot03} suggest further
that it implies uniform magnetic fields within the emitting
regions and the first even conclude that this implies that the
relativistic flow is Poynting flux dominated and that the
dissipation is in the form of external plasma instability.
\textcite{Waxman03} and \textcite{NakarPiranWaxman03} show however
that: (i) Random magnetic field in shock's plane could produce
almost as high polarization as a uniform field (provided that the
emitting jet is narrow and one is looking along the edge of the
jet). (ii) Even if the magnetic field is uniform the flow does not
have to be Poynting flux dominated. They also stress that while in
the uniform field case we expect high polarization in almost every
burst in the random field one we can  expect high polarization
only in very few bursts. The different time dependence of the
polarization \cite{NakarPiranWaxman03} could also enable us to
distinguish between the two possibilities.

\textcite{Lazzatietal03} and \textcite{DarDeRujula03} suggest that
this polarization implies IC (which can have in principle higher
intrinsic polarization). This shows that even the simplest
conclusion (that the polarization confirms synchrotron as the
emission mechanism) is uncertain. My overall conclusion is that
without further data on other bursts (which is, unfortunately,
quite unlikely in the nearby future) not much can be learnt from
this tentative detection.

\section{THE AFTERGLOW }
\label{sec:afterglow}

It is generally accepted that the afterglow is produced when the
relativistic ejecta is slowed down by the surrounding matter
\cite{MeszarosRees97}. The afterglow begins at $R_{ext}$  where
most of the energy of the ejecta is transferred to the  shocked
external medium. For a long burst this takes place while the burst
is still going on (see \textcite{Sari97} and \S
\ref{sec:transition}). Initially the process might be radiative,
namely a significant fraction of the kinetic energy is dissipated
and the radiation process effects the hydrodynamics of the shock.
I discuss  this phase in \S \ref{sec:rad-synch}. Later the
radiation processes become less efficient and an adiabatic phase
begins during which the radiation losses are minor and do not
influence the hydrodynamics. The hydrodynamic evolution  at this
stage is adiabatic. If the ejecta is in the form of a jet with an
opening angle $\theta$ then  a jet transition will take place when
$\Gamma$ reaches $\theta^{-1}$. A transition into the Newtonian
regime takes place when $\Gamma -1 \approx 0.5$. I begin the
discussion of the afterglow with the hydrodynamics of the
adiabatic phase and with the resulting synchrotron light curve. I
continue with a discussion of the possible early radiative
evolution. Then I turn to the jet break and to the Newtonian
transition. I continue with various complications and variations
on these themes.

\subsection{Relativistic Blast Waves and the Blandford-McKee solution
} \label{sec:Blast}

The theory of relativistic blast waves has been worked out in a
classical paper by Blandford \& McKee  (BM) already in 1976. The
BM model is a self-similar spherical solution describing an
adiabatic ultra relativistic blast wave in the limit $\Gamma \gg
1$.    This solution is the relativistic analogue of the well
known Newtonian Sedov-Taylor solution. Blandford and McKee also
describe in the same paper a generalization for varying ambient
mass density, $\rho =\rho_0 (R/R_0)^{-k}$, $R$ being the distance
from the center. The latter case would be particularly relevant
for $k=2$, as expected in the case of wind from a progenitor,
prior to the GRB explosion.

The BM solution describes a narrow shell of width $\sim
R/\Gamma^2$, in which the shocked material is concentrated. For
simplicity I approximate the solution with a thin homogenous
shell.  Then the adiabatic energy conservation yields:
\begin{equation}
E =  {\Omega\over 3-k} (\rho_0 R_0^k) R^{3-k} \Gamma^2 c^2 \ ,
\label{ad}
\end{equation}
where $E$ is the energy of the blast wave and $\Omega$ is the
solid angle of the afterglow. For a full sphere $\Omega= 4\pi$,
but it can be smaller if the expansion is conical with an opening
angle $\theta$: $\Omega = 4 \pi (1-cos \theta) \approx 2 \pi
\theta^2$ (assuming a double sided jet). This expression can be
simplified using a generalized Sedov scale:
\begin{equation}
l=\left[(3-k)E/ \rho_0 R_0^k c^2\right]^{1/(3-k)} . \label{lSedov}
\end{equation}
If $\Omega$ does not change with time then the blast wave collects
ambient rest mass that equals its initial energy at $R=l$.If we
take into account sideway expansion (after the jet break) we find
that  $\Gamma \approx 1$  and the blast wave becomes Newtonian at:
\begin{equation}
R = l (\Omega/4 \pi)^{1/(3-k)} .
\end{equation}
Using the approximate (the numerical factor in this equation
assumes that the shell is moving at a constant velocity) time -
radius relation Eq. \ref{Rt} one can invert  Eq. \ref{ad} (using
the definition of $l$ , Eq. \ref{lSedov}) and obtain $R$ and
$\Gamma$ as a function of time:
\begin{eqnarray}
 R &=&  [ { 2 l^{3-k} \over \Omega} ]^{1/(4-k)} t^{1/(4-k)}  \\ \nonumber
 \Gamma &=& [ { l^{3-k} \over  2^{3-k} \Omega } ]^{1/2(4-k)}
 t^{-(3-k)/2(4-k)}
 \label{RGamma}
 \end{eqnarray}
The time in these expressions is the observer time - namely the
time that photons emitted at $R$ arrive to the observer (relative
to the time that photons emitted at $R=0$). For spherical (or
spherical like) evolution $\Omega$ in these expressions is a
constant. In general it is possible that $\Omega$ varies with $R$
or with $\Gamma$ (as is the case in a sideways expansion of a
jet). This will produce, of course, a different dependence of $R$
and $\Gamma$ on $t$.

The values of $R$ and $\Gamma$ from Eq. \ref{RGamma} can be
plugged now into the typical frequencies $\nu_c$, $\nu_m$ and
$\nu_{sa}$ as well into the different expression for $F_{\nu,max}$
to obtain the light curve of the afterglow.

Alternatively, one can calculate the light curve using a more
detailed integration over the BM density and energy profiles. To
perform such integration recall that the radius of the front of
the shock is:
\begin{equation}
R=\hat t \{1-[2(4-k)\Gamma ^{2}]^{-1} \},
\end{equation}
where $\Gamma(t)$  is the shock's Lorentz factor and $\hat t $ is
the time since the explosion in its rest frame. The different
hydrodynamic parameters behind the shock can be expressed as
functions of a dimensionless parameter $\chi $:
\begin{equation}
 \chi \equiv [1+2(4-k)\Gamma ^{2}](1-R/\hat t) ,
\end{equation}
as:
\begin{eqnarray}
 n&=&2\sqrt{2}n_{1}\Gamma {\chi^{-(10-3k)/[2(4-k)]}}, \cr
 \gamma^{2}&=&\frac{1}{2}\Gamma ^{2}\chi ^{-1} \cr
 p&=&\frac{2}{3}w_{1}\Gamma ^{2}\chi^{-(17-4k)/(12-3k)},
\end{eqnarray}
where $ n_{1} $ and $ w_{1} $ are the number density and enthalpy
density of the undisturbed circumburst material and  $n$ and $p$
are measured in the fluid's rest frame.

The BM solution is self-similar and assumes $\Gamma \gg 1$.
Obviously, it breaks down when $R\sim l$. This
Relativistic-Newtonian transition should take place around
\begin{equation}
t_{NR}=l/c \approx 1.2 \, {\rm yr} (E_{\rm iso,52}/n_1)^{1/3} \ ,
\label{tNR}
\end{equation}
where the scaling is for $k=0$, $E_{52}$ is the isotropic
equivalent energy, $E_{\rm iso}=4\pi E/\Omega$, in units of
$10^{52} {\rm ergs}$ and $n_1$ is the external density in ${\rm
cm}^{-3}$. After this transition the solution will turn into the
Newtonian Sedov-Taylor solution with:
\begin{eqnarray}
R & = & R_{NR} (t/t_{NR})^{2/5} , \\ \nonumber
 v & = & v_{NR}(t/t_{NR})^{-3/5} , \\
e & = & e_{NR} (t/t_{NR})^{-6/5} . \label{SedovNR}
\end{eqnarray}

The adiabatic approximation is valid for most of the duration of
the afterglow. However, during the first hour or so (or even for
the first day, for $k = 2$), the system could be radiative
(provided that $\epsilon_e \approx 1$) or partially radiative.
During a radiative phase the evolution can be approximated as:
\begin{equation}
E =  {\Omega\over 3-k} A R^{3-k} \Gamma \Gamma_0 c^2 \ ,
\label{rad}
\end{equation}
where $\Gamma_0$ is the initial Lorentz factor. \textcite{CPS98}
derived an analytic self-similar solution describing this phase.

\textcite{CP99} describe a solution for the case when energy is
continuously added to the blast wave by the central engine, even
during the afterglow phase. A self-similar solution arises if the
additional energy deposition behaves like a power law.  This
would arise naturally in some models, e.g. in the pulsar like
model \cite{Usov94}.

\subsection{Light Curves for the "Standard" Adiabatic Synchrotron Model }
\label{sec:ad-synch}

In \S \ref{sec:synch-spec} I discussed the instantaneous
synchrotron spectrum. The light curve that corresponds to this
spectrum depends simply on the variation of the $F_{\nu,max}$ and
the break frequencies as a function of the observer time
\cite{MeszarosRees97,SPN98}. This in turn depends on the variation
of the physical quantities along the shock front. For simplicity I
approximate here the BM solution as a spherical homogeneous shell
in which the physical conditions are determined by the shock jump
between the shell and the surrounding matter. Like in \S
\ref{sec:synch-spec} the calculation is divided to two cases: fast
cooling and slow cooling.

\textcite{SPN98} estimate the observed emission as a series of
power law segments in time and in frequency\footnote{The
following notation appeared in the astro-ph version of
\cite{SPN98}. Later during the proofs that author realized that
$\alpha$ is used often in astrophysics to denote a spectral index
and in the Ap. J. version of \cite{SPN98} the notations have been
changed to $F_\nu \propto  t^{-\beta} \nu^{-\alpha}$. However, in
the meantime the astro-ph notation became generally accepted. I
use these notations here.}:
\begin{equation}
 F_\nu \propto  t^{-\alpha} \nu^{-\beta} \ ,
\end{equation}
that are separated by break frequencies, across which the
exponents of these power laws change: the cooling frequency,
$\nu_c$, the typical synchrotron frequency $\nu_m$ and the self
absorption frequency $\nu_{sa}$. To estimate the rates one plugs
the expressions for $\Gamma$ and $R$ as a function of the observer
time (Eq. \ref{RGamma}), using for a homogenous external matter
$k=0$:
\begin{eqnarray}
R(t) \cong   (17Et/4\pi m_p n c)^{1/4},
 \cr \Gamma(t) \cong
(17E/1024\pi n m_p c^5 t^3)^{1/8} , \label{RGammaISM}
\end{eqnarray}
to the expressions of the  cooling frequency, $\nu_c$, the
typical synchrotron frequency $\nu_m$ and the self absorption
frequency $\nu_{sa}$ (Eqs. \ref{numc}) and to the expression of
the maximal flux (Eq. \ref{spectrumslow} for slow cooling and Eq.
\ref{spectrumfast} for fast cooling). Note that the numerical
factors in the above expressions arise from an exact integration
over the BM profile.  This procedure results in:
\begin{eqnarray}
\label{abreaks} \nu_c & = & 0.85 \times 10^{14}{\rm \ Hz}
(\epsilon_B/0.1)^{-3/2} E_{52}^{-1/2} n_1^{-1} t_d^{-1/2} , \cr
 \nu_m & = & 1.8 \times 10^{12} {\rm \ Hz}(\epsilon_B/0.1)^{1/2}
 (\epsilon_e/0.1)^2 E_{52}^{1/2} t_d^{-3/2}, \cr
 F_{\nu,max}& = & 0.35 \times 10^5 \mu {\rm J}
(\epsilon_B/0.1)^{1/2} E_{52} n_1^{1/2} D_{28}^{-2} \ \ .
\end{eqnarray}
A nice feature of this light curve is that  the peak flux is
constant and does not vary with time \cite{MR97} as it moves to
lower and lower frequencies.

At sufficiently early times  $\nu_c<\nu_m$, i.e. fast cooling,
while at late times  $\nu_c>\nu_m$, i.e., slow cooling. The
transition between the two occurs when $\nu_c=\nu_m$.  This
corresponds  (for adiabatic evolution) to:
\begin{equation}
\label{tfc} t_0= 0.5 ~{\rm hours} (\epsilon_B/0.1)^2
(\epsilon_e/0.1)^2 E_{52} n_1 ~.
\end{equation}

Additionally one can translate Eqs. \ref{abreaks} to the time in
which a given break frequency passes a given band. Consider a
fixed frequency $\nu=\nu_{15}10^{15}$Hz. There are two critical
times, $t_c$ and $t_m$, when the break frequencies, $\nu_c$ and
$\nu_m$, cross the observed frequency $\nu$:
\begin{eqnarray}
 t_c= 0.2 ~{\rm hours} (\epsilon_B/0.1)^{-3} E_{52}^{-1} n_1^{-2}
\nu_{15}^{-2} ~, \cr
 t_m= 0.2 ~{\rm hours} (\epsilon_B/0.1)^{1/3} (\epsilon_e/0.1)^{4/3}
E_{52}^{1/3} \nu_{15}^{-2/3}  \ .
\end{eqnarray}

In the Rayleigh-Jeans part of the black body radiation
$I_\nu=kT(2\nu^2/c^2)$
 so that $F_\nu \propto kT\nu^2$. Therefore, in the part of the synchrotron
spectrum that is optically thick to synchrotron self absorption,
we have $F_\nu\propto kT_{eff}\nu^2$. For slow cooling $kT_{eff}
\sim \gamma_m m_e c^2 = const.$ throughout the whole shell of
shocked fluid behind the shock, and therefore $F_\nu \propto
\nu^2$ below $\nu_{sa}$ where the optical depth to synchrotron
self absorption equals one, $\tau_{\nu_{as}}=1$. For fast cooling,
as we go down in frequency, the optical depth to synchrotron self
absorption first equals unity due to absorption over the whole
shell of shocked fluid behind the shock, most of which is at the
back of the shell and has $kT_{eff}\sim\gamma_c$. The observer is
located in front of the shock, and the radiation that escapes and
reaches the observer is from $\tau_\nu \sim 1$. As $\nu$ decreases
below $\nu_{sa}$ the location where $\tau_\nu \sim 1$ moves from
the back of the shell toward the front of the shell, where the
electrons suffered less cooling so that $kT_{eff}(\tau_\nu=1)
\propto \nu^{-5/8}$. Consequently $F_nu\propto\nu^{11/8}$. At a
certain frequency $\tau_\nu \sim 1$ at the location behind the
shock where electrons with $\gamma_m$ start to cool significantly.
Below this frequency,
 $(\nu_{ac})$, even though $\tau_\nu \sim 1$ closer and closer to
the shock with decreasing $\nu$, the effective temperature at that
location is constant: $kT_{eff} \sim \gamma_m m_e c^2 = const.$,
and therefore $F_\nu \propto \nu^2$ for $\nu<\nu_{ac}$, while
$F_\nu \propto \nu^{11/8}$ for $\nu_{ac} < \nu < \nu_{sa}$.
Overall  the expression for the self absorption frequency depends
on the cooling regime. It divides to two cases, denoted $\nu_{sa}$
and $\nu_{ac}$, for fast cooling and both expression are different
from the slow cooling \cite{GPS00}. For fast cooling:
\begin{equation}
 \nu _{ac} = 1.7\times 10^{9}\ {\rm Hz}\ (\epsilon_B/0.1)^{-2/5}
 (\epsilon_e/0.1)^{-8/5}E_{52}^{-1/10}n_{1}^{3/10}(t/100{\rm sec})^{3/10}\ ,
\end{equation}
\begin{equation}
\label{ES_ISM}  \nu _{sa} = 1.8\times 10^{10}\ {\rm Hz}\
(\epsilon_B/0.1)^{6/5} E_{52}^{7/10}n_{1}^{11/10}(t/100{\rm
sec})^{-1/2}\ .
\end{equation}
For slow cooling:
\begin{equation}
\nu _{sa}= 1.24 \times 10^{9} \ {\rm Hz}
{(p-1)^{3/5}\over(3p+2)^{3/5}} (1+z)^{-1}\bar\epsilon_{e}^{\;
-1}\epsilon_B^{1/5} n_0^{3/5} E_{52}^{1/5}
\end{equation}

\begin{figure}[htb]
\begin{center}
\epsfig{file=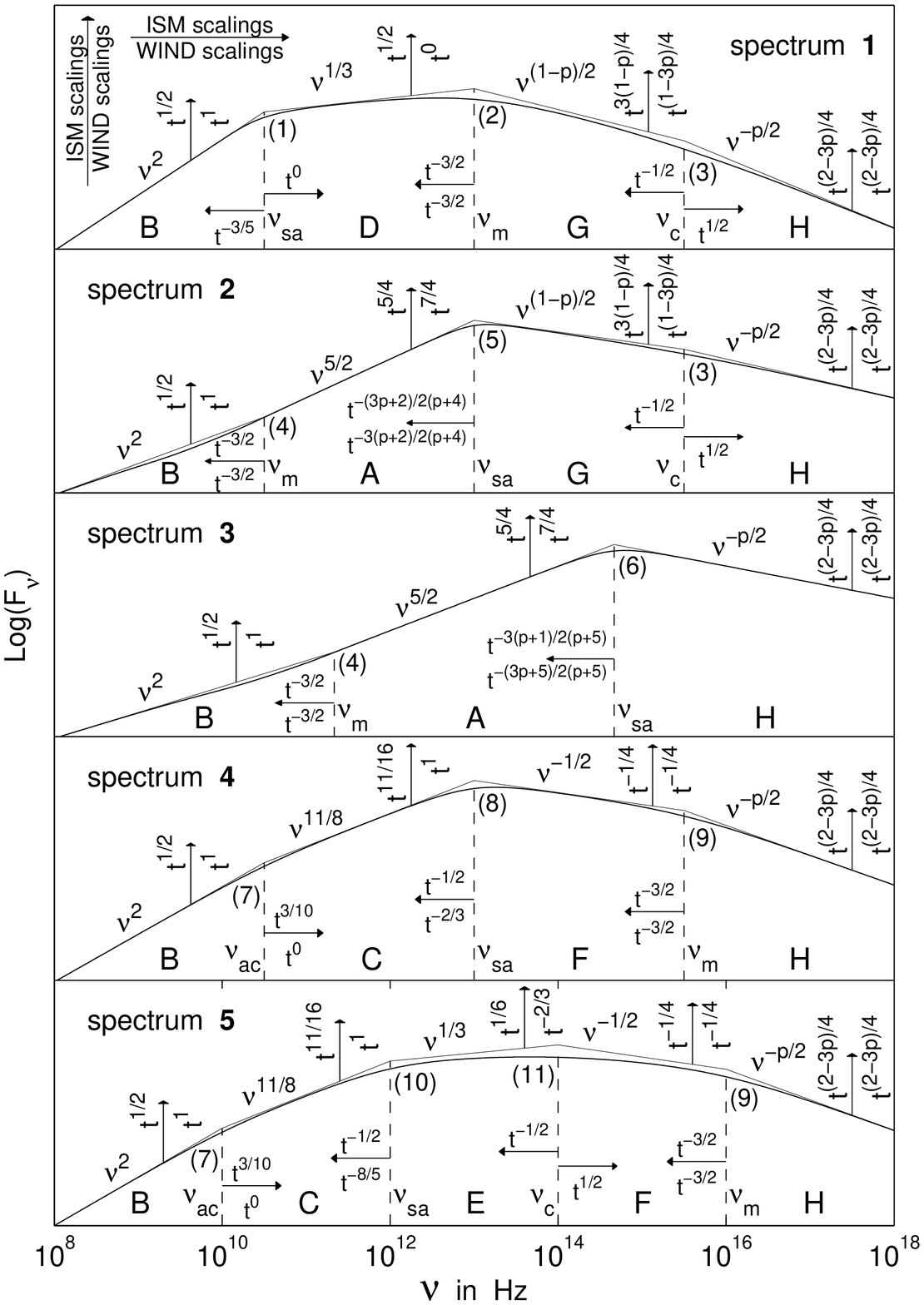,height=5in} \caption{The different
possible broad band synchrotron spectra from a
  relativistic blast wave, that accelerates the electrons to a power
  law distribution of energies. The thin solid line shows the
  asymptotic power law segments, and their points of
  intersection, where the break frequencies, $\nu_b$, and the
  corresponding flux densities, $F_{\nu_b,{\rm ext}}$, are defined.
  The different PLSs are labelled A through H, while the different
  break frequencies are labelled 1 through 11. The temporal scalings of
  the power law segments and the break frequencies, for an ISM ($k=0$) or stellar
  wind ($k=2$) environment, are indicated by the arrows.
  The different spectra are labelled 1 through
  5, from top to bottom. The relevant spectrum is determined by the
  ordering of the break frequencies. The top two panels (spectra 1 and
  2) correspond to slow cooling ($\nu_m<\nu_c$).  Spectrum 1 applies
  when $\nu_{sa}<\nu_m$, while spectrum 2 applies when
  $\nu_m<\nu_{sa}<\nu_c$.  The two bottom panels (spectra 4 and 5)
  correspond to fast cooling ($\nu_c<\nu_m$).  Spectrum 5 applies when
  $\nu_{sa}<\nu_c$, and spectrum 4 applies when
  $\nu_c<\nu_{sa}<\nu_m$. Spectrum 3 (middle panel) applies when
  $\nu_{sa}>\nu_m,\nu_c$, where in this case the relative ordering of
  $\nu_c$ and $\nu_m$ is unimportant (i.e. spectrum 3 may apply both
  to slow cooling or fast cooling). From \cite{GranotSari02}.}
\label{fig:full_spectrum}
\end{center}
\end{figure}

For a given frequency  either  $t_0>t_m>t_c$ (which is typical for
high frequencies) or $t_0<t_m<t_c$ (which is typical for low
frequencies). The results are summarized in two tables
\ref{table:Fast_ISM} and \ref{table:Slow-ISM} describing $\alpha$
and $\beta$ for fast and slow cooling. The different light curves
are depicted in Fig. \ref{fig:full_spectrum}.

\begin{table}
\centering
\begin{tabular}{|c|c|c|} \hline
  $ $ & $\alpha$ & $\beta$ \\ \hline
  $\nu < \nu_a$ & 1 & 2 \\ \hline
  $\nu_a < \nu < \nu_c $ & 1/6 & 1/3 \\\hline
  $\nu_c< \nu < \nu_m $ & -1/4   & -1/2 \\\hline
   $\nu_m < \nu $ & -(3p-2)/4 & $-p/2=(2\alpha-1)/3$ \\ \hline
\end{tabular}
\caption{$\alpha$ and $\beta$ for fast cooling ($\nu_a< \nu_c <
\nu_m$) into a constant density ISM}\label{table:Fast_ISM}
\end{table}

\begin{table}
\centering
\begin{tabular}{|c|c|c|} \hline
  $ $ & $\alpha$ & $\beta$ \\ \hline
  $\nu < \nu_a$ & 1/2 & 2 \\ \hline
  $\nu_a < \nu < \nu_m $ & 1/2 & 1/3 \\ \hline
  $\nu_m < \nu < \nu_c $ & -3(p-1)/4   & $-(p-1)/2=2\alpha/3$ \\ \hline
   $\nu_c < \nu $ & -(3p-2)/4 & $-p/2=(2\alpha-1)/3$ \\ \hline
\end{tabular}
\caption{$\alpha$ and $\beta$ for slow cooling ($\nu_a< \nu_m <
\nu_c $) into a constant density ISM}\label{table:Slow-ISM}
\end{table}

These results are valid only for $p>2$ (and for $\gamma_{max}$,
the maximal electron energy,  much  higher than $\gamma_{min}$).
If $p<2$ then $\gamma_{max}$ plays a critical role. The resulting
temporal and spectral indices for slow cooling with $1<p<2$ are
given by \textcite{DaiCheng01} and by \textcite{Bhattacharya01}
and summarized in table \ref{table:Slow_lowP} below.   For
completeness I include in this table also the cases of
propagation into a wind (see \S \ref{sec:wind}) and a jet break
(see \S \ref{sec:jets}).

\begin{table}
  \centering
\begin{tabular}{|c|c|c|c|} \hline
 & ISM & wind & Jet \\ \hline
  $\nu < \nu_a$ & (17p-26)/16(p-1) & (13p-18)/18(p-1) & 3(p-2)/4(p-1)
  \\ \hline
  $\nu_a < \nu < \nu_m $ & (p+1)/8(p-1) & 5(2-p)/12(p-1) & (8-5p)/6(p-1)
  \\ \hline
  $\nu_m < \nu < \nu_c $ & -3(p+2)/16  & -(p+8)/8 & -(p+6)/4 \\ \hline
   $\nu_c < \nu $ & -(3p+10)/16  & -(p+6)/8 & -(p+6)/4 \\ \hline
\end{tabular}
\caption{$\alpha$  for slow cooling ($\nu_a< \nu_m < \nu_c $)
into a constant density ISM, wind and jet for electron
distribution  with $1<p<2$.}\label{table:Slow_lowP}
\end{table}

The simple solution, that is based on a homogeneous shell
approximation, can be modified by using the full BM solution and
integrating over the entire volume of shocked fluid \cite{GPS99a}.
Following \cite{NakarPiran03b} I discuss  \S \ref{sec:BMlight} a
simple way to perform this integration. The detailed integration
yields a smoother spectrum and light curve near the break
frequencies, but the asymptotic slopes, away from the break
frequencies and the transition times, remain the same as in the
simpler theory. \textcite{GranotSari02} describe a detailed
numerical analysis of the smooth afterglow spectrum including a
smooth approximation for the spectrum over the transition regions
(see also \cite{GruzinovWaxman99}). They also describe additional
cases of ordering of the typical frequencies which were not
considered earlier.

A final note on this ``standard" model is that it assumes
adiabaticity. However, in reality a fraction of the energy is lost
and this influences over a long run the hydrodynamic behavior.
This could be easily corrected by an integration of the energy
losses and an addition a variable energy to Eq. \ref{ad}, followed
by the rest of the procedure described above
\cite{KumarPanaitescu00c}.

\subsection{Light Curves for the early radiative phase }
\label{sec:rad-synch}

If the electrons' energy is large (namely if $\epsilon_e$ is not
far from unity), then early on during the first few hours of the
afterglow there will be a radiative phase in which a significant
fraction of the kinetic energy is lost via the radiative
processes. One can generalize the BM solution to this radiative
stage (see \textcite{CPS98} and \S \ref{sec:Blast}). The essence
of the radiative phase is that in this case the energy varies as
$E\propto\Gamma$, where $\Gamma \cong (R/L)^{-3}$. Note that $L$
is calculated in terms of $M$ and the initial energy of the
explosion, $E_0$, via $M=E_0/\Gamma_0 c^2$, where $\Gamma_0$ is
the initial Lorentz factor of the ejecta:
\begin{eqnarray}
R(t) \cong (4ct/L)^{1/7} L, \cr
 \Gamma(t) \cong  (4ct/L)^{-3/7}
\end{eqnarray}
The transition time from the radiative to the adiabatic  phase
takes place when the radiation losses become negligible. This
happens at:
\begin{equation}
\label{trad} t_{rad}= 0.17 ~{\rm hours} ~(\epsilon_B/0.1)^{7/5}
(\epsilon_e/0.1)^{7/5} E_{52}^{4/5}(\Gamma/100)^{-4/5} n_1^{3/5}
~.
\end{equation}

Following \textcite{SPN98} one can use the above expressions to
express the different typical frequencies and fluxes as:
\begin{eqnarray}
\label{rbreaks} \nu_c  & = & 4.1 \times 10^{14} ~{\rm \ Hz}~
(\epsilon_B/0.1)^{-3/2}
 E_{52}^{-4/7} (\Gamma/100)^{4/7}  n_1^{-13/14} t_d^{-2/7} , \cr
\nu_m              & = & 3.8 \times 10^{11} ~{\rm \
Hz}~(\epsilon_B/0.1)^{1/2} (\epsilon_e/0.1)^2
 E_{52}^{4/7} (\Gamma/100)^{-4/7} n_1^{-1/14}  t_d^{-12/7}, \cr
F_{\nu,max}& = & 1.4 \times 10^3 ~\mu {\rm J}~ \epsilon_B^{1/2}
E_{52}^{8/7}(\Gamma/100)^{-8/7} n_1^{5/14} D_{28}^{-2} t_d^{-3/7}
\ .
\end{eqnarray}
Like in the adiabatic case  this can be translated to  the times
of passage of the break frequencies at a given observed frequency:
\begin{eqnarray}
t_c= 0.05 \times 10^{-7} ~{\rm days} (\epsilon_B/0.1)^{-21/4}
E_{52}^{-2} \Gamma_2^{2} n_1^{-13/4} \nu_{15}^{-7/2} ,
         \cr
t_m= 0.01 ~{\rm days} (\epsilon_B/0.1)^{7/24}
(\epsilon_e/0.1)^{7/6} E_{52}^{1/3} \Gamma_2^{-1/3}
\nu_{15}^{-7/12} n_1^{-1/24} ~.
\end{eqnarray}
Unlike the adiabatic case, here  $\nu_c$ must be below $\nu_m$.
Otherwise the bulk of the electrons do not cool and the system
won't be radiative. Indeed at $t_{rad}$ (given by Eq. \ref{trad}
above) $\nu_c=\nu_m$.

\subsection{Light Curve During  the Newtonian transition}
\label{sec:Newtonian}.

 At
$t \approx t_{NR}$ (see Eq. \ref{tNR}) the afterglow reaches the
Newtonian Sedov-Taylor phase. During this phase the adiabatic
hydrodynamic is described by Eq. \ref{SedovNR}.
\textcite{FrailWaxmanKulkarni00} calculate the synchrotron
spectrum and light curve of the afterglow in this stage.  The
energy scaling implies that $B \propto t^{-3/5}$ and $\ga_{e,min}
\propto t^{-6/5}$. Combined together this yields $\nu_m \propto
t^{-3}$. Using the standard assumptions of equipartition and of a
power law electron's distribution they find :
\begin{eqnarray}
\label{Sedov_light}
 \nu_c & = &  10^{13} {\rm
\ Hz} (\epsilon_B/0.3)^{-3/2} E_{51}^{-2/3} n_1^{-5/6}
  (t/t_{NR})^{-1/5} , \cr
 \nu_m & = & 1 {\rm GHz} (\epsilon_B/0.3)^{1/2} (\epsilon_e/0.3)^2
  n_1^{-1/2}, \cr
F_{\nu_m<\nu<\nu_c}& = & 1 {\rm mJ} (\epsilon_B/0.3)^{3/4}
(\epsilon_e/0.3) n_1^{3/4} E_{51} D_{28}^{-2} \nu^{-(p-1)/2}_{GHz}
(t_/t_{NR})^{-3(p-1)/2+3/5} \ .
\end{eqnarray}
This late time light curve provides a simple ``calorimetric"
estimate of the afterglow energy at this stage (see \S
\ref{sec:Energetics}). Additionally as the radio flux is rather
large and as it varies on a scale of several month it can be used
for search of orphan radio afterglows (see
\textcite{Levinsonetal02} and \S \ref{sec:orphan_radio}).

\subsection{Generalizations: I. Winds }
\label{sec:wind}

The simplest generalization of the previous models is to allow a
variable circuburst density with $n(R) \propto R^{-k}$. The
hydrodynamic evolution of a relativistic blast wave in such a
medium was considered already in the original paper of
\textcite{BLmc1}. The synchrotron light curve was considered
first by \textcite{MeszarosReesWijers98} and by
\textcite{DaiLu99}.

\textcite{ChevalierLi99,ChevalierLi00} stressed the importance of
the  $n(R) \propto R^{-2}$  case which arises whenever there is a
stellar wind ejected by the GRB's progenitor prior to the burst.
This arises naturally in  the Collapsar model that is based on
the collapse of a massive star. The calculations follow those
outlines in the previous sections, with the only difference that
the relations determining $R(t)$ and $\Gamma(t)$ for homogeneous
circumburst medium, Eqs. \ref{RGammaISM}, should be replaced by
Eqs. \ref{RGamma} with $k=2$

The high initial densities in a wind density profile implies a
low initial cooling frequency. Unlike the constant density case
the cooling frequency here increase with time
\cite{ChevalierLi99}. This leads to a different temporal
relations between the different frequencies and cooling regimes.
For example it is possible that the cooling frequency will be
initially below the synchrotron self absorption frequency.
\textcite{ChevalierLi00} consider five different evolution of the
light curves for different conditions and observed frequencies. We
list below the two most relevant cases, the first fits the \xr and
optical afterglows while the second is typical for the lower radio
frequencies.

\begin{table}
  \centering
\begin{tabular}{|c|c|c|} \hline
  & $\alpha$ & $\beta$ \\ \hline
  $\nu_c < \nu < \nu_m$ & -1/4 & -1/2 \\ \hline
  $\nu_m,\nu_c < \nu $  & -(3p-2)/4 & $-p/2=(2\alpha-1)/3$  \\ \hline
  $\nu_m < \nu < \nu_c $ & -(3p-1)/4   & $-(p-1)/2=(2\alpha+1)/3$ \\
  \hline
\end{tabular}
\caption{$\alpha$ and $\beta$ for \xr and optical frequencies from
a blast wave into a wind profile when $\nu_a < \nu_c,\nu_m,\nu$
\cite{ChevalierLi00}. Note that the order of the table is
according to the evolution of the light curve at a fixed high
observed frequency.}  \label{table:XR_wind1}
\end{table}

Note that for $\nu_m,\nu_c < \nu $ both the spectral slop and the
temporal evolution are similar for a wind and for a constant
density profile. This poses, of course,  a problem in the
interpretation of afterglow light curves.

\begin{table}
  \centering
\begin{tabular}{|c|c|c|} \hline
   & $\alpha$ & $\beta$ \\ \hline
   $\nu_c < \nu < \nu_a < \nu_m$ & 7/4 & 5/2 \\ \hline
   $\nu   < \nu_c < \nu_a < \nu_m$ & 2 & 2 \\ \hline
   $\nu   < \nu_a < \nu_m < \nu_c$ & 1 & 2 \\ \hline
   $\nu_a < \nu   < \nu_m < \nu_c$ & 0 & 1/3 \\ \hline
   $\nu_a < \nu_m < \nu  < \nu_c $ &  -(3p-1)/4   & $-(p-1)/2=(2\alpha+1)/3$ \\
  \hline
\end{tabular}
\caption{$\alpha$ and $\beta$ for radio frequencies from a blast
wave into a wind profile \cite{ChevalierLi00}. Note that the order
of the table is according to the evolution of the light curve at a
fixed low observed frequency.}  \label{table:XR_wind2}
\end{table}

\subsection{Generalizations: II. Energy injection and refreshed shocks
} \label{sec:energy}

The simple adiabatic model assumes that the energy of the GRB is
constant. However, the energy could change if additional slower
material is ejected behind the initial matter. This would be
expected generically in the internal shock model. In this model
the burst is produced by a series of collisions between shells
moving at different velocities. One naturally expect here also
slower moving matter that does not catch up initially with the
faster moving one. However, as the initially faster moving matter
is slowed down by the circum-burst matter this slower matter
eventually catch up  and produces refreshed shocks
\cite{ReesMeszaros98,KP00a,SariMeszaros00}.

There are two implications for the refreshed shocks. First the
additional energy injection will influence the dynamics of the
blast wave \cite{ReesMeszaros98,SariMeszaros00}. This effect can
be modelled by modifying $E$ in Eq. \ref{ad}, but the effect of
additional mass carrying the slower energy must be included in
some cases. This would change the decay slope from the canonical
one and produce a slower decay in the light curve. In the
following section \S \ref{sec:density} I describe a scheme for
calculating the light  curve resulting from a variable blast wave
energy. If the additional matter is emitted sporadically then the
shell collision could produce initial temporal variability in the
early afterglow signal. \textcite{Foxetal03_021004}, for example,
suggest that refreshed shocks are the origin of the variability in
the early afterglow of GRB 021004.

A second effect is the production of a reverse shock propagating
into the slower material  when it catches up with the faster one
\cite{KP00a}. This is of course in addition to the forward shock
that propagates into the outer shell.  This reverse shock could be
episodal or long lasting depending on the profile of the
additional matter. \textcite{KP00a} consider two shells with
energies $E_1$ and $E_2$ in the outer and the inner shells
respectively. The outer shell is moving with a  bulk Lorentz
factor $\Gamma_{0c}\sim 5 (t/day)^{3/8}$ at the (observed) time,
t, of the collision. As the inner shell catches up with the outer
one when both shells have comparable Lorentz factors the reverse
shocks is always mildly relativistic. The calculation of the shock
is slightly different than the calculation of a shell propagating
into a cold material (another shell or the ISM) discussed earlier.
Here the outer shell has already collided with the ISM. Hence it
is hot with internal energy exceeding the rest mass energy. The
reverse shock produces emission at a characteristic frequency that
is typically much lower than the peak of the emission from the
outer shell by a factor of $\sim 7 \Gamma_{0c}^2 (E_2/E_1)^{1.1}$,
and the observed flux at this frequency from the reverse shock is
larger compared to the flux from the outer shell by a factor of
$\sim 8 (\Gamma_{0c} E_2/E_1)^{5/3}$. This emission is typically
in the radio or the FIR range.

\textcite{KP00a} suggest that due to angular spreading the
refreshed shocks produce an enhancement with a typical time scale
$\delta t \sim t$. \textcite{GranotNakarPiran03} stress that the
fact that energy necessarily increases in refreshed shocks, the
overall light curve must have a step-wise shape (above the
continues power-law decline) with a break at the corresponding
shocks. This behavior was seen in GRB 030329. However there the
transitions are fast with $\delta t < t$.
\textcite{GranotNakarPiran03} point out that if the refreshed
shocks take place {\it after} the jet break (as is likely the case
in GRB 030329) then if the later shells remain cold and do not
spread sideways we would have $\delta t \sim t_{jet} < t$. This
explains nicely the fast transitions seen in  this burst.

\subsection{Generalizations: III. Inhomogeneous density profiles }
\label{sec:density}

An interesting possibility that arose with the observation of the
variable light curve of the afterglow of GRB 021004 is that the
ejecta encounters surrounding matter with an irregular density
profile \cite{Lazzati02,NakarPiranGranot03,HeylPerna03}. To
explore this situation one can resort to numerical simulation of
the propagation of the blast wave into a selected density profile
\cite{Lazzati02}. Instead one can attempt to model this
analytically or almost analytically \cite{NakarPiran03b}. The key
for this analytic model is the approximation of the light curve
from an inhomogeneous density profile as a series of emission from
instantaneous BM solutions, each with its own external density.

\subsubsection{The light curve of a BM solution}
\label{sec:BMlight}

The observed flux, at an observer time $ t $, from an arbitrary
spherically symmetric emitting region is given by \cite{GPS99a}:
\begin{equation} \label{eq Fnu1}
F_{\nu }(t)=\frac{1}{2D^{2}}\int _{0}^{\infty }dt' \int
_{0}^{\infty }r^{2}dr\int _{-1}^{1}d(cos\theta
)\frac{n'(r)P'_{\nu }(\nu \Lambda ,r)}{\Lambda ^{2}}\delta
(t'-t-\frac{r\, \, cos\theta }{c}),
\end{equation}
where $ n' $ is the emitters density and $ P'_{\nu } $ is the
emitted spectral power per emitter, both are measured in the
fluid frame; $ \theta  $ is the angle relative to the line of
sight, and $ \Lambda ^{-1}=1/\gamma (1-v\, \, cos\theta /c) $ ($ v
$ is the emitting matter bulk velocity) is the blue-shift factor.

\textcite{NakarPiran03b} show\footnote{See \cite{GPS99a} for an
alternative method for integrating Eq. \ref{eq Fnu1}.} that using
the self-similar nature of the BM profile (with an external
density $\propto r^{-k}$) one can reduce Eq. \ref{eq Fnu1} to:
\begin{equation} \label{eq Fnu general}
F_{\nu }(t)=\frac{1}{D^{2}}\int _{0}^{R_{max}(t)}A_{\nu
}(R)g_{\beta }(\widetilde{t},k)dR \ .
\end{equation}
The integration over $R$ is over the shock front of the BM
solution. The upper limit $ R_{max} $ corresponds to the shock
position from  where photons leaving along the line of sight reach
the observer at $t$. The factor $ D $ is the distance to the
source (neglecting cosmological factors). $ \beta  $ is the local
spectral index.

The factor $ g_{\beta } $ is a dimensionless factor that describes
the observed pulse shape of an instantaneous emission from a BM
profile. The instantaneous emission from a thin shell produces a
finite pulse (see \S \ref{sec:Temporal} and Fig.
\ref{fig:thinshell}). This is generalized now to a pulse from an
instantaneous emission from a BM profile. Note that even though
the BM profile extends from $0$ to $R$ most of the emission arise
from a narrow regions of width $\sim R/\Ga^2$ behind the shock
front. $ g_{\beta } $ is obtained by integration Eq. \ref{eq Fnu1}
over $ cos\theta $ and $ r $, i.e. over the volume of the BM
profile. It depends only on the radial and angular structure of
the shell. The self-similar profile of the shell enables us to
express $ g_{\beta } $ as a general function that depends only on
the dimensionless parameter $ \widetilde{t}\equiv
{(t-t_{los}(R))}/{t_{ang}(R)}$, with $t_{los}(R)$ is the time in
which a photon emitted at R along the line of sight to the center
reaches the observer and $t_{ang} \equiv R/2c\Gamma^{2}$. The
second function, $ A_{\nu }$, depends only on the conditions of
the shock front along the line-of-sight. It includes only
numerical parameters that remain after the integration over the
volume of the shell.

When all the significant emission from the shell at radius $ R $
is within the same power-law segment, $ \beta  $, (i.e $ \nu  $
is far from the break frequencies) then $ A_{\nu } $ and $
g_{\beta } $ are given by:
\begin{equation} \label{eq Anu
general} A_{\nu }(R)=H_{\nu }\left\{ \begin{array}{c} R^{2}\,
n^{4/3}_{ext,0}\, E_{52}^{1/3}\, M_{29}^{-1/3}\quad \nu <\nu
_{m}\\
R^{2}\, n^{(5+p)/4}_{ext,0}\, E_{52}^{p}\, M_{29}^{-p}\quad \nu
_{m}<\nu <\nu _{c}\\
R\, n^{(2+p)/4}_{ext,0}\, E_{52}^{p}\, M_{29}^{-p}\quad \nu >\nu
_{c}
\end{array}\right. \frac{erg}{sec\cdot cm\cdot Hz},
\end{equation}
where $ R $ is the radius of the shock front, $ n_{ext}(R) $ is
the external density, $ E $ is the energy in the blast-wave, $
M(R) $ the total collected mass up to radius $ R $ and $ H_{\nu }
$ is a numerical factor which depends on the observed power law
segment (see \cite{NakarPiran03b} for the numerical values.

\begin{equation}
\label{eq general pulse} g(\widetilde{t},\beta ,k)=\left\{
\begin{array}{c} \frac{2}{(4-k)}\int
^{1+2(4-k)\widetilde{t}}_{1}\chi ^{-\mu (\beta ,k)}\left(
1-\frac{1}{2(4-k)}+\frac{2(4-k)\widetilde{t}+1}{2(4-k)\chi
}\right) ^{-(2-\beta )}d\chi \quad \nu <\nu _{c}\\
(1+\widetilde{t})^{-(2-\beta )}\quad \nu >\nu _{c}
\end{array}\right. ,
\end{equation}
 where
 \begin{equation}
\label{eq mu} \mu (\beta ,k)\equiv 3\cdot (71-17k)/(72-18k)-\beta
\cdot (37+k)/(24-6k).
\end{equation}
 This set of equations is completed with the relevant relations
between the different variables of the blast wave, the observer
time and the break frequencies.

These equations describe the light curve within one power law
segment of the light curve. Matching between different power laws
can be easily done  \cite{NakarPiran03b}. The overall formalism
can be used to calculate the complete light curve of a BM blast
wave.

\subsubsection{The light curve with a variable density or energy}
\label{sec:variable_light}

The results of the previous section can be applied to study the
effect of variations in the external density or in the energy of
the blast-wave by approximating the solution as a series of
instantaneous BM solutions whose parameters are determined by the
instantaneous external density and the energy. Both can vary with
time. This would be valid, of course, if the variations are not
too rapid. The light curve can be expressed as an integral over
the emission from a series of instantaneous BM solutions.

When a blast wave at radius $ R $ propagates into the circumburst
medium, the emitting matter behind the shock is replenished within
$ \Delta R\approx R(2^{1/(4-k)}-1) $. This is the length scale
over which an external density variation relaxes to the BM
solution. This approximation is valid as long as the density
variations are on a larger length scales than $ \Delta R $. It
fails when there is a sharp density increase over a range of $
\Delta R $. However, the contribution to the integral from the
region on which the solution breaks is small ($ \Delta R/R\ll 1 $)
and the overall light curve approximation is acceptable.
Additionally  the density variation must be mild enough so that it
does not give rise to a strong reverse shock that destroys the BM
profile.

A sharp density decrease is more complicated. Here the length
scale in which the emitting matter behind the shock is
replenished could be of the order of $ R $. As an example we
consider a sharp drop at some radius $ R_{d} $ and a constant
density for $ R>R_{d} $. In this case the external density is
negligible at first, and the hot shell cools by adiabatic
expansion. Later the forward shock becomes dominant again.
\textcite{KumarPanaitescu00a} show that immediately after the drop
the light curve is dominated by the emission during the adiabatic
cooling. Later the the observed flux is dominated by emission
from $ R\approx R_{d} $, and at the end the new forward shock
becomes dominant. Our approximation includes the emission before
the density drop and the new forward shock after the drop, but it
ignores the emission during the adiabatic cooling phase.

\begin{figure}[htb]
\begin{center}
\epsfig{file=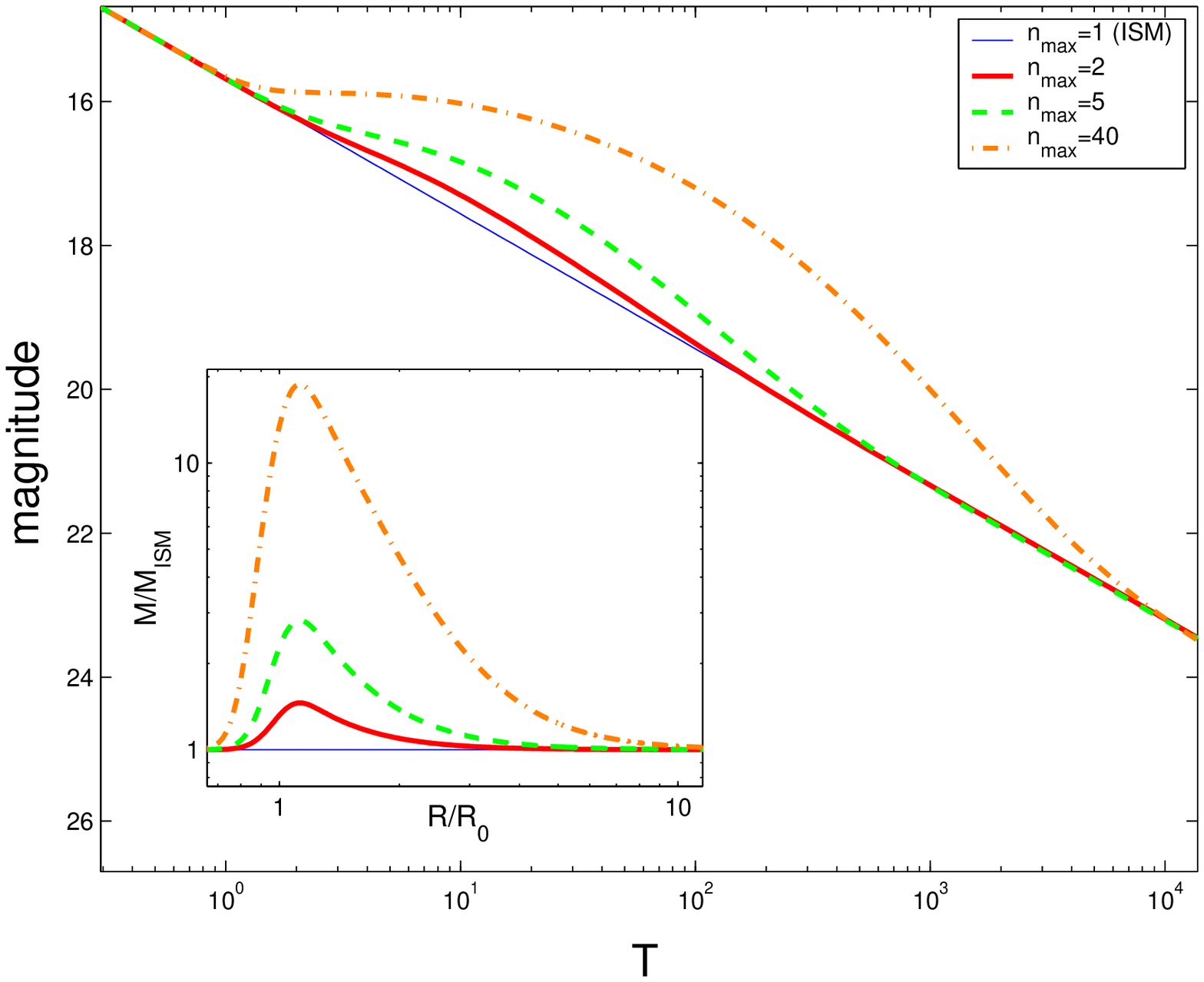,height=2in} \caption{The light curves
results from a Gaussian ($\Delta R/R=0.1$) over-dense region in
the ISM. The different thick lines are for a maximal
over-densities of $40$ ({\it dashed-dot}), $5$ ({\it dashed}) and
$2$ ({\it solid}). The thin line is the light curve for a constant
ISM density. The inset depicts the ratio of the mass, $M(R)$ over
the mass of an ISM (without the over-dense region), $M_{ISM}(R)$
(from \textcite{NakarPiran03b})} \label{fig:Gaussian over-density}
\end{center}
\end{figure}

As an example for this method  Fig \ref{fig:Gaussian over-density}
depicts the $ \nu _{m}<\nu <\nu _{c} $ light curve for a Gaussian
($\Delta R/R=0.1$) over-dense region in the ISM. Such a density
profile may occur in a clumpy environment. The emission from a
clump is similar to the emission from a spherically over-dense
region as long as the clump's angular size is much larger than $
1/\Gamma $. Even a mild short length-scale, over-dense region
(with a maximal over-density of 2)  influences the light curve for
a long duration (mainly due to the angular spreading). This
duration depends strongly on the magnitude of the over-density.

The calculations presented so far do not account, however, for the
reverse shock resulting from density enhancement and its effect on
the blast-wave. Thus the above models are limited to slowly
varying and low contrast density profiles. Now, the observed flux
depends on the external density, $n$, roughly as $n^{1/2}$. Thus,
a large contrast is needed to produce a significant
re-brightening. Such a large contrast will, however, produce a
strong reverse shock which will sharply decrease the Lorentz
factor of the emitting matter behind the shock, $\Gamma_{sh}$,
causing a sharp drop in the emission below $\nu_c$ and a long
delay in the arrival time of the emitted photons (the observer
time  is $\propto \Gamma_{sh}^{-2}$). Both factors combine to
suppresses the flux and to set a strong limit on the steepness of
the re-brightening events caused by density variations.

The method can be applied also to variations in the blast wave's
energy.  Spherically symmetric energy variations are most likely
to occur due to refreshed shocks, when new inner shells arrive
from the source and refresh the blast wave
\cite{ReesMeszaros98,KP00a,SariMeszaros00}. Once more, this
approximation misses the effect of the reverse shock that arise in
this case \cite{KP00a}. However it enables a simple calculation of
the observed light curve for a given energy profile.

\subsection{Generalizations: IV. Jets }
\label{sec:jets}
 The afterglow theory becomes much more
complicated if the relativistic ejecta is not spherical. The
commonly called ``jets" corresponds to relativistic matter ejected
into a cone of opening angle $\theta$. I stress that unlike other
astrophysical jets this ejecta is non steady state and generally
its width (in the direction parallel to the motion)  is orders of
magnitude smaller than the radius where the jet is. A ``flying
pancake" is a better description for these jets.

The simplest implication of a jet geometry, that exists regardless
of the hydrodynamic evolution, is that once $\Gamma \sim
\theta^{-1}$ relativistic beaming of light will become less
effective. The radiation was initially beamed locally into a cone
with an opening angle $\Gamma^{-1}$ remained inside the cone of
the original jet. Now with $\Gamma^{-1}> \theta$ the emission is
radiated outside of the initial jet. This has two effects: (i)  An
``on axis" observer, one that sees the original jet, will detect a
jet break due to the faster spreading of the emitted radiation.
(ii) An ``off axis" observer, that could not detect the original
emission will be able to see now an ``orphan afterglow", an
afterglow without a preceding GRB (see \S \ref{sec:orphan}). The
time of this transition is always give by Eq. \ref{tjet} below
with $C_2=1$.

Additionally the hydrodynamic evolution of the source changes when
$\Gamma \sim \theta^{-1}$. Initially, as long as $\Gamma \gg
\theta^{-1}$ \cite{Pi94} the motion would be almost conical. There
isn't enough time, in the blast wave's rest frame, for the matter
to be affected by the non spherical geometry, and the blast wave
will behave as if it was a part of a sphere. When $\Gamma = C_2
\theta^{-1}$, namely at\footnote{The exact values of the uncertain
constants $C_2$ and $C_1$ are extremely important as they
determine the jet opening angle (and hence the total energy of the
GRB) from the observed breaks, interpreted as $t_{\rm jet}$, in
the afterglow light curves.}:
\begin{equation}
t_{\rm jet}  =   {1\over C_1}\left( l\over c \right )
\left({\theta\over C_2}\right)^{2(4-k)\over (3-k)}
\end{equation}
sideways propagation begins. The constant $C_1$ expresses the
uncertainty at relation between the Lorentz factor and the
observing time and it depends on the history of the evolution of
the fireball. The constant $C_2$ reflects the uncertainty in the
value of $\Gamma$, when the jet break begins vs. the value of
opening angle of the jet $\theta$. For the important case of
constant external density $k=0$ this transition takes place at:
\begin{equation}
t_{\rm jet}=  {1 \, {\rm day} \over C_1 C_2^{8/3}} \left({E_{\rm
iso,52}\over n_1}\right)^{1/3} \left({\theta\over
0.1}\right)^{8/3} \ . \label{tjet}
\end{equation}

The sideways expansion continues with $\theta \sim \Gamma^{-1}$.
Plugging this relations in Eq. \ref{RGamma} and letting $\Omega$
vary like $\Ga^{-2}$ one finds that:
\begin{eqnarray}
R \approx {\rm const} \ ; \\ \nonumber \Ga \approx (R/2 t)^{1/2}
\ . \label{Rgammajet}
\end{eqnarray}
A more detailed analysis
\cite{Rhoads97,Rhoads99,P00,KumarPanaitescu00a} reveals that
according to the simple one dimensional analytic models $\Gamma$
decreases exponentially with $R$ on a very short length
scale.\footnote{Note that the exponential behavior is obtained
after converting Eq. \ref{ad} to a differential equation and
integrating over it. Different approximations used in deriving the
differential equation lead to slightly different exponential
behavior, see \cite{P00}.}

Table \ref{table:Slowjet} describes the parameters $\alpha$ and
$\beta$ for a post jet break  evolution \cite{SPH99}, The jet
break usually takes place rather late, after the radiative
transition. Therefore, I include in this table only the slow
cooling parameters.
\begin{table}
\label{table:Slowjet}
  \centering
\begin{tabular}{|c|c|c|} \hline
  & $\alpha$ & $\beta$ \\ \hline
  $\nu < \nu_a$ & 0 & 2 \\ \hline
  $\nu_a < \nu < \nu_m $ & -1/3 & 1/3 \\ \hline
  $\nu_m < \nu < \nu_c $ & -p   & $-(p-1)/2=(\alpha+1)/2$ \\ \hline
   $\nu_c < \nu $ & -p & $-p/2=\alpha/2$ \\ \hline
\end{tabular}
\caption{$\alpha$ and $\beta$ for slow cooling ($\nu_a< \nu_m <
\nu_c $) after a jet break.}
\end{table}

\begin{figure}[htb]
\begin{center}
\epsfig{file=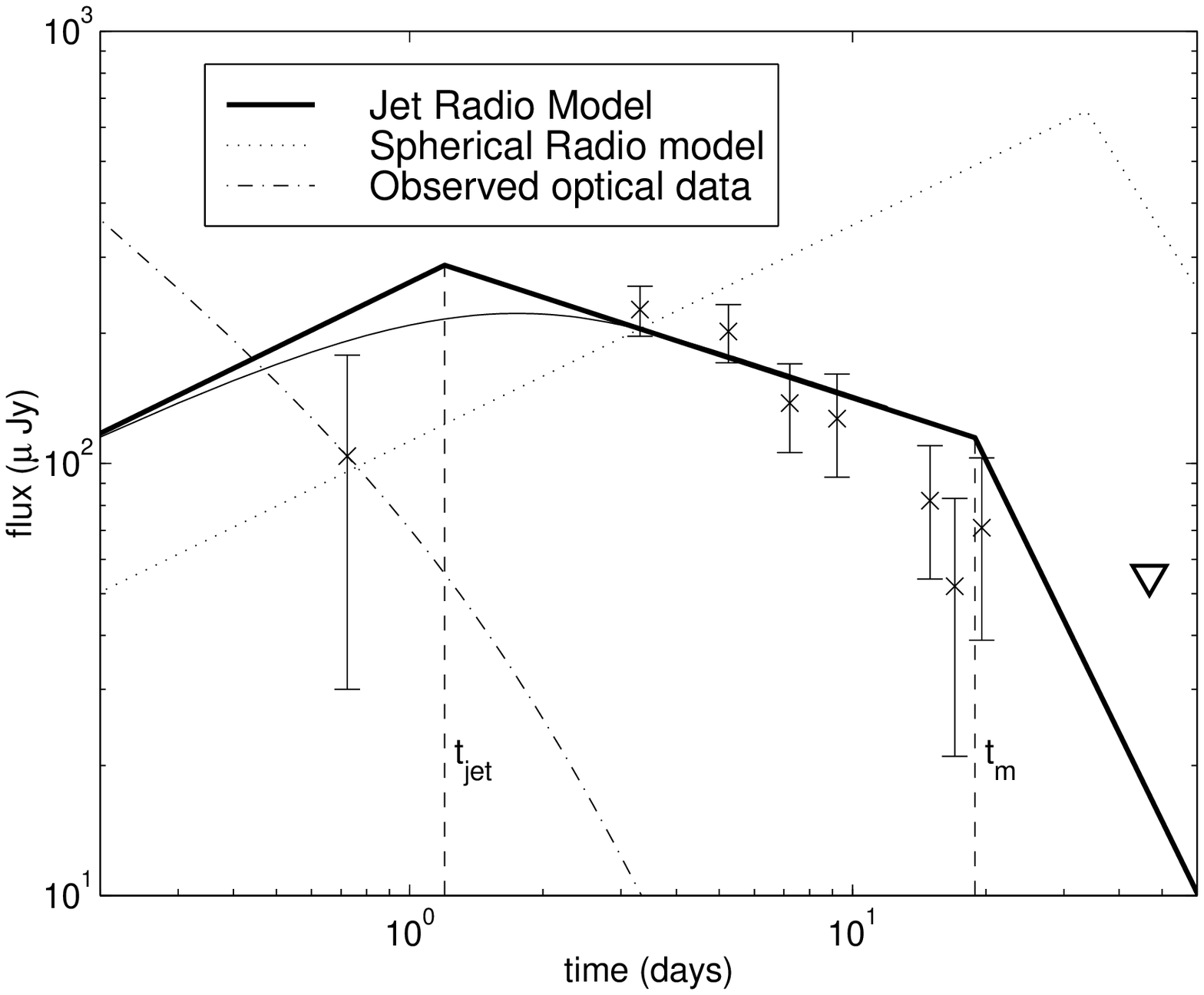,height=2in} \caption{Observed and
predicted light curve at 8.6Ghz  light curves of GRB990510 (from
\textcite{Harrisonetal99}). The different behavior of the optical
and radio light curves after the jet break is clearly seen.}
\label{fig:990510_radio}
\end{center}
\end{figure}

An important feature of the post jet-break evolution is that
$\nu_c$, the cooling frequency  becomes constant in time . This
means that the high frequency (optical and \xr) optical spectrum
does not vary after the jet-break took place. On the other hand
the radio spectrum varies (see Fig. \ref{fig:990510_radio}),
giving an additional structure that confirms the interpretation of
break as arising due to a sideways expansion of a jet (see e.g.
\cite{Harrisonetal99}).

\textcite{KumarPanaitescu00c} find that the jet break transition
in a wind profile will be very long (up to four decades in time)
and thus it will be hard to observe a jet break in such a case. On
the other hand it is interesting to note that for typical values
of $\alpha$ seen after a jet break ($\alpha \approx -2$) the high
frequency spectral index, $\beta=\alpha/2 \approx -1$, is similar
to the one inferred from a spherically symmetric wind $\beta =(2
\alpha + 1) /3 \approx -1$ \cite{Halpernetal99}. Note however,
that the wind interpretation requires a high ($\approx 3$) p value
(which may or may not be reasonable).  Still from the optical
observations alone  it is difficult to distinguish between these
two interpretations. Here the radio observations play a crucial
role as the radio behavior is very different \cite{Frail00}.

The sideways expansion causes a change in the hydrodynamic
behavior and hence a break in the light curve. The beaming outside
of the original jet opening angle also causes a break. If the
sideways expansion is at the speed of light than both transitions
would take place at the same time \cite{SPH99}. If the sideways
expansion is at the sound speed then the beaming transition would
take place first and only later the hydrodynamic transition would
occur \cite{PanaitescuMeszaros99}.  This would cause a slower and
wider transition with two distinct breaks, first a steep break
when the edge of the jet becomes visible and later a shallower
break when sideways expansion becomes important.

The  analytic or semi-analytic calculations of synchrotron
radiation from jetted afterglows
\cite{Rhoads99,SPH99,PanaitescuMeszaros99,ModerskiEtal00,KumarPanaitescu00a}
have led to different estimates of  the jet break time $t_{\rm
jet}$ and of the duration of the transition.  \textcite{Rhoads99}
calculated the light curves assuming emission from one
representative point, and obtained a smooth `jet break', extending
$\sim 3-4$ decades in time, after which $F_{\nu>\nu_m}\propto
t^{-p}$. \textcite{SPH99} assume that the sideways expansion is at
the speed of light, and not at the speed of sound ($c/\sqrt{3}$)
as others assume, and find a  smaller value for $t_{\rm jet}$.
\textcite{PanaitescuMeszaros99} included the effects of
geometrical curvature and finite width of the emitting shell,
along with electron cooling, and obtained a relatively sharp
break, extending $\sim 1-2$ decades in time, in the optical light
curve. \textcite{ModerskiEtal00} used a slightly different
dynamical model, and a different formalism for the evolution of
the electron distribution, and obtained that the change in the
temporal index $\alpha$ ($F_{\nu}\propto t^{-\alpha}$) across the
break is smaller than in analytic estimates ($\alpha=2$ after the
break for $\nu>\nu_m$, $p=2.4$), while the break extends over two
decades in time.

\begin{figure}
\begin{center}
\includegraphics[width=4.97cm]{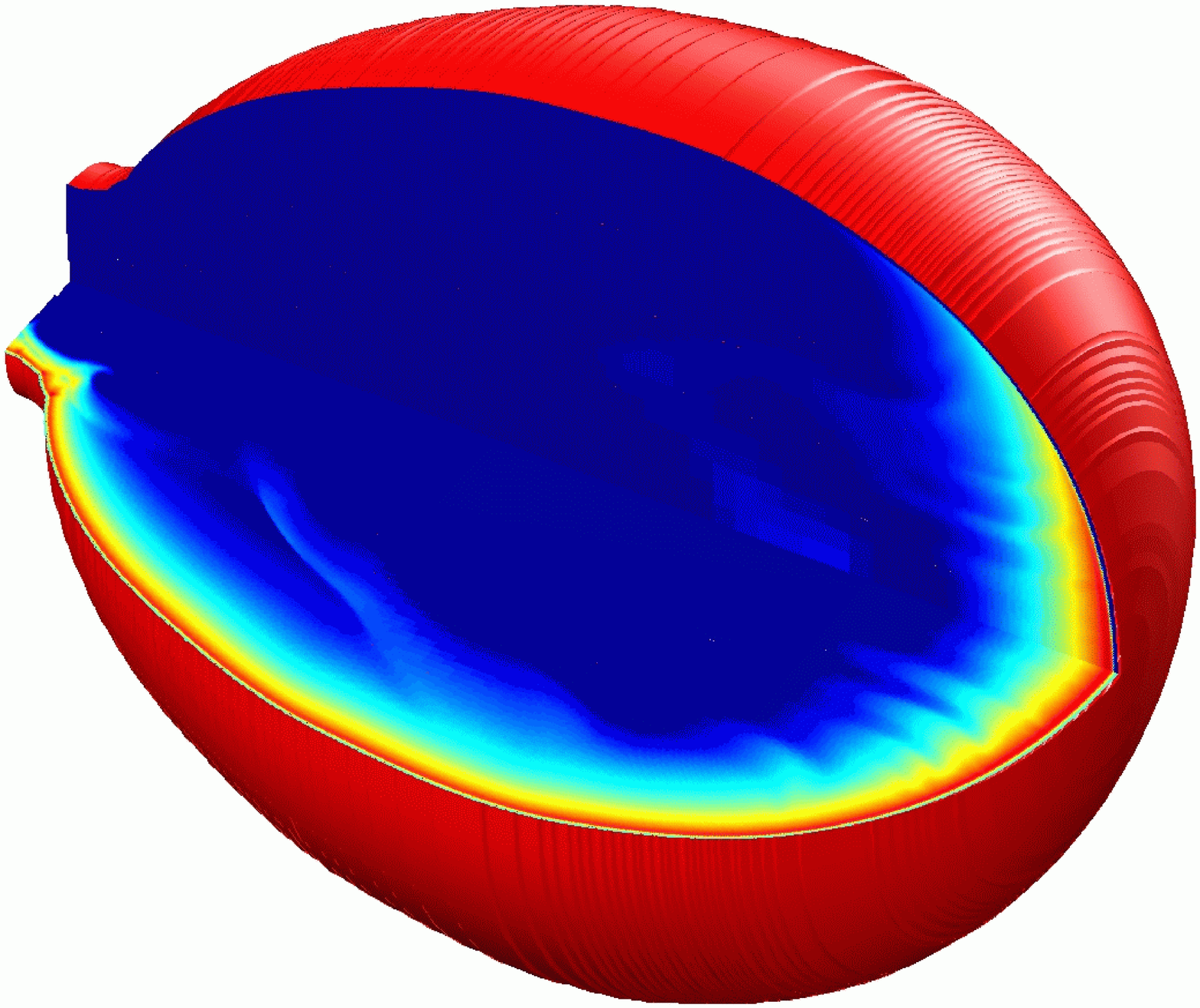}
\includegraphics[width=7.0cm]{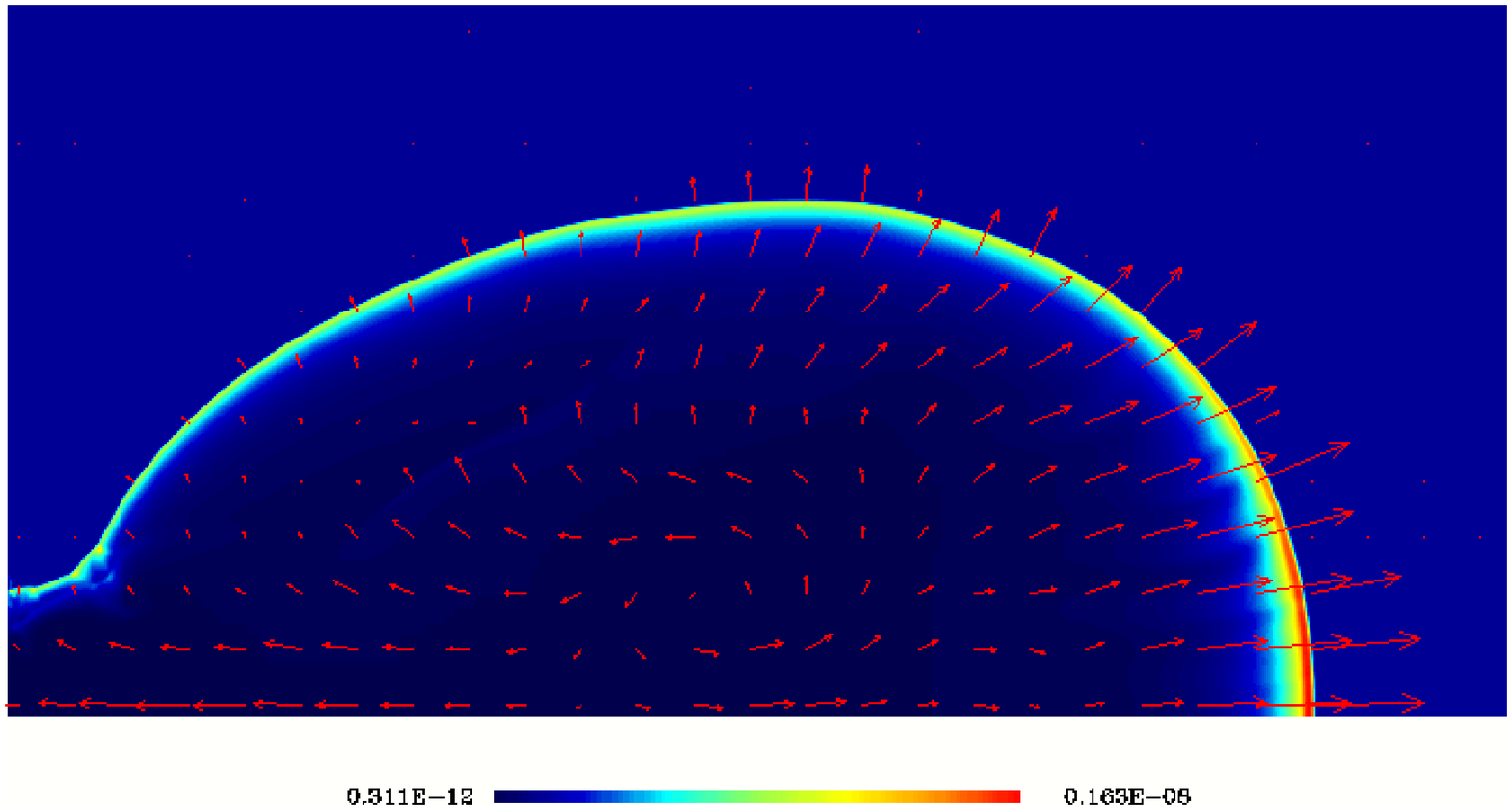}
\end{center}
\caption[]{A relativistic jet at the last time
  step of the simulation \cite{Granot01}.  ({\bf left}) A 3D view
  of the jet. The outer surface represents the shock front while the
  two inner faces show the proper number density ({\it lower face})
  and proper emissivity ({\it upper face}) in a logarithmic color
  scale. ({\bf right}) A 2D 'slice' along the jet axis, showing the
  velocity field on top of a linear color-map of the lab frame
  density.}
\label{3Djet}
\end{figure}

The different analytic or semi-analytic models have different
predictions  for the sharpness of the `jet break', the change in
the temporal decay index $\alpha$ across the break and its
asymptotic value after the break, or even the very existence a
`jet break' \cite{HDL00}. All these models rely on some common
basic assumptions, which have a significant effect on the dynamics
of the jet: (i) the shocked matter is homogeneous (ii) the shock
front is spherical (within a finite opening angle) even at
$t>t_{\rm jet}$ (iii) the velocity vector is almost radial even
after the jet break.

\begin{figure}
\begin{center}
\includegraphics[width=3.97cm]{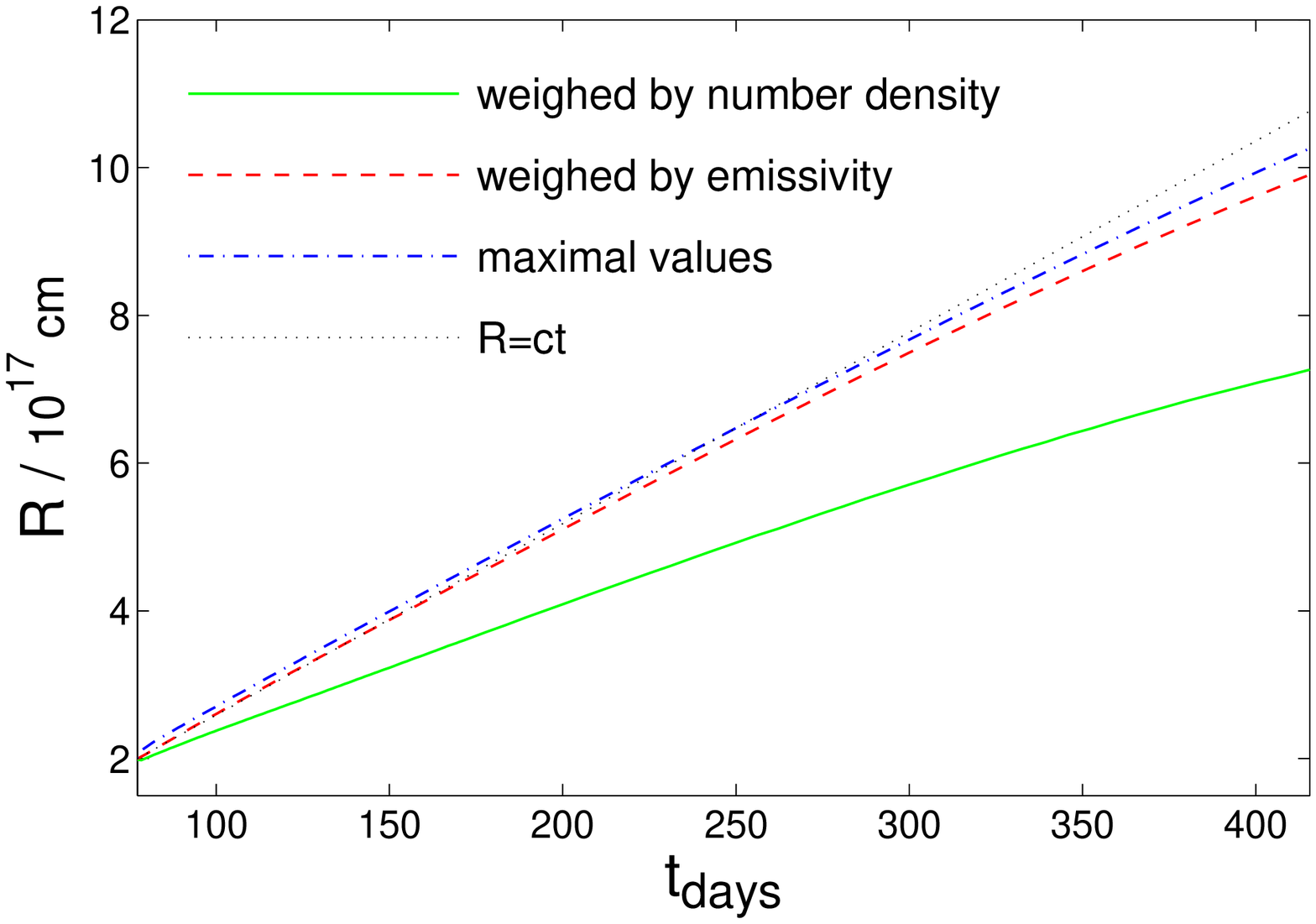}
\includegraphics[width=4.05cm]{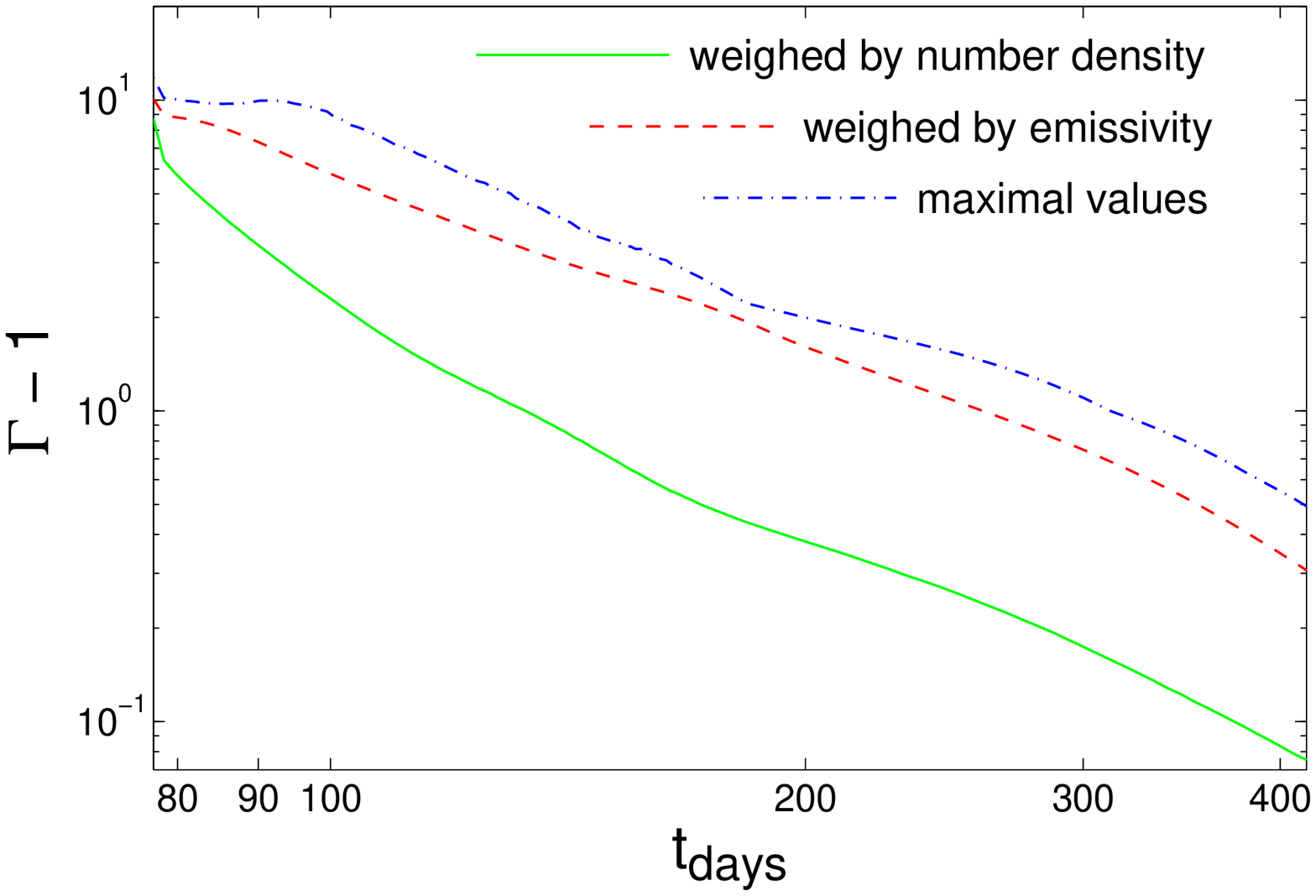}
\includegraphics[width=3.97cm]{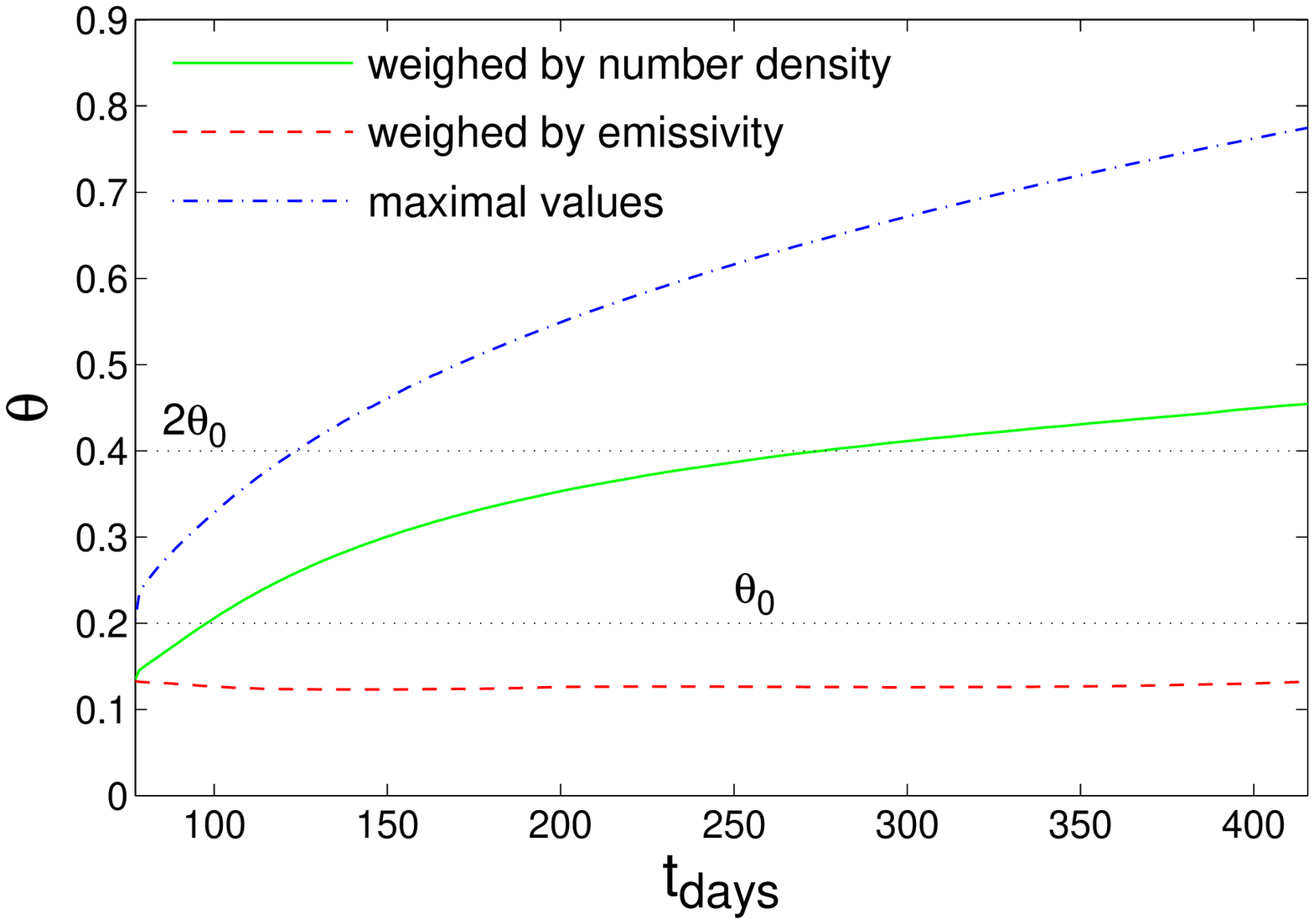}
\end{center}
\caption[]{The radius $R$ ({\it left frame}), Lorentz factor
$\Gamma-1$
  ({\it middle frame}) and opening angle $\theta$ of the jet ({\it
    right frame}), as a function of the lab frame time in days \cite{Granot01}.}
\label{averages}
\end{figure}

However, recent 2D hydrodynamic simulations \cite{Granot01} show
that these assumptions are not a good approximation of a realistic
jet. Using a very different approach \textcite{Cannizzoetal04}
find in another set of numerical simulations a similar result -
the jet does not spread sideways as much. Figure \ref{3Djet} shows
the jet at the last time step of the simulation of
\textcite{Granot01}. The matter at the sides of the jet is
propagating sideways (rather than in the radial direction) and is
slower and much less luminous compared to the front of the jet.
The shock front is egg-shaped, and quite far from being spherical.
Figure \ref{averages} shows the radius $R$, Lorentz factor
$\Gamma$, and opening angle $\theta$ of the jet, as a function of
the lab frame time. The rate of increase of $\theta$ with
$R\approx ct_{\rm lab}$, is much lower than the exponential
behavior predicted by simple models
\cite{Rhoads97,Rhoads99,P00,KumarPanaitescu00a}. The value of
$\theta$ averaged over the emissivity is practically constant, and
most of the radiation is emitted within the initial opening angle
of the jet. The radius $R$ weighed over the emissivity is very
close to the maximal value of $R$ within the jet, indicating that
most of the emission originates at the front of the
jet\footnote{This may imply that the expected rate of orphan
afterglows should be smaller than estimated assuming significant
sideways expansion!}, where the radius is largest, while $R$
averaged over the density is significantly lower, indicating that
a large fraction of the shocked matter resides at the sides of the
jet, where the radius is smaller. The Lorentz factor $\Gamma$
averaged over the emissivity is close to its maximal value, (again
since most of the emission occurs near the jet axis where $\Gamma$
is the largest) while $\Gamma$ averaged over the density is
significantly lower, since the matter at the sides of the jet has
a much lower $\Gamma$ than at the front of the jet. The large
differences between the assumptions of simple dynamical models of
a jet and the results of 2D simulations, suggest that great care
should be taken when using these models for predicting the light
curves of jetted afterglows. Since the light curves depend
strongly on the hydrodynamics of the jet, it is very important to
use a realistic hydrodynamic model when calculating the light
curves.

\textcite{Granot01} used 2D numerical simulations of a jet running
into a constant density medium to calculate the resulting light
curves, taking into account the emission from the volume of the
shocked fluid with the appropriate time delay in the arrival of
photons to different observers. They obtained an achromatic jet
break for $\nu>\nu_m(t_{\rm jet})$ (which typically includes the
optical and near IR), while at lower frequencies (which typically
include the radio) there is a more moderate and gradual increase
in the temporal index $\alpha$ at $t_{\rm jet}$, and a much more
prominent steepening in the light curve at a latter time when
$\nu_m$ sweeps past the observed frequency. The jet break appears
sharper and occurs at a slightly earlier time for an observer
along the jet axis, compared to an observer off the jet axis (but
within the initial opening angle of the jet). The value of
$\alpha$ after the jet break, for $\nu>\nu_m$, is found to be
slightly larger than $p$ ($\alpha=2.85$ for $p=2.5$). Due to the
fact that a significant fraction of the jet break occurs due to
the relativistic beaming effect (that do not depend on the
hydrodynamics) in spite of the different hydrodynamic behavior the
numerical simulations show a jet break at roughly the same time as
the analytic estimates.

\subsection{Generalizations: V. Angular Dependent Jets and the Structured
Jet Model } \label{sec:structured}

In a realistic jet one can expect either a random or regular
angular dependent structure. Here there are two dominant effects.
As the ejecta slows down its Lorentz factor decreases and an
observer will detect radiation from an angular region of size
$\Gamma^{-1}$ (see \S \ref{sec:patchy-shell}). At the same time
the mixing within the ejecta will lead to an intrinsic averaging
of the angular structure. Thus, both effects lead to an averaging
over the angular structure at later times.

Several authors \cite{Lipunov_Postnov_Pro01,Rossi02,Zhang02}
suggested independently a  different interpretation to the
observed achromatic breaks in the afterglow light curves. This
interpretation is based on a jet with a regular angular structure.
According to this model all GRBs are produced by a jet with a
fixed angular structure and the break corresponds to the viewing
angle. \textcite{Lipunov_Postnov_Pro01} considered a ``universal"
jet with  a 3-step profile: a spherical one, a $20^o$ one and a
$3^o$ one. \textcite{Rossi02} and \textcite{Zhang02} considered a
special profile where the energy per solid angle,
$\varepsilon(\theta) $, and the Lorentz factor
$\Gamma(t=0,\theta)$ are:
\begin{equation}
\varepsilon=\left\{\begin{array}{ll}
             \varepsilon_{c} & \;\;\;0 \leq \theta \leq \theta_{c}\\
             \varepsilon_{c}\big(\frac {\theta}{\theta_{c}})^{-a} &
\;\;\;\theta_{c}
             \leq \theta \leq \theta_{j}
    \end{array}\right.
\label{eq:Etheta}
\end{equation}
and
\begin{equation}
\Gamma=\left\{\begin{array}{lll}
       \Gamma_{c}& &\;\;\;0 \leq \theta \leq \theta_{c}\\
       \Gamma_{c}\big(\frac
{\theta}{\theta_{c}})^{-b},&b>0 & \;\;\; \theta_{c} \leq \theta
\leq \theta{j},
     \end{array}\right.
\label{eq:Gtheta}
\end{equation}
where $\theta_j$ is a maximal angle and core angle, $\theta_c$, is
introduced  to avoid a divergence at $\theta=0$ and the parameters
$a$ and $b$ here define the energy and Lorentz factor angular
dependence. This core angle can be taken to be smaller than any
other angle of interest. The power law index of $\Gamma$, $b$, is
not important for the dynamics of the fireball and the computation
of the light curve as long as
$\Gamma(t=0,\theta)\equiv\Gamma_{0}(\theta)>\theta^{-1}$ and
$\Gamma_{0}(\theta) \gg 1$.

\begin{figure}[htb]
\begin{center}
\epsfig{file=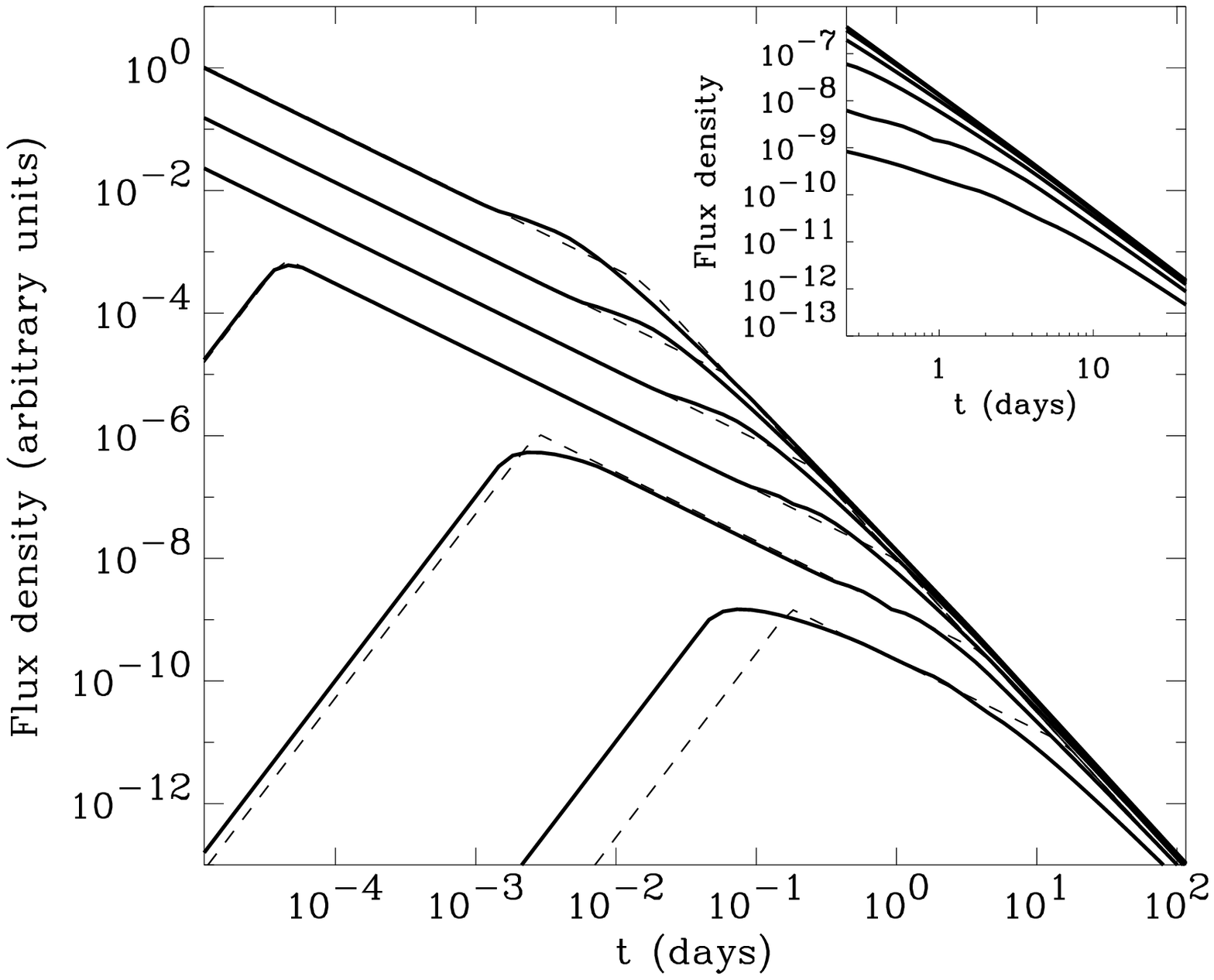,height=2in} \caption{{The light curves
of an inhomogeneous jet observed from different angles (From
\textcite{Rossi02}). From the top
$\theta_{o}=0.5,\;1,\;2,\;4,\;8,\;16^{\circ}$.
 The break time is related only to the observer angle:
$t_{b}\propto \theta_{o}^{2}$. The dashed line is the on-axis
light curve of an homogeneous jet with an opening angle
$2\theta_{o}$ and an energy per unit solid angle
$\varepsilon(\theta_{o})$. The blow up is the time range between 4
hours and 1 month, where most of the optical observations are
performed.  Comparing the solid and the dashed lines for a fixed
$\theta_{o}$, it is apparent that one can hardly distinguish the
two models by fitting the afterglow data.}} \label{fig:Rossi}
\end{center}
\end{figure}

To fit the constant energy result
\cite{Frail01,PanaitescuK01,Piranetal01} \textcite{Rossi02}
consider a specific angular structure with $a=2$.
\textcite{Rossi02} approximate the evolution assuming that at
every angle the matter behaves as if it is a part of a regular BM
profile (with the local $\varepsilon$ and $\Gamma(t,\theta)$)
until $\Gamma(t,\theta) =\theta^{-1}$. Then the matter begins to
expand sideways. The resulting light curve is calculated by
averaging the detected light resulting from the different angles.
They find that an observer at an angle $\theta_o$ will detect a
break in the light curve that appears around the time that
$\Gamma(t,\theta_o)=\theta_o^{-1}$ (see Fig. \ref{fig:Rossi}). A
simple explanation of break is the following: As the evolution
proceeds and the Lorentz factor decreases an observer will detect
emission from a larger and larger angular regions. Initially the
additional higher energy at small angles, $\theta<\theta_o$
compensates over the lower energies at larger angles $\theta>
\theta_o$. Hence the observer detects a roughly constant energy
per solid angle and the resulting light curve is comparable to the
regular pre-jet break light curve.  This goes on until
$\Gamma^{-1}(0)=\theta_{(o)}$. After this stage an further
increase in the viewing angle $\Gamma^{-1}$ will result in a
decrease of the energy per unit solid angle within the viewing
cone and will lead to a break in the light curve.

This interpretation of the breaks in the light curves in terms of
the viewing angles of a standard structured jets implies a
different understanding of total energy within GRB jets and of the
rate of GRBs. The total energy in this model is also a constant
but now it is larger as it is the integral of Eq. \ref{eq:Etheta}
over all viewing angles. The distribution of GRB luminosities,
which is interpreted in the uniform jet interpretation as a
distribution of jet opening angles is interpreted here as a
distribution of viewing angles. As such this distribution is fixed
by geometrical reasoning with $P(\theta_o) d\theta_o \propto
\sin\theta_o d\theta_o$ (up to the maximal observing angle
$\theta_j$). This leads to an implied isotropic energy
distribution of
\begin{equation}
P(log(E_{iso})) \propto E^{-1}_{iso} \ .
\end{equation}
\textcite{GuettaPiranWaxman03} and \textcite{NakarGranotGuetta03}
find that these two distributions are somewhat inconsistent with
current observations. However the present data that suffers from
numerous observational biases  is insufficient to reach a definite
conclusions.

\begin{figure}[htb]
\begin{center}
\epsfig{file=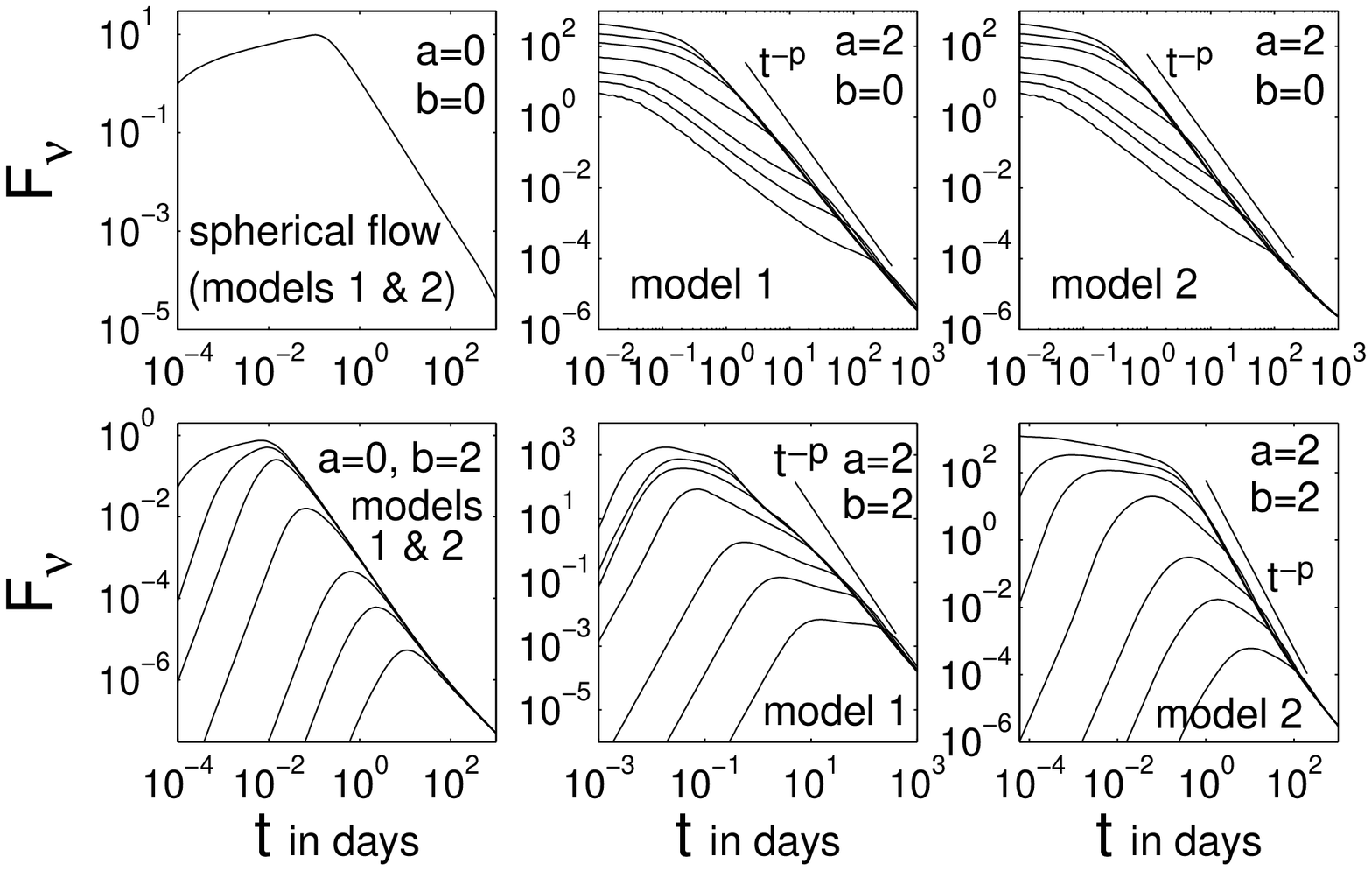,height=2in} \caption{Light curves for
structured jets (initially $\epsilon\propto\theta^{-a}$ and
$\Gamma\propto\theta^{-b}$), for models 1 and 2, in the optical
($\nu=5\times 10^{14}\;$Hz), for a jet core angle $\theta_c=0.02$,
viewing angles $\theta_{\rm obs}=0.01,0.03,0.05,0.1,0.2,0.3,0.5$,
$p=2.5$, $\epsilon_e=\epsilon_B=0.1$, $n=1\;{\rm cm}^{-3}$,
$\Gamma_0=10^3$, and $\epsilon_0$ was chosen so that the total
energy of the jet would be $10^{52}\;$erg (GK). A power law of
$t^{-p}$ is added in some of the panels, for comparison. From
\textcite{GranotKumarP03}} \label{fig:GranotKumar}
\end{center}
\end{figure}

In order to estimate better the role of the hydrodynamics on the
light curves of a structured jet
\textcite{GranotKumar02,GranotKumarP03} considered two simple
models for the hydrodynamics. In the first  (model 1) there is no
mixing among matter moving at different angles i.e.
$\varepsilon(\theta,t)=\varepsilon(\theta,t_0)$. In the second
(model 2) $\varepsilon$ is a function of time and it is averaged
over the region to which a sound wave can propagate (this
simulates the maximal lateral energy transfer that is consistent
with causality). They consider various energy and Lorentz factors
profiles and calculate the resulting light curves (see Fig.
\ref{fig:GranotKumar}).

\textcite{GranotKumar02} find that the light curves of models 1
and 2 are rather similar in spite of the different physical
assumptions. This suggests that the widening of the viewing angle
has a more dominant effect than the physical averaging.  For
models with a constant energy and a variable Lorentz factor
($(a,b)=(0,2)$) the light curve initially rises and there is no
jet break, which is quite different from observations for most
afterglows. For $(a,b)=(2,2),\,(2,0)$ they find a jet break at
$t_j$ when $\Gamma(\theta_o)\sim\theta_o^{-1}$. For $(a,b)=(2,2)$
the value, $\alpha_1$, of the temporal decay slope at $t < t_{j}$
increases with $\theta_o$, while $\alpha_2=\alpha(t>t_j)$
decreases with $\theta_o$. This effect is more prominent in model
1, and appears to a lesser extent in model 2. This suggests that
$\delta\alpha=\alpha_1-\alpha_2$ should increase with $t_j$,
which is not supported by observations. For $(a,b)=(2,0)$, there
is a flattening of the light curve just before the jet break
(also noticed by \textcite{Rossi02}), for $\theta_o > 3\theta_c$.
Again, this effect is larger in model 1, compared to model 2 and
again this flattening is not seen in the observed data.

Clearly a full solution of an angular dependent jet requires full
numerical simulations.  \textcite{KumarGranot03} present a simple
1-D model for the hydrodynamics that is obtained by assuming
axial symmetry and integrating over the radial profile of the
flow, thus considerably reducing the computation time. The light
curves that they find resemble those of models 1 and 2 above
indicating that these crude approximations are useful. Furthermore
they find relatively little change in $\epsilon(\theta)$ within
the first few days, suggesting that model 1 is an especially
useful approximation for the jet dynamics at early times, while
model 2 provides a better approximation at late times.

\subsection{Afterglow Polarization - a tool that distinguished between the different
jet models} \label{sec:pol_after}

Synchrotron emission from a jet (in which the spherical symmetry
is broken) would naturally produce polarized emission
\cite{Gruzinov99,GhiselliniLazzati99,Sari99}. Moreover, the level
and the direction of the polarization are expected to vary with
time and to give observational clues on the geometrical structure
of the emitting jet and our observing angle with respect to it.

\begin{figure}
\centerline{ \epsfxsize15pc \epsfbox{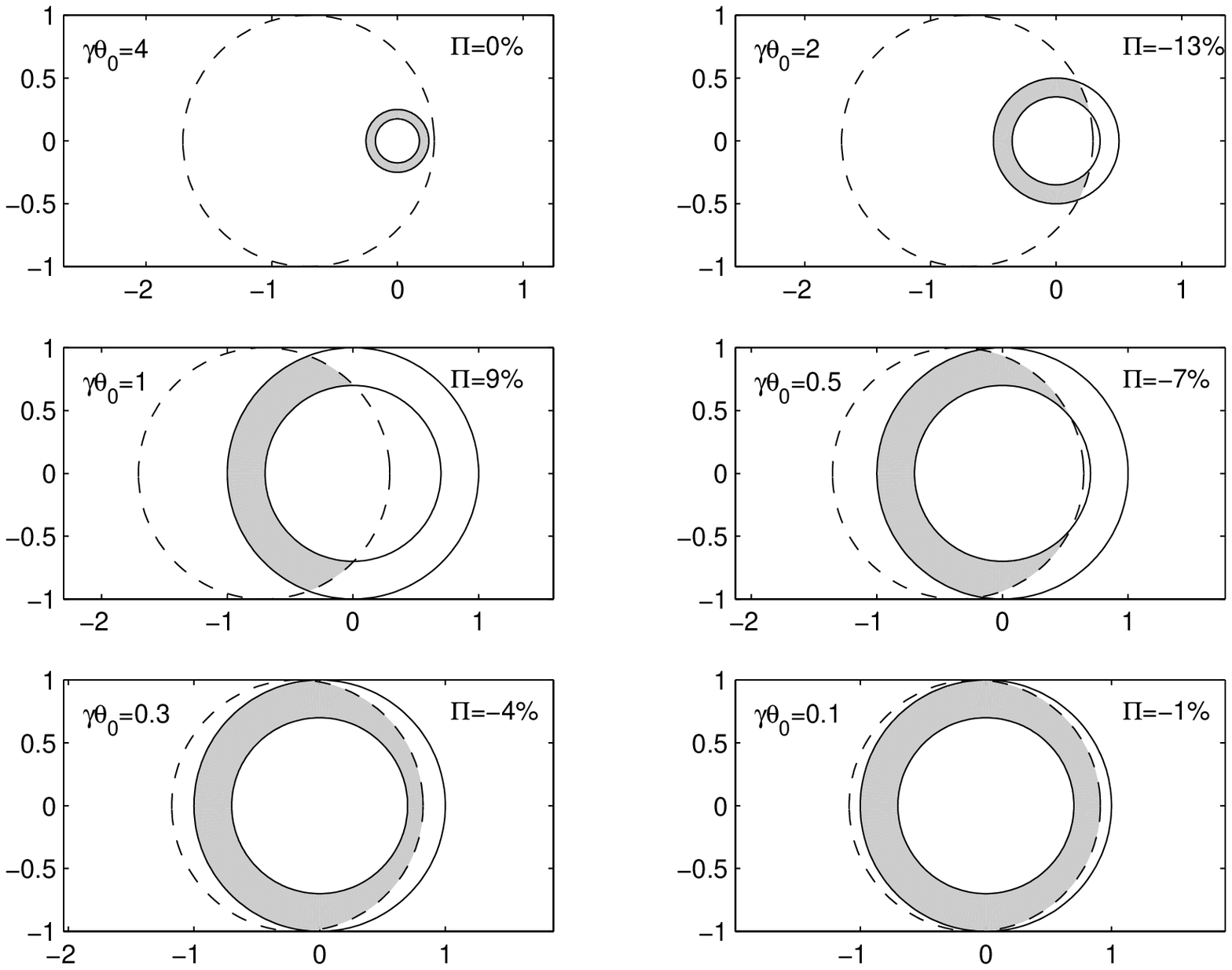}
\epsfxsize15pc \epsfbox{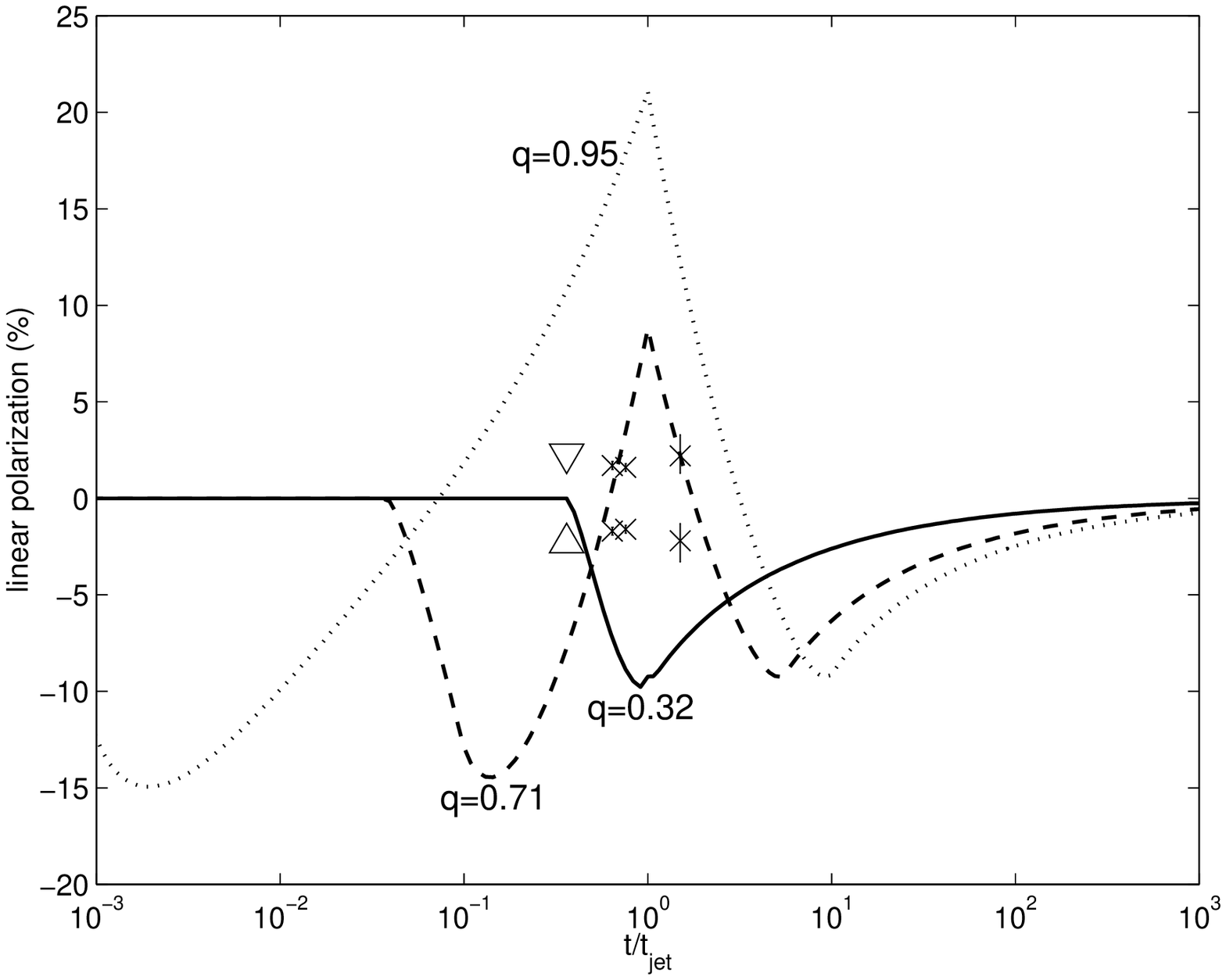}} \vspace{10pt} \caption{
Left: Shape of the emitting region.  The dashed line marks the
physical extent of the jet, and solid lines give the viewable
region $1/\gamma$. The observed radiation arises from the gray
shaded region.  In each frame, the percentage of polarization is
given at the top right and the initial size of the jet relative to
$1/\Gamma$ is given on the left.  The frames are scaled so that
the size of the jet is unity.  Right: Observed and theoretical
polarization light curves for three possible offsets of the
observer relative to the jet axis Observational data for
GRB\,990510 is marked by crosses (x), assuming
$t_{jet}=1.2$\,days.  The upper limit for GRB\,990123 is given by
a triangle, assuming $t_{jet}=2.1$\,days (from \cite{Sari99}).}
\label{fig:polfig}
\end{figure}

The key feature in the determination of the polarization during
the afterglow is the varying Lorentz factor and (after jet break)
varying jet width. This changes changes the overall geometry (see
Fig. \ref{fig:pol_random})   and hence the observer sees different
geometries  \cite{Sari99,Hurleyetal02}. Initially,  the
relativistic beaming angle $1/\Gamma$ is narrower than the
physical size of the jet $\theta_0$, and the observer see a full
ring and therefore the radial polarization averages out (the first
frame, with $\Gamma\theta_0=4$ of the left plot in Fig.
\ref{fig:polfig}). As the flow decelerates, the relativistic
beaming angle $1/\Gamma$ becomes comparable to $\theta_0$ and only
a fraction of the ring is visible; net polarization is then
observed. Assuming, for simplicity, that the magnetic field is
along the shock then the synchrotron polarization will be radially
outwards.  Due to the radial direction of the polarization from
each fluid element, the total polarization is maximal when a
quarter ($\Gamma\theta_0=2$ in Figure \ref{fig:polfig}) or when
three quarters ($\Gamma\theta_0=1$ in Figure \ref{fig:polfig}) of
the ring are missing (or radiate less efficiently) and vanishes
for a full and a half ring. The polarization, when more than half
of the ring is missing, is perpendicular to the polarization
direction when less than half of it is missing.

At late stages the jet expands sideways and since the offset of
the observer from the physical center of the jet is constant,
spherical symmetry is regained. The vanishing and re-occurrence of
significant parts of the ring results in a unique prediction:
there should be three peaks of polarization, with the polarization
position angle during the central peak rotated by $90^{\circ }$
with respect to the other two peaks. In case the observer is very
close to the center, more than half of the ring is always
observed, and therefore only a single direction of polarization is
expected. A few possible polarization light curves are presented
in Fig. \ref{fig:polfig}.

The predicted polarization from a structured jet is drastically
different from the one from a uniform jet,  providing an excellent
test between the two models \cite{Rossietalpolarization02}. Within
the structured jet model the polarization arises due to the
gradient in the emissivity. This gradient has a clear orientation.
The emissivity is maximal at the center of center of the jet and
is decreases monotonously outwards. The polarization will be
maximal when the variation in the emissivity within the emitting
beam are maximal. This happens around the jet break when
$\theta_{obs} \sim \Gamma^{-1}$ and the observed beam just reaches
the center. The polarization expected in this case is around 20\%
\cite{Rossietalpolarization02} and it is slightly larger than the
maximal polarization from a uniform jet. As the direction of the
gradient is always the same (relative to a given observer) there
should be no jumps in the direction of the polarization.

According to the patchy shell model \cite{KP00b} the jet can
includes variable emitting hot spots. This could lead to a
fluctuation in the light curve (as hot spots enter the observed
beam) and also to corresponding fluctuations in the polarization
\cite{Granot03,NakarOren03}. There is a clear prediction
\cite{NakarPiranGranot03,NakarOren03} that if the fluctuations are
angular fluctuations have a typical angular scale $\theta_f$ then
the first bump in the light curve should take place on the time
when $\Gamma^{-1} \sim \theta_f$ (the whole hot spot will be
within the observed beam). The following bumps in the light curve
should decrease in amplitude (due to statistical fluctuations).
\textcite{NakarOren03} show analytically and numerically that the
jumps in the polarization direction should be random, sharp and
accompanied by jumps in the amount of polarization.

\subsection{Orphan Afterglows}
\label{sec:orphan}

Orphan afterglows arise as a natural prediction of GRB jets. The
realization that GRBs are collimated with rather narrow opening
angles, while the following afterglow could be observed over a
wider angular range, led immediately to the search for orphan
afterglows: afterglows which are not associated with observed
prompt GRB emission. While the GRB and the early afterglow are
collimated to within the original opening angle, $\theta_j$, the
afterglow can be observed, after the jet break, from a viewing
angle of $\Ga^{-1}$. The Lorentz factor, $\Ga$, is  a rapidly
decreasing function of time. This means that an observer at
$\theta_{obs}>\theta_j$ couldn't see the burst but could detect an
afterglow once $\Ga^{-1}=\theta_{obs}$. As the typical emission
frequency and the flux decrease with time, while the jet opening
angle $\theta$ increases, this implies that observers at larger
viewing  angles will detect weaker and softer afterglows. \xr
orphan afterglows can be observed several hours or at most a few
days after the burst (depending of course on the sensitivity of
the detector). Optical afterglows (brighter than 25th mag ) can be
detected in R band for a week from small ($\sim 10^\circ$) angles
away from the GRB jet axis.  On the other hand, at very late
times, after the Newtonian break, radio afterglows could be
detected by observers at all  viewing angles.

The search for orphan afterglows is an observational challenge.
One has to search for a $~10^{-12}$ergs/sec/cm$^2$ signal in the
\xr, a 23th or higher magnitude in the optical or a mJy  in radio
(at GHz) transients.  Unlike afterglow searches that are
triggered by a well located GRB for the orphan afterglow itself
there is no information where to search and confusion with other
transients is rather easy. So far there was no detection of any
orphan afterglow in any wavelength.

\textcite{Rhoads97} was the first to suggest that observations of
orphan afterglows would enable us to estimate the opening angles
and the true rate of GRBs. \textcite{DalalGriestPruet02} have
pointed out that as the post jet-break afterglow light curves
decay quickly, most orphan afterglows will be dim and hence
undetectable. They point out that if  the maximal observing angle,
$\theta_{max}$, of an orphan afterglow will be a constant factor
times $\theta_j$ the ratio of observed orphan afterglows,
$R_{orph}^{obs}$, to that of GRBs, $R_{GRB}^{obs}$, will not tell
us much about the opening angles of GRBs and the true  rate of
GRBs, $R_{GRB}^{true} \equiv f_b R_{GRB}^{obs}$. However as we see
below this assumption is inconsistent with the constant energy of
GRBs that suggests that all GRBs will be detected to up to a fixed
angle which is independent of their jet opening angle.

\subsubsection{Optical Orphan Afterglow}
\label{sec:orphan-optical}

\begin{figure}[htb]
\begin{center}
\epsfig{file=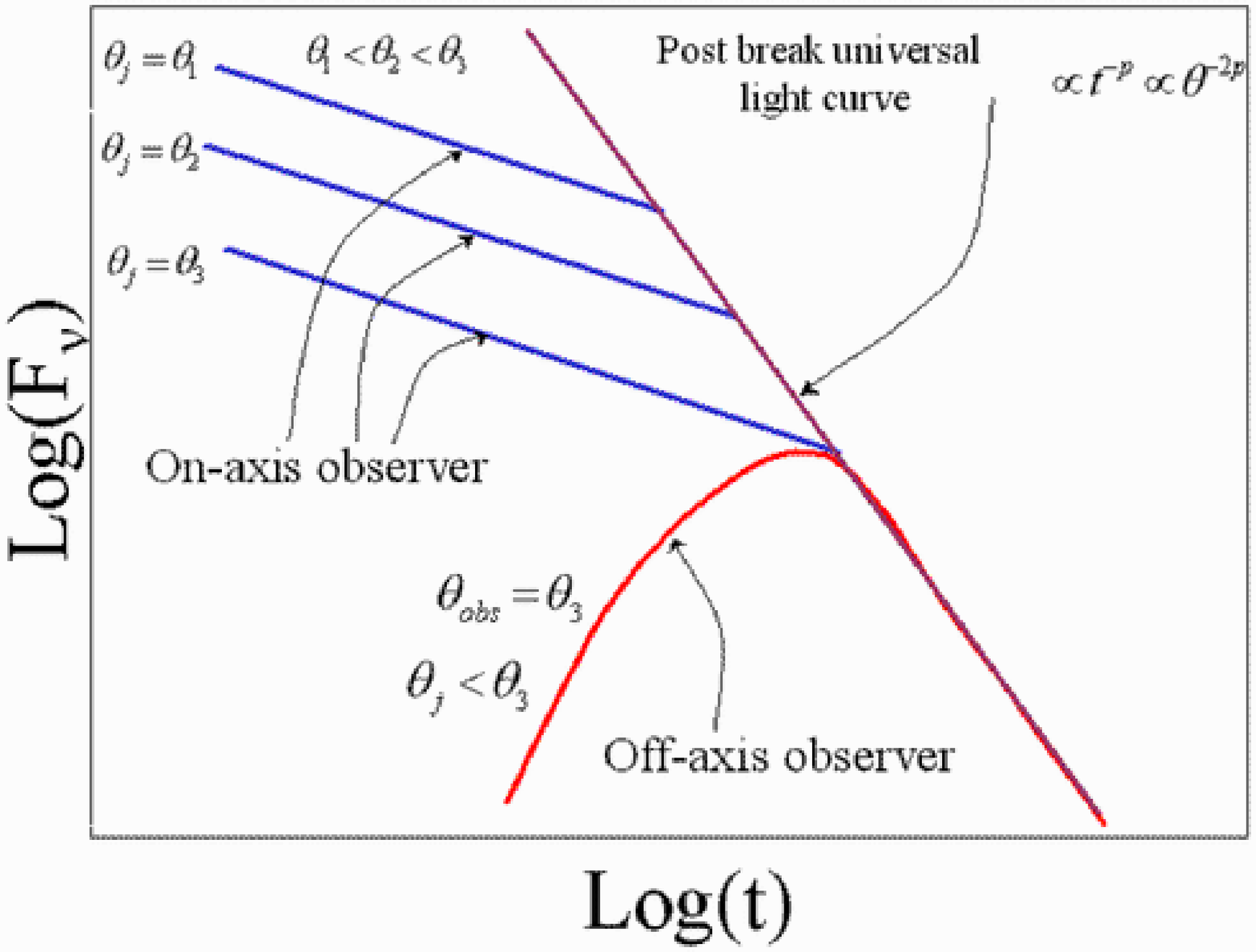,height=2in} \caption{A schematic
afterglow light curve. While the burst differ before the jet break
(due to different opening angles, the light curve coincide after
the break when the energy per unit solid angle is a constant.}
\label{fig:orphan_schmatic}
\end{center}
\end{figure}

Optical orphan afterglow is emitted at a stage when the outflow
is still relativistic.  The observation that GRBs have a roughly
constant total energy \cite{Frail01,PanaitescuK01,Piranetal01}
and that the observed variability in the apparent luminosity
arises mostly from variation in the jet opening angles leads to a
remarkable result: The post jet-break afterglow light curve is
universal \cite{GranotEtal02}. Fig. \ref{fig:orphan_schmatic}
depicts this universal light curve. This implies that for a given
redshift, $z$, and a given limiting magnitude, $m$, there will be
a fixed $\theta_{max}(z,m)$ (independent of $\theta_j$, for
$\theta_j < \theta_{max}$) from within which orphan afterglow can
be detected.

This universal post jet-break light curve can be estimated from
the observations \cite{TotaniPanaitescu02} or alternatively from
first principles \cite{Nakar_P_Granot02} . An observer at
$\theta_{obs}> \theta_j$ will (practically) observe the afterglow
emission only at $t_\theta$ when $\Gamma = \theta_{obs}^{-1}$.
Using Eq. \ref{tjet} and the fact that $\Gamma \propto t^{-1/2}$
after the jet break (Eq. \ref{Rgammajet}) one can estimate the
time, $t_{\theta}$ when a emission from a jet would be detected at
$\theta_{obs}$:
\begin{equation}
t_\theta=   A (\theta_{obs}/\theta_j)^2 t_{jet} \ , \label{ttheta}
\end{equation}
where $A$ is a factor of order unity, and $t_{jet}$ is the time
of the jet break (given by Eq. \ref{tjet}). The flux at this time
is estimated by substitution of this value into the post-jet-break
light curve (see \textcite{Nakar_P_Granot02} for details):
\begin{equation}
F(\theta_{obs}) = F_0 f(z) \theta_{obs}^{-2p} \ , \label{Fnumax2a}
\end{equation}
where $F_0$ is a constant and $f(z)=(1+z)^{1+\beta}D_{L28}^{-2}$
includes all the cosmological effects and $D_{L28}$ is the
luminosity distance in units of $10^{28}$cm. One notices here a
very strong dependence on $\theta_{obs}$. The peak flux drops
quickly when the observer moves away from the axis. Note also
that this maximal flux is independent of the opening angle of the
jet, $\theta_j$. The observations of the afterglows with a clear
jet break (GRB990510 \cite{Harrisonetal99,Staneketal99}, and
GRB000926 \cite{Harrisonetal01}) can be used to calibrate $F_0$.

Now, using Eq. \ref{Fnumax2a}, one can estimate
$\theta_{max}(z,m)$ and more generally the time,
$(t_{obs}(z,\theta,m)$ that a burst at a redshift, $z$, can be
seen from an angle $\theta$ above a limiting magnitude, $m$:
\begin{equation}
t_{obs}(z,\theta,m) \approx  {A t_{jet} \over \theta_j^2}
(\theta_{max}^2 - \theta_{obs}^2) \ . \label{tobs}
\end{equation}
One can then proceed and integrate over the cosmological
distribution of bursts (assuming that this follows the star
formation rate) and obtain an estimate of the number of orphan
afterglows that would appear in a single snapshot of a given
survey with a limiting sensitivity $F_{lim}$:
\begin{equation}
N_{orph}= \int_0^{\infty} {n(z)\over (1+z)} {dV(z) \over dz} dz
\times\int_{\theta_j}^{\theta_{max}(z,m)}   t_{obs}(z,\theta,m)
\theta d\theta \propto (F_0/F_{lim})^{2/p} \ . \label{Rate2}
\end{equation}
where n(z) is the rate of GRBs per unit volume and unit proper
time  and  dV(z) is the differential volume element at  redshift
$z$. Note that modifications of this simple model may arise with
more refined models of the jet propagation
\cite{GranotEtal02,Nakar_P_Granot02}.

\begin{figure}[htb]
\begin{center}
\epsfig{file=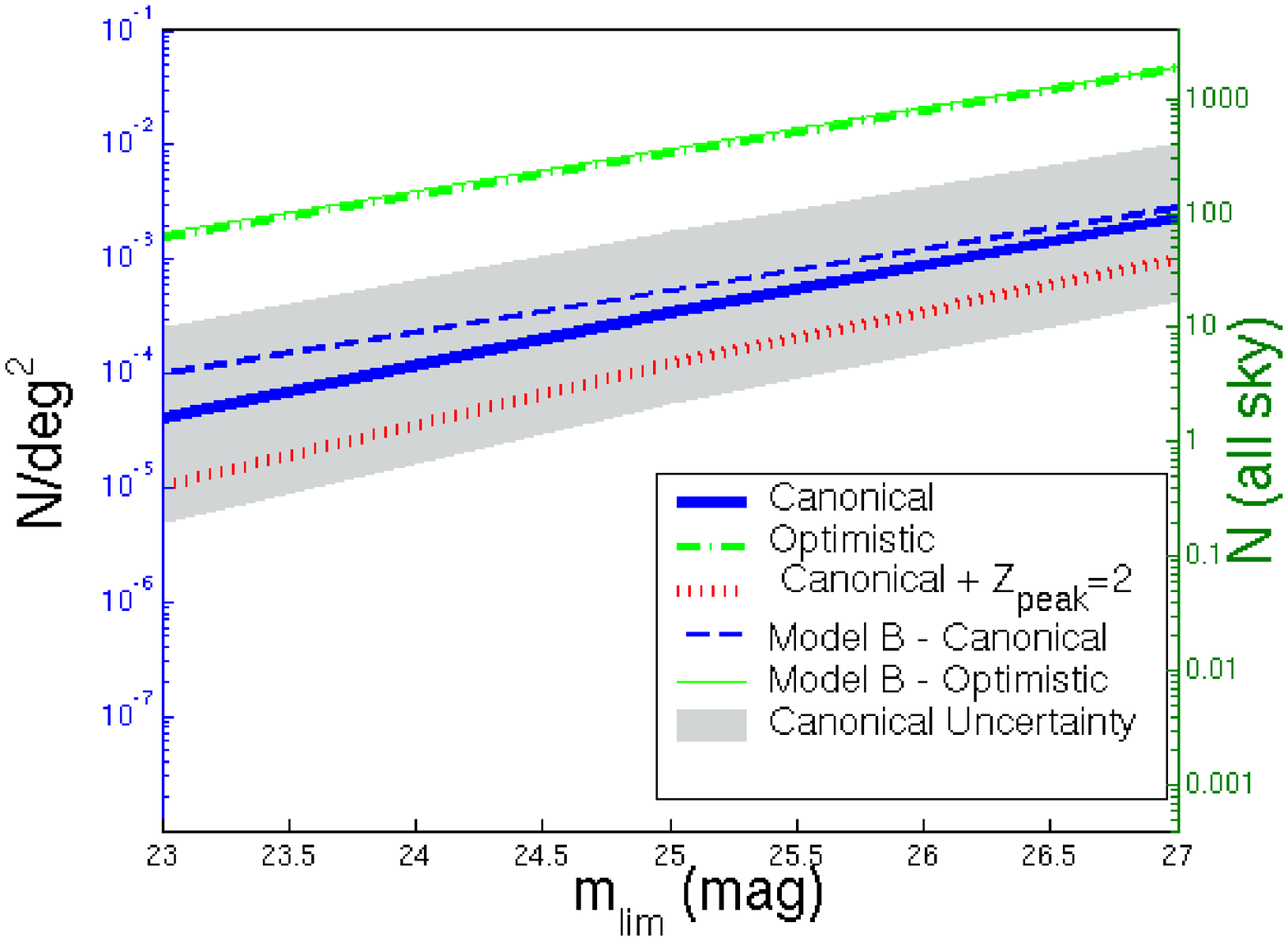,height=2in} \caption{The number of
observed orphan afterglows per square degree (left vertical scale)
and in the entire sky (right vertical scale), in a single
exposure, as a function of the limiting magnitude for detection.
The  {\it thick lines} are for model A with three different  sets
of parameters: i) Our ``canonical" normalization
$F_0=0.003\,\mu{\rm Jy}$, $z_{peak}=1$, $\theta_j=0.1$ ({\it solid
line}). The gray area around this line corresponds an uncertainty
by a factor of 5 in this normalization. ii) Our most optimistic
model with a relatively small $\theta_j=0.05$ and a large
$F_0=0.015\,\mu{\rm Jy}$ ({\it dashed-dotted line}).   iii) The
same as our ``canonical" model, except for $z_{peak}=2$ ({\it
dotted line}). The {\it thin lines} are for model B, where the
{\it solid line} is for our ``optimistic" parameters, while the
{\it dashed line} is for our ``canonical" parameters. Both models
are similar for the ``optimistic" parameters while model B
predicts slightly more orphan afterglows then model A for the
``canonical" parameters. From \cite{Nakar_P_Granot02}}
\label{fig:orphan_rate}
\end{center}
\end{figure}

The results of the intergration of Eq. \ref{Rate2} are depicted in
Fig. \ref{fig:orphan_rate}.
Clearly the rate of a single detection with a given limiting
magnitude increases with a larger magnitude. However, one should
ask  what will be the optimal strategy for a given observational
facility: short and shallow exposures that cover a larger solid
angle or long and deep ones over a smaller area. The exposure time
that is required in order to reach a given limiting flux,
$F_{lim}$, is proportional to $F_{lim}^{-2}$. Dividing  the number
density of observed orphan afterglows (shown in Fig.
\ref{fig:orphan_rate}) by this time factor results in   the rate
per square degree per hour of observational facility. This rate
increases for a shallow surveys that cover a large area. This
result can be understood as follows. Multiplying Eq. \ref{Rate2}
by $F_{lim}^2$ shows that the rate per square degree per hour of
observational facility $\propto F_{lim}^{2-2/p}$.  For  $p>1$ the
exponent is positive and  a shallow survey is preferred. The
limiting magnitude should not be, however, lower than  $\sim$ 23rd
as in this case more transients from on-axis GRBs will be
discovered than orphan afterglows.

Using these estimates \textcite{Nakar_P_Granot02} find that with
their most optimistic parameters  15 orphan afterglows will be
recorded  in the Sloan Digital Sky Survey (SDSS) (that covers
10$^4$ square degrees at 23rd mag) and 35 transients will be
recorded in a dedicated 2m class telescope operating full time for
a year in an orphan afterglow search.
\textcite{TotaniPanaitescu02} find a somewhat higher rate (a
factor $\sim 10$ above the optimistic rate). About 15\% of the
transients could be discovered with a second exposure of the same
area provided that it  follows after 3, 4 and 8 days for
$m_{lim}=23$, 25 and 27. This estimate does not tackle the
challenging problem of identifying the afterglows within the
collected data. \textcite{Rhoads01} suggested to identify
afterglow candidates by comparing the multi-color SDSS data to an
afterglow template. One orphan afterglow candidate was indeed
identified using this technique \cite{VandenBerketal02}. However,
it turned out that it has been a variable AGN \cite{Gal-Yam02}.
This event demonstrates  the remarkable observational challenge
involved in this project.

\subsubsection{Radio Orphan Afterglow}
\label{sec:orphan_radio}

After the Newtonian transition the afterglow is expanding
spherical. The velocities are at most mildly relativistic so
there are no relativistic beaming effects and the afterglow will
be observed from all viewing angles. This implies that
observations of the rate of orphan GRB afterglows at this stage
will give a direct measure of the beaming factor of GRBs. Upper
limits on the rate of orphan afterglows will provide a limit on
the beaming of GRBs \cite{PernaLoeb98}. However, as I discuss
shortly, somewhat surprisingly, upper limits on the rate of
orphan radio afterglow (no detection of orphan radio afterglow)
provide a lower (and not upper) limit on GRB beaming
\cite{Levinsonetal02}.

\textcite{FrailWaxmanKulkarni00} estimate the radio emission at
this stage using the Sedov-Taylor solution for the hydrodynamics
(see \S \ref{sec:Newtonian}). They find that the radio emission at
GHz will be around 1 mJy at the time of the Newtonian transition
(typically three month after the burst) and it will decrease like
$t^{-3(p-1)/2+3/5}$ (see Eq. \ref{Sedov_light}). Using this limit
one can estimate the rate of observed orphan radio afterglow
within a given limiting flux. The beaming factor $f^{-1}_b$ arises
in two places in this calculations. First, the overall rate of
GRBs: $R_{GRB}^{true} \equiv f_b R_{GRB}^{obs}$, increases with
$f_b$. Second the total energy is proportional to $f_b^{-1}$ hence
the flux will decrease when $f_B$ increases. The first factor
implies that the rate of orphan radio afterglows will increase
like $f_b$. To estimate the effect of the second factor
\textcite{Levinsonetal02} use the fact that (for a fixed observed
energy) the time that a radio afterglow is above a given flux is
proportional to $E^{10/9}$ in units of the NR transition time
which itself is proportional to $E^{1/3}$. Overall this is
proportional to $E^{13/9}$ and hence to $f_b^{-13/9}$. To obtain
the overall effect of $f_b$ \textcite{Levinsonetal02} integrate
over the redshift distribution and obtain the total number of
orphan radio afterglow as a function of $f_b$. For a simple limit
of a shallow survey (which is applicable to current surveys)
typical distances are rather ``small", i.e. less than 1Gpc and
cosmological corrections can be neglected. In this case it is
straight forwards to carry the integration analytically and obtain
the number of radio orphan afterglows in the sky at any given
moment \cite{Levinsonetal02}:
\begin{equation}
N_R\simeq 10^{4}f_b^{5/6} (R/0.5)\left(\frac{f_{\nu min}}{5 mJy}
\right)^{-3/2}\left(\frac{\epsilon_e}{0.3}
\right)^{3/2}\left(\frac{\epsilon_B}{0.03}\right)^{9/8}n_{-1}^
{19/24}E_{\rm iso,54}^{11/6}\nu_9^{-3/4}(t_i/3 t_{NR})^{-7/20}.
\label{NR}
\end{equation}
where $R$ is the observed rate of GRBs per Gpc$^3$ per year, and
$t_i$ is the time in which the radio afterglow becomes isotropic.

\textcite{Levinsonetal02} search the FIRST and NVSS surveys  for
point-like radio transients with flux densities greater than 6
mJy. They find 9 orphan candidates. However, they argue that the
possibility that most of these candidates are radio loud AGNs
cannot be ruled out without further observations. This analysis
sets an upper limit for the all sky number of radio orphans, which
corresponds to a lower limit $f_b^{-1}>10$  on the beaming factor.
Rejection of all candidates found in this search would imply
$f_b^{-1}>40$ \cite{GuettaPiranWaxman03}.

\subsection{Generalizations: VI. Additional Physical Processes }
With the development of the theory of GRB afterglow it was
realized that several additional physical ingredients may
influence the observed afterglow light emission. In this section
I will review two such processes: (i) Pre acceleration of teh
surrounding matter  by the prompt \gr emission and (ii) Decay of
neutrons within the outflow.

\subsubsection{Pre-acceleration}
\label{sec:pre-acceleration}

The surrounding regular ISM or even stellar wind is optically thin
to the initial \gr pulse. Still the interaction of the pulse and
the surrounding matter may not be trivial.
\textcite{ThompsonMadau00} pointed out that a small fraction of
the  \gr radiation will be Compton scattered on the surrounding
electrons. The backscattered photons could now interact with the
outwards going \gr flux and produce pairs. The pairs will increase
the rate of backscattering and this could lead to an instability.
When sufficient number of pairs will be produced the surrounding
matter will feel a significant drag by the \gr flux and it will be
accelerated outwards \cite{MadauThompson00}. These
pre-acceleration of the ambient medium could have several
implications to the early afterglow
\cite{Meszaros-Ramirez-Rees01,Beloborodov_front_02}.

The key issue is that while the optical depth of the surrounding
medium (as ``seen" the  \gr photons) is very small, the mean free
path of an ambient electron within the \gr photons is large (at
small enough radius) and each electron  scatters many photons.
While the medium absorbs only a small fraction of the prompt \gr
energy, the effect of this energy can be significant.
\textcite{Beloborodov02a} characterizes the interaction of the \gr
radiation front with the surrounding medium  by a dimensionless
parameter\footnote{Note that \textcite{Beloborodov02a} uses the
notation $\xi$ for this parameter}:
\begin{equation}
\eta = \frac{\sigma_T E_{iso}}{4 \pi R^2 m_ec^2}=6.5
E_{52}R_{16}^{-2} \ ,
\end{equation}
the energy that a single electron scatters relative to its rest
mass. \textcite{Beloborodov_front_02} calculates the Lorentz
factor  of the ambient medium  and the number of pairs per
initial electron as functions of $\eta$.
where $\eta_{load}=20-30$, depending on the spectrum of the
gamma-rays, $\eta_{acc}=5\eta_{load}=100-150$, and
$f_{acc}=[\exp(\eta_{acc}/\eta_{load})+\exp(-\eta_{acc}
/\eta_{load})]/2=74$.

If $\eta<\eta_{load} \approx 20-30$, depending on the spectrum of
the gamma-rays, the medium remains static and $e^\pm$-free. When
the front has $\eta>\eta_{load}$, a runaway $e^\pm$ loading
occurs. The number of loaded pairs depends exponentially on $\eta$
as long as $\eta<\eta_{acc} =5\eta_{load}=100-150$. The medium is
accelerated if $\eta>\eta_{acc}$. $\eta_{acc}$ is around 100
because the electrons are coupled to the ambient ions, and and
the other hand the loaded $e^\pm$ increase the number of scatters
per ion. At $\eta=\eta_{gap}\approx 3\times 10^3$, the matter is
accelerated to a Lorentz factor $\Gamma_{ambient}$ that exceeds
the Lorentz factor of the ejecta. It implies that the radiation
front pushes the medium away from the ejecta and opens a gap.

As the GRB radiation front expands, the energy flux and hence
$\eta$ decreases $\propto R^{-2}$. $\eta$ passes through
$\eta_{gap}$, $\eta_{acc}$, and $\eta_{load}$ at $R_{gap}$,
$R_{acc}$, and $R_{load}$, respectively. These three
characteristic radii define four stages:

{\bf i.} $R<R_{gap}\approx R_{acc}/3$: The ejecta moves in a
cavity produced by the radiation front with
$\Gamma_{ambient}>\Ga_{ejecta}$.

{\bf II.} $R_{gap}<R<R_{acc} \approx  3 \times 10^{15} {\rm ~cm}
E_{52}^{1/2}{\rm ~cm}$: The ejecta sweeps the $e^\pm$-rich medium
that has been preaccelerated to
$1\ll\Gamma_{ambient}<\Ga_{ejecta}$.

{\bf III.} $R_{acc}<R<R_{load}\approx 2.3 R_{acc}$. The ejecta
sweeps the ``static'' medium ($\Gamma_{ambient}\approx 1$) which
is still dominated by loaded $e^\pm$.

{\bf IV.} $R>R_{load}$. The ejecta sweeps the static pair-free
medium.

\begin{figure}[htb]
\begin{center}
\epsfig{file=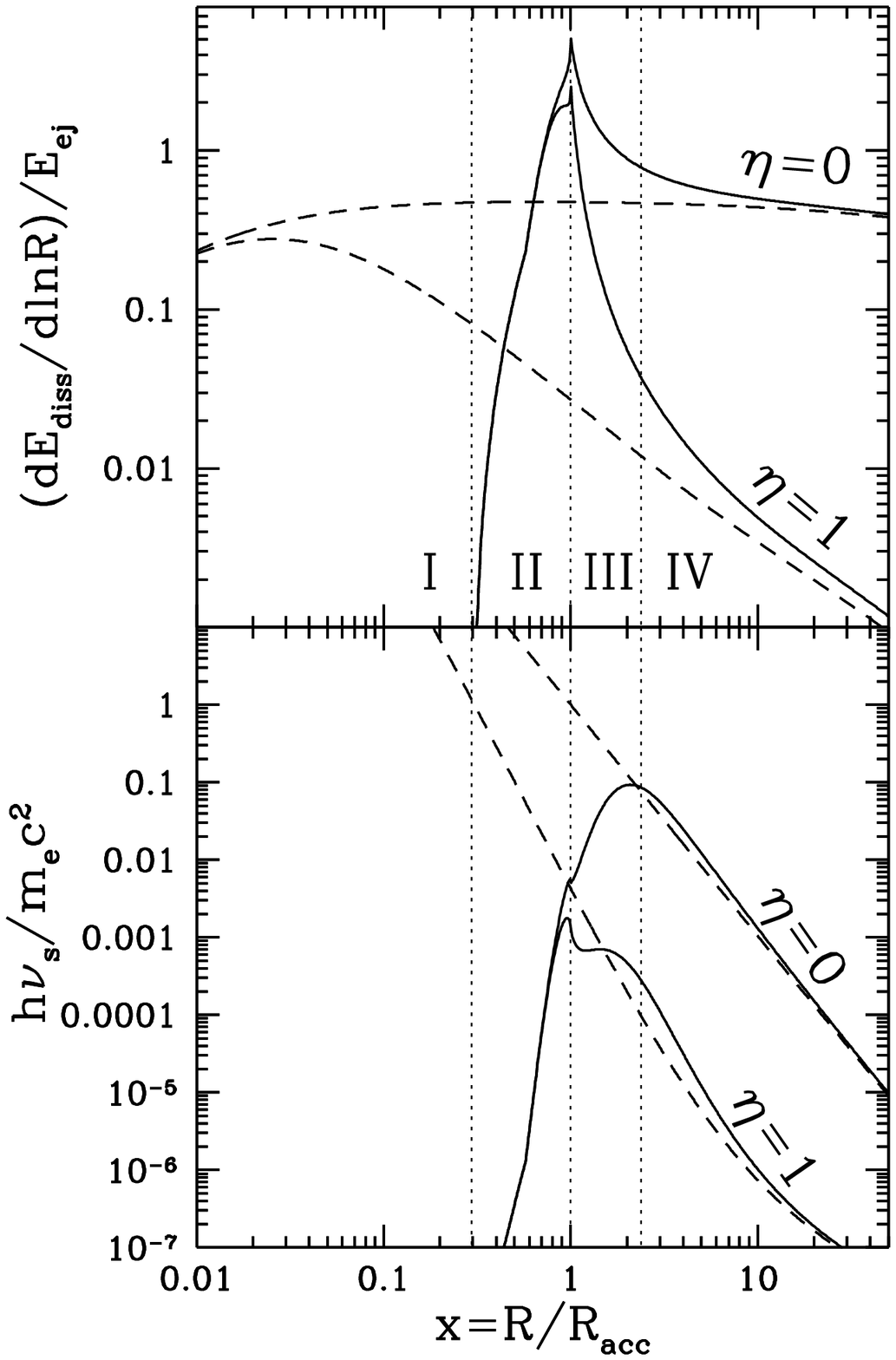,height=2in} \caption{Afterglow from a
GRB ejecta decelerating in a wind of a Wolf-Rayet progenitor with
$\dot{M}=2\times 10^{-5}M_\odot$~yr$^{-1}$ and
$w=10^{3}$~km~s$^{-1}$. The burst has an isotropic energy
$E=10^{53}$~erg) and a thin ejecta shell with kinetic energy
$E=10^{53}$~erg and a Lorentz factor $\Ga=200$. Dashed curves show
the prediction of the standard model that neglects the impact of
the radiation front, and solid curves show the actual behavior.
Two extreme cases are displayed in the figure: $\eta=0$ (adiabatic
blast wave) and $\eta=1$ (radiative blast wave). Four zones are
marked: I - $R<R_{gap}$ (the gap is opened), II -
$R_{gap}<R<R_{acc}$ (the gap is closed and the ejecta sweeps the
relativistically preaccelerated $e^\pm$-loaded ambient medium),
III - $R_{acc}<R<R_{load}$ ($e^\pm$-loaded ambient medium with
$\Gamma_{ambient}\approx 1$), and IV - $R>R_{load}$ (pair-free
ambient medium with $\Gamma_{ambient}\approx 1$). The radie are
measured in units of $R_{acc}\approx 10^{16}$~cm. {\it Top panel}:
dissipation rate. {\it Bottom panel}: synchrotron peak frequency
(assuming $\epsilon_B=0.1$) in units of $m_ec^2/h$. From
\textcite{Beloborodov02a}} \label{fig:flash}
\end{center}
\end{figure}

This influence of the \gr on the surrounding matter  may modify
the standard picture of interaction of external shocks with the
surrounding medium (see \S \ref{sec:Ex-hydro}. This depends
mostly on the relation between   $R_{ext}$ and $R_{gap} \approx
 10^{15}E_{52}^{1/2}$~cm. If $R_{ext} > R_{gap}$ this effect
won't be important. However, if $R_{ext} < R_{gap}$ then
effective decceleration will begin only at $R_{gap}$.  At
$R<R_{gap}$ the ejecta freely moves in a cavity cleared by the
radiation front and only at $R=R_{gap}$ the blast wave gently
begins to sweep the preaccelerated medium with a small relative
Lorentz factor. With increasing $R>R_{gap}$, $\Gamma_{ambient}$
falls off quickly, and it approaches $\Gamma_{ambient}=1$ at
$R=R_{acc} \approx 3 R_{gap}$ as
$\Gamma_{ambient}=(R/R_{acc})^{-6}$. Thus, after a delay, the
ejecta suddenly ``learns'' that there is a substantial amount of
ambient material on its way. This resembles a collision with a
wall and results in a sharp pulse (see Fig. \ref{fig:flash}).

While $R_{gap}$ does not depend on the external density $R_{ext}$
does (see Eq. \ref{Rext}). The condition $R_{ext}<R_{gap}$
implies:
\begin{equation}
 E_{52}^{1/6}n_1^{1/3} \Ga^{2/3}_{100} > 0.02
\end{equation}
Thus it requires a dense external medium and large initial Lorentz
factor. Otherwise $R_{gap}$ is too large and the deceleration
takes place after the gap is closed. Hence the conditions for
pre-acceleration will generally occur if the burst takes place in
a dense circumburst regions, like in a Wolf-Rayet progenitor
\cite{Beloborodov_front_02}. \textcite{KumarPanatescu03} elaborate
on this model and find that the observational limits by LOTIS and
ROTSE on prompt emission from various burst limit the ambient ISM
density (within $10^{16}$cm to less than $10^{3}{rm cm}^{-3}$.
Similarly the find that in case of a wind the progenitors mass
loss to wind's velocity ratio is below
$10^{-6}M_\odot$/yr/(10$^3$km/sec).

\subsubsection{Neutron decoupling and decay}
\label{sec:neutrons}

\textcite{DerishevKocharovskyKocharovsky01a,DerishevKocharovskyKocharovsky99b}
pointed out that neutrons that are included initially in the
fireball will change its dynamics  and modify the standard
afterglow evolution. While the protons slow down due to the
interaction with the surrounding matter the neutrons will coast
freely after they decouple with $\Gamma_n$, which equals to the
Lorentz factor while decoupling took place.

At
\begin{equation}
R_{decay} \approx 0.3 \times 10^{16} {\rm cm} (\Gamma_n/100)
\end{equation}
the neutrons decay. A new baryonic shell forms ahead of the
original fireball shell, with energy comparable to the initial
energy of the protons' shell (this depends, of course, on the
initial ratio of neutrons to protons). At this stage the neutrons
front that is not slowed down like the rest of the fireball is at
a distance:
\begin{equation}
 \Delta R= R (1/ 2\Gamma^2
- 1/2Gamma^2_n) \ ,
\end{equation}
from the fireball front, where $\Gamma$ is the current Lorentz
factor of the fireball.

Once more the situation depends on whether $R_{decay}$ is smaller
or larger than $R_{ext}$, the original deceleration radius. If
$R_{decay}< R_{ext}$:
\begin{equation}
E_52^{1/3}n_1^{-1/3} \Ga^{-2/3}_{100} (\Gamma_n/300)^{-1} < 0.06 \
,
\end{equation}
the decaying neutron products will mix with the original protons
and won't influence the evolution significantly (apart from
adding their energy to the adiabatic fireball energy).  Otherwise,
they situation depends on $\Gamma_n$ the Lorentz factor at
decoupling.

\textcite{PruetDalal02} consider a situation in which the neutron
decouple with a low $\Gamma_n$. In this case one will get a
delayed shock scenario when the neutronic decay produce will
eventually catch up with the slowing down protons (when their
Lorentz factor is of order $\Gamma_n$. Along the same line of
thought \textcite{DalalGriestPruet02} suggest that a large
neutronic component that may exist within the initial fireball
material may help to eliminate the baryon load problem
\cite{SP90}.

\textcite{Beloboradov03neutron} considers a situation when
$\Gamma_n \approx \Ga_0$, the initial Lorentz factor of the
protons. In this case the decaying neutrons' products will be
ahead of the shell of the protons. The decaying products will
interact with the surrounding matter and will begin to slow down.
There will be a triple interaction between the two shells and the
surrounding ambient medium (resembling to some extend the
pre-acceleration scenario described earlier) . This will take
place at radii of a few times $R_{decay}$ and at an observed time
of  a few $\times R_{decay}/2 c \Gamma^2 \approx {\rm a few
seconds} /(\Gamma_n/300)$, i.e. extremely early. This will produce
brightening when the fronts pass $R_{decay}$.

The neutrons could also influence the behavior of the relativistic
flow during the prompt (internal shocks) phase. Specifically
inelastic collisions between differentially streaming protons and
neutrons can produce pions and eventually $\nu_\mu$ of ~10 GeV as
well as $\nu_e$ of ~5 GeV \cite{BahcallMeszaros00,MeszarosRees00}.
These neutrino fluxes could produce $\sim 7$ events/year in km3
neutrino detectors. GeV photons will also be produced but it is
unlikely that they could be detected.

 \label{sec:Afterglow-IC}

\section{Additional Emission from GRBs }
\label{sec:Other}
\subsection{TeV \gr }
\label{sec:TeV}

\textcite{Hurley94} reported on detection of  18 GeV photons from
GRB 940217.  Milagrito - A  TeV detector - discovered a possible
TeV signal coincident with  GRB 970417 \cite{Milagrito_970417}.
\textcite{Gonzalez03} discovered  a high energy tail that extended
up to 200 MeV from GRB 941017.

A natural source for high energy \gr is the SSC (Synchrotron self
Compton) component produced by IC from the burst itself or from
the afterglow \cite{MR94a,Meszaros-Rees-Papathanassiou94}. The SSC
photons energy  should be $\ga_e^2$ higher than the synchrotron
photons. Typical random Lorentz factors of electrons, $\gamma_e$,
within internal shocks are of order a thousand (in the fluid's
rest frame). This implies that if the observed \gr emission is
produced by synchrotron in internal shocks then the IC emission
would produce a second peak around a few hundred GeV. This would
be the analogue of the high energy component observed in Blazars.
Note that emission above $\sim 10-100$GeV might be self absorbed
by pair production with the source
\cite{Papathanassiou-Meszaros96,Pilla-Loeb98,GuettaGranot03z}.

The SSC component would be even higher from the early afterglow.
The synchrotron emission from the forward shock is expected to be
around 10 keV (if the observed early afterglow is indeed produced
by the external shocks). With a Lorentz factor of a typical
electron around $10^5$ the expected SSC component should be around
$100$TeV. Finally the reverse shock emission is expected to
produce 100 eV photons \cite{SP99b}. With typical electrons
Lorentz factor of a few thousand this should correspond to SSC
photons with typical energy of 100 MeV. Depending on the relevant
Y parameter the fluxes of these high energy components should be
comparable or even larger than the prompt GRB \gr fluxes. This
emission should be simultaneous with the GRB emission. It is also
possible that the forward shock electrons will IC reverse shocks
photons. It is likely that this is the cause of the high energy
emission seen in GRB 941017
\cite{PeerWaxman04,PiranNakarGranot03}.

Other mechanisms can produce high energy emission as well.
\textcite{Vietri97} suggested that as GRB can accelerate protons
up to $10^{20}$eV (see \S \ref{sec:UHECRs} below). These protons
can emit 0.01 of the GRB energy as high energy \gr with energies
up to 300 GeV. \textcite{Bottcher_Dermer98} considered the
synchrotron spectrum resulting from high energy protons and
leptons produced in cascade initiated by photo-pion production.
They predict a significant flux of 10Mev-100GeV photons.

While the high energy photons flux could be significant these
photons might not be detectable on earth.   The high energy photon
flux above 1 TeV would be attenuated significantly due to pair
production of such high energy  photons with the intergalactic NIR
flux \cite{GouldSchreder67}. \textcite{DaiLu02} suggest that
secondary emission produced via these interaction (upscattering of
the CMB  by the produced pairs) would still point towards the
initial direction and hence might be detectable as a delayed GeV
emission. However, even a tiny intergalactic magnetic field ($>
10^{-22}$G would be sufficient to deflect the electrons and dilute
these signal \cite{GuettaGranot03z}.

\subsection{Neutrinos }
\label{sec:neutrinos}

Neutrinos can be produced in several regions within GRB sources.
First some models, like the Collapsar model or the neutron star
merger model predict ample ($\sim 10^{53}$ ergs) production of
low (MeV) neutrinos. However, no existing or planned detector
could see those from cosmological distances. Furthermore, this
signal will be swamped in rate by the much more frequent SN
neutrino signals which would typically appear closer.

However, GRBs could be detectable sources of high energy
neutrinos, with energies ranging from $10^{14}$eV to $10^{17}$eV.
These neutrinos are produced by internal or external shocks of
the GRB process itself and hence are independent of the nature of
the progenitor.

To understand the process of  neutrino emission recall that
neutrinos are ``best" produced in nature following  pions
production in proton-photon or proton-proton collisions. The
proton-photon process requires that the photon's energy is around
the $\Delta$ resonance in the proton's energy frame: namely at
$\sim 200$MeV. The resulting pion decays emitting neutrinos with a
typical energy of $\sim 50$ MeV  in the proton's rest frame. If
the proton is moving relativistically, with a Lorentz factor
$\gamma_p$ within the laboratory frame the required photon energy
in the lab frame is smaller by a factor of $\gamma_p$ and the
resulting neutrino energy is larger by a factor of $\gamma_p$.
Depending on the surrounding environment very energetic pions may
lose some of their energy before decaying producing a ``cooling
break" in the neutrino spectrum. In this case the resulting
neutrinos' energy will be lower than this naive upper limit.

Within GRBs protons are accelerated up to $10^{20}$eV
\cite{Waxman95,Vietri95}. The relevant Lorentz factors of these
protons range from $\Gamma$ up to $10^{11}$ (at the very high
energy tail of the protons distribution). Thus we expect neutrinos
up to $10^{19}$eV provided that there is a sufficient flux of
photons at the relevant energies so that the pions can be
produced and there are no energy loses to the pions.

\textcite{PaczynskiXu94} and \textcite{WaxmanBahcall97} calculated
the flux of VHE neutrinos from internal shocks. They found that a
significant flux of $\sim 10^{14}$eV neutrinos can be produced by
interaction of the accelerated internal shocks protons with the
GRB photons. \textcite{GuettaSpadaWaxman01} estimate that on
average each GRB produces a flux of $\sim 10^{-9}$ GeV/cm$^2$ sec
sr corresponding to 0.01 events in a km$^3$ detector.
Calculations of specific fluxes from individual bursts (that
factor in the observed \gr spectrum) were performed by
\textcite{Guettaetal03}. \textcite{WaxmanBahcall00} suggest that
protons accelerated at the reverse shock (that arises at the
beginning of the afterglow) would interact with the optical - uv
flux of the afterglow and produce $10^{18}$eV neutrinos.

Within the Collapsar model
\textcite{WaxmanMeszaros00,Razzaque-Meszaros-Waxman03} suggested
that as the jet punches the stellar shell it can produce a flux of
TeV neutrinos. Within the Supranova model the internal shock
protons \cite{GuettaGranot03} or external shocks protons
\cite{Dermer03} can also interact with external, pulsar wind
bubble, photons producing $10^{16}$eV neutrinos with a comparable
detection rate to the one obtained form interaction of the
internal shock protons with \gr photons. If the external magnetic
field is sufficiently large (as in the pulsar wind bubble)
external shocks can also accelerate protons to high energy
\cite{VietriDeMarcoGuetta03}. In this case the protons can
interact with afterglow photons and can produce neutrinos up to
$10^{17}$eV \cite{LiDaiLu02}.

\subsection{Cosmic Rays and Ultra High Energy Cosmic Rays }
\label{sec:UHECRs}

Already in 1990 \textcite{SP90} noticed that a fireball may
produce cosmic rays. However the flux of ``low" energy (up to
$10^{14}$ eV) that they considered was smaller by several orders
of magnitude than the observed flux of cosmic rays that are
accelerated in SNRs. Hence this component isn't important.

\textcite{Waxman95} and independently \textcite{Vietri95} noticed
that protons can be accelerated up to $10^{20}$eV within the
relativistic shocks that take place in GRBs. Namely internal
shocks or the reverse shock in GRBs are among the few locations
in the Universe where the shock acceleration condition (Eq.
\ref{emax_acc} needed to accelerate protons up to $10^{20}$eV,
the Hillas criterion can be satisfied. Moreover to within an
order of magnitude the flux of \gr reaching earth from GRBs  is
comparable to the observed flux of UHECRs (Ultra High Energy
Cosmic Rays) \cite{Waxman95}. Thus, if GRBs produce a comparable
energy in \gr and in UHECRs they could be the source of the
highest energy Cosmic rays.

\textcite{Greisen66} and \textcite{ZK66} (GZK) pointed out that
the highest energy CR (above $10^{19.5}$ eV) are attenuated as
they propagate via the Cosmic Microwave background (CMBR). This
happens because at this high energies the protons can interact
with the CMBR photons and produce pions. The typical mean free
path of a ultra high energy proton in the CMBR decreases rapidly
with energy and for a $10^{20}$eV proton it is only several tens
Mpc. Thus, the observed UHECRs at energies above the GZK energy
($\sim 10^{19.5}$eV must arrive from relatively nearby (on
cosmological scale) sources. However, there are no known steady
state sources within this distance (see however
\textcite{FarrarPiran00}). GRBs as a transient phenomenon could be
a ``hidden" source of UHECRs. There won't be a direct association
between GRBs and arrival of UHECRs as the later are deflected by
the intergalactic magnetic field. This leads to an angular
deflection as well as a long time delay. If GRBs are sources of
UHECRs then we expect a break in the UHECR spectrum at the GZK
energy - below the GZK energy we will detect UHECRs from the whole
universe. Above the GZK energy we will detect only ``local" UHECRs
from within the nearest several dozen Mpc.
\textcite{BahcallWaxman03} suggested that recent observations
imply that such a break has been seen. However, the observational
situation is not clear as yet and a final resolution would most
likely require the Auger UHECR detector.

\subsection{Gravitational Radiation }
\label{sec:Gravitational-rad}

Like GRBs, typical sources of gravitational radiation involve the
formation of  compact objects. Hence it is reasonable to expect
that gravitational waves will accompany GRBs. This association is
indirect: the gravitational waves are not directly related to the
GRB.  Additinionally, GRBs have their own, albeit weak,
gravitational radiation pulse which arises during the
acceleration of the jets to relativistic velocities.
Unfortunately this signal is weak and moreover it is
perpendicular to the GRB signal.

To estimate the rates of observed gravitational radiation events
associated with GRB we use the rate of long GRBs. The nearest
(long) GRB detected within a year would be at 1Gpc. As GRBs are
beamed the nearest (long) event would be at would be much nearer,
at $135 \theta_{0.1}^2$Mpc. However, this burst would be directed
away from us. Still a GRB that is beamed away from us is expected
to produce an ``orphan" afterglow.

The rate of short bursts is less certain. \textcite{Schmidt01a}
estimates that the rate of short GRBs is smaller by a factor of
two than the rate of long ones. In this case the distances
mentioned above should be revised up by a factor of 1.25.
However, if the rate of short GRBs is larger by a factor 10 than
the rate of long ones then the corresponding distances should be
revised downwards by a factor of $10^{-1/3}$ This would put one
event per year  at $\sim 80\theta_{0.1}^2$Mpc, but once again
this burst won't be pointing towards us. The nearest event with a
burst in our direction would be at  $\sim 450$Mpc.

\subsubsection{Gravitational Radiation from Neutron Star Mergers }
\label{sec:GRBNS}

Binary neutron star mergers are the "canonical" sources of
gravitational radiation emission.  LIGO and VIRGO both aim in
detecting these sources. Specifically the goal of these detectors
is to detect the characteristic``chirping" signals arising from
the in-spiraling phase of these events. The possibility of
detection of such signals has been extensively discussed (see
e.g. \cite{LIGO-merger}). Such events could be detected up to a
distance of several tens of Mpc with LIGO I and up to $\sim 100
Mpc$ with LIGO II.

Comparing with  GRB rates we find that if, as some expect, neutron
star mergers are associated with short GRBs and if the rate of
short GRBs is indeed large, then we have one event per year within
the sensitivity of LIGO II and marginally detectable by LIGO I.
However, this burst will be pointing away from us.

The detection of the chirping merger signal is based on fitting
the gravitational radiation signal to pre-calculated templets.
\textcite{Kochaneck_Piran93} suggested that the detection of a
merger gravitational radiation signal would require a lower S/N
ratio if this signal coincides with a GRB. This would increase
somewhat the effective sensitivity of LIGO and VIRGO to such
events. \textcite{Finnetal99} suggest using the association of
GRBs with sources of gravitational waves in a statistical manner
and propose to search for enhanced gravitational radiation
activity towards the direction of a GRB during the short periods
when GRBs are detected.  Given the huge distances of observed GRBs
it is not clear if any of these techniques will be useful.

\subsubsection{Gravitational Radiation from Collapsars }
\label{sec:GRCol}

The Collapsar model \cite{Woosley93,Pac98,MacFadyen_W99} is based
on the collapse of the core of a massive star to a black hole
surrounded by a thick massive accretion disk. As far as
gravitational radiation is concerned this system is very similar
to a regular supernova. Rotating gravitational collapse has been
analyzed by \textcite{Stark_Piran85}. They find that the
gravitational radiation emission emitted in a rotating collapse
to a black hole is dominated by the black hole's lowest normal
modes, with a  typical frequency of $~20c^3/GM$. The total energy
emitted is:
\begin{equation}
{\Delta E_{GW}} = \epsilon M c^2 = {\rm min}(1.4 \cdot 10^{-3}
a^4, \epsilon_{max}) M c^2 \ ,
\end{equation}
where $a$ is the dimensionaless specific angular momentum and
$\epsilon_{max}$ is a maximal efficiency which is of the order
${\rm a few} \times 10^{-4}$.  The expected amplitude of the
gravitational radiation signal, $h$, would be of the order of
$\sqrt{\epsilon} GM/c^2 d$ where $d$ is the distance to the
source. Even LIGO II won't be sensitive enough to detect such a
signal from a distance of 1Gpc or even from 100 Mpc.

\subsubsection{Gravitational Radiation from Supranova }
\label{sec:GRSUP}

According to the Supranova model a GRB arises after a neurton star
collapse to a black hole. This collapse takes PLACE several weeks
or months after the Supernova that formed the neutron star (see
\ref{sec:Supranova}). The expected gravitational waves signal from
a Supranova \cite{VietriStella98} includes two components. First
the signal from the initial supernova is similar to the
gravtitational waves from the collapsar model. However, here the
first collapse (the Supernova) takes place several weeks or months
before the GRB. Thus, there won't be any correlation between the
gravitational waves emitted by the first collapse and the GRB. A
second component may arise from the second collapse from the
supramassive neutron star to a black hole.  This signal should
coincide with the GRB.

\subsubsection{Gravitational Radiation from the GRB}
\label{sec:GRGRB}

The most efficient generation of gravitational radiation could
take place here during the acceleration phase, in which the mass
is accelerated to a Lorentz factor $\Gamma$. To estimate this
emission I follow  \textcite{Weinberg73} analysis of gravitational
radiation emitted from a relativistic collision between two
particles.  Consider the following simple toy model: two particles
at rest with a mass $M$ that are accelerated instantly at $t=0$ to
a Lorentz factor $\Gamma$ and energy $E$. Conservation of energy
requires that some (actually most) of the rest mass is converted
to kinetic energy during the acceleration and the rest mass of
the accelerated particle is $m = E/\Gamma = M/\Gamma$. The
energy  emitted per unit frequency per unit solid angle in the
direction at an angle $\alpha$ relative to $\vec \beta$ is:
\begin{equation}
{d E \over d \Omega d \omega} = {G M^2 \beta^2 \over c \pi^2}
\big[   {\Gamma^2 (\beta^2 - \cos^2\alpha) \over  (1 - \beta^2
\cos^2\alpha)^2} + { \cos^2\alpha \over \Gamma^2 (1 - \beta^2
\cos^2\alpha)^2} \big]\ . \label{fluxgrav}
\end{equation}
The result is independent of the frequency, implying that the
integral over all frequency will diverge. This nonphysical
divergence arises from the nonphysical assumption that the
acceleration is instantaneous. In reality this acceleration takes
place over a time $\delta t$, which is of order 0.01sec. This
would produce a cutoff $\omega_{max} \sim 2 \pi / \delta t$ above
which Eq. \ref{fluxgrav} is not valid. The angular distribution
found in Eq. \ref{fluxgrav} is disappointing. The EM emission from
the ultrarelativistic source is beamed forwards into a small angle
$1/\Gamma$, enhancing the emission in the forwards direction by a
large factor ($\Gamma^2$). The gravitational radiation from this
relativistic ejecta is spread rather uniformly  in almost all
$4\pi$ steradians. Instead of beaming there is ``anti-beaming"
with no radiation at all emitted within the forward angle
$1/\Gamma$ along the direction of the relativistic motion.

Integration of the energy flux over different directions yields:
\begin{equation}
{d E \over d \omega} = {G M^2 \over c \pi^2} [ 2 \Gamma^2 + 1+ {(
1 - 4 \Gamma^2) \over \Gamma^2 \beta} \arctan(\beta)] \
.\label{energy_flux}
\end{equation}
As expected the total energy emitted is proportional to $m^2
\Gamma^2$. Further integration over frequencies up to the cutoff
$2 \pi / \delta t$ yields:
\begin{equation}
E  \approx { 2 G M^2 \Gamma^2 \over c \pi \delta t } \ .
\end{equation}

In reality the situation is much more complicated than the one
presented here. First, the angular width of the emitted blobs is
larger than $1/\Gamma$. The superposition of emission from
different directions washes out the no emission effect in the
forward direction. Additionally according to the internal shocks
model the acceleration of different blobs go on independently.
Emission from different blobs should be combined to get the actual
emission. Both effects {\it reduce} the effective emission of
gravitational radiation and makes the above estimate an upper
limit to the actual emission.

The gravitational signal is spread in all directions (apart from
a narrow beam along the direction of the relativistic motion of
the GRB). It ranges in frequency from $0$ to $f_{max} \approx
100$Hz. The amplitude of the gravitational radiation signal at the
maximal frequency, $f_{max} \approx 100$Hz, would be: $ h \approx
(GM\Gamma^2 /c^2 d) $.  For typical values of $E=M\Gamma =
10^{51}$ ergs, $\delta t = 0.01$ sec and a distance of $500$ Mpc,
$ h \approx .5 \cdot 10^{-25}$, it is far below the sensitivity
of planned gravitational radiation detectors. Even if we a burst
is ten times nearer this "direct" gravitational radiation signal
would still be undetectable .

Some specific models for GRBs' inner engine predict additional
amount of energy. For example
\textcite{vanPutten01,vanPuttenLevinson01} suggest a model of a
black hole - accretion torus in which a large fraction of the
emitted energy of the black hole - accretion torus system escapes
as gravitational radiation. The radiation arises due to
instabilities within the torus that break down the axial
symmetry. They estimate that as much as $10^{53}$ergs would be
emitted as gravitational radiation which will have a
characteristic signature corresponding to the normal mode of the
black hole - accretion torus system with typical frequencies
around few hundred Hz, conveniently within the frequency range of
LIGO/VIRGO. If correct than GRBs are the most powerful
burst-sources of gravitational waves in the Universe
\cite{vanPutten01}.

\section{MODELS OF INNER ENGINES }
\label{sec:inner-engine}

The Fireball model tells us how GRBs operate. However, it does not
answer the most interesting astrophysical question: what produces
them? which astrophysical process generates the energetic
ultrarelativistic flows needed for the Fireball model? Several
observational clues  help us  answer these questions.
\break\indent {\bf $\bullet$ Energy:} The total energy involved is
large $\sim 10^{51}$ergs,
  a significant fraction of the binding energy of a
  stellar compact object. {\it The ``inner engine" must be able to
  generate this  energy and
  accelerate  $\sim 10^{-5}M_\odot$ (or the equivalent in terms of Poynting flux)
  to relativistic velocities.}
\break\indent {\bf $\bullet$ Collimation:} Most GRBs are
  collimated with typical opening angles $1^o<\theta<20^o$.
   {\it The ``inner engine" must be
  able to collimate the relativistic flow.}
\break\indent {\bf $\bullet$ Long and Short Bursts:} The bursts
are divided to two
  groups according to their overall duration. Long bursts with $T>2$sec
  and short ones with $T<2$sec. As the duration is determined by
  the inner engine this may imply that there are two different
  inner engines.
\break\indent{\bf $\bullet$ Rates:} GRBs take place once per $ 3
\cdot 10^5$ yr per
  galaxy.  {\it GRBs are very rare at about 1/3000 the rate of
  supernovae.}
\break\indent{\bf $\bullet$ Time Scales:} The variability time
scale,
  $\delta t$, is at times as short
  as 1ms. The overall duration (of long GRBs), $T$, is of the order of 50sec.
According to the internal shocks model these time scales are
determined by the activity of the ``inner engine". {\it $\delta t
\sim 1$ ms suggests a compact object. $T \sim 50$sec  is much
longer than the dynamical time scale, suggesting a prolonged
activity.\footnote{The ratio $\delta t/T \ll 1$ for short bursts
as well \cite{NakarPiran02b}}. This requires two (or possibly
three \cite{Ramirez-Ruiz_Merloni01,NakarPiran02a}) different time
scales operating within the ``inner engine". This rules out any
``explosive" model that release the energy in a single explosion.}

These clues, most specifically the last one suggest that GRBs
arise due to accretion of a massive ($\sim 0.1 m_\odot$) disk
onto a compact object, most likely a newborn black hole. A
compact  object is required because of the short time scales.
Accretion is needed to produce the two different time scales, and
in particular the prolonged activity. A massive ($\sim 0.1
m_\odot$) disk is required because of the energetics.  Such a
massive disk can form only simultaneously with the formation of
the compact object. This leads to the conclusions that GRBs
accompany the formation of black holes. This model is supported
by the observations of relativistic (but not as relativistic as
in GRBs) jets in AGNs, which are powered by accretion onto black
holes.

An important alternative to accretion is Usov's model
\cite{Usov92,Usov94} in which the relativistic flow is mostly
Poynting flux and it  is driven  by the magnetic and rotational
energies of a newborn rapidly rotating neutron star.

\subsection{Black hole accretion}
\label{sec:accretion}

 Several scenarios could lead to a black
hole - massive accretion disk system. This could include mergers
(NS-NS binaries \cite{Eichler_LPS89,NPP92}, NS-BH  binaries
\cite{Pac91}  WD-BH binaries \cite{Fryer_WHD99}, BH-He-star
binaries \cite{Fryer_Woosley98}) and models based on ``failed
supernovae'' or ``Collapsars''
\cite{Woosley93,Pac98,MacFadyen_W99}. \textcite{Narayan_P_Kumar01}
have recently shown that accretion theory suggests that from all
the above scenarios only Collapsars could produce long bursts and
only NS-NS (or NS-BH) mergers could produce short bursts. The
basic idea is that the duration of the accretion depends on the
size of the disk. So short burst must be produced by small disks
and those are natureally produced in mergers. On the other hand
long burst require large disks. However, those are inefficient.
One can overcome this if we have a small disk that is fed
continuously. In this case the efficiency can be large and the
duration long. This happens naturally within the collapsar model.

\subsection{The Pulsar Model}
\label{sec:Pulsar}

Several ``inner engine" models involve pulsar like activity of the
inner engine which is directly connected to a Poynting flux
decimated relativistic flow (in a contrast to a baryonic flux
dominated flow).  Energy considerations require an extremely large
magnetic fields of the order of $10^{15}$G within such sources.

\textcite{Usov92} suggested that GRB arise during the formation
of rapidly rotating highly magnetized neutron stars.  Such
objects could form by the gravitational collapse of accreting
white dwarfs with anomalously high magnetic fields in binaries,
as in magnetic cataclysmic binaries. The rapidly rotating and
strongly magnetized neutron stars would lose their rotational
kinetic energy on a timescale of seconds or less In a pulsar like
mechanism. The energy available in this case is the rotational
and magnetic energies of the neutron star that are of the order
of a few $\times 10^{51}$ergs for a neutron star rotating near
breakup. The rotation of the magnetic field creates a strong
electric field and an electron-positron plasma which is initially
optically thick and in quasi-thermodynamic equilibrium.
Additionally  a very strong magnetic field forms. The pulsar
produces a relativistic Poynting flux dominated flow.

While a Poynting flux dominated flow may be dissipated in a
regular internal shocks. \textcite{Usov94} and
\textcite{Thompson94} discuss a scheme in which  the energy is
dissipated from the magnetic field to the plasma and then via
plasma instability to the observed \gr outside the \gr
photosphere, which is at around $10^{13}$cm. At this distance the
MHD approximation of the pulsar wind breaks down and intense
electromagnetic waves are generated. The particles are
accelerated by these electromagnetic waves to Lorentz factors of
$10^6$ and produce the non thermal spectrum.
\textcite{SmolskyUsov96,SmolskyUsov00} and
\textcite{SpruitDaigneDrenkhahn01,DrenkhahnSpruit02} discuss
various aspects of the conversion of the Poynting flux energy to
\gr but these issues are more related to the nature of the
emitting regions and only indirectly to the nature of the inner
engine.

Usov's model is based on rotating highly magnetized neutron star
and from this point of view it indeed resembles to a large extend
a regular pulsar. Other authors consider pulsar like activities
in other contexts.  \textcite{Katz97}, for example, considers a
black hole - thick disk model in which the electromagnetic
process turn rotational energy to particle energy in a pulsar
like mechanism. \textcite{MR97b} discuss related idea on the
formation of a Poynting flux dominated flow within a black hole
accretion disk model.

\subsection{Rotating black holes and the Blandford Znajek
mechanism}

It is possible and even likely that the process of energy
extraction involves the Blandford-Znajek mechanism
\cite{BlandfordZnajek} in which the black hole - torus system is
engulfed in a magnetic field and the rotational energy of the
black hole is extracted via this magnetic field. The exploration
of the Blandford-Znajek mechanism involves relativistic MHD
consideration which are beyond the scope of this review. I refer
the reader to  several recent extensive reviews on this subject
(see e.g. \textcite{LeeWijersBrown00}).

\subsection{The Collapsar Model }
\label{sec:Collapsar}

The evidence for the association of (long) GRBs with supernovae
(see \textcite{Bloom02} and \S \ref{sec:obs-SN})  provides a
strong support for the Collapsar model. \textcite{Woosley93}
proposed that GRB arise from the collapse of a single Wolf-Rayet
star endowed with fast rotation ('failed' Type Ib supernova).
\textcite{Pac98} pointed out that there is tentative evidence that
the GRBs 970228, 970508, and 970828 were close to star-forming
regions and that this suggests that GRBs are linked to cataclysmic
deaths of massive stars. \textcite{MacFadyen_W99} begun a series
of calculations
\cite{AloyEtal00,MacFadyenWoosleyHeger01M,ZhangWoosleyMacFadyen03}
of a relativistic jet propagation through the stellar envelope of
the collapsing star which is the most important ingredient unique
to this model (other features like the accretion process onto the
black hole, the corresponding particle acceleration and to some
extend the collimation process are common to other models). The
collimation of a jet by the stellar mantle was shown to occur
analytically by \textcite{Meszaros-Rees01}.
\textcite{ZhangWoosleyMacFadyen03} numerically confirmed and
extended the basic features of this collimation process.

According to the Collapsar model the massive iron core of a
rapidly rotating massive star, of mass $M>30M_\odot$, collapses to
a black hole (either directly or during the accretion phase that
follows the core collapse). An accretion disk form around this
black hole and a funnel forms along  the rotation axis, where the
stellar material has relatively little rotational support. The
mass of the accretion disk is around 0.1 $M_\odot$. Accretion of
this disk onto the black hole takes place several dozen seconds
and powers the GRB. Energy can be extracted via neutrino
annihilation \cite{MacFadyen_W99} or via the Bladford-Znajek
mechanism. The energy deposited in the surrounding matter will
preferably leak out along the rotation axis producing jets with
opening angles of $<10^o$. If the jets are powerful enough they
would penetrate the stellar envelope and produce the GRB.

\textcite{ZhangWoosleyMacFadyen03} find that relativistic jets are
collimated by their passage through the stellar mantle. Starting
with an initial half-angle of up to $20^o$, the jet emerges with
half-angles that, though variable with time, are around $5^o$.
The jet becomes very hot in this phase and it has only a moderate
Lorentz factor, modulated by mixing, and a very large internal
energy (more than $80\%$ of the total energy). As the jet escapes,
conversion of the remaining internal energy into kinetic energy
gives terminal Lorentz factors along the axis of $\sim 150$
(depending, of course, on the initial conditions considered).
Because of the large ratio of internal to kinetic energy in both
the jet and its cocoon, the opening angle of the final jet is
significantly greater than at breakout. A small amount of
material emerges at large angles, but with a Lorentz factor still
sufficiently large to make a weak GRB. When the jet breaks out
from the star  it may produce a thermal precursor (seen in
several GRBs)
\cite{Pac98,Ramirez-RuizMacFadyenLazzati02,WaxmanMeszaros03}.
Instabilities in the accretion process, or in the passage of the
jet through the stellar envelope
\cite{AloyEtal02,ZhangWoosleyMacFadyen03} can produce the
required variability in the Lorentz factor that is needed to
produce internal shocks.

The processes of core collapse, accretion along the polar column
(which is essential in order to create the funnel) and the jet
propagation through the stellar envelope take together $\sim
10$sec \cite{MacFadyen_W99}. The duration of the accretion onto
the black hole is expected to take several dozen seconds. These
arguments imply that Collapsars are expected to produce long GRBs
(see however, \textcite{ZhangWoosleyMacFadyen03} for a suggestion
that the breakout of a relativistic jet and its collision with
the stellar wind will produce a brief transient with properties
similar to the class of ``short-hard'' GRBs.).

\subsection{The Supranova Model}
\label{sec:Supranova}

\textcite{VietriStella98} suggested that GRBs take place when a
``supermassive" (or supramassive as \textcite{VietriStella98} call
it)  neutron star (namely a neutron star that is above the maximal
cold nonrotating neutron star mass) collapses to a black hole. The
collapse can take place because  the neutron star losses angular
momentum via a pulsar wind and it looses the extra support of the
centrifugal force. Alternatively the supramassive neutron star can
simply cool and become unstable if rotation alone is not enough to
support it. The neutron star could also become over massive and
collapse if it accretes slowly matter from a surrounding accretion
disk \cite{VietriStella99}. In this latter case the time delay
from the SN could be very large and the SNR will not play any role
in the GRB or its afterglow.

The Supranova model is a two step event. First, there is a
supernova, which may be more energetic than an average one, in
which the supermassive neutron star forms. Then a few weeks or
months later this neutron star collapses producing the GRB. While
both the Supranova and the Collapsar (or hypernova) events are
associated with Supernovae or Supernovae like event the details of
the model are very different. First, while in the Collapsar model
one expect a supernova bump on the afterglow light curve, such a
bump is not expected in the Supranova model unless the time delay
is a few days. On the other hand while it is not clear in the
Collapsar model how does the Fe needed for the Fe \xr lines reach
the implied large distances form the center, it is  obvious in
this model, as the supernova shell was ejected to space several
month before the GRB. As mentioned earlier (see \S
\ref{sec:obs-SN}) the association of GRB 030329 with SN 2003dh
\cite{Stanek03SN,Hjorth03SN} is incompatible with the Supranova
model. Proponents of this model, argue however, that there might
be a distribution of delay times between the first and second
collapses.

The models are also very different in their physical content.
First in the Supranova model the GRB jet does not have to punch a
whole through the stellar envelope. Instead the ejecta propagates
in almost free space polluted possibly by a pulsar wind
\cite{GranotKonigl,GuettaGranot03}. In both models, like in many
other models,  the GRB is powered by accretion of a massive
accretion disk surrounding the newborn black hole. This accretion
disk forms, from the debris of the collapsing neutron star at the
same time that the black hole is formed. Again, the time scale of
the burst is determined by the accretion time of this disk.
\textcite{Narayan_P_Kumar01} (see also \S \ref{sec:accretion})
point however that long lived (50 sec) accretion disks must be
large and hence  extremely inefficient. This may pose a problem
for this model.

\textcite{GranotKonigl}, \textcite{GuettaGranot03} and
\textcite{GuettaInue} considered the effects a strong pulsar wind
(that may exist after the SN and before the second collapse) on
this scenario. The pulsar wind can have several effects. First it
would produce a denser highly magnetized medium into which the GRB
jet propagates. The strong magnetic field will be amplified by the
afterglow shock. This resolves the problem of the source of the
strong magnetic field needed for the synchrotron afterglow model.
This can also explain the high energy emission detected by EGRET
in GRB 940217 (\textcite{Hurley94} and \S \ref{sec:spec-obs}) by
Inverse Compton scattering on the pulsar wind bubble photons.  On
the other hand the density of this wind matter ($\sim
10^3$cm$^{-3}$) might be too high for the spherical model. Note
however, that this depends on the time delay as $t^{-3}$. However,
the pulsar wind won't be spherical and one would expect that it
will form an elongated supernova shell cavity within which the
pulsar wind is bounded. If, as expected, the pulsar jet coincides
with the GRB jet then the relativistic ejecta will move along the
elongated direction of this shell.

\subsection{Merging neutron stars}
\label{sec:NSmergers}

Neutron star binary mergers \cite{Eichler_LPS89,NPP92} or neutron
star - black hole binary mergers \cite{Pac91} (hereafter called
mergers) also produce a black hole - accretion disk system and
are candidates for the inner engines of GRBs, specifically of
short GRBs. These mergers take place because of the decay of the
binary orbits due to gravitational radiation emission as was
beautifully demonstrated in the famous binary pulsar PSR 1913+16
\cite{TaylorWeisberg82}.

These mergers take place at a rate of $\approx 10^{-6}$ events
per year per galaxy \cite{NPS91,Phinney91,vandenHeuvelLorimer96}.
This is the rate of merger of binaries of the type of PSR 1913+16
whose life time is or order of several $10^8$ years. Various
population synthesis calculations suggest that there is also
another population of short lived binaries
\cite{TutukovYungelson93,TutukovYungelson94,BelczynskiBulikKalogera02,PernaBelczynski02}.
These binaries form with very close orbits and hence with short
lifetimes of the order of $10^5$yrs. Even thought the overall
rate of such mergers could be comparable to those of the PSR
1913+16 type  one cannot expect to catch a binary in our galaxy
in such a stage. Similarly unlike the long lived mergers that may
be kicked from their host galaxy within their long life time
\cite{NPP92,BulikBelczynskiZbijewski99} this short lived
population remains within the galaxy when they merge
\cite{BelczynskiBulikKalogera02}.

Earlier simulations of mergers focused on the gravitational
radiation from this system.  \textcite{DaviesEtal94} begun a
series of numerical simulation of neutron star merger that focused
on GRB related aspects
\cite{RosswogEtal99,RosswogEtal00,AyalEtal01,RosswogDavies02}.
Using a SPH scheme they followed NS mergers  under different
assumptions (Newtonian with ad hoc addition of gravitational
radiation back reaction or Post Newtonian), with different
equations of state (adiabatic or realistic) and with different
initial spin axis and mass rations and different estimates of the
effects of neutrino cooling. A parallel set of simulations was
carried out by
\textcite{RuffertJankaSchafer95,JankaRuffert96,RuffertJanka98,RuffertJanka99,RuffertJanka01}
who used particle in cell methods. Both kinds of simulations yield
comparable results. The merger results in a black hole - accretion
disk system. The mass of the accretion disk is of order
0.1$M_\odot$ and it depends, of course somewhat on the orientation
of the spins and the relative masses of the two neutron stars.

A merger releases  $\sim 5 \times 10^{53}$ergs but most of this
energy is in the form of low energy neutrinos and gravitational
waves. Still there is  enough energy available to power a GRB but
is not clear how the GRB is produced.  A central question is, of
course, how does a merger generate the relativistic wind required
to power a GRB.  \textcite{Eichler_LPS89} suggested that about
one thousandth of these neutrinos annihilate and produce pairs
that in turn produce gamma-rays via $\nu \bar \nu \rightarrow e^+
e^- \rightarrow \gamma\gamma$.  This idea was criticized on
several grounds by different authors the main problem is that it
does not produce enough energy. For example
\textcite{jaroszynksi96} pointed out that a large fraction of the
neutrinos will be swallowed by the black hole that forms. An
alternative source of energy within the merger model is the
accretion power of a disk that forms around the black hole. This
brings us back to the canonical black hole - accretion disk
scenario.

\section{Open Questions and Future Prospects}

I believe that overall we have a basic understanding of the GRB
phenomenon. As usual some aspects are understood better than
others.

There is a very good understanding of the afterglow. Here there
are numerous observations over a wide range of wavelengths with
which the theory can be confronted. The overall picture, of a
slowing down relativistic flow and of synchrotron emission fit
the data to a large extent (see e.g.
\textcite{WijersGalama99,PanaitescuK01} and many other fits of the
observations to the model). We have already learned that the ``cow
is not spherical", namely that the relativistic flow is
collimated. New observations, like those of GRB 021004 and GRb
030329, poses at times new puzzles and suggest that the basic
simple picture has to be refined. It seems however, that the
answers are within the scope of the current model, such as:
refreshed shocks, patchy shells and variable external densities.
All these phenomena are fairly reasonable in a realistic
environment. Within the afterglow, I believe that the \xr lines
pose the greatest current puzzle, in terms of their energy
requirements and other implications on the source (see
\textcite{Lazzati02}). Another interesting open question is what
distinguished between GHOSTs and OTGRBs - an environmental
(extinction or ``improper conditions within the circum-burst
matter) or an intrinsic mechanism?

The main observational challenges concerning  the afterglow are
the determination whether short GRBs have afterglow. A wealth of
information on long GRBs arises from the information on hosts,
environments and redshifts, that are determined from the afterglow
observations. All these are missing for short GRBs. If short GRBs
don't have afterglows, then an immediate  theoretical question is
why? Is it possible that they are produced in a very different
environment than long ones (such as outside their original
galaxies) in a region with no circum-burst matter suitable for
producing the afterglow? At the moment the observational situation
is not clear. The coming Swift satellite may resolve this mystery.

Another important observational question involves the search for
orphan afterglows  (either in radio or in optical). Their
detection will establish the collimated jets picture. But even
firm upper limits will  set independent limits on the rates of
GRBs. However, as mentioned in \S \ref{sec:orphan} this is a very
challenging observational task. This has important implication for
the nature of the jets - are GRB jets standard with a fixed
angular structure \cite{Lipunov_Postnov_Pro01,Rossi02,Zhang02}?
This question is related both to the overall energetics and to the
rate of GRBs.

Another interesting challenge will be the resolution of the
afterglow image (see \textcite{GPS99c}). This may be possible in
radio for a nearby burst and the afterglow of GRB 030329 provides
an excellent candidate for that. Polarization measures could pave
the way for understanding of the collimation geometry and for a
confirmation of the synchrotron emission process.

As we move backwards with time towards the burst we encounter the
very early afterglow and the optical flash that coincides with the
burst itself. Here a great progress was made with recent
observations triggered by HETE II (e.g. the almost complete light
curve of GRB 021004 \cite{Foxetal03}). SWIFT may contribute a lot
to this issue. These observations could shed a light on issues
like the role of pre-acceleration and neutrons that are  unclear
as yet. Here, I stress the importance of early and continuous
radio observations, which could determine whether there are
refreshed shocks during the early afterglow, that have a clear
radio signature \cite{KP00a}.

The understanding of the \gr emitting regions is less clear. Here,
within the internal shocks model there is a reasonable
understanding of the temporal structure (in terms of activity of
the inner engine). However, it is not clear how is the observed
spectrum produces and it seems that the simple synchrotron
spectrum has to be modified (see e.g.
\textcite{LloydPetrosian00,Medvedev00} for ideas on such
modifications). Another possibly related puzzle is the origin of
the narrow $E_p$ distributions (see however, e.g.
\textcite{DaigneMochkovitch98,Guetta_Spada_Waxman01,DaigneMochkovitch03}).
Another set of open questions is what is the origin of the
intrinsic correlation between luminosity (which in fact reflects
the collimation angle \cite{Frail01,PanaitescuK01}) discovered by
\textcite{Fenimore_Ramirez-Ruiz01} or the lag-luminosity relation
discovered by \textcite{Norris_lags00}. Similarly or even more
puzzling are the implied correlations between redshift and
intrinsic luminosity \cite{Lloyd-RonningFryerRamirez-Ruiz02} and
between redshift and intrinsic hardness \cite{Schmidt01} (note
that this later correlation is essential in view of  the narrow
$E_p$ distribution of GRBs). Here  pairs
\cite{Ghisellini_Celotti99} and IC can play an important role.
Theoretical open basic physical questions that arise here (as well
as in the theory of the afterglow) deal with the processes of the
behavior of collisionless shocks (see e.g.
\textcite{Medvedev01,NiktoMedvedev01}, particle acceleration (see
\S \ref{sec:acc}) and the generation of strong magnetic field (see
\cite{MedvedevLoeb99}). Issues like relativistic turbulence and
relativistic plasma instabilities might play an important role
here (see e.g. \textcite{LyutikovBlandford02}).

>From an observational point, it will be a challenge to beat the
statistical power of the BATSE data in terms of number of bursts.
Somewhat surprisingly, the questions what is the luminosity
function of GRBs and what is the rate of GRBs as a function of
redshift and to what extend GRBs follow the star formation rate
are still open. Detectors with better spectral resolutions could
shed some additional light on the spectrum. Another hope for new
data, or at least for upper limits, arises from observational
windows in higher \gr bands. On the low energy side it seems that
there is a continuum between XRFs and GRBs
\cite{BATSE_XRF,BarraudEtal03}. This result still has to  be fully
understood in the context of the narrow $E_p$ distribution.

Looking far into the future one can hope to observe neutrinos or
gravitational radiation correlated to GRBs. UHE neutrinos (Fluxes
of  MeV neutrinos would be too weak to be detected from
cosmological distances) could confirm that protons are accelerated
to UHE energies within GRBs. In turn this would proof (or
disprove) the possible role of GRBs as sources of  UHECRs.
Gravitational radiation could give a direct clue on the activity
of the inner engine (see \S \ref{sec:GRGRB} and identify, for
example, merger events.

There is a lot of observational evidence associating long GRBs
with core collapse SNes. This gives a clear clue on what is the
inner engine of long GRBs. There is no direct or indirect evidence
on the progenitors of short GRBs. Even with this clue the
situation is far from clear  when we turn to the inner engine.
Here most models assume some variant of a black-hole - torus
system with various energy extraction mechanisms ranging from
neutrino annihilation (which is less likely) to variants on the
theme of electromagnetic extraction (magnetic turbulence within
the accretion disk; the Blandford-Znajek mechanism which involves
a disk-black hole-magnetic field interaction; pulsar like
activity). Here there are open questions all around: What is the
content of the ultrarelativistic flow - baryonic or Poynting flux?
How is the flow accelerated and collimated? What determines the
variability of the flow (required for internal shocks) and the
different time scales? This part of the model seems to be in a
rather poor shape - but this is understandable as we don't have
any direct observations of this inner engine. One hope that arises
is that there seem to be an emerging similarity between GRBs,
galactic micro quasars and AGNs. All these systems  accelerate
collimated flows to relativistic velocities and they all seem to
involve accretion onto black holes. Hopefully,  this similarity
could lead to a common resolution of how inner engines operate in
all those systems.

\section*{Acknowledgments}
I would like to thank J. Granot, D. Guetta, P. Kumar, E. Nakar and
R. Sari for many helpful discussions and J. Bloom, J. Hjorth, P.
M{\' e}sz{\' a}ros, E. Pian, K. Stanek, P. Vreeswijk and an
anonymous referee for remarks. This research was supported by a
grant from the US-Israel Binational Science Foundation.


\end{document}